\documentclass[3p,times]{elsarticle}

\usepackage{epsfig}
\usepackage{graphicx}
\usepackage{epstopdf}
\usepackage{pifont}
\usepackage{natbib}
\usepackage{geometry}
\usepackage{amssymb}
\usepackage{fleqn}
\usepackage{framed}
\usepackage{txfonts}
\usepackage{morefloats}
\usepackage{amsmath}
\usepackage{xcolor}
\usepackage{bm}
\usepackage[normalem]{ulem}

\newenvironment{enumerate-roman}{
  \begin{enumerate}}{\end{enumerate}}

\renewcommand{\BibitemShut}[1]{}

\newcommand{\vek}[1]{\bm{\mathrm{#1}}}
\DeclareMathOperator{\Imag}{Im}
\DeclareMathOperator{\Real}{Re}
\DeclareMathOperator{\Tr}{Tr}

\begin{document}

\begin{frontmatter}

\title{The BCS-BEC crossover: From ultra-cold Fermi gases to nuclear systems}

\author[unicam,infn]{Giancarlo Calvanese Strinati}
\author[unicam,infn]{Pierbiagio Pieri} \author[rostock]{Gerd
  R\"{o}pke} \author[ipno,lpmmc]{Peter Schuck} \author[ipno]{Michael
  Urban} \address[unicam]{School of Science and Technology, Physics Division,
  Universit\`{a} di Camerino, 62032 Camerino (MC), Italy}
\address[infn]{INFN, Sezione di Perugia, 06123 Perugia (PG), Italy}
\address[rostock]{Institut f\"{u}r Physik, Universit\"{a}t Rostock,
  18051 Rostock, Germany} \address[ipno]{Institut de Physique
  Nucl\'eaire, CNRS-IN2P3 and Universit\'e~Paris-Sud,
  91406~Orsay~cedex, France} \address[lpmmc]{Laboratoire de Physique
  et de Mod\'elisation des Milieux Condens\'es, CNRS and
  Universit\'e~Joseph~Fourier, BP~166, 38042~Grenoble~cedex 9, France}

\begin{abstract} 
This report adresses topics and questions of common interest in the fields of ultra-cold gases and nuclear physics in the context of the BCS-BEC crossover. 
By this crossover, the phenomena of Bardeen-Cooper-Schrieffer (BCS) superfluidity and Bose-Einstein condensation (BEC), which share the same kind of spontaneous symmetry breaking, are smoothly connected through the progressive 
reduction of the size of the fermion pairs involved as the fundamental entities in both phenomena. 
This size ranges, from large values when Cooper pairs are strongly overlapping in the BCS limit of a weak inter-particle attraction, to small values when composite bosons are non-overlapping
in the BEC limit of a strong inter-particle attraction, across the intermediate unitarity limit where the size of the pairs is comparable with the average inter-particle distance.

The BCS-BEC crossover has recently been realized experimentally, and essentially in all of its aspects, with ultra-cold Fermi gases. 
This realization, in turn, has raised the interest of the nuclear physics community in the crossover problem, since it represents an unprecedented tool to test fundamental and 
unanswered questions of nuclear many-body theory. 
Here, we focus on the several aspects of the BCS-BEC crossover, which are of broad joint interest to both ultra-cold Fermi gases and nuclear matter, and which will likely help
to solve in the future some open problems in nuclear physics (concerning, for instance, neutron stars). 
Similarities and differences occurring in ultra-cold Fermi gases and nuclear matter will then be emphasized, not only about the relative phenomenologies
but also about the theoretical approaches to be used in the two contexts. 
Common to both contexts is the fact that at zero temperature the BCS-BEC crossover can be described at the mean-field level with reasonable accuracy. 
At finite temperature, on the other hand, inclusion of pairing fluctuations beyond mean field represents an essential ingredient of the theory, especially in the normal phase
where they account for precursor pairing effects.

After an introduction to present the key concepts of the BCS-BEC crossover, this report discusses the mean-field treatment of the superfluid phase, both for homogeneous and 
inhomogeneous systems, as well as for symmetric (spin- or isospin-balanced) and asymmetric (spin- or isospin-imbalanced) matter. 
Pairing fluctuations in the normal phase are then considered, with their manifestations in thermodynamic and dynamic quantities. 
The last two Sections provide a more specialized discussion of the BCS-BEC crossover in ultra-cold Fermi gases and nuclear matter, respectively. 
The separate discussion in the two contexts aims at cross communicating to both communities topics and aspects which, albeit arising in one of the two fields, share a strong common interest.
\end{abstract}

\end{frontmatter}

\tableofcontents

\section{Introduction}
\subsection{Historical background}
\label{Historical-background}

The idea behind the BCS-BEC crossover dates back just after the birth
of the BCS theory in 1957
\cite{Bardeen-1957}. The authors of this theory made a point to
emphasize the differences between their theory for superconductors
based on strongly-overlapping Cooper pairs and the
Schafroth-Butler-Blatt theory \cite{Schafroth-1957} resting on
non-overlapping composite bosons which undergo Bose-Einstein condensation at
low temperature. Subsequently, the interest in these two different
situations has been kept disjoint for some time, until theoretical
interest arose for unifying them as two limiting (BCS and BEC)
situations of a single theory where they share the same kind of broken
symmetry. In this way, the physical system passes through an
intermediate situation, where the size of the pairs is comparable to
the average inter-particle spacing. (For a short-range (contact)
interaction, this intermediate situation is nowadays referred to as
the ``unitary'' regime.)  Pioneering work in this sense was done by
the Russian school, motivated by the exciton condensation in
semiconductors \cite{Keldysh-1965} or simply by intellectual curiosity
\cite{Popov-1966}. The theory of the BCS-BEC crossover took shape
initially through the work by Eagles \cite{Eagles-1969} with possible
applications to superconducting semiconductors, and later through the
works by Leggett \cite{Leggett-1980} and Nozi\`{e}res and Schmitt-Rink
\cite{Nozieres-1985} where the formal aspects of the theory were
developed at zero temperature and above the critical temperature,
respectively.

The interest in the BCS-BEC crossover grew up with the advent of
high-temperature (cuprate) superconductors in 1987, in which the size
of the pairs appears to be comparable to the inter-particle spacing
\cite{Randeria-1989,Randeria-1990,Micnas-1990,Randeria-1992,Drechsler-1992,Haussmann-1993,Pistolesi-1994,Casas-1994}. 
Related interest in the BCS-BEC crossover soon spread to some problems in nuclear physics
\cite{Baldo-1995}, but a real explosion of this activity appeared
starting from 2003 with the advent of the fully controlled
experimental realization essentially of all aspects of the BCS-BEC
crossover in ultra-cold Fermi gases (see
Refs.~\cite{Inguscio-2007} and \cite{Zwerger-2012} for an
experimental and theoretical overview, respectively). This
realization, in turn, has raised the interest in the crossover problem
especially of the nuclear physics community, as representing an
unprecedented tool to test fundamental and unanswered questions of
nuclear many-body theory. The Fermi gas at the \emph{unitary limit}
(UL), where fermions of opposite spins interact via a contact
interaction with infinite scattering length, was actually introduced
as a simplified model of dilute neutron matter
\cite{Baker-2000,Baker-1999}, and the possibility to realize this
limit with ultra-cold atoms was hence regarded as extremely important
for this field of nuclear physics.

In nuclear physics there is the paradigmatic example of a
proton-neutron bound state (the deuteron) for which in symmetric or
asymmetric nuclear matter, as a function of density, one may find a
transition from BEC to BCS 
\cite{Alm-1993,Baldo-1995,Stein-1995,Lombardo-2001b}
(see also Ref.\cite{Andrenacci-1999} for the density-induced BCS-BEC crossover).
Such a scenario may be realized in expanding
nuclear matter generated from heavy ion collisions \cite{Baldo-1995}
or in proto-neutron stars \cite{Heckel-2009}. More recently, this
density driven crossover has been studied together with the competing
liquid-gas phase transition \cite{Jin-2010}. Deuteron condensation
also heavily competes with alpha-particle condensation, which is
somehow related to \emph{pairing of pairs} discussed also in
condensed-matter physics. Nuclear physics is a precursor of the theory
for quartet condensation \cite{Roepke-1998}, but theoretical
studies of quartets came up later also in the area of ultra-cold atoms
\cite{Capponi-2007}. Alpha-particle condensation is presently
very much discussed in finite nuclei (Hoyle state)
\cite{Yamada-2012}. It was also predicted that in the non-condensed
phase, the deuterons give rise to a pseudo-gap formation
\cite{Schnell-1999}. With today's ultra-cold atoms experiments, it has
become possible to test such theories
\cite{Stewart-2008,Gaebler-2010,Feld-2011}. For neutrons, no bound
state exists, but rather a virtual state at almost zero energy. As a
consequence, a dilute gas of neutrons, as it exists in the inner crust
of neutron stars, is almost in the unitary limit mentioned above.
Recent studies of the dilute neutron gas, from (almost) unitarity at
low density to the BCS limit at high density, were done within BCS
theory \cite{Matsuo-2006}, Quantum-Monte-Carlo calculations
\cite{Abe-2009,Gezerlis-2010}, and the Nozi\`{e}res-Schmitt-Rink
approach \cite{Ramanan-2013}. One can say that the equation of state
and pairing properties of very dilute neutron matter, although
inaccessible in experiments, are now known thanks to the analogy with
ultra-cold atoms.

As these examples show, there are several aspects of the BCS-BEC
crossover which are of broad joint interest to both ultra-cold atoms
and nuclear communities. This paper is thus meant to provide a 
comprehensive review which focuses mainly on these common aspects
of ultra-cold atoms and nuclear physics. Along these views, this
paper provides also a pedagogical review of the main essential aspects
of the BCS-BEC crossover. 
In the process of writing, we have mostly adopted the quantum many-body diagrammatic techniques in line with our own technical expertise, and we have mainly focused on the topics to which we have provided original contributions over the last several years. Accordingly, for readers interested in complementary theoretical approaches to the problem we refer to other reviews which cover various aspects of the BCS-BEC crossover. In particular, we can refer to reviews on the application to strongly-interacting Fermi gases (in particular, at unitarity) of Quantum Monte Carlo methods~\cite{Bulgac-2012,Carlson-2012}, functional-renormalization-group techniques~\cite{Diehl-2010}, epsilon \cite{Nishida-2012} and virial \cite{Liu-2013} expansions.  For the application of functional-integral approaches (in particular, to the superfluid phase) we refer instead to the original research works of Refs.~\cite{Pistolesi-1996,Hu-2006,Diener-2008}. 

In addition, it should be mentioned that further reviews cover a number of aspects on the BCS-BEC crossover which share a partial overlap with the material discussed here. Specifically, the unitary limit of Section \ref{sec:unitarylimit} and the Tan contact of Section \ref{sec:Tancontact} have been of most interest in other reviews owing to the widespread recent interest in these topics, which have been treated in Refs.~\cite{Zwerger-2012,Randeria-2014,Pitaevskii-2016,Zwerger-2016}.  The Fano-Feshbach resonances of Section~\ref{sec:Feshbach}, which are at the heart of the interaction-induced crossover, have been discussed in Refs.~\cite{Chin-2010,Gubbels-2013,Randeria-2014,Pitaevskii-2016,Zwerger-2016}. Polarized Fermi gases (considered here in Sections \ref{sec:polarized} and \ref{sec:nsr-polarized}) have also been treated in Refs.~\cite{Giorgini-2008,Radzihovsky-2010,Chevy-2010,Zwerger-2012,Gubbels-2013,Pitaevskii-2016}. The topic of the single-particle spectral function and pseudo-gap of Section~\ref{sec:pseudo-gap} can be found discussed also in Refs.~\cite{Chen-2005,Randeria-2014,Zwerger-2016}. Also some aspects of pairing in nuclear systems discussed in Section 5 can be found in the reviews of Refs.~\cite{Brink-2005,Dean-2003,Gezerlis-2014}. Finally, Refs.~\cite{Pitaevskii-2016} and \cite{Gezerlis-2014} cover also some introductory material treated here in Sections \ref{sec:BCSwavefunction}, \ref{sec:BCS_homogeneous}, \ref{sec:bdg-hfb}, \ref{sec:gmb}, \ref{sec:collective}, \ref{sec:vortices}, and \ref{sec:deuteron}. However, it should be remarked that no other reference thus far has emphasized the aspects of the BCS-BEC crossover common to ultra-cold atoms and nuclear physics as we have done here.

\subsection{The BCS wave function and its BEC limit}
\label{sec:BCSwavefunction}

The starting point is the BCS ground-state wave function, of the form \cite{Bardeen-1957,Schrieffer-1964}
\begin{equation}
|\Phi_{\mathrm{BCS}} \rangle = \prod_{\vek{k}} \left(u_{\vek{k}} + \varv_{\vek{k}} c^{\dagger}_{\vek{k} \uparrow} c^{\dagger}_{-\vek{k} \downarrow}\right) |0 \rangle \, .
\label{BCS-wave-function}    
\end{equation} 
\noindent
In this expression, $|0 \rangle$ is the vacuum state,
$c^{\dagger}_{\vek{k} \sigma}$ is a fermionic creation operator for
wave vector $\vek{k}$ and spin projection $\sigma
=(\uparrow,\downarrow)$, and $u_{\vek{k}}$ and $\varv_{\vek{k}}$ are
probability amplitudes given by:
\begin{equation}
\varv_{\vek{k}}^2 = 1 - u_{\vek{k}}^2 = \frac{1}{2} \left( 1 - \frac{\xi_{\vek{k}}}{E_{\vek{k}}} \right)
\label{v-square}
\end{equation}
where $\xi_{\vek{k}}=\vek{k}^2/(2m) - \mu$ ($m$ being the fermion
mass and $\mu$ the chemical potential) and $E_{\vek{k}}
=\sqrt{\xi_{\vek{k}}^2 + |\Delta_0|^2}$ where $\Delta_0$ is
the BCS order parameter (sometimes referred to as the superconducting
gap) here taken at zero temperature. \ For simplicity, we have assumed
that $\Delta_0$ is independent of the wave vector, as it is the case for a
contact interaction. More generally, $\Delta_0$ in the expression of
$E_{\vek{k}}$ will be replaced by the wave-vector (and temperature)
dependent value $\Delta_{\vek{k}}$.
[Throughout this paper, we use units where the reduced Planck constant
$\hbar$ and the Boltzmann constant $k_B$ are set equal to unity.]

The BCS wave function (\ref{BCS-wave-function}) has the important
property that it is the vacuum to the so-called quasi-particle
operators $\alpha_{\vek{k}} = u_{\vek{k}} c_{\vek{k}\uparrow} -
\varv_{\vek{k}} c^{\dag}_{-{\vek{k}}\downarrow}$ (see
Eq.~(\ref{canonical-transformation}) below and Ref.~\cite{Ring-1980}),
that is, $\alpha_{\vek{k}}|\Phi_{\mathrm{BCS}} \rangle=0$.  This
relation facilitates considerably the evaluation of expectation
values.  One can readily show that $\langle \Phi_{\mathrm{BCS}}|
c^{\dag}_{\vek{k}\sigma} c_{\vek{k}\sigma} | \Phi_{\mathrm{BCS}}
\rangle = \varv_{\vek{k}}^2$ is the occupation number $n_{\vek{k}}$
which goes over to the Fermi step in the limit as $\Delta_0$ tends to
zero, and that $\langle \Phi_{\mathrm{BCS}}|
c_{-\vek{k}\downarrow}c_{\vek{k}\uparrow} | \Phi_{\mathrm{BCS}}
\rangle = u_{\vek{k}}^{*} \varv_{\vek{k}}$ is the so-called 
anomalous density $\phi_{\vek{k}}$ (known also as the ``pairing tensor" $\kappa_{\vek{k}}$ in nuclear physics) which characterizes the BCS wave
function.

With reference to the BCS-BEC crossover, it has long been known that
the BCS wave function (\ref{BCS-wave-function}) contains the
Bose-Einstein condensation of composite bosons as a limit.  This is
because, upon setting $g_{\vek{k}} = \varv_{\vek{k}}/u_{\vek{k}}$, the
expression (\ref{BCS-wave-function}) can be rewritten in the form
(see, e.g., Ref.~\cite{Ring-1980}):
\begin{equation}
|\Phi_{\mathrm{BCS}} \rangle = \left(\prod_{\vek{k}'} u_{\vek{k}'}\right) \exp\left[\sum_{\vek{k}} g_{\vek{k}}  §c^{\dagger}_{\vek{k} \uparrow} c^{\dagger}_{-\vek{k} \downarrow}\right]|0 \rangle       
\label{BCS-wave-function-rewritten} 
\end{equation}
\noindent
since $(c^{\dagger}_{\vek{k} \sigma})^2=0$ owing to Pauli principle.
Here, the operator $b_0^{\dagger} \equiv \sum_{\vek{k}} g_{\vek{k}}
c^{\dagger}_{\vek{k} \uparrow} c^{\dagger}_{-\vek{k} \downarrow}$
contains fermion pairs but it is not a truly bosonic operator, to the
extent that the commutator $[b_0, b_0^{\dagger}] = \sum_{\vek{k}}
|g_{\vek{k}}|^2 (1 - \hat{n}_{\vek{k} \uparrow} - \hat{n}_{-\vek{k}
  \downarrow})$ is not a c-number but explicitly contains the
fermionic operators $\hat{n}_{\vek{k} \sigma} = c^{\dagger}_{\vek{k}
  \sigma} c_{\vek{k} \sigma}$.  However, under some circumstances one
may consider that $[b_0, b_0^{\dagger}] \cong 1$ for all practical
purposes, provided $\langle \Phi_{\mathrm{BCS}}|
\hat{n}_{\vek{k}\sigma} | \Phi_{\mathrm{BCS}} \rangle =
\varv_{\vek{k}}^2 \ll 1$ for all $\vek{k}$ of physical relevance.
As a consequence, $|\Phi_{\mathrm{BCS}} \rangle = \exp (b_0^{\dagger})
|0 \rangle$ represents a bosonic coherent state (that is, a
Bose-Einstein condensate) with a non-vanishing broken-symmetry average
$\langle \Phi_{\mathrm{BCS}} | b_0 | \Phi_{\mathrm{BCS}} \rangle =
\sum_{\vek{k}} \, |g_{\vek{k}}|^2 \ne 0$.

\begin{figure}[h]
\begin{center}
\includegraphics[width=7.0cm,angle=0]{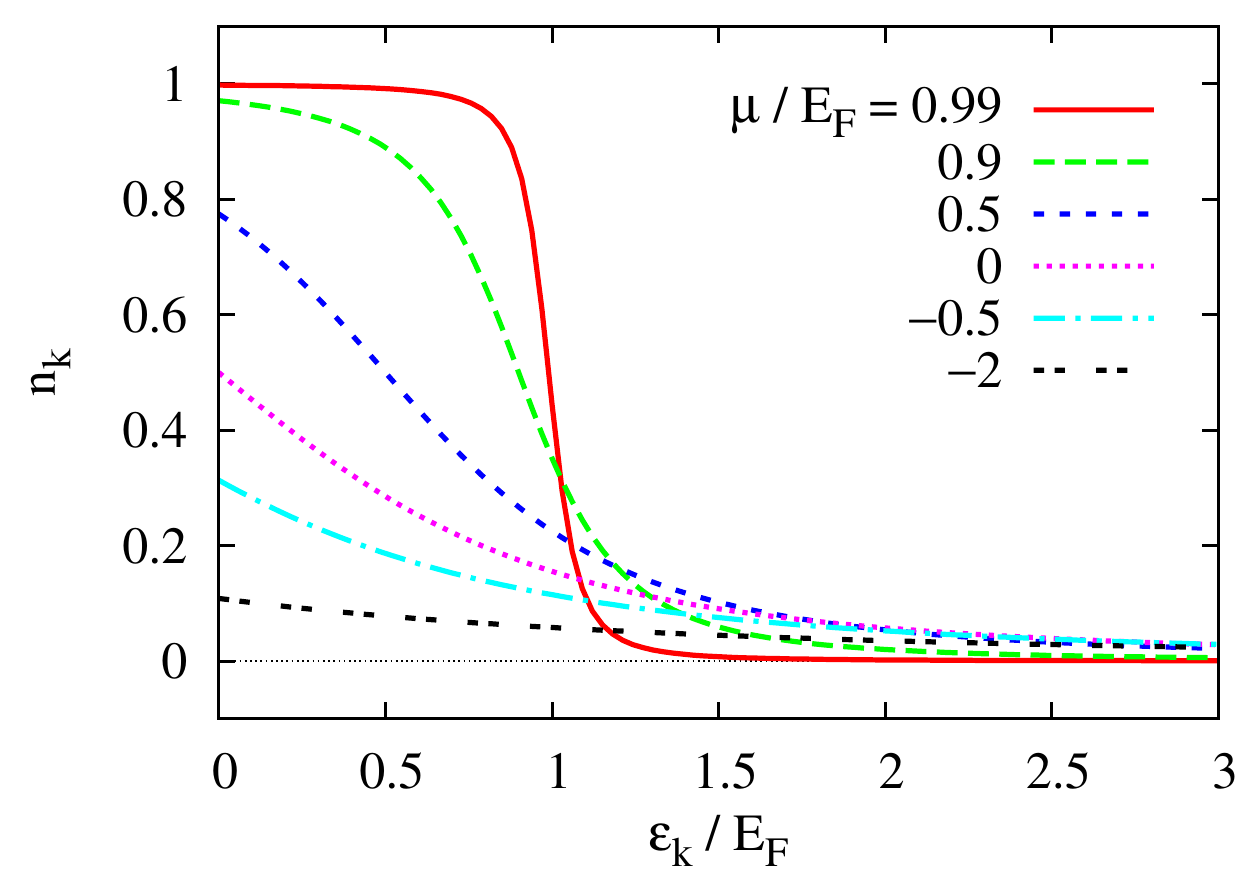}
\caption{BCS occupation number $n_{\vek{k}} = \varv_{\vek{k}}^{2}$ at zero temperature vs the energy $\varepsilon_{\vek{k}} = k^2/(2m)$ in units of the Fermi energy $E_F = (3\pi^2n)^{2/3}/(2m)$
where $n$ is the density, for various values of the chemical potential ranging from $\mu = 0.99 E_F$ to $\mu = -2 E_F$ [the corresponding values of the gap are 
$\Delta_0/E_F = ($ 0.09, 0.32, 0.76, 1.05, 1.24, and 1.59), from top to bottom].}
\label{Figure-1}
\end{center}
\end{figure} 

It is clear from Eq.~(\ref{v-square}) that the condition $\varv_{\vek{k}}^2 \ll 1$ can be satisfied for \emph{al\/l} $\vek{k}$ provided the fermionic chemical potential 
$\mu$ becomes large and negative.  
This condition can be achieved when a bound-state with
binding energy $\varepsilon_0 = (m a_F^2)^{-1}$ occurs for the
two-body problem \emph{in vacuum} with positive
scattering length $a_F$, and the coupling parameter $(k_Fa_F)^{-1}$
becomes large such that $\varepsilon_0/E_F \gg 1$.  In this limit,
$\mu$ approaches the value $-\varepsilon_0/2$, which amounts to saying
that all fermions are paired up in tight (composite) bosons with a
vanishing residual interaction among the bosons.  This result for the
BEC limit of the fermionic chemical potential $\mu$ can directly be
obtained from the mean-field gap equation (at zero temperature), which
in the case of a wave-vector dependent interaction
$V_{\mathrm{eff}}(\vek{k},\vek{k}')$ reads:
\begin{equation}
\Delta_{\vek{k}} = - \int \! \frac{d\vek{k}'}{(2 \pi)^3} \, V_{\mathrm{eff}}(\vek{k},\vek{k}') \, \frac{\Delta_{\vek{k}'}}{2E_{\vek{k}'}}
\label{gap-eq}
\end{equation}
\noindent
where now $E_{\vek{k}} =\sqrt{\xi_{\vek{k}}^2 +
  |\Delta_{\vek{k}}|^2}$.  (In the case of a contact potential,
$V_{\mathrm{eff}}(\vek{k},\vek{k}')$ and $\Delta_{\vek{k}}$ reduce,
respectively, to the coupling constant $\varv_0$ of
Eq.~(\ref{regularization}) and to the constant gap $\Delta_0$ given by
the gap equation (\ref{gap-homogeneous}) below.)  Using
$\phi_{\vek{k}} = \Delta_{\vek{k}}/(2E_{\vek{k}})$, $n_{\vek{k}} =
\varv_{\vek{k}}^2$, and Eq.~(\ref{v-square}), one can rewrite
Eq.~(\ref{gap-eq}) in the form:
\begin{equation}
2 \xi_{\vek{k}} \phi_{\vek{k}} + (1-2 n_{\vek{k}}) \, \int \! \frac{d\vek{k}'}{(2 \pi)^3} \, V_{\mathrm{eff}}(\vek{k},\vek{k}') \ \phi_{\vek{k}'} = 0 \, .
\label{alternative-gap-equation}
\end{equation}
\noindent
Provided $n_{\vek{k}} \ll 1$ for all $\vek{k}$,
Eq.~(\ref{alternative-gap-equation}) is just the Schr\"{o}dinger
equation for the relative motion of two particles of equal mass $m$
which are mutually interacting via the potential $V_{\mathrm{eff}}$.
The negative eigenvalue $2\mu$ of this equation thus corresponds to
(minus) the two-body binding energy $\varepsilon_0$ as stated above.

More generally, as we shall see below, it is the fermionic chemical potential $\mu$ that takes the key role of the driving field which enables the system to pass from the BCS to the BEC limits of the BCS-BEC crossover. 
As an illustration, Fig.~\ref{Figure-1} shows the occupation number $n_{\vek{k}} = \varv_{\vek{k}}^2$ at zero temperature for different (from positive to negative) values of the chemical potential $\mu$,
which correspond to increasing values of the gap parameter $\Delta_0$. 
Note that only when $\mu > 0$ the curves have an inflection point at $\varepsilon_{\vek{k}} = \mu$, which highlights the presence of an underlying Fermi surface even for a system with attractive inter-particle interaction.
As a consequence, when $\mu$ becomes negative, the Fermi sea gets completely dissolved and the occupation number becomes quite small for all $\vek{k}$.

\subsection{Pairing correlations}
\label{sec:pairingcorrelations}

The BCS wave function (\ref{BCS-wave-function}), or its equivalent
form (\ref{BCS-wave-function-rewritten}), treats all fermion pairs on
the same footing to the extent that a single wave function
$g_{\vek{k}}$ is assigned to each pair.  This \emph{mean-field} type
of approach is appropriate to describe a system of fermions with a
mutual attractive interaction when the inter-particle correlations
extend much beyond the average inter-particle distance, in such a way
that many pairs are contained within the size of a given pair and
different pairs strongly overlap with each other.  This is definitely
the case for the (BCS) weak-coupling limit, to which the BCS theory of
superconductivity was originally meant to apply \cite{Bardeen-1957}.
When considering the BCS-BEC crossover, however, the range of the
inter-particle correlations can decrease down to the size of the bound
pair, which, in turn, can be much smaller than the inter-particle
distance.  Under these circumstances, the gas of composite bosons
becomes dilute and one accordingly anticipates that \emph{pairing
  fluctuations} beyond mean field can acquire a major role, especially
at finite temperature when they are accompanied by thermal
fluctuations.

\begin{figure}[h]
\begin{center}
\includegraphics[width=8.0cm,angle=0]{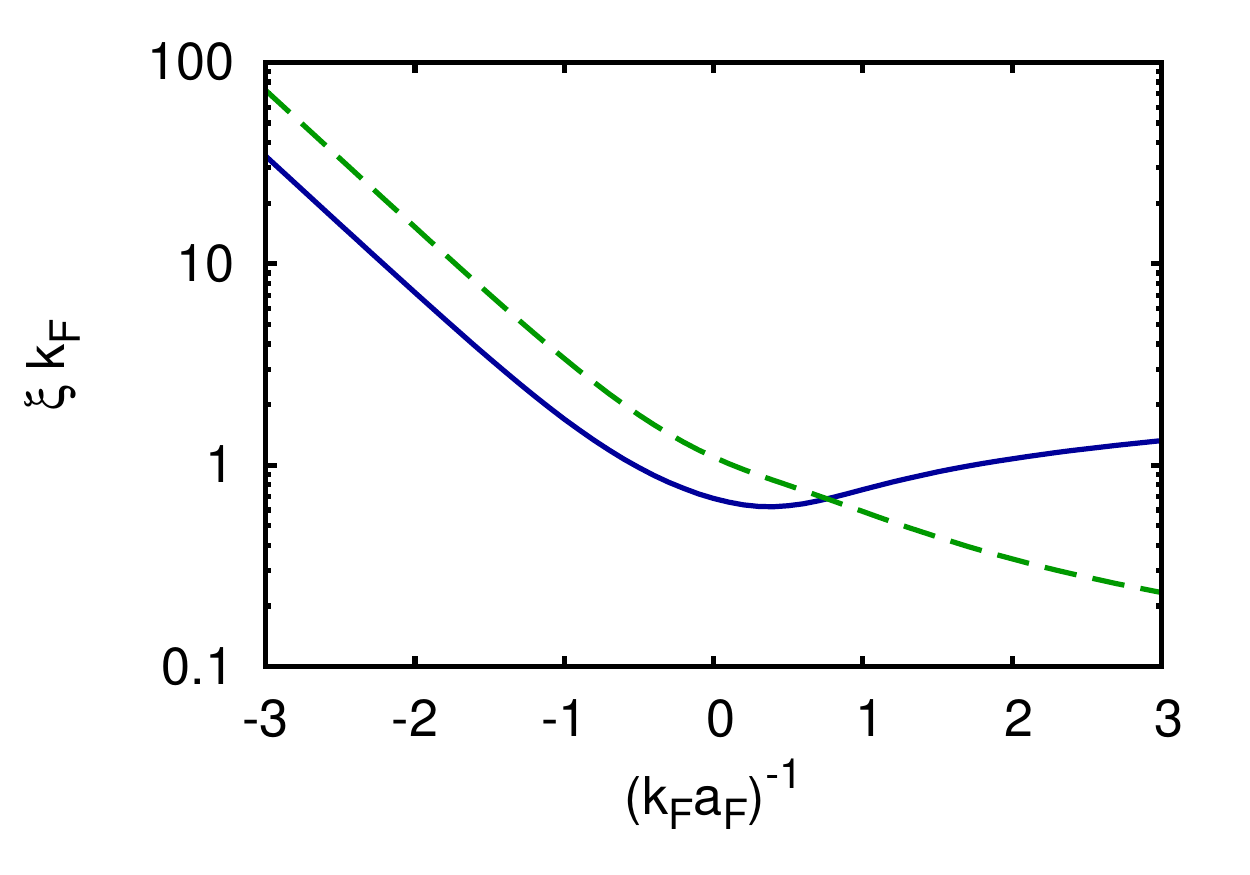}
\caption{Zero-temperature intra-pair coherence length $\xi_{\mathrm{pair}}$ at the mean-field level (dashed line) vs the coupling parameter $(k_F a_F)^{-1}$.  
The full line gives the corresponding evolution of the inter-pair coherence (healing) length $\xi$, to be discussed extensively in Section~\ref{sec:intra-inter-pair}.  
Note that the two lengths differ by an irrelevant numerical factor in the BCS limit owing to their independent definitions. 
[Reproduced from Ref.~\cite{Spuntarelli-2010}.]}
\label{Figure-2}
\end{center}
\end{figure} 

Figure~\ref{Figure-2} shows the evolution throughout the BCS-BEC
crossover of the zero-temperature \emph{pair coherence length}
$\xi_{\mathrm{pair}}$ (dashed line) calculated for the BCS ground
state (\ref{BCS-wave-function}).  This quantity provides information
about the intra-pair correlation established between fermions of
opposite spins \cite{Pistolesi-1994} and, at the mean-field level,
corresponds to the spatial extension of the order parameter
$\phi_{\vek{k}}$.  Since $\phi_{\vek{k}}$ can, in turn, be interpreted
as a wave function as shown by Eq.~(\ref{alternative-gap-equation}),
the r.m.s. radius $\xi_{\mathrm{pair}}$ of the pairs can be obtained
from the expression:
\begin{equation}
\xi_{\mathrm{pair}}^{2} = \frac{\int\! d\vek{r}\,\vek{r}^2\,|\phi(\vek{r})|^2} {\int\! d\vek{r}\, |\phi(\vek{r})|^2}
\label{CL}
\end{equation}
\noindent
where $\phi(\vek{r})$ is the Fourier transform of $\phi_{\vek{k}}$.
[For later convenience, Fig.~\ref{Figure-2} shows also the corresponding evolution of the \emph{healing length} $\xi$ (full line), that instead provides information about the inter-pair
correlation \cite{Pistolesi-1996} and requires the inclusion of pairing fluctuations beyond mean field - see Section~\ref{sec:intra-inter-pair}].  
In Fig.~\ref{Figure-2}, the
average inter-particle distance $k_F^{-1}$ is used as the unit for
$\xi_{\mathrm{pair}}$, where $k_F$ is the Fermi wave vector related
to the particle density via the relation $n = k_F^3/(3 \pi^2)$.
In Fig.~\ref{Figure-2} and throughout this paper, the quantity
$(k_F a_F)^{-1}$ plays the role of the \emph{coupling parameter}
of the theory.  Depending on the sign of
$a_F$, this parameter ranges from $(k_F\, a_F)^{-1} \lesssim -1$
characteristic of the weak-coupling BCS regime when $a_F < 0$, to
$(k_F\, a_F)^{-1} \gtrsim +1$ characteristic of the
strong-coupling BEC regime when $a_F > 0$, across the value
$(k_F\, a_F)^{-1}=0$ at unitarity when $|a_F|$ diverges.  In
practice, the ``crossover region'' of most interest, to be discussed
throughout this paper, turns out to be approximately limited to the
interval $-1 \lesssim (k_F\, a_F)^{-1} \lesssim +1$.

Pairing fluctuations play a particularly important role in the normal
phase above the critical temperature $T_c$ where the order parameter
vanishes (the generalization of the BCS approach to finite temperature
will be considered in Section~\ref{sec:BCS_meanfield}).  This is
because, a ``local'' order is expected to survive above $T_c$ if the
system is fluctuating, even though the long-range order characteristic
of the superfluid phase is lost.  These considerations have led people
to associate pairing fluctuations with the occurrence of a
\emph{pseudo-gap} above $T_c$ (as seen, for instance,
by a depression of the single-particle density of states about the
chemical potential), in a similar way to what occurs in the superfluid
phase below $T_c$ when the order parameter is instead nonvanishing.
Such an analogy (between the order parameter below $T_c$ and the
pseudo-gap above $T_c$) has been carried to the point that even the
pseudo-gap state occurring in copper-oxide superconductors has been
interpreted from the point of view of a BCS-BEC crossover scenario
\cite{Chen-2005}.

In this context, it can be useful to consider the similarity between
the pseudo-gap physics resulting from pairing fluctuations above
$T_c$ and the persistence of (damped) spin waves, which are present
over regions of limited extent in the normal phase of ferromagnetic
(or antiferromagnetic) materials when a strict long-range order is
absent \cite{Mook-1973}.  Through this analogy, pseudo-gap phenomena
in a Fermi system with an attractive inter-particle interaction are
attributed to the persistence of a ``local pairing order'' above the
superfluid temperature $T_c$, which occurs even though the
(off-diagonal) long-range order is absent.  This local order, which is
built up by pairing fluctuations above $T_c$, makes the
single-particle excitation spectrum to resemble that below $T_c$,
although with a frequency broadening due to the decay of the local
excitations which cannot propagate over long distances in the absence
of long-range order.

In addition, it is relevant to point out that, when raising the temperature and approaching $T_c$ from below, pairing correlations turn out to persist
over a finite distance even within mean field, without the need of considering pairing fluctuations beyond mean field.  
Indeed, explicit calculations of the temperature dependence of the pair correlation function within mean field show that $\xi_{\mathrm{pair}}$ maintains a finite value at $T_{c}^{-}$ \cite{Marsiglio-1990}, 
a result which remains true irrespective of coupling across the BCS-BEC crossover \cite{Palestini-2014}).  
This remark, too, points out the importance of including pairing fluctuations in the normal phase above $T_c$, in order to obtain a meaningful finite value of $\xi_{\mathrm{pair}}$
even when approaching $T_c$ from above.
Otherwise, $\xi_{\mathrm{pair}}$ would be discontinuous when passing from $T_{c}^{-}$ to $T_{c}^{+}$, since above $T_{c}$ fermions become non-interacting within the BCS (mean-field) approximation.
We shall return to this point more extensively in Section~\ref{sec:intra-inter-pair}.

\section{BCS mean field}
\label{sec:BCS_meanfield}

\subsection{The homogeneous infinite matter case for a contact interaction}
\label{sec:BCS_homogeneous}

The BCS theory of superconductivity considers an effective
\emph{attractive} interaction $V_{\mathrm{eff}}(\vek{r}-\vek{r}')$
between otherwise free fermions of two different species
(conventionally referred to as spin $\uparrow$ and $\downarrow$),
which are embedded in a continuous medium at a distance
$|\vek{r}-\vek{r}'|$ \cite{Bardeen-1957,Schrieffer-1964}.  This
attractive interaction, albeit weak, is responsible for the formation
of Cooper pairs in the medium \cite{Cooper-1956}.  In condensed matter
(like in metallic superconductors), detailed knowledge of the form of
the attractive interaction is not required for most purposes and one
may accordingly consider the simple form of a ``contact'' (zero-range)
potential $V_{\mathrm{eff}}(\vek{r}) = \varv_0 \,\, \delta (\vek{r})$,
where $\varv_0$ is a negative constant.  This model interaction fully
applies to ultra-cold Fermi gases (at least in the presence of a broad
Fano-Feshbach resonance as discussed in Section~\ref{sec:Feshbach}),
but does not apply to nuclear systems discussed in
Section~\ref{sec:nuclearsystems} for which a finite-range interaction
should in principle be retained.  In order to reduce the numerical
difficulties, however, even in the treatment of nuclear superfluids
the finite-range interaction is often approximated by a zero-range
one, which requires one to introduce a density-dependent coupling
constant (see, e.g., Ref.~\cite{Garrido-2001}).  This approach has
further been elaborated (for instance, in
Refs.~\cite{Bulgac-2002,Yu-2003}) with the introduction of an
efficient regularization scheme for finite systems (as discussed in
Section~\ref{sec:bdg-hfb}). Here, like in most part of the paper, we
consider the case of equal populations $N_{\uparrow} =
N_{\downarrow}$, while the new features arising under the more general
condition $N_{\uparrow} \ne N_{\downarrow}$ will be explicitly
considered in Section~\ref{sec:polarized}.

The price to pay for the use of a contact interaction is that, when dealing with a homogeneous system, integrals over the wave vector
$\vek{k}$ may diverge in the ultraviolet since the Fourier transform $V_{\mathrm{eff}}(\vek{k}) = \varv_0$ is a constant (in what follows,
we shall consider a three-dimensional system).  
This difficulty can be simply overcome by introducing a cut-off, given for metallic superconductors by the Debye frequency and for nuclei by a
phenomenologically adjusted cut-off energy that reflects the range of the effective force. 
For ultra-cold gases, on the other hand, one can exploit the fact that a similar divergence affects also the two-body problem \emph{in vacuum}, whereby the
fermionic scattering length $a_F$ is obtained from the relation \cite{Sa-de-Melo-1993}:
\begin{equation}
\frac{m}{4\pi a_F} = \frac{1}{\varv_0} + \int_{|\vek{k}| \le k_0} \! \frac{d\vek{k}}{(2 \pi)^3} \, \frac{m}{\vek{k}^2}
\label{regularization}
\end{equation}
\noindent
with an ultraviolet cutoff $k_0$.  This regularization procedure
entails the limits $\varv_0 \rightarrow 0^{-}$ and $k_0 \rightarrow
\infty$ to be taken simultaneously, in such a way that $a_F$ is kept
at the desired value. Replacing the parameter $\varv_0$ with the
physical quantity $a_F$ is especially relevant in the context of
ultra-cold gases, where the natural cut-off given by the inverse of
the interaction range would be orders of magnitude larger than all
wave vectors of physical interest and where $a_F$ can be
experimentally controlled \cite{Regal-2003a}.  As we shall see in
Section~\ref{sec:nsr}, the regularization (\ref{regularization}) is
also of help when dealing with the many-body problem based on this
two-body interaction, to the extent that it greatly reduces the number
of many-body diagrams that survive in the limit $\varv_0 \rightarrow
0$.
 
In this way, the BCS mean-field equations for the order (gap)
parameter $\Delta$ at the temperature $T$ and for the density $n$
become \cite{Schrieffer-1964}:
\begin{gather}
- \frac{m}{4\pi a_F}\, = \, \int \! \frac{d\vek{k}}{(2 \pi)^3} \, \left( \frac{1-2f(E_{\vek{k}})}{2 E_{\vek{k}}} \, - \, \frac{m}{\vek{k}^2} \right)
\label{gap-homogeneous} \\
n \, = \, \int \! \frac{d\vek{k}}{(2 \pi)^3} \, \left(1 \, - \, \frac{\xi_{\vek{k}}}{E_{\vek{k}}} \left( 1-2f(E_{\vek{k}}) \right) \right)
\label{density-homogeneous}
\end{gather}
\noindent
where $f(E) =(e^{E/T} + 1)^{-1}$ is the Fermi function and
$E_{\vek{k}} =\sqrt{\xi_{\vek{k}}^2 + |\Delta|^2}$.  Note that the
quantity $\Delta$ that enters Eqs.~(\ref{gap-homogeneous})
and (\ref{density-homogeneous}) plays the dual role of the order
parameter (which is non-vanishing between $T=0$ and $T=T_c$) and of
the minimum value of the single-particle excitation energy
$E_{\vek{k}}$ at $|\vek{k}| = \sqrt{2 m \mu}$ when $\mu > 0$. 
[The extension to finite temperature of the gap
  equation (\ref{gap-eq}) with a finite-range interaction will be
  considered in Section~\ref{sec:deuteron} - cf. Eq.~(\ref{gapeqn}).]

For a \emph{weak} attractive interaction, $a_F$ is small and
negative such that $(k_F a_F)^{-1} \ll -1$.  This limit
characterizes what is called the ``conventional'' BCS theory
\cite{Bardeen-1957}.  In this limit, the Fermi surface of
non-interacting fermions is only slightly perturbed by the presence of
the interaction, and the chemical potential $\mu$ at $T=0$ coincides
with the Fermi energy $E_F = k_F^2/(2m)$.  This is actually the
limit where the mean-field approximation is expected to work best, to
the extent that the size $\xi_{\mathrm{pair}}$ of a Cooper pair is
much larger than the average inter-particle distance $k_F^{-1}$ and
a large number of pairs is contained within the size of
$\xi_{\mathrm{pair}}$ \cite{Schrieffer-1964}.

\subsection{Solution at $T=0$ for a contact potential}
\label{sec:BCS_solution}

For a generic value of the coupling $(k_F a_F)^{-1}$ that ranges
between the BCS and BEC limits, the values of $\Delta_0$ and $\mu_0$
at $T=0$ are determined by solving the coupled equations
(\ref{gap-homogeneous}) and (\ref{density-homogeneous}), where one
sets $f(E_{\vek{k}})=0$ since $E_{\vek{k}}$ is always positive. 
Although a numerical solution of the
ensuing equations is possible \cite{Carter-1995}, it was found independently that the
three-dimensional integrals over the wave vector occurring in these
equations can be expressed either in terms of the complete elliptic
integrals \cite{Marini-1998} or in terms of the Legendre function
\cite{Papenbrock-1999}, in such a way that closed-form expressions for
$\Delta_0$ and $\mu_0$ are obtained.  In Ref.~\cite{Marini-1998}, the
results were expressed in terms of the parameter $x_0 =
\mu_0/\Delta_0$ that ranges from $- \infty$ (BEC limit) to $+ \infty$
(BCS limit), while in Ref.~\cite{Papenbrock-1999} the inverse of this
parameter was used.

In Ref.~\cite{Marini-1998}, the following results were obtained:
\begin{gather}
\frac{\Delta_0}{E_F} = \frac{1} {\left( x_0 J(x_0) + I(x_0) \right)^{2/3}} \, ,    
\label{Delta-exact}\\
\frac{\mu_0}{E_F} = x_0 \, \frac{\Delta_0}{E_F} \, , 
\label{mu-exact}\\
\frac{1}{k_F a_F}  =  - \frac{4}{\pi} \, \frac{ x_0 I(x_0) - J(x_0)}{\left( x_0 J(x_0) + I(x_0) \right)^{1/3}} \, ,
\label{k_Fa_F-exact}
\end{gather}
\noindent
where
\begin{equation}
I(x_0) = \frac{1}{2(1+x_0^2)^{1/4}} \, F\left(\frac{\pi}{2},\kappa\right)
\label{integral-I}
\end{equation}
\noindent
and
\begin{equation}
J(x_0) = (1+x_0^2)^{1/4} \, E\left(\frac{\pi}{2},\kappa\right) - \frac{1}{4y_0 (1+x_0^2)^{1/4}}\, F\left(\frac{\pi}{2},\kappa\right)\, .
\label{integral-J}
\end{equation}
\noindent
In these expressions, $F\left(\frac{\pi}{2},\kappa\right)$ and
$E\left(\frac{\pi}{2},\kappa\right)$ are the complete elliptic
integrals of the first and second kind, respectively, and
\begin{equation}
y_0 = \frac{\sqrt{1+x_0^2} + x_0}{2}\, ,\qquad
\kappa^2 = \frac{y_0}{\sqrt{1+x_0^2}} \, .         
\label{yo-kappa}
\end{equation}
\noindent
In Ref.~\cite{Papenbrock-1999}, on the other hand,
Eq.~(\ref{k_Fa_F-exact}) was equivalently expressed in the form:
\begin{equation}
\frac{1}{k_Fa_F} = \bigg(\frac{1+x_0^2}{x_0^2}\bigg)^{1/4} P_{1/2}\left(-|x_0| \, (1+x_0^2)^{-1/2} \right) 
\label{Papen}
\end{equation}
where $P_{\alpha}$ denotes the Legendre function.
In addition, analytic expressions for the pair coherence length $\xi_{\rm pair}$ and the condensate fraction were obtained in Refs.~\cite{Marini-1998} and \cite{Salasnich-2005}, respectively.

\begin{figure}[t]
\begin{center}
\includegraphics[width=14.0cm,angle=0]{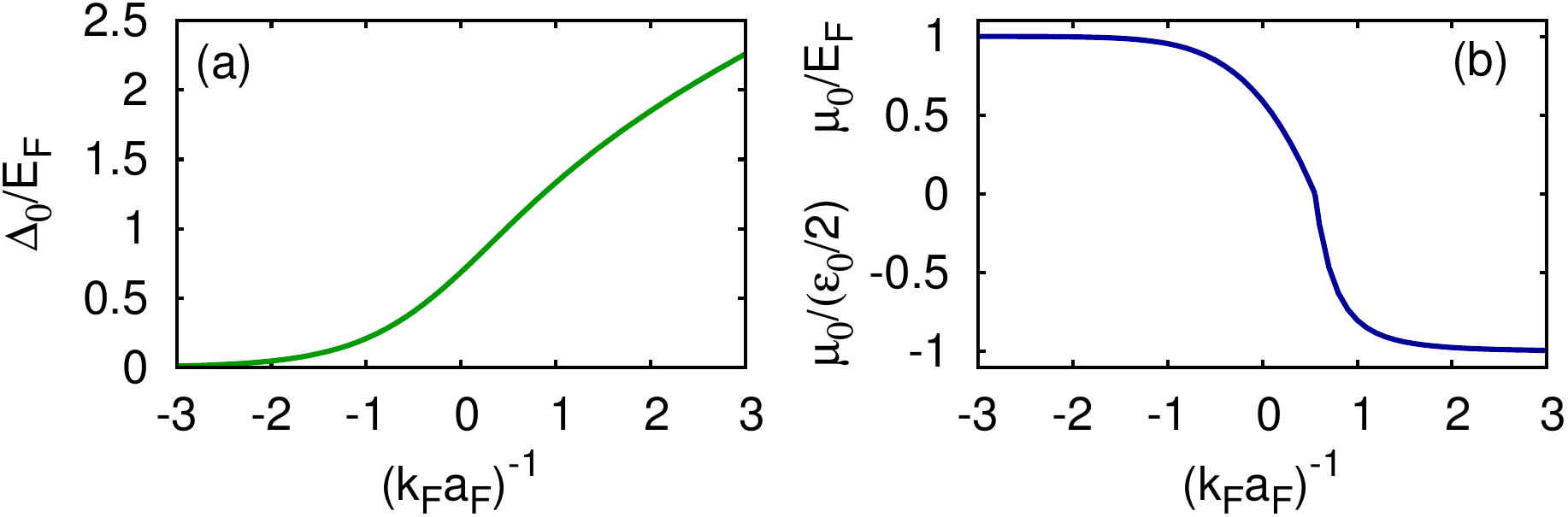}
\caption{(a) Gap $\Delta_0$ and (b) chemical potential $\mu_0$ at the mean-field level vs $(k_F a_F)^{-1}$ for a homogeneous system at $T=0$. 
[Reproduced from Ref.~\cite{Spuntarelli-2010}.]}
\label{Figure-3}
\end{center}
\end{figure} 

Analytic approximations can be readily obtained in the BCS and BEC
limits.  In the BCS limit where $x_0 \gg 1$, one obtains $\kappa^2
\simeq 1 - 1/(4x_0^2) \approx 1$ in Eq.(\ref{yo-kappa}).  One can
thus approximate $\Delta_0/E_F \simeq 1/x_0$, $\mu_0/E_F
\simeq 1$, and $1/(k_F a_F) \simeq - (2/\pi) \ln{(8x_0/e^2)}$,
in such a way that:
\begin{equation}
\Delta_0= \frac{8 E_F}{e^2}\,\exp \bigg (\frac{\pi}{2k_F a_F} \bigg) \, .
\label{PF}
\end{equation}
\noindent
In the BEC limit, on the other hand, $x_0<0$ and $|x_0| \gg 1$
such that $\kappa^2 \simeq 1/(4x_0^2) \approx 0$.  One thus
finds:
\begin{gather}
\frac{1}{k_F a_F} \simeq \left(\frac{16}{3\pi}\right)^{1/3} |x_0|^{2/3} \,,
\label{approximate-k_Fa_F}\\
\frac{\Delta_0}{E_F}  \simeq \left(\frac{16}{3\pi}\right)^{2/3} |x_0|^{1/3} \simeq \,  \left(\frac{16}{3\pi}\right)^{1/2} \left(\frac{1}{k_F a_F}\right)^{1/2},            
\label{approximate-Delta}\\
\frac{\mu_0}{E_F} \simeq - \left(\frac{16}{3\pi}\right)^{2/3} |x_0|^{4/3} \simeq - \frac{1}{2} \frac{\varepsilon_0}{E_F} \, . 
\label{approximate-mu}
\end{gather}
\noindent
The full expressions (\ref{Delta-exact}), (\ref{mu-exact}), and
(\ref{k_Fa_F-exact}) interpolate \emph{smoothly} between the above
(BCS and BEC) limits through the crossover region $k_F|a_F|
\gtrsim1$.  Plots of $\Delta_0$ and $\mu_0$ as functions of
$(k_Fa_F)^{-1}$ are readily obtained from these equations, as
shown in Figs.~\ref{Figure-3}(a) and \ref{Figure-3}(b).
Note, in particular, that the chemical potential: (i) In the BCS limit
equals the Fermi energy $E_F$; (ii) In the crossover region $-1
\lesssim (a_F k_F)^{-1} \lesssim +1$ gradually decreases and
eventually changes its sign; (iii) In the BEC limit reaches the
asymptotic value $-\varepsilon_0/2$.  It is just this behaviour of
the chemical potential that drives the BCS-BEC crossover, as it was
anticipated in Section~\ref{sec:BCSwavefunction}.

\subsection{Extension to finite temperature}
\label{sec:BCS_finite_T}

At finite temperature, the presence of the Fermi function in the gap
equation (\ref{gap-homogeneous}) and density equation
(\ref{density-homogeneous}) makes it impossible to find an analytic
solution and one has then to resort to a numerical solution of these
equations.  An exception occurs, however, upon approaching the
critical temperature $T_c$ from below, where analytic results can be
obtained both in BCS and BEC limits (see, e.g., Ref.~\cite{Strinati-2000}).

At the mean-field level, the critical temperature $T_c$ is defined
by the vanishing of the BCS gap parameter $\Delta$.
Equations~(\ref{gap-homogeneous}) and (\ref{density-homogeneous}) are
thus replaced by:
\begin{gather}
- \, \frac{m}{4 \pi a_F} \, = \, \int \! \frac{d\vek{k}}{(2 \pi)^3} \, \left( \frac{\tanh{(\xi_{\vek{k}}/2T_c)}}{2 \xi_{\vek{k}}} \right. \left. \, - \, \frac{m}{\vek{k}^2} \right)              
\label{regular-T-c}\\
n \, = \, 2 \, \int \! \frac{d\vek{k}}{(2 \pi)^3} \, \frac{1}{\exp{(\xi_{\vek{k}}/T_c)} \, + \, 1} \, .       
\label{density-T-c}
\end{gather}
\noindent
In the weak-coupling limit where $(k_F a_F)^{-1} \ll -1$,
Eq.~(\ref{density-T-c}) yields the value of the chemical potential of
a Fermi gas, which is only slightly smaller than the zero-temperature
result $\mu=E_F$ provided $T_c \ll E_F$.  Correspondingly,
Eq.~(\ref{regular-T-c}) gives the following expression for $T_c$:
\begin{equation}
T_c \, = \, \frac{8 e^{\gamma} E_F}{\pi e^2} \, \exp\bigg(\frac{\pi}{2k_Fa_F}\bigg)
\label{T-c-w-c}
\end{equation}
\noindent
where $\gamma$ is Euler's constant (such that $\Delta_0/T_c = \pi
/ e^{\gamma} \simeq 1.764$ with reference to the result (\ref{PF})
that holds in the same limit).  Since in the weak-coupling regime
$a_F$ is negative and $k_F |a_F| \ll 1$, Eq.~(\ref{T-c-w-c})
yields $T_c \ll E_F$ consistently with our assumptions.

In the strong-coupling limit where $(k_F a_F)^{-1} \gg +1$, the role of the two equations (\ref{regular-T-c}) and (\ref{density-T-c}) is reversed.  
If we assume that $T_c \ll |\mu|$, we may set $\tanh{(\xi_{\vek{k}}/2T_c)}
\simeq 1$ in Eq.~(\ref{regular-T-c}), making it to reduce to the
bound-state equation and yielding $\mu \simeq -\varepsilon_0/2$
at the leading order.  The same kind of
approximation, however, cannot be used in Eq.~(\ref{density-T-c})
since $n$ has to remain finite.  In this equation, we set instead
$(\exp{(\xi_{\vek{k}}/T_c)}+1)^{-1} \simeq
\exp{(-\xi_{\vek{k}}/T_c)}$ and obtain
\begin{equation}
\frac{\mu}{T_c} \, \simeq \, \ln\left[\frac{n}{2} \left(\frac{2\pi}{mT_c}\right)^{3/2}\right]  \, ,     
\label{classical-mu}
\end{equation}
\noindent
a result which coincides with the expression of the classical chemical
potential at temperature $T_c$.  Using at this point the value
$\mu(T_c) \simeq -\varepsilon_0/2$ as determined from
Eq.~(\ref{regular-T-c}), we can solve the expression
(\ref{classical-mu}) iteratively for $T_c$ yielding
\cite{Sa-de-Melo-1993}:
\begin{equation}
T_c \, \simeq \, \frac{\varepsilon_0} {2 \ln \left(\frac{\varepsilon_0}{E_F}\right)^{3/2}}
\label{classical-T-c}
\end{equation}
\noindent
at the leading order in $E_F/\varepsilon_0 \ll 1$.  Although
$T_c \ll |\mu|$ consistently with our assumptions, the expression of
$T_c$ given by Eq.~(\ref{classical-T-c}) diverges in the BEC limit
\emph{at fixed density}, instead of approaching as expected the finite
value $T_{\mathrm{BEC}} = \frac{\pi}{m} \left(\frac{n/2}{\zeta(3/2)}
\right)^{2/3}$ at which the Bose-Einstein condensation of an ideal gas
of composite bosons with mass $2m$ and density $n/2$ occurs
($\zeta(3/2) \approx 2.612$ being the Riemann $\zeta$ function of
argument $3/2$).  Physically, the mean-field temperature
(\ref{classical-T-c}) corresponds to the \emph{dissociation} of
composite bosons, with the $\ln$ factor in the denominator originating
from entropy effects.  It is thus appropriate to associate the
mean-field result (\ref{classical-T-c}) with the ``pair dissociation
temperature'' of the composite bosons, which is completely
unrelated to the BEC temperature $T_{\mathrm{BEC}}$ at which quantum
coherence is established among the composite bosons.  The reason for
this failure is that only the internal degrees of freedom of the
composite bosons are taken into account by the mean-field expressions
(\ref{regular-T-c}) and (\ref{density-T-c}), thereby leaving aside the
translational degrees of freedom of the composite bosons.  To include
these, pairing fluctuations beyond mean field need to be considered.
The corresponding analysis will be discussed in Section~\ref{sec:nsr}.

\subsection{Bogoliubov-de Gennes and Hartree-Fock-Bogoliubov equations for
  inhomogeneous Fermi systems}
\label{sec:bdg-hfb}

The BCS wave function (\ref{BCS-wave-function}), for the ground state
of a \emph{homogeneous} Fermi gas with an attractive inter-particle
interaction, is a variational wave function with an associated gap
parameter $\Delta_0$ at zero temperature, which extends to the
superfluid phase the Hartree-Fock approximation for the normal phase by
retaining also operator averages of the type $\langle
c^{\dagger}_{\vek{k} \uparrow} c^{\dagger}_{-\vek{k} \downarrow}
\rangle$ that do not conserve the particle number.

At finite temperature, the ``mean-field'' character of the BCS gap
equation (\ref{gap-homogeneous}) for $\Delta$ and of the associated
coherence factors $u_{\vek{k}}$ and $\varv_{\vek{k}}$ given by the
expressions (\ref{v-square}), is better captured by an alternative
procedure, whereby a mean-field (or Hartree-Fock-type) decoupling is
directly performed at the level of the Hamiltonian by including
particle non-conserving averages.  Accordingly, one replaces the
grand-canonical Hamiltonian $K = H - \mu N$ by its mean-field
approximation as follows
\begin{equation}
K = \sum_{\vek{k}\sigma} \xi_{\vek{k}} \, c^{\dagger}_{\vek{k}\sigma} c_{\vek{k}\sigma} + \varv_0 \sum_{\vek{k} \vek{k}' \vek{q}} 
c^{\dagger}_{\vek{k}+\vek{q}/2\uparrow} c^{\dagger}_{-\vek{k}+\vek{q}/2\downarrow} c_{-\vek{k}'+\vek{q}/2\downarrow} c_{\vek{k}'+\vek{q}/2\uparrow}
\rightarrow 
\sum_{\vek{k}\sigma} \xi_{\vek{k}} \, c^{\dagger}_{\vek{k}\sigma} c_{\vek{k}\sigma} - \sum_{\vek{k}} \left( \Delta^{*} \, c_{-\vek{k}\downarrow} c_{\vek{k}\uparrow} 
+ \Delta \, c^{\dagger}_{\vek{k}\uparrow} c^{\dagger}_{-\vek{k}\downarrow} \right) - \frac{\Delta^2}{\varv_0}               
\label{BCS-HF-decoupling}                         
\end{equation}
\noindent
where, for simplicity, we have considered the case of a zero-range
interaction with strength $\varv_0$ (the generalization to
finite-range interactions will be considered below).  The parameter
$\Delta$ in Eq.(\ref{BCS-HF-decoupling}) thus satisfies the condition
\begin{equation}
\Delta = - \varv_0 \sum_{\vek{k}} \langle \, c_{-\vek{k}\downarrow} c_{\vek{k}\uparrow} \, \rangle
\label{Delta-mf-decoupling}
\end{equation}
\noindent
where the thermal average $\langle \dots \rangle$ is self-consistently
determined through the mean-field Hamiltonian itself.  The quadratic
form (\ref{BCS-HF-decoupling}) can be readily diagonalized by the
canonical transformation
\begin{equation}
c_{\vek{k}\uparrow} = u_{\vek{k}}^{*} \, \alpha_{\vek{k}} \, + \, \varv_{\vek{k}} \, \beta^{\dagger}_{-\vek{k}} \,\, , \qquad
c_{\vek{k}\downarrow} = - \varv_{\vek{k}} \, \alpha^{\dagger}_{-\vek{k}} \, + \, u_{\vek{k}}^{*} \, \beta_{\vek{k}} \,\, ,             
\label{canonical-transformation}
\end{equation}
\noindent
where $u_{\vek{k}} = u_{-\vek{k}}$ and $\varv_{\vek{k}} =
\varv_{-\vek{k}}$ are constrained by $|u_{\vek{k}}|^2 +
|\varv_{\vek{k}}|^2 = 1$ at any given $\vek{k}$.  In terms of the
new operators $\alpha_{\vek{k}}$ and $\beta_{\vek{k}}$, the mean-field
Hamiltonian (\ref{BCS-HF-decoupling}) becomes (see, e.g., Chapt. 6 of
Ref.~\cite{Ring-1980})
\begin{multline}
K \rightarrow \sum_{\vek{k}} \left( \alpha^{\dagger}_{\vek{k}} \alpha_{\vek{k}} 
  + \beta^{\dagger}_{\vek{k}} \beta_{\vek{k}} \right)
  \left[ \xi_{\vek{k}} \left( u_{\vek{k}}^2 - \varv_{\vek{k}}^2 \right)
  + 2 \Delta u_{\vek{k}} \varv_{\vek{k}} \right] 
  + \sum_{\vek{k}} \left( \alpha^{\dagger}_{\vek{k}} \beta^{\dagger}_{-\vek{k}} 
  + \beta_{-\vek{k}} \alpha_{\vek{k}} \right)
  \left[ 2 \xi_{\vek{k}} u_{\vek{k}} \varv_{\vek{k}}
  - \Delta \left( u_{\vek{k}}^2 - \varv_{\vek{k}}^2 \right) \right]
\\
 + 2 \sum_{\vek{k}} 
  \left( \xi_{\vek{k}} \varv_{\vek{k}}^2 - \Delta u_{\vek{k}} \varv_{\vek{k}} \right) 
- \frac{\Delta^2}{\varv_0}  \label{quadratic_form} 
\end{multline}
\noindent
(here, for simplicity, all quantities are assumed to be real).
Provided that $u_{\vek{k}}$ and $\varv_{\vek{k}}$ are solutions to the
equation
\begin{equation}
\left( 
\begin{array}{cc}
\xi_{\vek{k}} & \Delta\\
\Delta & - \xi_{\vek{k}}
\end{array} 
\right)
\left( \begin{array}{c}
u_{\vek{k}} \\
\varv_{\vek{k}} 
\end{array} 
\right) 
= E_{\vek{k}}
\left( \begin{array}{c}
u_{\vek{k}} \\
\varv_{\vek{k}}
\end{array} 
\right)  \, ,                                          
\label{BdG-equations-homogeneous} 
\end{equation}
\noindent
whereby $u_{\vek{k}}^2 = 1 - \varv_{\vek{k}}^2 = (1+\xi_{\vek{k}}/E_{\vek{k}})/2$ 
and $E_{\vek{k}} = \sqrt{\xi_{\vek{k}}^2+\Delta^2}$, the coefficient 
$[2\xi_{\vek{k}}u_{\vek{k}}\varv_{\vek{k}} - \Delta ( u_{\vek{k}}^2
  - \varv_{\vek{k}}^2 )]$ of the second term on the right-hand side
of Eq.~(\ref{quadratic_form}) vanishes identically, while the
coefficient $[ \xi_{\vek{k}} ( u_{\vek{k}}^2 - \varv_{\vek{k}}^2 )
  + 2 \Delta u_{\vek{k}} \varv_{\vek{k}} ]$ of the first term equals
$E_{\vek{k}}$.  Correspondingly, the self-consistent condition
(\ref{Delta-mf-decoupling}) for $\Delta$ reduces to:
\begin{equation}
\Delta = - \varv_0 \sum_{\vek{k}}  u_{\vek{k}} \varv_{\vek{k}} \left[ 1 - 2 f(E_{\vek{k}}) \right]
\label{Delta-self-consistency}
\end{equation}
\noindent
with $u_{\vek{k}} \varv_{\vek{k}} = \Delta/(2 E_{\vek{k}})$.  Note
further that the last term of Eq.~(\ref{quadratic_form}), which can be
written as $\sum_{\vek{k}} ( \xi_{\vek{k}} -E_{\vek{k}})  -
  \Delta^2/\varv_0 $ with the help of
Eq.~(\ref{BdG-equations-homogeneous}), corresponds to the
(grand-canonical) ground-state energy associated with the BCS state.

What is relevant here is that the above procedure can be readily generalized to situations in which $\Delta \rightarrow \Delta(\vek{r})$ depends on the spatial position $\vek{r}$, owing, for
instance, to the presence of an external potential (like a trapping potential or a barrier), or of a magnetic field, or of the self-consistent mean field in finite nuclei, which introduce spatial 
inhomogeneities in the system.  
In this case, the algebraic matrix equation (\ref{BdG-equations-homogeneous}) is replaced by a pair of coupled Schr\"{o}dinger-like equations for a two-component single-particle fermionic eigenfunction, 
of the form \cite{de-Gennes-1966}:
\begin{equation}
\left( 
\begin{array}{cc}
\mathcal{H}(\vek{r}) & \Delta(\vek{r})            \\
\Delta^{*}(\vek{r})  & - \mathcal{H}(\vek{r})  
\end{array} 
\right)
\left( \begin{array}{c}
u_{\nu}(\vek{r}) \\
\varv_{\nu}(\vek{r}) 
\end{array} 
\right) 
= E_{\nu}
\left( \begin{array}{c}
u_{\nu}(\vek{r}) \\
\varv_{\nu}(\vek{r}) 
\end{array} 
\right)                                           
\label{BdG-equations} 
\end{equation}
\noindent
where $\mathcal{H}(\vek{r}) = - \nabla^2/(2m) + V_{\mathrm{ext}}(\vek{r}) - \mu$ contains the external potential $V_{\mathrm{ext}}(\vek{r})$ (in addition, in the presence of an
external vector potential $\vek{A}(\vek{r})$ the gradient operator $\vek{\nabla}$ in the kinetic energy is replaced as usual by $\nabla - i \vek{A}(\vek{r})$). 
 Note that the Hartree term is absent in the diagonal elements of the matrix
(\ref{BdG-equations}), owing to the use of a contact inter-particle
interaction which makes this term to vanish in the limit $\varv_0 \rightarrow 0$.
[In weak coupling, however, one often introduces in $\mathcal{H}(\vek{r})$ a Hartree-like-term of the form $\frac{4 \pi a_{F}}{m} \, \frac{n}{2}$ \cite{Bruun-1999,Grasso-2003}.]
These equations, known as the Bogoliubov-de Gennes (BdG) equations, have to be solved up to self-consistency for the local gap parameter $\Delta(\vek{r})$, which acts like an
off-diagonal pair potential and satisfies the local condition:
\begin{equation}
\Delta(\vek{r}) = - \varv_0 \sum_{\nu} u_{\nu}(\vek{r}) 
\varv_{\nu}^{*}(\vek{r})
\left[ 1 - 2 f(E_{\nu}) \right] \, .                               
\label{self-consistency}
\end{equation}
\noindent
In addition, the eigenfunctions
$\{u_{\nu}(\vek{r}),\varv_{\nu}(\vek{r})\}$ obey the orthonormality condition:
\begin{equation}
\int \! d\vek{r} \, \left[ u_{\nu}^{*}(\vek{r}) u_{\nu'}(\vek{r})
\, + \, \varv_{\nu}^{*}(\vek{r}) \varv_{\nu'}(\vek{r}) \right] \, 
  = \, \delta_{\nu\nu'}  
\label{normalization-condition}                       
\end{equation}
\noindent
where the Kronecker delta on the right-hand side is readily generalized to account for continuous eigenvalues.  
Physical quantities like the local number density, the current density, and the energy density can all be expressed in terms of the eigenfunctions and eigenvalues of Eq.~(\ref{BdG-equations}) 
\cite{de-Gennes-1966}.

Note that the complex character of $\Delta(\vek{r})$ (as well as of
the eigenfunctions $\{u_{\nu}(\vek{r}),\varv_{\nu}(\vek{r})\}$) has
been restored in
Eqs.~(\ref{BdG-equations})--(\ref{normalization-condition}), to allow
for the presence of a particle current.   Note also that it is the particle-hole mixing
characteristic of the BCS pairing to be responsible for the coupling
of the two components $(u_{\nu}(\vek{r}),\varv_{\nu}(\vek{r}))$ of the
eigenfunctions of Eq.~(\ref{BdG-equations}), in contrast to the normal
Fermi gas where particle- and hole-excitations are separately good
quasi-particles.

When solving the BdG equations for a contact inter-particle interaction with coupling constant $\varv_{0}$ (which is the case of ultra-cold Fermi gases), technical problems arise 
from the need to regularize the self-consistency condition (\ref{self-consistency}) for the gap parameter $\Delta(\vek{r})$ since the sum over $\nu$ therein diverges in the ultraviolet. 
In the homogeneous case with a uniform gap parameter $\Delta_0$ this regularization can be readily implemented with the help of Eq.(\ref{regularization}), by expressing
the bare coupling constant $\varv_0$ that enters the gap equation in terms of the scattering length $a_F$ of the two-body problem [cf. Eq.(\ref{gap-homogeneous})].  
This procedure, however, cannot be utilized when the gap parameter $\Delta(\vek{r})$ has a spatial dependence, and a new strategy is required.  
A first generalization of the regularization procedure for fermions on the BCS side of the crossover trapped in a harmonic potential was given in Ref.~\cite{Bruun-1999} in terms 
of a pseudo-potential method in real space, but it turns out that this method is not easily implemented in numerical calculations. 
A numerically more efficient regularization procedure was described in Ref.~\cite{Grasso-2003} for the solution of the BdG equations (\ref{BdG-equations}) in a harmonic
oscillator basis, which for practical purposes has to be truncated at some energy cutoff $E_c$. Following Refs.~\cite{Bulgac-2002,Yu-2003}, the contribution of states above the cutoff 
is included within the Thomas-Fermi approximation. 
In this way, one obtains the following \emph{regularized version of the gap equation}:
\begin{equation}
\Delta(\vek{r}) = -g_{\mathrm{eff}}(\vek{r}) \, \sum_{\nu}^{E_{\nu} < E_{c} - \mu} \! u_{\nu}(\vek{r}) \varv_{\nu}(\vek{r})^{*} \left[ 1 - 2 f(E_{\nu}) \right] \,,
\label{regularized-gap-equation1}
\end{equation}
\noindent
where the numerical solution of the BdG equations (\ref{BdG-equations}) is explicitly performed only for eigenvalues $E_{\nu}$ up to $E_{c} - \mu$ that appear on the right-hand side
of Eq.(\ref{regularized-gap-equation1}).
The pre-factor $g_{\mathrm{eff}}(\vek{r})$ plays the role of a (cutoff dependent) effective coupling constant and is given by
\begin{equation}
\frac{1}{g_{\mathrm{eff}}(\vek{r})} = \frac{m}{4\pi a_F} - \mathcal{R}(\vek{r})
\label{geff1}
\end{equation}
\noindent
where
\begin{equation}
\mathcal{R}(\vek{r}) = \frac{m}{2\pi^2}\bigg [k_c(\vek{r}) - \frac{k_{\mu}(\vek{r})}{2} \ln \frac{k_c(\vek{r}) + k_{\mu}(\vek{r})} {k_c(\vek{r}) - k_{\mu}(\vek{r})}\bigg ] \, .
\label{value-R(kc)}
\end{equation}
The position-dependent wave vectors $k_{\mu}(\vek{r})$ and $k_c(\vek{r})$ are related to the chemical potential $\mu$ and energy cutoff $E_c$ by 
$\mu = k_{\mu}^2(\vek{r})/(2m) + V_{\mathrm{ext}}(\vek{r})$ and $E_c = k_c^2(\vek{r})/(2m) + V_{\mathrm{ext}}(\vek{r})$. 
As noticed in Refs.~\cite{Bulgac-2002,Grasso-2003}, Eq.~(\ref{value-R(kc)}) can also be used for negative values of $\mu-V_{\mathrm{ext}}(\vek{r})$ by
allowing for imaginary values of $k_{\mu}(\vek{r})$. 

This approach was extended in Ref.~\cite{Simonucci-2013} to the whole BCS-BEC crossover and generic inhomogeneous situations,
by combining the introduction of the cutoff $E_{c}$ in the quasi-particle energies such that $E_{\nu} < E_{c} - \mu$ with the derivation of the Gross-Pitaevskii equation for composite bosons 
in the BEC limit that was done in Ref.~\cite{Pieri-2003} (cf. Section \ref{sec:GLandGP}). 
Under the assumption that the energy cutoff $E_{c}$ is the largest energy scale in the problem (such that $\Delta(\vek{r})/E_{c} \ll 1$, $T/E_{c} \ll 1$, and $E_F/ E_{c} \ll 1$, where $E_F$ is the
Fermi energy associated with the mean density - conditions that can be somewhat relaxed in the weak-coupling limit), the regularized version of the gap equation reads eventually \cite{Simonucci-2013}:
\begin{multline}
\left\{ - \frac{m}{4 \pi a_F} + \mathcal{R}(\vek{r}) - \left[ \frac{1}{2} \, \mathcal{I}_{02}(\vek{r}) - \frac{1}{3} \, \mathcal{I}_{13}(\vek{r}) \right] \frac{\nabla^2}{4m} \right.
+  \left. 2 V_{\mathrm{ext}}(\vek{r}) \frac{1}{4} \mathcal{I}_{02}(\vek{r}) +  \frac{1}{4} \, \mathcal{I}_{03}(\vek{r}) |\Delta(\vek{r})|^2 \right\} \Delta(\vek{r}) \\
= \sum_{\nu}^{E_{\nu} < E_{c} - \mu} \! u_{\nu}(\vek{r}) \varv_{\nu}(\vek{r})^{*}  \left[ 1 - 2 f(E_{\nu}) \right] \, .
\label{regularized-gap-equation}
\end{multline}
\noindent
The quantities $\mathcal{I}_{ij}(\vek{r})$ that enter the left-hand side of this equation are defined as follows:
\begin{equation}
\mathcal{I}_{ij}(\vek{r}) \, = \,  \int_{|\vek{k}|>k_{c}(\vek{r})} \! \frac{d\vek{k}}{(2 \pi)^3} \, \frac{\left( \frac{\vek{k}^2}{2m} \right)^{i}} {\left( \frac{\vek{k}^2}{2m}+V_{\mathrm{ext}}(\vek{r})-\mu\right)^{j}} \, .
\label{definition-I_i_j-integrals}
\end{equation}
The right-hand side of Eq.(\ref{regularized-gap-equation}), which results from an explicit numerical integration of the BdG equations over the reduced energy range $E_{\nu} < E_{c} - \mu$, 
acts as a \emph{source term} on the non-linear differential equation for $\Delta(\vek{r})$ associated with the left-hand side. 
By implementing the numerical calculation of the BdG equations for an isolated vortex embedded in a uniform superfluid, it was shown in Ref.~\cite{Simonucci-2013} that the inclusion of the various
terms on the left-hand side of Eq.(\ref{regularized-gap-equation}) [from the linear ($\Delta$) term, to the linear plus cubic ($\Delta + \Delta^3$) terms, and finally to the linear plus cubic plus
Laplacian ($\Delta + \Delta^3 + \nabla^2$) terms] plays an increasingly important role in reducing the total computational time and memory space at any coupling throughout the BCS-BEC
crossover, by decreasing the value of the cutoff $E_{c}$ up to which the eigenfunctions of the BdG equations have to be explicitly calculated \cite{Simonucci-2013}.

In practice, the BdG equations have been solved numerically up to self-consistency only for relatively simple geometries since their numerical solution becomes rapidly 
too demanding due to computational time and memory space. 
For ultra-cold Fermi gases, the BdG equations have been solved to account for the spatial dependence of the order parameter in balanced \cite{Bruun-1999,Grasso-2003,Ohashi-2005} 
and imbalanced \cite{Castorina-2005,Jensen-2007} trapped gases, as well as for the microscopic structure of a single vortex at zero temperature on the BCS side \cite{Nygaard-2003} 
and across the BCS-BEC crossover \cite{Sensarma-2006}, and also at finite temperature across the BCS-BEC crossover \cite{Simonucci-2013}.  
The microscopic structure of a single vortex for the unitary Fermi gas has been analyzed in Ref.~\cite{Bulgac-2003} with BdG-like equations resulting from the superfluid local-density 
approximation of Ref.~\cite{Bulgac-2002b}, which has later been used also to calculate the profile of the order parameter in balanced trapped gases \cite{Bulgac-2007} or in the FFLO phase 
of imbalanced Fermi gases throughout the BCS-BEC crossover \cite{Bulgac-2008b} (see Section \ref{sec:polarized}).
In addition, the BdG equations have been used to determine the occurrence of vortex lattices in a rotating trap on the BCS side of the crossover \cite{Feder-2004,Tonini-2006}. 
A study of the Josephson effect with a one-dimensional barrier throughout the BCS-BEC crossover at zero temperature \cite{Spuntarelli-2010} will be reported in Section \ref{sec:Josephson}.

In nuclear physics, where in the most sophisticated cases finite-range
(instead of zero-range) forces are used in the mean-field and gap
equations (a famous example in this context being the Gogny-force
\cite{Decharge-1980}), both matrix elements in (\ref{BdG-equations}) become
non-local. The corresponding equations, usually referred to as the
Hartree-Fock-Bogoliubov (HFB) equations, are thus more general than
the BdG equations and read (for simplicity, we do not write down the spin and isospin structure):
\begin{equation}
\int d\vek{r}'
\begin{pmatrix}\mathcal{H}(\vek{r}, \vek{r}')&\Delta(\vek{r}, \vek{r}')\\
\Delta^*(\vek{r}, \vek{r}')& -\mathcal{H}^*(\vek{r}, \vek{r}')\end{pmatrix}
\begin{pmatrix} u_{\nu}(\vek{r}')\\ \varv_\nu(\vek{r}')\end{pmatrix}
= E_\nu \begin{pmatrix} u_{\nu}(\vek{r})\\ \varv_\nu(\vek{r})\end{pmatrix}
\label{nonlocal-HFB}
\end{equation}
where
\begin{equation}
\mathcal{H}(\vek{r}, \vek{r}')= \bigg [-\frac{\nabla_{\vek{r}}^2}{2m} + \Gamma_{\mathrm{H}}(\vek{r}) \bigg ] \, \delta(\vek{r}-\vek{r}') + \Gamma_{\mathrm{F}}(\vek{r}, \vek{r}') 
\label{HF}
\end{equation}
\noindent
and
\begin{equation}
\Delta(\vek{r}, \vek{r}') = -V_{\mathrm{eff}}(\vek{r}-\vek{r}',n(\tfrac{\vek{r}+\vek{r}'}{2})) \sum_\nu u_\nu(\vek{r})\varv_\nu(\vek{r}')^*[1-2f(E_\nu)] \, .
\label{nonlocal-gap-eq}
\end{equation}
In the above expressions, $\Gamma_{\mathrm{H}}(\vek{r})= \int d\vek{r}' V_{\mathrm{eff}}(\vek{r},\vek{r}') \, n(\vek{r}')$ and 
$\Gamma_{\mathrm{F}}(\vek{r}, \vek{r}') = - V_{\mathrm{eff}}(\vek{r},\vek{r}') \, n(\vek{r}, \vek{r}')$ are the Hartree and Fock fields, respectively, with the density matrix given
by $n(\vek{r}, \vek{r}')= \sum_\nu \varv_\nu(\vek{r})\varv_\nu^*(\vek{r}')$ (such that $n(\vek{r}) = n(\vek{r}, \vek{r})$ is the number density).  
In nuclear physics, the phenomenological effective forces are generally density dependent, also in the pairing channel, as indicated in the argument of $V_{\mathrm{eff}}$ in Eq.(\ref{nonlocal-gap-eq}).

The non-local HFB equations (\ref{nonlocal-HFB}), that hold for finite nuclei, are numerically much harder to solve than the local BdG equations (\ref{BdG-equations}).  
Yet, they are almost routinely solved even for deformed and rotating nuclei, in general the results of the calculations being in excellent agreement with experiments 
(see Refs.~\cite{Bender-2003,Vretenar-2005} for reviews).
Owing to the high numerical cost to solve these equations, however, approximate local approximations have been developed for the mean-field terms of Eq.(\ref{HF}) 
(such as the Skyrme Energy Density Functionals (EDF) \cite{Erler-2011}) and for the pairing interaction of Eq.(\ref{nonlocal-gap-eq}).

The need to regularize the gap equation arises also for the nuclear problem. 
In this context, it is worth mentioning that the renormalization scheme of Eq.~(\ref{regularized-gap-equation1}) discussed above was originally developed in
Refs.~\cite{Bulgac-2002,Yu-2003} for the nuclear problem, in the case when the gap equation reduces to a local form. 
In this case, Eqs.~(\ref{geff1}) and (\ref{value-R(kc)}) are replaced by an effective coupling constant which takes the form:
\begin{equation}
\frac{1}{g_{\mathrm{eff}}(\vek{r})} = \frac{1}{g(\vek{r})} - \frac{m^*(\vek{r})}{2\pi^2}\bigg [k_c(\vek{r}) - \frac{k_{\mu}(\vek{r})}{2} \ln \frac{k_c(\vek{r}) + k_{\mu}(\vek{r})} {k_c(\vek{r}) - k_{\mu}(\vek{r})}\bigg ] \, .
\label{geff}
\end{equation}
Here, in contrast to Eqs.~(\ref{geff1}) and (\ref{value-R(kc)}), the effective mass $m^*(\vek{r})$ and the bare coupling constant $g(\vek{r})$ are now position dependent through the value of the local density, which can be fitted such that $\Delta_0$ as a function of $k_F$ reproduces the gap obtained with the Gogny force \cite{Garrido-2001} in infinite matter. 
In addition, the wave vectors $k_{\mu}(\vek{r})$ and $k_c(\vek{r})$ are now related to the chemical potential $\mu$ and energy cutoff $E_c$ by 
$\mu = \frac{k_{\mu}^2(\vek{r})}{2m^*(\vek{r})} + U(\vek{r})$ and $E_c = \frac{k_c^2(\vek{r})}{2m^*(\vek{r})} + U(\vek{r})$, where $U(\vek{r})$ is the self-consistent mean-field potential 
(given, for instance, by the Skyrme EDF) which takes the place of the trapping potential $V_{\mathrm{ext}}(\vek{r})$ of Eqs.~(\ref{geff1}) and (\ref{value-R(kc)})

Finally, it is worth mentioning that the information contained in the
BdG and HFB equations [Eqs.~(\ref{BdG-equations}) and
  (\ref{nonlocal-HFB}), respectively] can be cast in an integral form
in terms of the (single-particle) Gor'kov propagators, from which one can better
appreciate also the particle-hole mixing characteristic of the pairing
theory. These propagators read:
\begin{equation}
\hat{\mathcal{G}}(\vek{r},\vek{r}';\omega_{n}) =  \sum_{\nu}  
\left[ \!\! \begin{array}{c}
u_{\nu}(\vek{r}) \\
\varv_{\nu}(\vek{r}) \end{array}
\!\! \right]  \frac{1}{i \omega_{n} \, - \, E_{\nu}}
\left[ u_{\nu}^{*}(\vek{r}'), \varv_{\nu}^{*}(\vek{r}') \right]    
\, + \, \sum_{\nu}
\left[ \!\! \begin{array}{c} 
- \varv_{\nu}^{*}(\vek{r}) \\
u_{\nu}^{*}(\vek{r}) \end{array}
\!\! \right] \frac{1}{i \omega_{n} \, + \, E_{\nu}}
\left[ - \varv_{\nu}(\vek{r}') , u_{\nu}(\vek{r}') \right] 
\label{Green's-functions-BdG}
\end{equation}
\noindent
where $\omega_{n}=(2 n + 1) \pi T$ ($n$ integer) is a fermionic
Matsubara frequency and the sums are limited to $E_{\nu} > 0$.  In the
matrix (\ref{Green's-functions-BdG}), the diagonal (off-diagonal)
elements correspond to the normal (anomalous) Gor'kov propagator
\cite{Fetter-1971}.  Upon taking the analytic continuation $i
\omega_{n} \rightarrow \omega + i \eta$ ($\eta = 0^{+}$) to the real
frequency $\omega$, the two terms on the right-hand side of the
expression (\ref{Green's-functions-BdG}) give rise to the particle and
hole components of the spectral function, which are peaked about the
poles at $\omega = +E_{\nu}$ and $\omega = -E_{\nu}$, respectively,
and have their weights expressed in terms of $u_{\nu}(\vek{r})$ and
$\varv_{\nu}(\vek{r})$.  We shall return to this issue in Section
\ref{sec:pseudo-gap}.

\subsection{Ginzburg-Landau and Gross-Pitaevskii equations}
\label{sec:GLandGP}

There are cases when the BdG equations (\ref{BdG-equations}) can be
replaced by suitable non-linear differential equations for the gap
parameter $\Delta(\vek{r})$, which are somewhat easier to solve
numerically and conceptually more appealing than the BdG equations
themselves.  These non-linear differential equations for
$\Delta(\vek{r})$ are the Ginzburg-Landau (GL) equation for the
Cooper-pair wave function and the Gross-Pitaevskii (GP) equation for
the condensate wave function of composite bosons.  As a matter of
fact, it turns out that the GL and GP equations can be microscopically
derived from the BdG equations in two characteristic limits; namely,
the GL equation in the weak-coupling (BCS) limit close to $T_c$
\cite{Gorkov-1959} and the GP equation in the strong-coupling (BEC)
limit at $T=0$ \cite{Pieri-2003}.  In both cases, the integral form
(\ref{Green's-functions-BdG}) of the (single-particle) Gor'kov
propagators associated with the solutions of the BdG equations
provides the starting point for the microscopic derivation of these
non-linear differential equation for $\Delta(\vek{r})$.

In particular, one obtains the GL equation for strongly overlapping Cooper pairs through an expansion of the expression (\ref{Green's-functions-BdG}) in terms of the small parameter
$|\Delta(\vek{r})|/T_c$ \cite{Gorkov-1959}, in the form:
\begin{equation}
\left\{  \frac{(i\vek{\nabla} + 2\vek{A}(\vek{r}))^2}{4 \, m} + \frac{6 \pi^2 T_c^2 }{7 \zeta(3) E_F} \left[ \left(1- \frac{\pi}{4k_Fa_F}\right) \frac{V_{\mathrm{ext}}(\vek{r})}{E_F} 
-  \left(1 - \frac{T}{T_c} \right) \right] \right\} \Delta(\vek{r}) + \frac{3}{4 E_F} \, |\Delta(\vek{r})|^2 \Delta(\vek{r}) = 0
\label{GL-equation} 
\end{equation}
\noindent
where $\zeta(3)\approx 1.202$ is the Riemann $\zeta$ function of
argument $3$.  [For the meaning of the vector potential {\bf A}, see below.] 
In the presence of the trapping (harmonic) potential
$V_{\mathrm{ext}}(\vek{r})$, the values of the critical temperature
$T_c$ and of the Fermi wave vector $k_F$ correspond to those of a
uniform system with density equal to that at the center of the trapped
system.  This extra term did not appear in the original Gor'kov
derivation of the GL equation \cite{Gorkov-1959} and was first
considered in Ref.~\cite{Baranov-1998}.  A simple way to derive it is
by following the method of Ref.~\cite{Simonucci-2014} and adopting a
local-density approximation in the expression of the regularized
particle-particle bubble given by Eq.(\ref{approximate-pp-bubble})
below.  Upon replacing in this expression $E_F \rightarrow (E_F -
V_{\mathrm{ext}}(\vek{r}))$ and expanding to linear order in the small
quantity $V_{\mathrm{ext}}(\vek{r}) / E_F$, one readily recovers
both terms within parentheses that multiply $V_{\mathrm{ext}}(\vek{r})
/ E_F$ in Eq.(\ref{GL-equation})

In addition, from the BdG equations one also obtains the GP equation for a gas of dilute composite bosons of mass $2m$ through an expansion of the expression (\ref{Green's-functions-BdG}) 
in terms of the small parameter $|\Delta(\vek{r})/\mu|$ \cite{Pieri-2003}, in the form:
\begin{equation}
\frac{(i\vek{\nabla} + 2\vek{A}(\vek{r}))^2}{4 \, m} \, \Phi(\vek{r}) + 2 V_{\mathrm{ext}}(\vek{r}) \, \Phi(\vek{r})  + \frac{8 \pi a_F}{2 \, m} \, |\Phi(\vek{r})|^2 \Phi(\vek{r}) = \mu_B \, \Phi(\vek{r}) \, .
\label{GP-equation} 
\end{equation}
\noindent
Here, $\Phi(\vek{r}) = \sqrt{\frac{m^2 a_F}{8 \pi}} \,
\Delta(\vek{r})$ is the condensate wave function expressed in terms of
the gap parameter $\Delta(\vek{r})$ and $\mu_B$ is the residual
chemical potential of the composite bosons, defined by $2 \mu = -
\varepsilon_0 + \mu_B$ with $\mu_B \ll \varepsilon_0 = (m
a_F^2)^{-1} $ in the relevant BEC limit ($a_F > 0$).  Note in this
context that the GP equation (\ref{GP-equation}) could also be
obtained from the regularized gap equation
(\ref{regularized-gap-equation}) discussed in Section
\ref{sec:bdg-hfb}, in the BEC limit when $|\mu|$ is the largest energy
scale in the problem such that $E_{c}$ (and thus $k_{c}$) can be taken
to vanish.  In this limit, the source term on the right-hand side of
Eq.(\ref{regularized-gap-equation}) can be neglected while the
coefficients on the left-hand side become 
$\mathcal{R} = \frac{m}{4 \pi a_F} - \frac{m^2 a_F}{8\pi} \, \mu_B$, $\mathcal{I}_{02} = \frac{m^2 a_F}{2\pi}$, $\mathcal{I}_{13} = \frac{3 m^2 a_F}{8\pi}$, and 
$\mathcal{I}_{03} = \frac{m^3 a_F^3}{4\pi}$, such that Eq.(\ref{GP-equation}) with $\vek{A}(\vek{r}) = 0$ is recovered from Eq.~(\ref{regularized-gap-equation}).

Since we are here concerned with superfluidity of neutral particles, in Eqs.~(\ref{GL-equation}) and (\ref{GP-equation}) the vector potential $\vek{A}(\vek{r})$ is meant to describe 
rotating systems and is accordingly given by $\vek{A}(\vek{r}) = m \, \vek{\Omega} \times \vek{r}$ where $\vek{\Omega}$ is the angular velocity of the rotating trap.  
[In this case, the quadratic term in $\vek{A}(\vek{r})$ should be omitted from the kinetic energy.]
At the same time, the factor of $2$ in front of the vector potential $\vek{A}(\vek{r})$ in Eqs.~(\ref{GL-equation}) and (\ref{GP-equation}) (as well as of the external (trapping) potential
$V_{\mathrm{ext}}(\vek{r})$ in Eq.~(\ref{GP-equation})) reflects the composite nature of Cooper pairs and bosons in the two (GL and GP) situations.  
In practice, the independent solutions of the GL and GP equations (\ref{GL-equation}) and (\ref{GP-equation}) represent useful benchmarks for the numerical results of the BdG equations for the
local gap parameter $\Delta(\vek{r})$, in the respective (BCS and BEC) limits of the BCS-BEC crossover.

\subsection{Spin-imbalanced (polarized) systems} 
\label{sec:polarized}

The effect of spin-population imbalance (``polarization'') on superconductivity was first addressed shortly after the formulation of BCS theory
\cite{Clogston-1962,Chandrasekhar-1962,Sarma-1963,Fulde-1964,Larkin-1964}.
In these works, the spin-population imbalance was assumed to be produced by an external magnetic field acting only on the spins of the electrons, while the orbital effects were completely neglected.  
This assumption requires quite stringent conditions on superconductors, which, in practice, have made the experimental study of this issue in condensed-matter systems an extremely difficult task 
(see Ref.~\cite{Matsuda-2007} for a review). 
In nuclear matter, the problem of pairing in population-imbalanced Fermi gases naturally arises because most of the heavier nuclei have more neutrons ($n$) than protons ($p$), and the same is true for nuclear matter in proto-neutron stars. 
In nuclear physics, one quite generally refers in this context to \emph{asymmetric nuclear matter}. In the context of deuteron-like Cooper pairs, several works deal with this problem also with respect to 
the BCS-BEC crossover \cite{Alm-1993,Sedrakian-1997,Lombardo-2001b}  (cf. Section~\ref{sec:deuteron-asy}).
This problem has also been studied in the context of dense quark matter \cite{Casalbuoni-2004,Alford-2008}.

Recently, great interest in population-imbalanced Fermi gases has been
prompted by experiments with two-component ultra-cold Fermi gases, for
which the populations of different ``spin'' states (that correspond to
different atomic hyperfine states) can be controlled independently of
orbital effects.  In addition, and similarly to the situation in
nuclear matter, these populations are separately conserved, owing to
the long relaxation time associated with the processes that change the
hyperfine state of an atom.  The first two experiments with imbalanced
ultra-cold Fermi gases \cite{Zwierlein-2006a,Partridge-2006} have
stimulated a large theoretical activity on these systems over the last
several years.  Detailed reviews of the recent theoretical and
experimental activity on population-imbalanced ultra-cold Fermi gases
can be found in Refs.~\cite{Radzihovsky-2010} and \cite{Chevy-2010}.
Below, we will first describe the historical approaches to the problem
of an imbalanced Fermi gas as related to conventional (weak-coupling)
superconductors, and then we will deal with the more recent connection
of this problem to the BCS-BEC crossover for arbitrary couplings.

\begin{figure}[t]
\begin{center}
\includegraphics[width=14cm,angle=0]{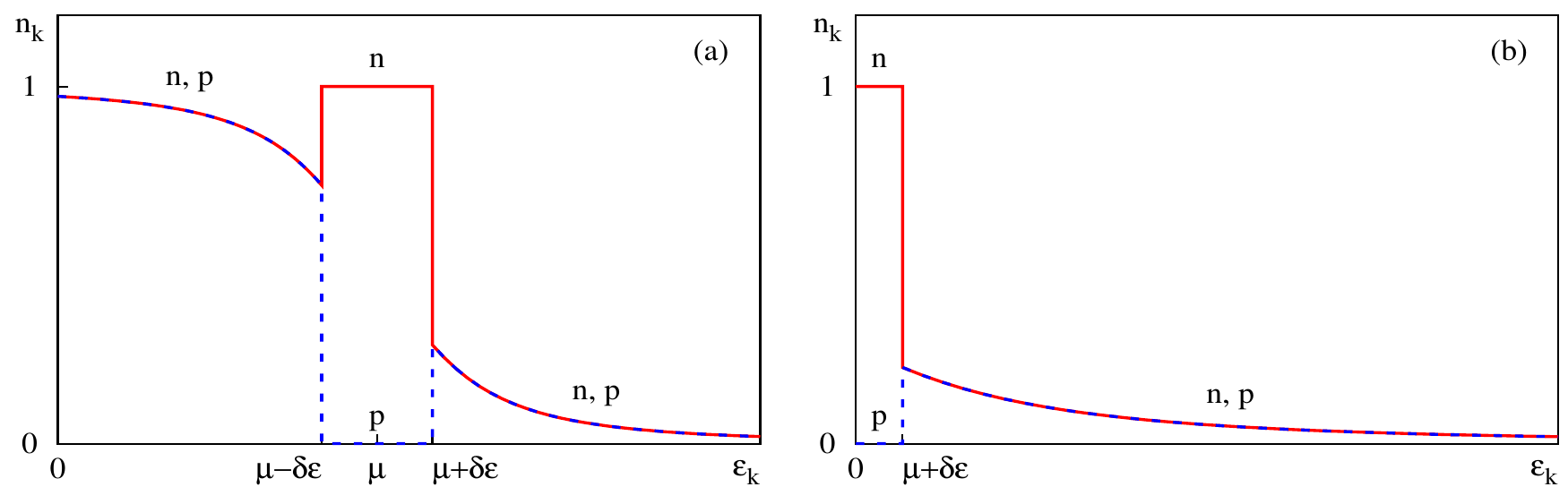}
\end{center}
\caption{Schematic representation of the occupation number of neutrons $n = \uparrow$ (full line) and protons $p = \downarrow$ (dashed line) as a function of the energy $\varepsilon _k$ 
in asymmetric ($n_n > n_p$) nuclear matter with proton-neutron pairing, for (a) $\mu >\delta\varepsilon$ and (b) $\mu < \delta\varepsilon$. 
Here, $\mu=(\mu_{\uparrow}+\mu_{\downarrow})/2$ and $\delta\varepsilon = \sqrt{h^{2}-\Delta^{2}}$, with $h=(\mu_{\uparrow}-\mu_{\downarrow})/2$.
}
\label{Figure-4}
\end{figure}
The BCS mean-field approximation, that was discussed in the previous
Sections for a balanced system, can be readily generalized to the
presence of different spin populations (and thus of different chemical
potentials).  For a contact interaction, the gap equation
(\ref{gap-homogeneous}) is modified as follows
\begin{equation}
-\frac{m}{4\pi a_F} = \int \!\frac{d\vek{k}}{(2 \pi)^3} \left( \frac{ 1 - f(E^{+}_{\vek{k}}) - f(E^{-}_{\vek{k}})}{2 E_{\vek{k}}} - \frac{m}{\vek{k}^2}  \right) \, ,
\label{Sarma-gap}
\end{equation}
while the number equation (\ref{density-homogeneous}) splits into an
equation for the total density
\begin{equation}
n_{\uparrow} + n_{\downarrow} =  \int \!\frac{d\vek{k}}{(2 \pi)^3}
\left( 1 - \frac{\xi_{\vek{k}}}{E_{\vek{k}}} \left[ 1 -  f(E^{+}_{\vek{k}}) - f(E^{-}_{\vek{k}}) \right] \right)
\label{total-density-Sarma}
\end{equation}
and an equation for the difference between the two densities:
\begin{equation}
n_{\uparrow} - n_{\downarrow} = \int \!\frac{d\vek{k}}{(2 \pi)^3} \left[ f(E^{+}_{\vek{k}})  - f(E^{-}_{\vek{k}}) \right] \, .
\label{difference-density-Sarma}
\end{equation}
Here, $E_{\vek{k}} = \sqrt{\xi_{\vek{k}}^2 + |\Delta|^2}$ with
$\xi_{\vek{k}} = (\xi_{\vek{k}\uparrow} + \xi_{\vek{k}\downarrow})/2$,
and $E^{\pm}_{\vek{k}} = E_{\vek{k}} \pm \delta \xi_{\vek{k}}$ with
$\delta \xi_{\vek{k}} = (\xi_{\vek{k}\uparrow} -
\xi_{\vek{k}\downarrow})/2$, where
$\xi_{\vek{k}\sigma}=k^2/(2m)-\mu_\sigma$ is the single particle
energy relative to the chemical potential $\mu_\sigma$ for the spin
$\sigma = (\uparrow,\downarrow)$.  One sees from
Eq.~(\ref{difference-density-Sarma}) that at $T=0$ the difference
$(n_{\uparrow} - n_{\downarrow})$ between the two densities is finite,
provided there is a region of $\vek{k}$-space where one of the two
single-particle excitation energies $E^{\pm}_{\vek{k}}$ becomes
negative.  In terms of the field $h=(\mu_{\uparrow}-\mu_{\downarrow})/2$ such
that $E^{\pm}_{\vek{k}}=E_{\vek{k}} \mp h$, one has that
$n_{\uparrow}> n_{\downarrow}$ provided $h > \min E_{\vek{k}}$ (for
definiteness, in the following we shall assume $h > 0$ such that
spin-$\uparrow$ fermions correspond to the majority species).  In
addition, it can be verified that at $T=0$ the above equations
(\ref{Sarma-gap})-(\ref{difference-density-Sarma}) correspond to the
following ground-state wave function \cite{Lombardo-2001b}:
\begin{equation}
\vert \Phi_{\mathrm{S}} \rangle=\prod_{\vek{k} \in \mathcal{R}} c^\dagger_{\vek{k}\uparrow} 
\prod_{\vek{k} \notin \mathcal{R}} (u_{\vek{k}}+\varv_{\vek{k}} c^\dagger_{\vek{k}\uparrow} c^\dagger_{-\vek{k}\downarrow}) \, \vert 0 \rangle
\label{Sarma-wave-function}
\end{equation}
where $u_{\vek{k}}$ and $\varv_{\vek{k}}$ are defined in terms of
$\xi_{\vek{k}}$ and $E_{\vek{k}}$ as in Eq.(\ref{v-square}) for the
balanced case. In the expression (\ref{Sarma-wave-function}), the
region $\mathcal{R}$ of $\vek{k}$-space with $E^{+}_{\vek{k}}<0$ is
fully occupied by the $\uparrow$-fermions of the majority species,
while outside this region Cooper pairing occurs. 
The fact that the states in the region  $\mathcal{R}$ are no longer available for pairing is called ``blocking effect" in nuclear physics.
The state $\vert \Phi_{\mathrm{S}}\rangle$ (referred to either as the ``Sarma'' state from the original work by Sarma \cite{Sarma-1963} or as the ``breached-pair'' phase from the more 
recent work of Ref.~\cite{Liu-2003,Gubankova-2003,Forbes-2005}) by Wilczek and co-workers accommodates at the same time superfluidity and a finite population imbalance, through a sort 
of phase separation in $\vek{k}$-space between one region which is unpolarized and superfluid and a second region which is fully polarized and normal.   
Depending on the value of $\mu=(\mu_{\uparrow}+\mu_{\downarrow})/2$, the normal region $\mathcal{R}$ is enclosed by two (when $\mu > \delta\varepsilon$), or one 
(when $\mu < \delta\varepsilon$) Fermi surfaces with gapless single-particle excitations, where $\delta\varepsilon = \sqrt{h^{2}-\Delta^{2}}$.  
The state (\ref{Sarma-wave-function}) thus corresponds to a gapless superfluid.
As an example of the breached-pair phase, Fig.~\ref{Figure-4} shows a schematic picture of the occupation number for protons and neutrons in asymmetric nuclear matter (which can be transferred to the case of a spin-imbalanced system by replacing the labels $n$ and $p$ with $\uparrow$ and $\downarrow$, respectively). 
The two panels of Fig.~\ref{Figure-4} distinguish the two cases with (a) $\mu > \delta\varepsilon$ and (b) $\mu < \delta\varepsilon$. 
In both cases, one sees that there is a window of energies with unpaired occupancies for the extra neutrons.
This problem will be further elaborated in Section~\ref{sec:deuteron-asy}.

\begin{figure}[t]
\begin{center}
\includegraphics[width=16cm]{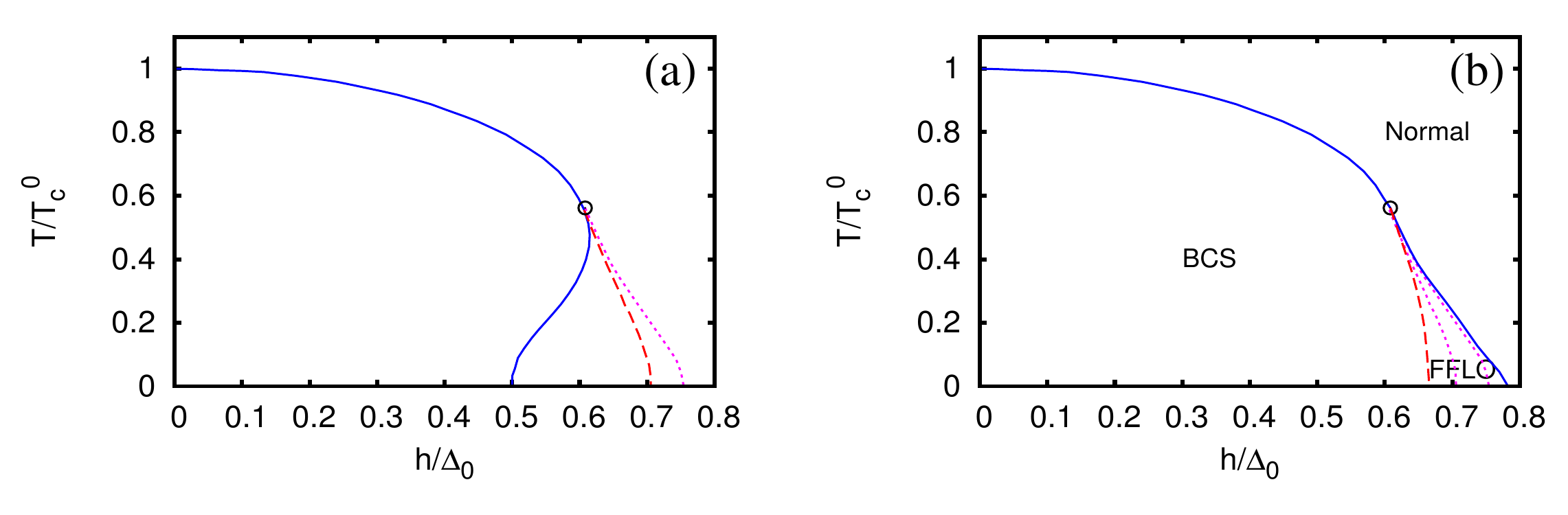}
\end{center}
\vspace{-0.9cm}
\caption{(a) Phase diagram for the superfluid/normal transition in the (extreme) weak-coupling limit. 
The temperature is in units of the  critical temperature $T_c^0$ for $h=0$, and $h=(\mu_{\uparrow}-\mu_{\downarrow})/2$ is in units of the zero-temperature gap $\Delta_0$ at $h=0$.  
Full line: solution of the equation for the critical temperature obtained by setting $\Delta=0$ in the BCS gap equation, which corresponds to a true  transition line only above the tricritical point (circle).  Dashed line: first-order transition line that separates the BCS superfluid from the normal phase.  
Dotted line: second-order transition line between the (single $\vek{q}$) FF and normal phase.  
(b) Refined phase diagram for the superfluid/normal transition in weak coupling, where more general solutions in the FFLO phase are considered.  
Full line: transition line separating the superfluid from the normal phase, which becomes of first order below the tricritical point (circle) and separates the FFLO from the normal phase.  
Dashed line: second-order transition line that separates the BCS and FFLO phases.
The transition lines BCS $\to$ FF and FF $\to$ normal obtained by the simple FF solution are also reported for comparison (dotted lines).}
\label{Figure-5}
\end{figure}

A problem with the Sarma state (\ref{Sarma-wave-function}) in weak coupling is that it
is energetically unstable, since it corresponds to a local maximum of
the variational energy at $T=0$ \cite{Sarma-1963}. Specifically, by
solving the gap equation (\ref{Sarma-gap}) in the weak-coupling limit
one finds that $\Delta(h)$ shows a re-entrant behaviour below a certain
temperature $T_0$, while at high enough temperature $\Delta(h)$
decreases monotonically with $h$ and vanishes continuously at a
critical value of $h$.  The Sarma state is then associated with this
re-entrant branch at $T=0$.  An analogous re-entrant behaviour is found
for the dependence of the critical temperature on $h$, which is
obtained by letting $\Delta\to 0$ in the gap equation
(\ref{Sarma-gap}) (as shown by the full line in Fig.~\ref{Figure-5}(a)).
Comparison between the (grand-canonical) free energies of the normal
and superfluid phases shows, however, that below the temperature $T_0
\, (=0.56 T_c^0$ in weak coupling) the transition becomes of first
order, with the gap dropping discontinuously to zero at the
transition.  Correspondingly, the transition curve eliminates
completely the re-entrant behaviour below $T_0$ (as shown by the dashed
curve in Fig.~\ref{Figure-5}(a)).  An analysis along these lines made
by Sarma at $T=0$ \cite{Sarma-1963} recovered the Clogston
\cite{Clogston-1962} and Chandrasekhar \cite{Chandrasekhar-1962}
prediction of a first-order transition at $h=\Delta_0/\sqrt{2}$, from
an unpolarized BCS state (when $h<\Delta/\sqrt{2}$) to a polarized
normal state (when $h>\Delta_0/\sqrt{2}$).
 
An alternative solution to the problem of superconductivity
(superfluidity) in the presence of spin imbalance was proposed by
Fulde and Ferrell (FF) \cite{Fulde-1964} and independently by Larkin
and Ovchinnikov (LO) \cite{Larkin-1964}.  The basic idea is that a
mismatch of the spin-up and spin-down Fermi surfaces due to spin
imbalance should favor Cooper pairing for a finite value of the
center-of-mass wave vector $\vek{Q}$.  This is because, by taking
$|\vek{Q}|$ of the order of ($k_{F\uparrow}-k_{F\downarrow}$), the
pairing states $|\vek{k}+\vek{Q}/2,\uparrow\rangle$ and
$|-\vek{k}+\vek{Q}/2,\downarrow\rangle$ remain in the vicinity of both
Fermi surfaces.  Specifically, this matching occurs only on one side
of the respective Fermi surfaces as determined by $\vek{Q}$ itself,
while on the other side pairing is completely suppressed.  A
non-vanishing value of the polarization is then obtained by making
these regions completely empty/filled with $\downarrow$/$\uparrow$
fermions, respectively.  In addition, pair condensation at finite
$\vek{Q}$ leads to a space-dependent order parameter
$\Psi(\vek{r})=\langle\psi_\uparrow(\vek{r})\psi_\downarrow(\vek{r})\rangle$
(where $\psi_{\sigma}(\vek{r})$ is a fermion field operator with spin
$\sigma$), which is given by a single plane wave $\Psi(\vek{r})=\psi_0
e^{i \vek{Q} \cdot \vek{r}}$ in the Fulde-Ferrell analysis and by a
superposition of plane waves with the same $|\vek{Q}|$ in the
Larkin-Ovchinnikov analysis (in particular, the solution
$\Psi(\vek{r})=\psi_0 \cos (\vek{Q} \cdot \vek{r})$ was considered).
  
The above mean-field equations
(\ref{Sarma-gap})$-$(\ref{difference-density-Sarma}) are readily
extended to take into account the FF pairing.  By setting
$\Delta_{\vek{Q}} e^{i\vek{Q}\cdot \vek{r}} \equiv -
\int\!\frac{d\vek{k}}{(2\pi)^3} \varv_0 \langle
c_{-\vek{k}+\vek{Q}/2\downarrow} c_{\vek{k}+\vek{Q}/2
  \uparrow}\rangle$, the gap equation is modified as follows:
\begin{equation}
-\frac{m}{4\pi a_F} = \int \!\frac{d\vek{k}}{(2 \pi)^3} \left( \frac{1 - f(E^{+}_{\vek{k},\vek{Q}}) - f(E^{-}_{\vek{k},\vek{Q}})}{2 E_{\vek{k},\vek{Q}}} - \frac{m}{\vek{k}^2} \right) 
\label{Delta-eqn-FFLO}
\end{equation}
where $E_{\vek{k},\vek{Q}} = \sqrt{\xi_{\vek{k},\vek{Q}}^2 +
  |\Delta_{\vek{Q}}|^2}$ with $\xi_{\vek{k},\vek{Q}} =
(\xi_{\vek{k}+\vek{Q}/2\uparrow} + \xi_{-\vek{k}+\vek{Q}/2\downarrow})/2$, and 
$E^{\pm}_{\vek{k},\vek{Q}} = E_{\vek{k},\vek{Q}} \pm \delta \xi_{\vek{k},\vek{Q}}$ with
$\delta\xi_{\vek{k},\vek{Q}} = (\xi_{\vek{k}+\vek{Q}/2\uparrow} -
\xi_{-\vek{k}+\vek{Q}/2\downarrow})/2$.  Similarly to the gap equation
(\ref{Delta-eqn-FFLO}), the number equations for the FF phase are
obtained from the corresponding equations (\ref{total-density-Sarma})
and (\ref{difference-density-Sarma}) of the Sarma phase via the
replacements $(\xi_{\vek{k}}, E_{\vek{k}},E^{\pm}_{\vek{k}}) \to
(\xi_{\vek{k},\vek{Q}}, E_{\vek{k},\vek{Q}},
E^{\pm}_{\vek{k},\vek{Q}})$.  And, analogously to the Sarma phase
[cf. Eq.(\ref{difference-density-Sarma})], a finite population
imbalanced is obtained at $T=0$ when (at least) one of the two
single-particle excitations $E^{\pm}_{\vek{k},\vek{Q}}$ becomes
negative in a region of $\vek{k}$-space.  Calling
$\mathcal{R_\uparrow}$ and $\mathcal{R_\downarrow}$ the regions where
$E^{+}_{\vek{k},\vek{q}}$ and $E^{-}_{\vek{k},\vek{q}}$ are negative,
the FF solution at $T=0$ then corresponds to the variational wave
function (see, e.g., Ref.~\cite{Takada-1969})
\begin{equation}
\vert \Psi_{\mathrm{FF}} \rangle 
  = \prod_{\vek{k} \in \mathcal{R_\uparrow}} \! c^\dagger_{\vek{k}+\frac{\vek{Q}}{2}\uparrow} 
    \prod_{\vek{k} \in \mathcal{R_\downarrow}} \! c^\dagger_{-\vek{k}+\frac{\vek{Q}}{2}\downarrow}
    \prod_{\vek{k} \notin \mathcal{R_{\uparrow,\downarrow}}} \!
      (u_{\vek{k}}+\varv_{\vek{k}} c^\dagger_{\vek{k}+\frac{\vek{Q}}{2}\uparrow} 
        c^\dagger_{-\vek{k}+\frac{\vek{Q}}{2}\downarrow})\vert 0 \rangle \, ,
\label{FFLO-wavefunction}
\end{equation}
where in the region $\mathcal{R_\uparrow}$ ($\mathcal{R_\downarrow}$)
the state $|\vek{k} + \vek{Q}/2, \uparrow\rangle$ is fully occupied
(empty) and the state $|-\vek{k}+\frac{\vek{Q}}{2},\downarrow\rangle$
is fully empty (occupied), with the remaining states being available
for pairing.  A finite population imbalance results when the regions
$\mathcal{R_\uparrow}$ and $\mathcal{R_\downarrow}$ have different
volumes.  Note, however, that no finite value of the current is
associated with a finite value of $|\vek{Q}|$.

The thermodynamic stability of the FF phase, together with the value
of $|\vek{Q}|$, are obtained by minimizing the mean-field free-energy.
In weak coupling, one obtains that at $T=0$ the Clogston-Chandrasekhar
first-order transition from the BCS state to the normal phase at
$h=\Delta_0/\sqrt{2}$ is replaced by a first-order transition BCS
$\to$ FF at essentially the same value of $h$, followed by a
second-order transition FF $\to$ normal when $h$ reaches the value
$h=0.754 \Delta_0$.  Correspondingly, $|\vek{Q}|$ changes from $1.28
\, (k_{F\uparrow}-k_{F\downarrow})$ at the transition BCS $\to$ FF, to
$1.2 (k_{F\uparrow}-k_{F\downarrow})$ at the transition FF $\to$
normal \cite{Takada-1969}.  The transition FF $\to$ normal remains of
second order also at finite temperature, while the transition BCS
$\to$ FF is of first order and corresponds, in practice, to the
transition BCS $\to$ normal, since the free energies of the FF and
normal phases are quite close in value.  The resulting phase diagram
is reported in Fig.~\ref{Figure-5}(a), where the three transition
lines meet at the tricritical point with $h=0.61 \Delta_0$ and $T=0.56
T_c^0$.  It turns out that the region of existence of the FF phase (in
the figure delimited by the dashed and dotted lines) is quite narrow.

The assumption of a single wave vector $\vek{Q}$ was overcome by the
LO approach \cite{Larkin-1964}, where a more general superposition of
plane waves with the same $|\vek{Q}|$ was considered, corresponding in
real space to a crystalline order for the order parameter
$\Psi(\vek{r})$ (or of the gap parameter $\Delta(\vek{r})$).  The LO
analysis was based on an expansion in powers of $\Delta(\vek{r})$ of
the Gor'kov equations for the normal and anomalous Green's functions
in the presence of a Zeeman splitting between the two spin species,
and thus holds only near the second-order transition where
$\Delta(\vek{r})$ vanishes.  It was found that near the transition
point the solution $\Delta(\vek{r})= 2 \Delta \cos(\vek{Q}\cdot
\vek{r})$ has lower energy than the single plane-wave FF solution, and
has also the lowest energy among the crystalline solutions.  It was
further found that the location of the transition to the normal state
is unchanged with respect to the FF analysis, since the degeneracy
among the possible superpositions of plane waves with the same
$|\vek{Q}|$ is lifted only inside the superfluid phase.

When dealing with a first-order transition, however, the power
expansion of the Gor'kov equations utilized by the LO approach is not
appropriate and one should resort to a full solution of the Gor'kov
equations (or, equivalently, of the BdG equations) for the polarized
Fermi gas.  Since this is rather a difficult task, in weak coupling
the problem was approached using the Eilenberger quasi-classical
equations which are appropriate to this limit
\cite{Eilenberger-1968,Larkin-1968}.  These equations have been
applied to a three-dimensional polarized Fermi gas in
Refs.~\cite{Matsuo-1998} and \cite{Mora-2005}, where a generic
combinations of $\vek{Q}$ with the same value of $|\vek{Q}|$ was
considered.  In particular, the analysis of Ref.~\cite{Mora-2005}
shows that the transition to the normal phase is of first order from
the tricritical point down to $T=0$, although it remains very close to
the standard FFLO second-order transition.  At $T=0$, the critical
field for the transition to the normal phase has now the value
$h=0.781 \Delta_0$, to be compared with the FFLO value $h=0.754
\Delta_0$ for which a second-order transition was assumed.  Similar
results were obtained by an alternative approach (valid, in principle,
only close to the tricritical point) based on a GL expansion up to
sixth order in $\Delta$ and fourth order in $|\vek{Q}|$
\cite{Buzdin-1997,Houzet-1999,Combescot-2002}.  An expansion up to
sixth order of the GL energy functional was also considered in
Ref.~\cite{Bowers-2002} to study the FFLO phase in the context of
color superconductivity in QCD (for a review, see
Refs.~\cite{Casalbuoni-2004} and \cite{Alford-2008}).  Owing to all
these results, in weak coupling the description of the FFLO phase near
the transition to the normal phase can be considered to be quite
accurately described.

A similar degree of accuracy is lacking for the description of the
FFLO phase away from the critical line.  The reason is that, as one
gets inside the superfluid phase, a single value of $|\vek{Q}|$ no
longer suffices: $\Delta(\vek{r})$ remains a periodic function in
space, but Fourier harmonics of higher order are required.  Insights
in this problem are provided by the analytic solution of the BdG
equations in one dimension
\cite{Machida-1984,Buzdin-1983,Buzdin-1987}.  This yields an
evolution, from a soliton lattice at the BCS-FFLO transition (for
which the gap parameter changes its sign over a region of the order of
the coherence length and then remains uniform over a region that gets
progressively wider upon approaching the BCS-FFLO transition), to the
ordinary LO solution at the FFLO-normal transition.  A similar kind of
evolution is expected to occur in higher dimensions
\cite{Matsuo-1998}.  In particular, in three dimensions the formation
of a one-dimensional soliton lattice, with a period that tends to
infinity at the transition point, makes the BCS-FFLO transition
continuous and shifts it to lower values of $h$ compared to the FF
solution.  For instance, at $T=0$ the BCS-FFLO transition occurs at
$h=0.666 \Delta_0$ \cite{Matsuo-1998}, in contrast with $h=0.707
\Delta_0$ of the FF solution.  The complete phase diagram, which takes
into account the formation of a soliton lattice at the BCS-FFLO
transition and the analysis of Ref.~\cite{Mora-2005} (as elaborated in \cite{Chevy-2010}) at the
FFLO-normal transition, is reported in Fig.~\ref{Figure-5}(b).  One
sees that, although the region for the occurrence of the FFLO phase is
now enlarged with respect to the FF solution, the FFLO phase still
remains confined to a small corner of the phase-diagram.

The discussion about the Sarma and FFLO phases and their stability was
thus far restricted to the weak-coupling limit of the interaction.
The question about what happens when this assumption is relaxed was
considered only recently, stimulated by the experiments with
ultra-cold Fermi gases that we now pass to consider.  The first
theoretical works on ultra-cold polarized Fermi gases focused on the
Sarma solution, initially for weak coupling
\cite{Bedaque-2003,Wu-2003} in line with the work on the breached-pair
phase of Refs.~\cite{Liu-2003,Gubankova-2003} and later on across the
BCS-BEC crossover \cite{Son-2006,Iskin-2006,Gubbels-2006,Chen-2006}.  Also in
this context, the Sarma phase turns out to be unstable against the
FFLO phase or the phase separation between the unpolarized BCS phase
and the polarized normal phase.  Since in ultra-cold gases the spin
populations are separately conserved, here phase separation is a
manifestation of the (Clogston-Chandrasehkhar) first-order transition
discussed above for superconductors as a function of the magnetic
field.  The stability of the Sarma phase was further checked by
looking at the sign of the superfluid density or of the
compressibility matrix, which yield the spinodal curves of local
stability against the FFLO phases and phase-separation
\cite{Wu-2003,Iskin-2006,Pao-2006,Chen-2006}.

\begin{figure}[t]
\begin{center}
\includegraphics[width=16cm]{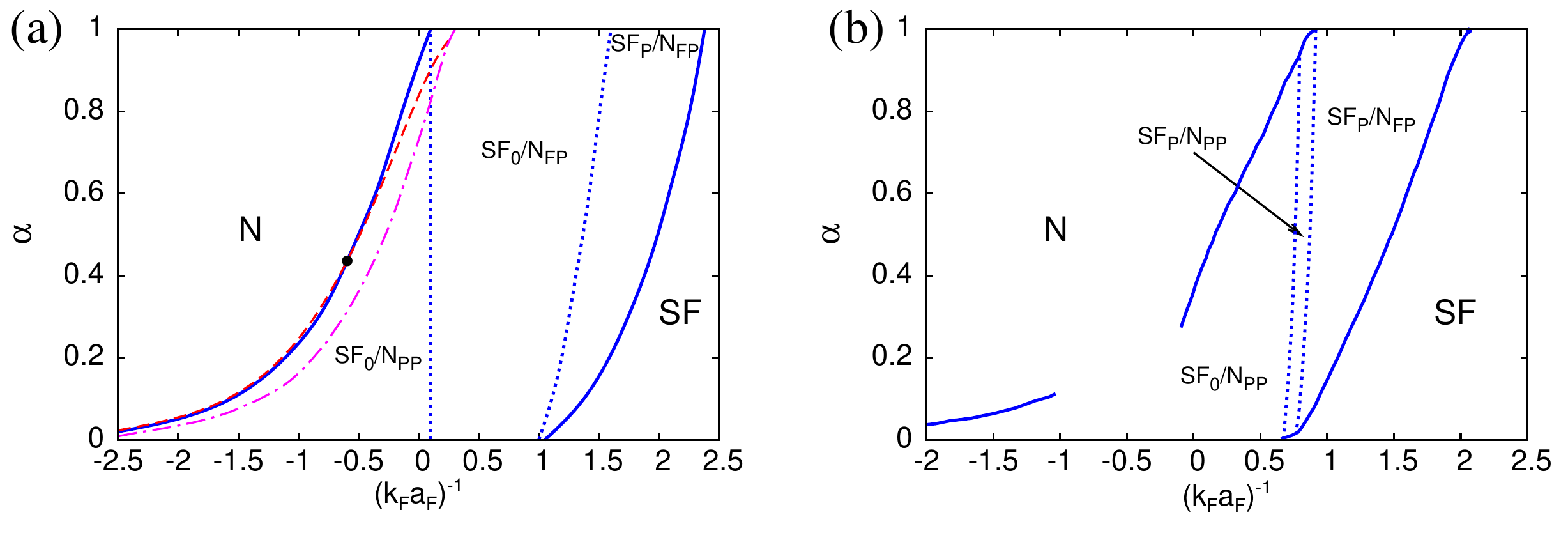}
\end{center}
\vspace{-0.7cm}
\caption{(a) Mean-field $T=0$ phase diagram for the population imbalance $\alpha= (n_{\uparrow}-n_{\downarrow})/(n_{\uparrow}+n_{\downarrow})$ vs the coupling parameter $(k_F a_F)^{-1}$, 
where $k_F$ is determined by the total density $n=n_{\uparrow}+n_{\downarrow}$.  
Full lines: boundaries between the phase separation region and the homogeneous normal (N) and superfluid (SF) phases.
Dotted lines: boundaries within the phase separation region between 
balanced superfluid and partially polarized normal state
(SF$_0$/N$_{\mathrm{PP}}$), balanced superfluid and fully
polarized normal state (SF$_0$/N$_{\mathrm{FP}}$), and polarized
superfluid and fully polarized normal state
(SF$_{\mathrm{P}}$/N$_{\mathrm{FP}}$).
Dashed line: transition line to the FFLO phase (which does not take into account phase separation).  
Dashed-dotted line: normal to superfluid transition line for the Sarma phase (which is reported here regardless of its stability).    
(b) Corresponding phase-diagram obtained in Ref.~\cite{Pilati-2008} by fixed-node Quantum Monte Carlo simulations. }
\label{Figure-6}
\end{figure}

A more complete mean-field analysis of the ground-state phase diagram
for the polarized Fermi gas throughout the BCS-BEC crossover was
carried out in Ref.~\cite{Sheehy-2007}, by taking into account the FF
phase with a single plane wave as well as the phase-separation, while
searching for the global minima of the grand-canonical potential
$\Omega$ at $T=0$.  This has the form \cite{Radzihovsky-2010}:
\begin{equation}
\frac{\Omega}{V} = -\frac{m}{4\pi a_F} \Delta_{\vek{Q}}^2 + \int \!\!\frac{d\vek{k}}{(2 \pi)^3}\left[\xi_{\vek{k},\vek{Q}} - E_{\vek{k},\vek{Q}} + \frac{m\Delta_{\vek{Q}}^2}{k^2} 
+ E^{+}_{\vek{k},\vek{Q}} \Theta(- E^{+}_{\vek{k},\vek{Q}}) + E^{-}_{\vek{k},\vek{Q}} \Theta(- E^{-}_{\vek{k},\vek{Q}})\right]
\end{equation}
\noindent
where $V$ is the volume.
 The resulting phase diagram is reported in Fig.~\ref{Figure-6}(a)
and shows that phase separation dominates, especially in the unitary
regime of interest for ultra-cold gases.  The polarized superfluid (Sarma) phase is stable
only on the BEC side of the crossover, where it corresponds to a
mixture of composite bosons and excess fermions \cite{Pieri-2006}.
The FFLO phase is instead confined to a small region on the BCS side
(in between the dashed and full lines on the left of the black dot in
Fig.~\ref{Figure-6}(a)).  The previous discussion about the FFLO phase
in weak coupling suggests, however, that also in this case the region
of stability of the FFLO phase could be enlarged by considering more
complex solutions than the FF ansatz.  Work in this direction was
reported in Refs.~\cite{Yoshida-2007} and \cite{Bulgac-2008b}.  In
particular, Ref.~\cite{Bulgac-2008b} suggested the occurrence of an
FFLO phase over an extended range of polarizations for the unitary
Fermi gas.  In addition, inside the region of phase-separation
Fig.~\ref{Figure-6}(a) distinguishes (dotted lines) the phases made by: (i) A balanced
superfluid and a partially polarized normal state
(SF$_0$/N$_{\mathrm{PP}}$); (ii) A balanced superfluid and a fully
polarized normal state (SF$_0$/N$_{\mathrm{FP}}$); (iii) A polarized
superfluid and a fully polarized normal state
(SF$_{\mathrm{P}}$/N$_{\mathrm{FP}}$).  All these phases were obtained
by minimizing the total energy with respect to different kinds of
phase separation and their relative volumes (a procedure equivalent to
the Maxwell construction of Ref.~\cite{Pao-2009}).

For comparison, Fig.~\ref{Figure-6}(b) shows the corresponding $T=0$
phase diagram obtained in Ref.~\cite{Pilati-2008} by fixed-node
Quantum Monte Carlo simulations (which, however, do not take into
account FFLO correlations in the nodal surface of the trial wave
functions).  This phase diagram is qualitatively similar to that at
the mean-field level of Fig.~\ref{Figure-6}(a), with the polarized
superfluid being stable as a homogeneous phase only on the BEC side of
the crossover (albeit now somewhat closer to unitarity).
  At unitarity, the critical polarization above which superfluidity
disappears is $\alpha_c = 0.39$, in agreement with the (extrapolated)
experimental value $\alpha_{c}\simeq 0.36$ obtained in
Ref.~\cite{Shin-2008b}, but in contrast with the mean-field result
$\alpha_c = 0.93$.  This reduction of the value of $\alpha_c$ is
expected, since the mean-field analysis neglects correlations in the
normal phase and thus overestimates the extension of the BCS phase in
the coexistence region.  A further difference with respect to the mean-field results is the prediction of a region of coexistence
between a partially polarized superfluid and a partially polarized normal phase  (SF$_{\mathrm{P}}$/N$_{\mathrm{PP}}$) at the center 
of the phase separation region, which replaces the SF$_{\mathrm{0}}$/N$_{\mathrm{FP}}$ region found at the mean-field
level.

\begin{figure}[t]
\begin{center}
\includegraphics[width=16cm]{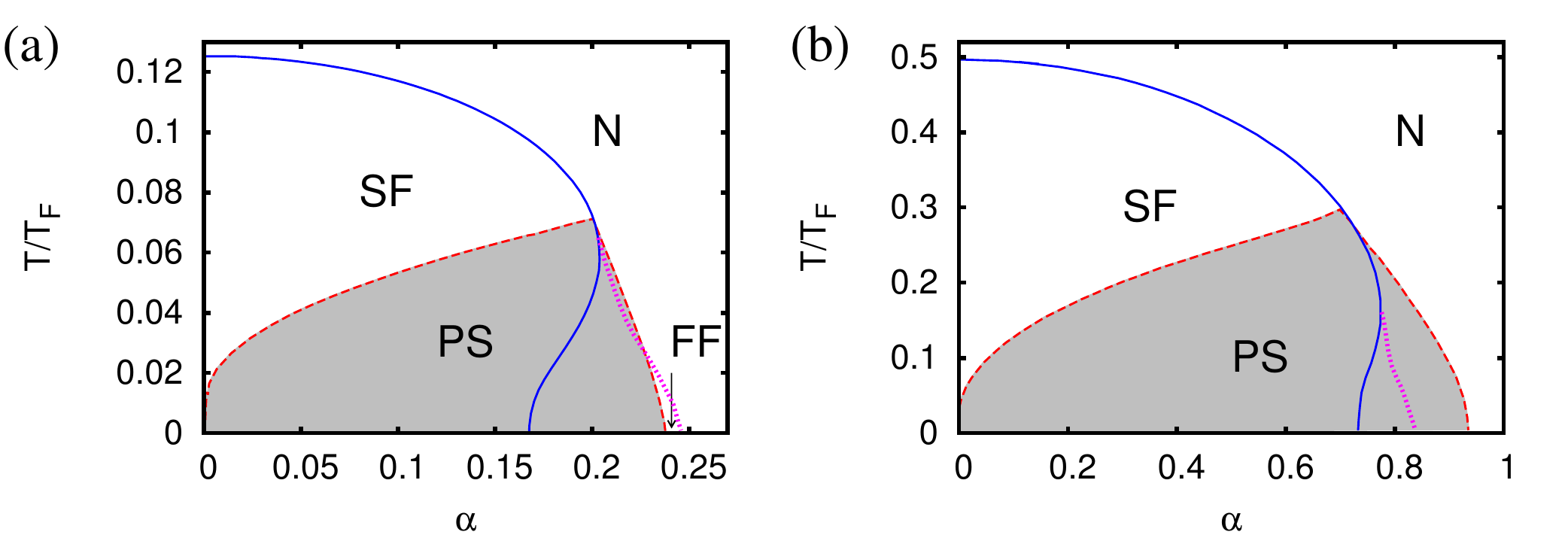}
\end{center}
\vspace{-0.7cm}
\caption{Mean-field phase diagram for temperature $T$ (in units of the Fermi temperature $T_F$) vs population imbalance $\alpha$ with the coupling values 
(a) $(k_F a_F)^{-1}= -1.0$ and (b) $(k_F a_F)^{-1}= 0$, where $k_F$ and $T_F$ are defined in terms of the total density $n=n_{\uparrow}+n_{\downarrow}$.  
Full line: second order transition line obtained by taking $\Delta \to 0$ in the gap equation (\ref{Sarma-gap}).  
Dotted line: second order transition line to the FFLO phase.  
Shaded region enclosed by dashed lines: phase separation (PS) region between the superfluid (SF) and normal (N) phases.  
Full and dotted lines correspond to true transition lines only outside the phase separation region. 
[Data reproduced from Refs.~\cite{Kashimura-2012,Gubbels-2013,Tartari-2011}.]}
\label{Figure-7}
\end{figure}

At finite temperature, the analysis of the phase diagram and of the
properties of polarized superfluid Fermi gases throughout the BCS-BEC
crossover requires one to include (like for a balanced system) pairing
fluctuation over and above mean field.  This extension will be
discussed in detail in Section~\ref{sec:nsr-polarized}.  Here, for
future reference, Fig.~\ref{Figure-7} shows the temperature vs
population imbalance phase diagram at the mean-field level for the two
coupling values $(k_F a_F)^{-1} = (-1.0,0)$ which are representative
of the evolution away from weak coupling. 
One sees that the curve for
$T_c$ obtained by taking $\Delta \to 0$ in the gap equation
(\ref{Sarma-gap}) (and solving simultaneously the number equations
(\ref{total-density-Sarma}) and (\ref{difference-density-Sarma}))
presents a re-entrant behaviour as a function of population imbalance
$\alpha$, in analogy to what was found in Fig.~\ref{Figure-5}(a) in
the weak-coupling limit as a function of the magnetic field $h$.
Allowing for the FFLO pairing, but still looking for a second-order
transition line by setting $\Delta_{\vek{Q}}\to 0$ in
Eq.~(\ref{Delta-eqn-FFLO}), one obtains the curves represented by
dotted lines which eliminate the re-entrant behaviour (as also found in
weak coupling).  By further allowing for phase separation between a
balanced superfluid phase and a polarized normal phase, the shaded
areas of Fig.~\ref{Figure-7} result
\cite{Parish-2007,Kashimura-2012,Gubbels-2013}.  One sees that, while
in the weak-coupling regime of Fig.~\ref{Figure-5}(a) the FFLO
instability anticipates the first-order normal-superfluid transition
line and merges with it at the tricritical point, as the coupling
increases the FFLO second-order transition line is progressively
replaced by phase separation, while the tricritical point no longer
coincides with the point where the FFLO second-order transition line
merges with the BCS transition line (full line in
Fig.~\ref{Figure-7}).

\section{Pairing fluctuations}
\label{sec:pairingfluctuations}

Quite generally, a consistent description of the BCS-BEC crossover
cannot proceed without proper inclusion of pairing fluctuations.  This
is true not only in the superfluid phase below $T_c$ where pairing
fluctuations are added on top of the BCS mean field, but it is
especially relevant in the normal phase above $T_c$ where pairing
fluctuations account for important precursor effects.  This Section
provides a concise overview about the physical phenomena relevant to
the BCS-BEC crossover, for which pairing fluctuations turn out to be
mostly important.

\subsection{Nozi\`{e}res-Schmitt-Rink approach and its extensions}
\label{sec:nsr}  

\begin{figure}[t]
\begin{center}
\includegraphics[width=16.5cm,angle=0]{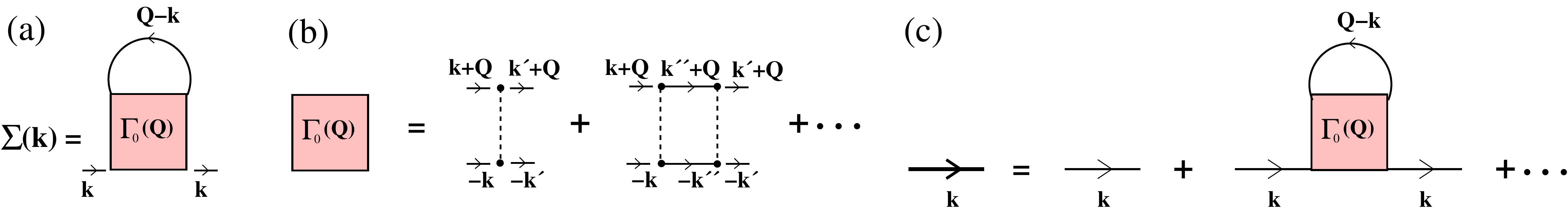}
\caption{Graphical representation of (a) the fermionic self-energy $\Sigma(k)$ given by Eq.(\ref{Sigma-NSR}), (b) the particle-particle propagator $\Gamma_0(Q)$ given by Eq.(\ref{Gamma-0}), 
and (c) the fermionic single-particle propagator $G(k)$ that enters the expression (\ref{density-1}) for the density.  
Thick and thin lines represent, respectively, dressed ($G(k)$) and bare ($G_0(k)$) fermionic single-particle propagators, while dashed lines stand for the inter-particle interaction.  
The colored box identifies the particle-particle propagator $\Gamma_0(Q)$.}
\label{Figure-8}
\end{center}
\end{figure}

It was already mentioned in Section~\ref{sec:BCS_finite_T} that, to
obtain the expected value $T_{\mathrm{BEC}}$ of the Bose-Einstein
critical temperature in the BEC limit of the BCS-BEC crossover, one
has to include the translational degrees of freedom of the composite
bosons and account for their dynamics.  This, in turn, requires one to
include pairing fluctuations beyond the BCS mean-field approximation.
Specifically, it was clear from the analysis of
Section~\ref{sec:BCS_finite_T} that, to obtain the correct value of
the critical temperature $T_c$ in the BEC limit of the BCS-BEC
crossover, one needs to modify the equation (\ref{density-T-c}) for
the density.  To this end, one can extend the diagrammatic description
of the dilute repulsive Fermi gas due to Galitskii
\cite{Galitskii-1958} to the case of the attractive Fermi gas which
undergoes the BCS-BEC crossover, along the lines of the original
approach by Nozi\`{e}res and Schmitt-Rink (NSR) \cite{Nozieres-1985}
(a firm basis for this extension was later discussed in
Ref.~\cite{Vagov-2007}).  In the normal phase above $T_c$, this
amounts to considering the effects of the fermionic self-energy
depicted in Fig.~\ref{Figure-8}(a) which is given by the following
expression:
\begin{equation}
\Sigma(k) = - \int \! \frac{d\vek{Q}}{(2 \pi)^3} \, T \sum_{\nu} \, \Gamma_0(Q) \, G_0(Q-k) \, .
\label{Sigma-NSR} 
\end{equation}
\noindent
Here, $k=(\vek{k},\omega_{n})$ is a four-vector with fermionic
Matsubara frequency $\omega_{n}=(2 n + 1) \pi T$ ($n$ integer),
$Q=(\vek{Q},\Omega_{\nu})$ is a four-vector with bosonic Matsubara
frequency $\Omega_{\nu}=2 \pi \nu T$ ($\nu$ integer), $G_0(k) = (i
\omega_{n} - \xi_{\vek{k}})^{-1}$ is the bare fermionic
single-particle propagator, and $\Gamma_0$ is the particle-particle
propagator (sometimes called the $t$-matrix or vertex of ladder
diagrams) which is depicted in Fig.~\ref{Figure-8}(b) and is given by
the expression
\begin{equation}
\Gamma_0(Q) = \frac{ - \varv_0}{1+\varv_0 \, \chi_{\mathrm{pp}}(Q)} = - \frac{1}{\frac{m}{4 \pi a_F}+R_{\mathrm{pp}}(Q)}
\label{Gamma-0}
\end{equation}
\noindent
where $\varv_{0}$ is the bare coupling constant appearing in Eq.(\ref{regularization}), 
\begin{equation}
\chi_{\mathrm{pp}}(Q) = \int \! \frac{d\vek{k}}{(2\pi)^3} \, T \, \sum_{n} \, G_0(\vek{k}+\vek{Q},\omega_{n}+\Omega_{\nu}) \, G_0(-\vek{k},-\omega_{n})
\label{particle-particle-bubble}
\end{equation}
is the particle-particle bubble, and
\begin{equation}
R_{\mathrm{pp}}(Q) = \chi_{\mathrm{pp}}(Q) - \int \!\frac{d\vek{k}}{(2 \pi)^3} \frac{m}{\vek{k}^2} = \int \!\frac{d\vek{k}}{(2 \pi)^3} 
\left( \frac{1 - f(\xi_{\vek{k}+\vek{Q}}) - f(\xi_{\vek{k}})} {\xi_{\vek{k}+\vek{Q}}+\xi_{\vek{k}}-i\Omega_{\nu}} - \frac{m}{\vek{k}^2} \right)
\label{bubble-pp-exact}  
\end{equation}
\noindent
is the regularized version of the particle-particle bubble obtained with the help of Eq.~(\ref{regularization}).
Note that the minus signs in front of the expressions (\ref{Sigma-NSR}) and (\ref{Gamma-0}) comply with a standard convention for the Matsubara Green's functions (cf., e.g., Ref.~\cite{Fetter-1971}), 
according to which the value $- \varv_0$ is associated with each diagrammatic potential line.  
Sometimes, however, a change of sign is adopted in the definition of $\Gamma_0$, such that at the leading order $\Gamma_0$ agrees instead with the bare potential $\varv_0$; 
this is, for instance, the convention used in nuclear physics (cf. Section~\ref{sec:nuclearsystems}).

The self-energy (\ref{Sigma-NSR}) enters the fermionic single-particle
propagator
\begin{equation}
G(k) = \frac{1}{G_0(k)^{-1} - \Sigma(k)} \,,
\label{Dyson}
\end{equation}
in terms of which the fermionic density can be calculated as follows ($\eta=0^+$):
\begin{equation}
n =  2  \int \! \frac{d\vek{k}}{(2 \pi)^3} \, T \, \sum_{n} \, e^{i\omega_{n}\eta} \, G(k) \, .
\label{density-1}
\end{equation}
In the original NSR approach \cite{Nozieres-1985}, on the other hand,
the density was instead derived from the thermodynamic potential that
is obtained by closing the ladder diagrams.  As shown in
Ref.~\cite{Sa-de-Melo-1993}, this corresponds to including Gaussian
fluctuations around the saddle point (the latter corresponding to the
BCS mean field approximation), or to calculating the thermodynamic potential to order $1/N$ in a large-$N$ expansion~\cite{Enss-2012}.  
The same result can also be obtained
in the present language, if one calculates the density from
Eq.~(\ref{density-1}) with the additional approximation of truncating
the Dyson equation (\ref{Dyson}) at first order in $\Sigma$ as
depicted in Fig. \ref{Figure-8}(c), thereby writing:
\begin{equation}
G(k) \simeq G_0(k) + G_0(k) \, \Sigma(k) \, G_0(k) \,.
\label{Dyson1storder}
\end{equation}
The two above approximations are equivalent when the self-energy
corrections are small; however, if the corrections are not small,
Eq. (\ref{density-1}) might yield a sensible result when the NSR
approach does not \cite{Serene-1989}.  We shall discuss below some
advantages as well as disadvantages of adopting the NSR (truncated)
approach (\ref{Dyson1storder}) with respect to its extension
(\ref{Dyson}) with the full single-particle propagator.

An argument in favour of the perturbative expansion
(\ref{Dyson1storder}) of the Dyson equation considers the shift of the
grand-canonical potential $\Omega(T,V,\mu)$ with respect to the value
$\Omega_0(T,V,\mu)$ for the non-interacting system, which quite
generally is given by the following expression (cf. Eq.~(25.27) of
Ref.~\cite{Fetter-1971}):
\begin{equation}
\Omega -\Omega_0 = V \int_0^1 \frac{d\lambda}{\lambda} \int \! \frac{d\vek{k}}{(2 \pi)^3} \, T  \sum_{n} \, e^{i\omega_{n}\eta} \, (i \omega_n - \xi_{\vek{k}}) \, G^{\lambda}(k)\, ,
\label{Omega_lambda}
\end{equation}
where $V$ is the volume and the single-particle propagator
$G^{\lambda}$ is evaluated with the variable coupling constant
$\lambda \varv_0$.  With the choice of the self-energy
(\ref{Sigma-NSR}) and the perturbative expansion
(\ref{Dyson1storder}), Eq.~(\ref{Omega_lambda}) reduces to the
following approximate expression
\begin{equation}
\Omega - \Omega_0 \cong V  \int \! \frac{d\vek{Q}}{(2 \pi)^3} \, T \sum_{\nu} \, e^{i\Omega_{\nu}\eta} \ln [1 + \varv_0 \, \chi_{\mathrm{pp}}(Q)] 
\label{Omega_lambda-approximate}
\end{equation}
where $\chi_{\mathrm{pp}}$ is given by
Eq.(\ref{particle-particle-bubble}), which coincides with the
expression derived in the original NSR approach on the basis of a
linked-cluster expansion for $\Omega$ which sums up ladder diagrams
\cite{Nozieres-1985}.  This shows that the results for the
grand-canonical potential, obtained from the single-particle
propagator and from the linked-cluster expansion, are consistent with
each other, a property which is not obvious for an approximate theory
(cf., e.g., the discussion in Section 3.6 of Ref.\cite{Mahan-2000}).
As a consequence of this property, the derivative of
Eq.~(\ref{Omega_lambda-approximate})Ê with respect to the chemical
potential yields:
\begin{equation}
- \frac{\partial}{\partial \mu}\left(\frac{\Omega-\Omega_0}{V}\right)_{T} = n - n_0 \cong 2  \int \! \frac{d\vek{k}}{(2 \pi)^3} \, T \sum_{n} \, G_0(k)^2 \, \Sigma(k) \, ,
\label{approxiate-density}
\end{equation}
which coincides with the expression (\ref{density-1}) for the density
obtained in terms of the single-particle propagator $G$ once this is
expanded as in Eq.(\ref{Dyson1storder}).

A similar argument was made in Ref.~\cite{Urban-2014} with the use of the zero-temperature formalism in the canonical ensemble \cite{Fetter-1971,Mahan-2000}.  
In this case, the NSR approach corresponds to what in nuclear physics is known as the particle-particle random-phase approximation (pp-RPA) \cite{Ring-1980}.  
Here, the correlation energy of the ground-state is obtained from the zero-temperature counterpart of Eq.~(\ref{Omega_lambda}) (cf., e.g., Eq.~(7.32) of
Ref.\cite{Fetter-1971}), where the expanded Dyson equation (\ref{Dyson1storder}) is again used.  
Note, however, that for an attractive inter-particle interaction, the use of the NSR approach at zero temperature is physically meaningful only in the presence of a
mechanism that suppresses superfluidity.  
In finite systems, like nuclei or ultra-small metallic grains, this occurs when the level spacing exceeds a critical value~\cite{Ring-1980}.  
For infinite systems, on the other hand, this may occur when the system is polarized above a critical value, as discussed in Section~\ref{sec:polarized}.  
In this respect, it was shown in Ref.~\cite{Urban-2014} that the NSR approach in the zero-temperature formalism, together with the truncation of the Dyson equation,
fulfils the Luttinger theorem (see Section~\ref{sec:nsr-polarized} for more details).  
Nevertheless, the truncation of the Dyson equation has some notable drawbacks.  
For example, within the pp-RPA at $T=0$ the quasi-particle residue $Z$ (associated with the discontinuity of the Fermi distribution at the Fermi surface) can
become negative \cite{Schuck-2003}, a feature that does not occur when the self-energy is summed up to all orders.  
Furthermore, the shape of the single-particle spectral function (cf. Section~\ref{sec:pseudo-gap}) crucially depends on the location of the poles of the single-particle propagator 
in the complex frequency plane, which is of course different when the full expression (\ref{Dyson}) or its truncation (\ref{Dyson1storder}) are adopted.

A definite virtue of the NSR approach (in both its variants) is that
it interpolates for the critical temperature between two limiting
cases, that is, the free Bose gas and the BCS weak-coupling limit
(albeit with the some provisions for the latter case).  This is
because, in the BEC (strong-coupling) limit whereby $|\mu| / T \gg 1$,
the particle-particle propagator $\Gamma_0$ acquires the polar form
\cite{Haussmann-1993,Haussmann-1994,Pieri-2000}:
\begin{equation}
\Gamma_0(Q) \simeq - \frac{8\pi}{m^2a_F} \, \frac{1}{i\Omega_{\nu} - \frac{\vek{Q}^2}{4m} + \mu_B} 
\label{Gamma-0-BEC_limit}
\end{equation}
which (apart from the presence of the residue $- 8\pi/(m^2a_F)$) is
equivalent to a bare bosonic propagator with mass $2m$ and chemical
potential $\mu_B=2\mu + \varepsilon_0$, where  $\varepsilon_0 = (m a_F^2)^{-1}$ is the dimer binding energy.
 In this limit, $|\mu|$ is the
largest energy scale in the fermionic propagator $G_0$, in such a way
that one can neglect the $Q$-dependence of $G_0$ in
Eq.~(\ref{Sigma-NSR}) and obtain:
\begin{equation}
\Sigma(k) \simeq - \, G_0(-k) \!\! \int \! \frac{d\vek{Q}}{(2 \pi)^3} \, T \, \sum_{\nu} \, e^{i\Omega_\nu\eta} \, \Gamma_0(Q) \qquad (\eta = 0^{+}) \, .
\label{approximate-Sigma-NSR} 
\end{equation}
With the approximation (\ref{Dyson1storder}), one thus ends up with
the following expression for the density (\ref{density-1}):
\begin{equation}
n \simeq 2 \! \int \! \frac{d\vek{k}}{(2 \pi)^3} \, T \, \sum_n \, e^{i\omega_n\eta} \, G_0(k) 
\, - 2  \! \int \! \frac{d\vek{k}}{(2\pi)^3} \, T \sum_n \, G_0^2(k) \, G_0(- k) \! \int \!  \frac{d\vek{Q}}{(2\pi)^3} \, T \sum_\nu \, e^{i\Omega_\nu\eta} \, \Gamma_0(Q)
\label{approximate-density-BEC_limit}
\end{equation}
\noindent
where the first term on the right-hand side can be neglected since
$|\mu| / T \gg 1$ (this holds true provided the value of $T_c$ we
are after is finite).  When substituting the polar form
(\ref{Gamma-0-BEC_limit}) for $\Gamma_0$ in the second term on the
right-hand side of Eq.(\ref{approximate-density-BEC_limit}), one sees
that the factor
\begin{equation}
\int \! \frac{d\vek{k}}{(2 \pi)^3} \, T \, \sum_n \,G_0^2(k) \, G_0(-k)
\simeq - \frac{m^2}{8\pi\sqrt{2m |\mu|}}
 \simeq - \frac{m^2a_F}{8\pi}
\label{triple-G-0}
\end{equation}
\noindent
cancels the residue of the expression (\ref{Gamma-0-BEC_limit}).  One
is thus left with the following Bose-like expression for the density:
\begin{equation}
n \, \simeq \, 2 \! \int \! \frac{d\vek{Q}}{(2 \pi)^3} \, \frac{1}{\exp{(\xi^B_{\vek{Q}}/T)} - 1} \equiv 2 \, n_B
\label{density-1-bosonic}
\end{equation}
\noindent
where $\xi^B_{\vek{Q}} = \vek{Q}^2/(4m) - \mu_B$, with $n_B$
playing the role of the density of a system of non-interacting
composite bosons with mass $2m$.  The correct value of the
Bose-Einstein temperature $T_{\mathrm{BEC}} = 3.31 \,
n_B^{2/3}/(2m)$ then results by letting $\mu_B \rightarrow 0^{-}$
in the expression (\ref{density-1-bosonic}).  In the BCS
(weak-coupling) limit, on the other hand, at the leading order the
self-energy (\ref{Sigma-NSR}) approaches the mean-field value
$\Sigma_0= \frac{4 \pi a_F}{m} \frac{n}{2}$ and the BCS result
(\ref{T-c-w-c}) for $T_c$ is recovered, {\em provided} the mean-field
shift $\Sigma_0$ is also self-consistently included in the fermion
propagators entering $\Gamma_0$ (see Refs.~\cite{Perali-2002,Pieri-2004b,Pieri-2005b} for a discussion about the importance of including
this shift in the weak-coupling limit).  It should be noted that the
NSR approach in its original formulation, which does not include the
mean-field shift $\Sigma_0$, recovers the BCS result (\ref{T-c-w-c})
for $T_c$ in the weak-coupling limit only with logarithmic accuracy
(specifically, it yields the result (\ref{T-c-w-c}) divided by a
``spurious" factor $e^{1/3}$).  In this limit, however, particle-hole
(rather than particle-particle) effects are known to be important and
modify the pre-factor in front of the exponential dependence in
(\ref{T-c-w-c}), as it will be discussed in Section \ref{sec:gmb}.

Quite generally, for generic values of the coupling parameter $(k_F
a_F)^{-1}$ throughout the BCS-BEC crossover, the density equation
(\ref{density-1}) with $\Sigma$ and $G$ given by
Eqs.~(\ref{Sigma-NSR}) and (\ref{Dyson}) can be solved numerically in
conjunction with Eq.~(\ref{regular-T-c}), to yield $T_c$ and
$\mu(T_c)$ as functions of $(k_Fa_F)^{-1}$.
\begin{figure}[t]
\begin{center}
\includegraphics[width=8cm,angle=0]{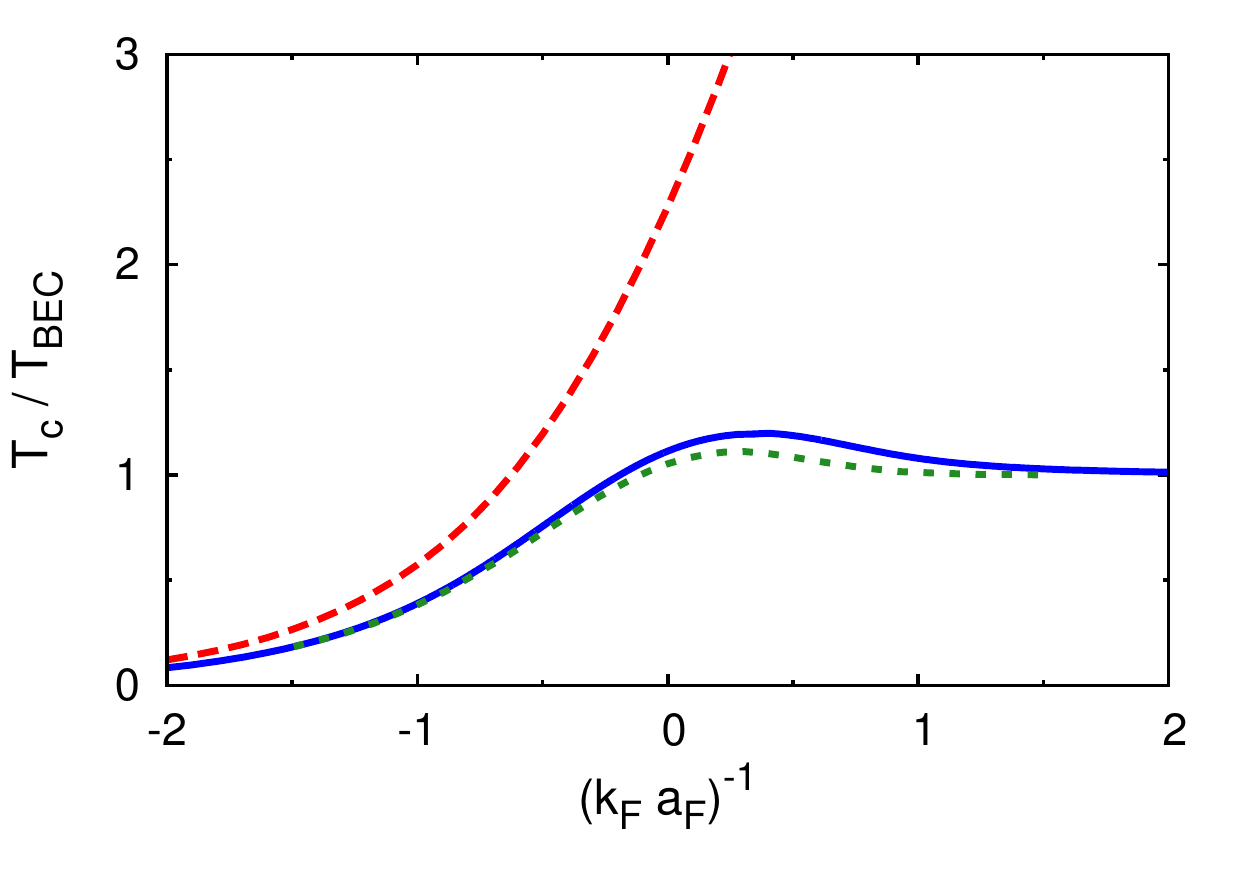}
\caption{Critical temperature $T_c$ for the superfluid transition vs the coupling parameter $(k_Fa_F)^{-1}$, as obtained by solving the density equation with the fermionic 
single-particle propagator that includes (full line) or neglects (dashed line) the self-energy (\ref{Sigma-NSR}).  
The NSR result, where the approximation (\ref{Dyson1storder}) is further adopted in the density equation, is also reported for comparison (dotted line).}
\label{Figure-9}
\end{center}
\end{figure}
Figure~\ref{Figure-9} compares the values of $T_c$ obtained by using
the full $G$ in the density equation that includes pairing
fluctuations through the self-energy $\Sigma$ of Eq.~(\ref{Sigma-NSR})
(full line), and the corresponding mean-field values of $T_c$ where
the bare $G_0$ is instead used in the density equation (dashed line).
The two results are seen to strongly deviate from each other on the
BEC side of the crossover where $(k_F a_F)^{-1} \gtrsim 1$.  If we
consider for all couplings (and not only in the BEC limit as we did in
Section~\ref{sec:BCS_finite_T}) the mean-field critical temperature to
represent a sort of ``pair dissociation temperature'' $T^{*}$, the
above result indicates that, while in weak coupling fermion pairs form
and condense at the same temperature (that is, $T^{*} \simeq T_c$),
in strong coupling pair formation occurs at a temperature higher than
the condensation temperature (that is, $T^{*} \gg T_c$).  The
difference between $T_c$ and $T^{*}$, which naturally arises within
a pairing-fluctuation scenario, points to the presence of precursor
pairing phenomena (like the pseudo-gap effects to be discussed below
in Section~\ref{sec:pseudo-gap}) at temperatures intermediate between
$T_c$ and $T^{*}$.

A few additional comments on the structure of the particle-particle
propagator $\Gamma_0(Q)$ are in order.  At any coupling, $\Gamma_0(Q)$
admits a spectral representation of the form
\cite{Pieri-2004a,Pieri-2004b}:
\begin{equation}
\Gamma_0(\vek{Q},\Omega_{\nu}) = - \, \int_{- \infty}^{+ \infty} \! \frac{d \omega}{\pi} \,\, \frac{\Imag\Gamma_0^{R}(\vek{Q},\omega)} {i\Omega_{\nu} - \omega} 
\label{spectral-representation-Gamma-0}
\end{equation}
\noindent
where the retarded particle-particle propagator (or vertex function)
$\Gamma_0^{R}(\vek{Q},\omega)$ is defined by
$\Gamma_0(\vek{Q},\Omega_{\nu})$ given by the expressions
(\ref{Gamma-0})--(\ref{bubble-pp-exact}) in which one makes the
replacement $i\Omega_{\nu} \rightarrow \omega + i \eta$.  With this
replacement, it appears that the expression (\ref{bubble-pp-exact})
has a branch cut in the complex $\omega$-plane for $\omega \ge
\frac{\vek{Q}^2}{4m} - 2 \mu$, whose branch point at $\omega =
\frac{\vek{Q}^2}{4m} - 2 \mu$ is pushed to large positive values in
the BEC limit when $-2 \mu \rightarrow \varepsilon_0 - \mu_B$.  In
this limit, $\Gamma_0^R(\vek{Q},\omega)$ shows a pole at $\omega =
\frac{\vek{Q}^2}{4m} - \mu_B$, such that $\Imag
\Gamma_0^R(\vek{Q},\omega) = \frac{8 \pi^2}{m^2a_F} \delta(\omega -
\frac{\vek{Q}^2}{4m} + \mu_B)$ and the spectral representation
(\ref{spectral-representation-Gamma-0}) reduces to the previous result
(\ref{Gamma-0-BEC_limit}).  Away from the BEC limit, however, the
branch cut of $\Gamma_0^R(\vek{Q},\omega)$ in the complex
$\omega$-plane plays in practice an important role, which is related
on physical grounds to the fermionic nature of the constituent
particles.  This fermionic nature actually manifests itself even in
the BEC limit, when calculating the residual interaction which acts
among the composite bosons (cf. Section \ref{sec:b-b-interaction}).

It is also worth emphasizing that, for a zero-range interaction
potential, it is the regularization (\ref{regularization}) that makes
the infinite summation of the particle-particle bubble
(\ref{particle-particle-bubble}) (which yields eventually the
particle-particle propagator $\Gamma_0$ of Eq.~(\ref{Gamma-0})) to
remain finite in the limit $\varv_0 \rightarrow 0$ and $k_0
\rightarrow \infty$, because the ultraviolet divergence of the
expression (\ref{particle-particle-bubble}) is compensated by the
simultaneous vanishing of $\varv_0$ \cite{Pieri-2000}.  An analogous
mechanism works also for the normal and anomalous particle-particle
bubbles that are present below $T_c$ \cite{Andrenacci-2003}.  As a
consequence, the only diagrammatic structures that survive the
regularization procedure (\ref{regularization}) are those which are
built directly on the particle-particle propagator $\Gamma_0$ and not
on the bare potential $\varv_0$ (the only exception to this rule
occurs below $T_c$ for the BCS diagram of the gap equation, where
the trace of the anomalous single-particle Green's function itself
diverges in the ultraviolet \cite{Pieri-2004a,Pieri-2004b}).  In
practice, this provides a strong simplification on the classification
of the many-body diagrams, and as such has recently been used also to
develop a sophisticated re-summation technique known as ``bold
diagrammatic'' Monte Carlo \cite{Prokofev-2008}.  This simplification,
however, cannot be exploited when the interaction potential has a
finite range, like in the case of the potentials used in nuclear
physics.

As already mentioned, the sum of ladder diagrams that defines the
particle-particle propagator $\Gamma_0$ of Eq.(\ref{Gamma-0})
(together with the ensuing fermionic self-energy $\Sigma$ of
(\ref{Sigma-NSR})) is sometimes referred to as the $t$-matrix
approximation.  The above expressions actually represent the
\emph{$G_0G_0$ version} of this approximation \cite{Perali-2002},
since \emph{all} the fermionic single-particle propagators (that is,
both those entering $\Gamma_0$ and the one closing the fermion loop in
$\Sigma$ - cf. Fig.~\ref{Figure-8}) are bare ($G_0$) ones.  Different
variants of the $t$-matrix approximation have actually been
introduced, which correspond to:
\begin{enumerate-roman}
\item The \emph{$G_0G$ version}, where one $G_0$ in the
  particle-particle bubble $\chi_{\mathrm{pp}}$ of
  Eq.~(\ref{particle-particle-bubble}) is replaced by a full $G$
  \cite{Janko-1997,Kosztin-1998,Chen-1998,Chen-2005};
\item The \emph{extended $t$-matrix approximation} (ETMA), where the
  propagator $G_0$ closing the loop in $\Sigma$
  [cf. Eq.~(\ref{Sigma-NSR})] is replaced by the full $G$, but not the
  propagators in $\chi_{\mathrm{pp}}$ \cite{Kashimura-2012};
\item The \emph{self-consistent Green's function method} (also called
  Luttinger-Ward method), where all $G_0$ (including that closing the
  loop in $\Sigma$) are replaced by full $G$
  \cite{Haussmann-1993,Haussmann-1994,Haussmann-2007};
\item A variant following the approach by Zimmermann and Stolz
  \cite{Zimmermann-1985}, where the bare $G_0$ are replaced by
  quasi-particle propagators (cf. Section~\ref{sec:ZS}).
\end{enumerate-roman}

\noindent
These different variants of the $t$-matrix approximation unavoidably produce different quantitative results when applied to specific problems in the context of the BCS-BEC crossover \cite{Hu-2008,Chien-2010}.  
Mostly for the purpose of illustration,  in what follows we shall mainly present the results obtained within the $G_0G_0$ version of the $t$-matrix approximation.  
In this context, one should anyway keep in mind that not always what is believed to be an improvement in the treatment of a given fermionic single-particle propagator necessarily
leads to a corresponding improvement in the comparison with the experimental data, nor even in comparison with results of exactly solvable models (see for instance Ref.~\cite{Combescot-2008}
for the description of a spin-$\downarrow$ impurity in a gas of spin-$\uparrow$ particles). 

\subsection{Intra- and inter-pair correlations}
\label{sec:intra-inter-pair}

In Section~\ref{sec:nsr} we went through the original argument due to
Nozi\`{e}res and Schmitt-Rink \cite{Nozieres-1985}, according to which
considering pairing fluctuations over and above mean field is
essential to recover the correct value of the Bose-Einstein critical
temperature in the BEC limit of the BCS-BEC crossover.  In addition,
in Section~\ref{sec:pairingcorrelations} we mentioned that the need to
include pairing fluctuations beyond mean field arises even in the BCS
limit of the BCS-BEC crossover, where already at the mean-field level
one finds that the size $\xi_{\mathrm{pair}}$ of the intra-pair
correlations between opposite-spin fermions remains finite when
approaching $T_c$ from below \cite{Marsiglio-1990}.  In
Section~\ref{sec:pairingcorrelations} we have further anticipated
that, even at zero temperature, the inclusion of pairing fluctuations
is required to calculate the inter-pair coherence (or healing) length
$\xi$, which strongly deviates from the intra-pair coherence length
$\xi_{\mathrm{pair}}$ upon approaching the BEC limit \cite{Pistolesi-1996,Engelbrecht-1997}, as shown in
Fig.~\ref{Figure-2}.  On physical grounds, the difference between
$\xi_{\mathrm{pair}}$ and $\xi$ has the same origin of that occurring
between the pair dissociation temperature $T^{*}$ and the critical
temperature $T_c$, as shown in Fig.~\ref{Figure-9}.

Here, we dwell further on the difference between the two lengths
$\xi_{\mathrm{pair}}$ and $\xi$.  At zero temperature this difference
lies at the very origin of the BCS-BEC crossover, while at finite
temperature it gives valuable information about the persistence of
pairing above $T_c$.  To this end, we shall rely on the results of
Ref.~\cite{Palestini-2014} where the $t$-matrix approximation of
Section~\ref{sec:nsr} (as well as its extension below $T_c$
\cite{Pieri-2004a,Pieri-2004b}) was employed to calculate the pair
correlation function for opposite-spin fermions (to get
$\xi_{\mathrm{pair}}$) and the correlation function of the order
parameter (to get $\xi$).

The pair correlation function for opposite-spin fermions is defined by:
\begin{equation}
g_{\uparrow \downarrow}(\vek{\rho},\vek{R}) = \left\langle
\psi^{\dagger}_{\uparrow}\left(\vek{R}+\frac{\vek{\rho}}{2}\right)
\psi^{\dagger}_{\downarrow}\left(\vek{R}-\frac{\vek{\rho}}{2}\right)
\psi_{\downarrow}\left(\vek{R}-\frac{\vek{\rho}}{2}\right)
\psi_{\uparrow}\left(\vek{R}+\frac{\vek{\rho}}{2}\right) \right\rangle
\, - \, \left( \frac{n}{2} \right)^2
\label{definition-pair-correlation-function}
\end{equation}
\noindent
where $\psi_{\sigma}(\vek{r})$ is a fermion field operator with spin
$\sigma$, and $\vek{\rho} = \vek{r} - \vek{r}'$ and $\vek{R} =
(\vek{r} + \vek{r}')/2$ are the relative and center-of-mass
coordinates of the pair (the dependence of $g_{\uparrow \downarrow}$
on $\vek{R}$ can be dropped for a homogeneous system).  The pair
coherence length $\xi_{\mathrm{pair}}$ is then obtained as follows
\begin{equation}
\xi_{\mathrm{pair}}^2 = \frac{\int \! d \vek{\rho} \, \vek{\rho}^2 \, g_{\uparrow \downarrow}(\vek{\rho})} {\int \! d \vek{\rho} \, g_{\uparrow\downarrow}(\vek{\rho})} \, ,
\label{xi-pair-definition}
\end{equation}
which generalizes Eq.(\ref{CL}) that is only valid at the mean-field level.
[The definition (\ref{xi-pair-definition}) for $\xi_{\mathrm{pair}}$ is meaningful only when the integrals therein converge.
This is the case at the mean-field level and with the further inclusion of pairing fluctuations within the $t$-matrix approximation, to be considered below.
When going beyond this approximation, a long-range (power-law) tail may appear in $g_{\uparrow \downarrow}(\vek{\rho})$ which is related to correlations between spin-$\uparrow$ and
spin-$\downarrow$ belonging to different pairs, such that the integrals in Eq.(\ref{xi-pair-definition}) may not converge.
In this case, it should anyway be possible to identify a short-range from a long-range part in $g_{\uparrow \downarrow}(\vek{\rho})$, in such a way that $\xi_{\mathrm{pair}}$ could be
isolated by inspection.]

\begin{figure}[t]
\begin{center}
\includegraphics[width=16.0cm,angle=0]{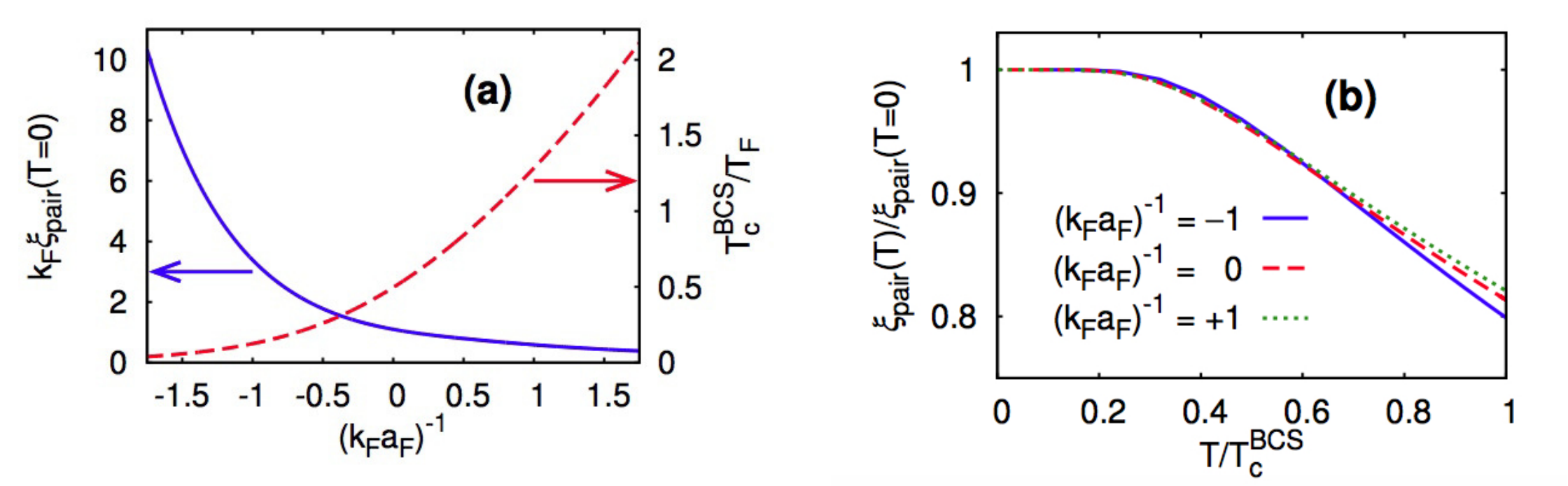}
\caption{(a) BCS pair coherence length $\xi_{\mathrm{pair}}(T=0)$ at zero temperature in units of the inverse Fermi wave vector $k_F^{-1}$ (full line, left scale) and BCS critical temperature
$T^{\mathrm{BCS}}_{c}$ in units of the Fermi temperature $T_F$ (dashed line, right scale) vs the coupling parameter $(k_F a_F)^{-1}$.  (b) BCS pair coherence length $\xi_{\mathrm{pair}}(T)$
in units of $\xi_{\mathrm{pair}}(T=0)$ vs the temperature $T$ in units of the respective BCS critical temperature $T^{\mathrm{BCS}}_{c}$ at various couplings.
[Adapted from Ref.~\cite{Palestini-2014}.]}
\label{Figure-10}
\end{center}
\end{figure}

Quite generally, the thermal average of
Eq.~(\ref{definition-pair-correlation-function}) can be related to the
two-particle Green's function by a suitable choice of its space
$\vek{\rho}$, (imaginary) time $\tau$, and spin arguments
\cite{Palestini-2014}.  The two-particle Green's function, in turn,
can be conveniently calculated in terms of many-body diagrams.  In
particular, at the mean-field level for any temperature below $T_c$,
$g_{\uparrow \downarrow}(\vek{\rho})$ is given by the square of the
anomalous BCS single-particle Green's function:
\begin{equation}
\mathcal{G}_{12}(\vek{\rho},\tau=0^{-}) = \Delta \! \int \! \frac{d\vek{k}}{(2 \pi)^3} \,\, e^{i \vek{k} \cdot \vek{\rho}} \, \frac{1 - 2 f(E_{\vek{k}})}{2 E_{\vek{k}}} \, .
\label{G12-BCS}
\end{equation}
\noindent
The results of this calculation are shown in Fig.~\ref{Figure-10}.
For convenience, Fig.~\ref{Figure-10}(a) reproduces the coupling dependence of $\xi_{\mathrm{pair}}$ at $T=0$ from Fig.~\ref{Figure-2} and of $T^{\mathrm{BCS}}_{c}$ from Fig.~\ref{Figure-9}.  
It turns out that, irrespective of coupling, the temperature dependence of $\xi_{\mathrm{pair}}$ follows approximately a kind of a ``law of corresponding states'' (that is, with a common behaviour when reduced variables are used as in Fig.~\ref{Figure-10}(b)), provided that $\xi_{\mathrm{pair}}(T)$ is expressed in units of $\xi_{\mathrm{pair}}(T=0)$ and $T$ in units of the BCS critical
temperature $T^{\mathrm{BCS}}_{c}$ at the given coupling.  
It turns further out that the temperature dependence of $\xi_{\mathrm{pair}}$ from $T=0$ to $T^{\mathrm{BCS}}_{c}$ is rather weak, not only in the BCS weak-coupling limit 
(where it was originally pointed out in Ref.~\cite{Marsiglio-1990}) but also at stronger couplings throughout the BCS-BEC crossover \cite{Palestini-2014}.  
Specifically, in all cases $\xi_{\mathrm{pair}}$ reaches a finite value at $T^{\mathrm{BCS}}_{c}$ which is about $80 \%$ of its value at $T=0$, as shown in Fig.~\ref{Figure-10}(b) for various couplings.
On physical grounds, as the temperature raises the spatial range of the correlations between spin-$\uparrow$ and spin-$\downarrow$ fermions is bound to decrease, because the effect of temperature 
is to disorder the system over a progressively shorter length scale which sets a limit on the value of $\xi_{\mathrm{pair}}$.

Although below $T_c$ pairing fluctuations can be added on top of
mean field to improve on the description of the pair correlation
function (\ref{definition-pair-correlation-function}) as well as to
obtain the correlation function of the order parameter, from a
physical point of view it is most interesting to consider the behaviour
of the intra-pair ($\xi_{\mathrm{pair}}$) and inter-pair ($\xi$)
coherence lengths above $T_c$.  In particular, the pair correlation
function (\ref{definition-pair-correlation-function}) can be obtained
above $T_c$ within the $t$-matrix approximation of
Section~\ref{sec:nsr} by the following expression
\cite{Palestini-2014}:
\begin{equation}
g_{\uparrow \downarrow}(\vek{\rho}) = \!\! \int \!\! \frac{d\vek{Q}}{(2 \pi)^3} \, T \! \sum_{\nu} \Gamma_0(\vek{Q},\Omega_{\nu}) \int \!\! \frac{d\vek{k}}{(2\pi)^3} \, e^{i\vek{k}\cdot\vek{\rho}} \!
\left( \frac{1 - f(\xi_{\vek{k}+\vek{Q}}) - f(\xi_{\vek{k}})} {\xi_{\vek{k}+\vek{Q}}+\xi_{\vek{k}}-i\Omega_{\nu}}\right)
\int\!\! \frac{d\vek{k}'}{(2 \pi)^3} \, e^{-i\vek{k}'\cdot\vek{\rho}} \!
\left( \frac{1 - f(\xi_{\vek{k}'+\vek{Q}}) - f(\xi_{\vek{k}'})} {\xi_{\vek{k}'+\vek{Q}}+\xi_{\vek{k}'}-i\Omega_{\nu}} \right)
\label{pair-correlation-function-above-Tc}
\end{equation}
\noindent
in terms of which $\xi_{\mathrm{pair}}(T)$ results again using the
definition (\ref{xi-pair-definition}).  Note that the expression
(\ref{pair-correlation-function-above-Tc}) has the expected
short-range behaviour:
\begin{equation}
g_{\uparrow \downarrow}(\vek{\rho}) \underset{(\vek{\rho} \rightarrow 0)}{\longrightarrow} \left( \frac{m \, \Delta_{\infty}}{4 \, \pi}
\right)^2 \left( \frac{1}{\rho^2} - \frac{2}{a_F \rho} + \cdots
\right)
\label{pair-correlation-function-short-range}
\end{equation}
\noindent
with $\rho = |\vek{\rho}|$ and the notation
\begin{equation}
\Delta_{\infty}^2 = \int \!\! \frac{d\vek{Q}}{(2 \pi)^3} \, T \, \sum_{\nu} \, e^{i \Omega_{\nu} \eta} \, \Gamma_0(\vek{Q},\Omega_{\nu}) \, .
\label{Delta-infinity}
\end{equation}
\noindent
In Section~\ref{sec:Tancontact} we shall identify the quantity
$(m\Delta_{\infty})^2$ with the form of the Tan contact within the
present theory.  [Note that below $T_c$ within mean field one ends
  up with an expression similar to
  (\ref{pair-correlation-function-short-range}) with the BCS gap
  $\Delta$ replacing $\Delta_{\infty}$.]

The correlation function of the order parameter can also be obtained
above $T_c$ within the $t$-matrix approximation of
Section~\ref{sec:nsr}, in the form \cite{Palestini-2014}:
\begin{equation}
F(\vek{R} -\vek{R}') = \frac{1}{2} \int \!\! \frac{d\vek{Q}}{(2 \pi)^3} \, e^{i \vek{Q} \cdot (\vek{R} -\vek{R}')} \, \Gamma_0(\vek{Q},\Omega_{\nu}=0)
\label{correlation-function-order-parameter-above-Tc}
\end{equation}
\noindent
where only the static ($\Omega_{\nu}=0$) component, which corresponds
to taking a time averaging of pairing fluctuations, is considered for
extracting the spatial dependence of the correlation function.  To
obtain the value of the inter-pair coherence length $\xi$ from the
above expression it is enough to expand
$\Gamma_0(\vek{Q},\Omega_{\nu}=0)^{-1} = a + b \, \vek{Q}^2 + \cdots$
up to quadratic order.  The Fourier transform in
(\ref{correlation-function-order-parameter-above-Tc}) then yields a
Yukawa potential form with range $\xi = \sqrt{b/a}$
\cite{Pistolesi-1996}.  In the limit $T \rightarrow T_c$ the
propagator $\Gamma_0(\vek{Q}, \Omega_{\nu})$ then develops the Cooper
singularity for ($\vek{Q} \rightarrow 0, \Omega_{\nu} = 0$), in
agreement with the Thouless criterion.  Accordingly, the coefficient
$a$ vanishes like $(T-T_c)$ in the limit $T \rightarrow T_c^{+}$.
\begin{figure}[t]
\begin{center}
\includegraphics[width=17.5cm,angle=0]{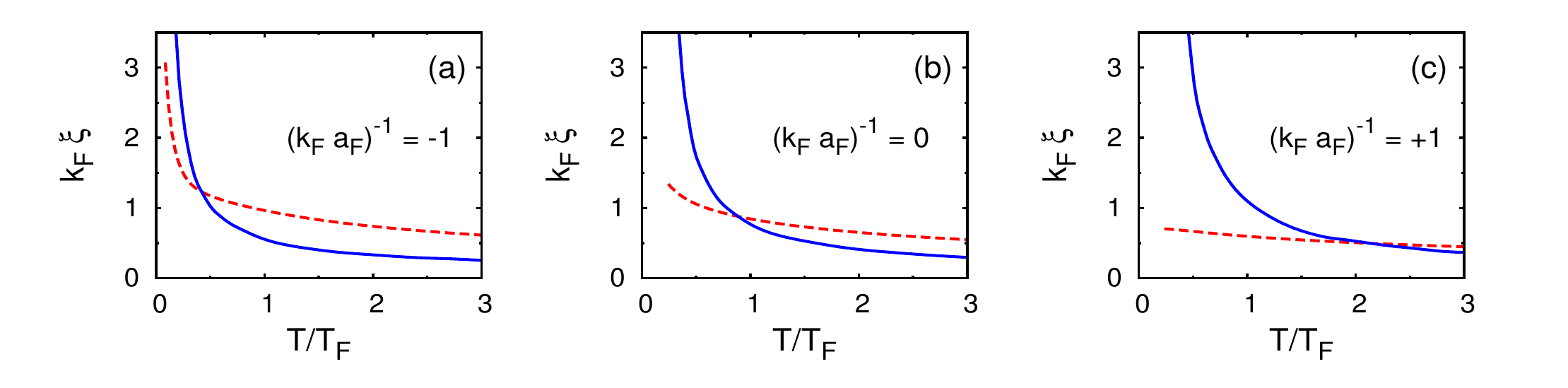}
\caption{The temperature dependence of the inter-pair coherence length $\xi$ (full lines) and of the intra-pair coherence length $\xi_{\mathrm{pair}}$ (dashed lines) is shown above $T_c$ for
three values of the coupling $(k_F a_F)^{-1}$.  
Each curve terminates at the corresponding value of $T_c$ obtained with the inclusion of pairing fluctuations (cf. the full line of Fig.~\ref{Figure-9}).
[Adapted from Ref.~\cite{Palestini-2014}.]}
\label{Figure-11}
\end{center}
\end{figure}

\begin{figure}[t]
\begin{center}
\includegraphics[width=8.0cm,angle=0]{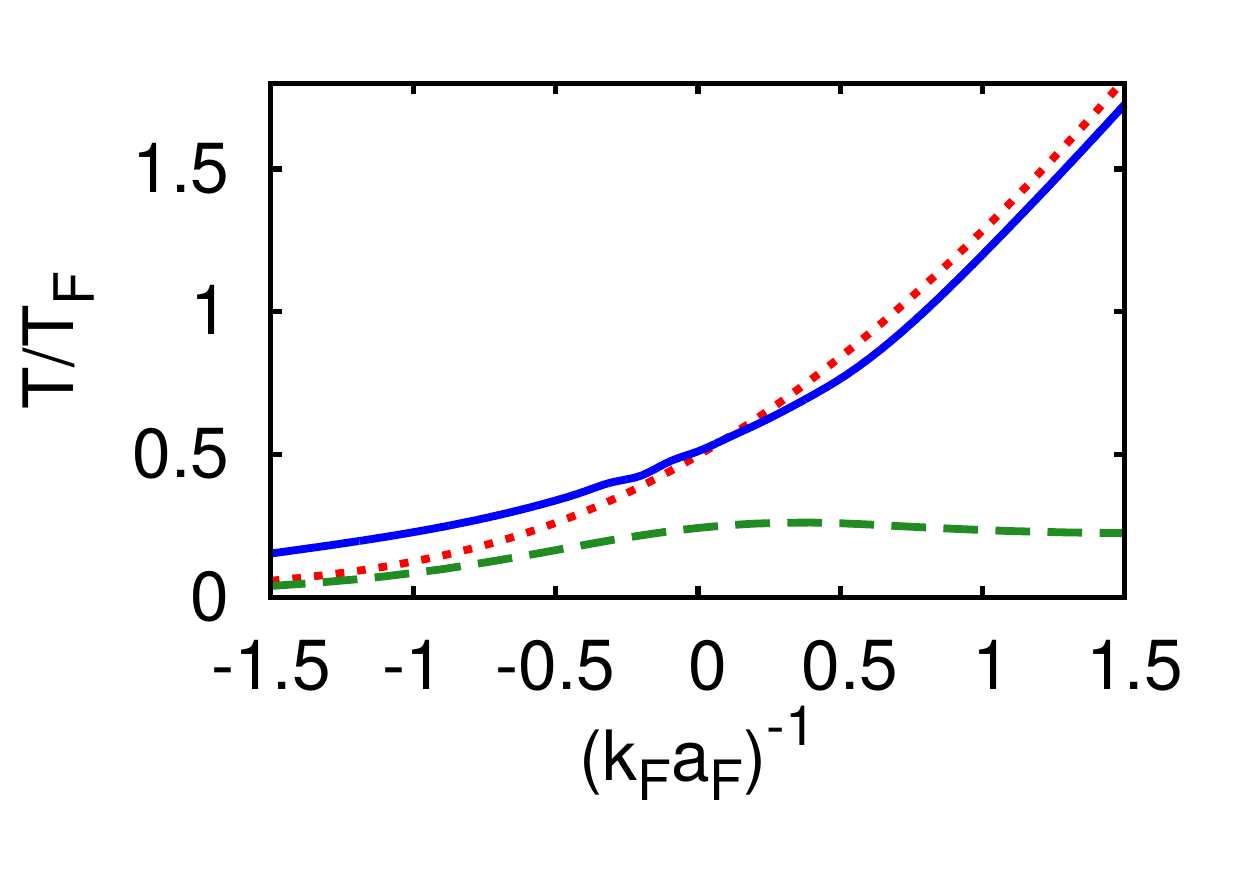}
\caption{The coupling dependence of the crossover temperature $T^{*}$ (full line) is compared with that of the ``pair dissociation temperature'' obtained at the mean-field level (dotted line).  
Also shown is the critical temperature $T_c$ obtained within the $t$-matrix approximation (dashed line).
[Reproduced from Ref.~\cite{Palestini-2014}.]}
\label{Figure-12}
\end{center}
\end{figure}

The temperature dependence of both lengths $\xi_{\mathrm{pair}}$ and $\xi$ is shown in Fig.~\ref{Figure-11} for three characteristic couplings throughout the BCS-BEC crossover.  
Owing to the steeper temperature dependence of $\xi$ with respect to $\xi_{\mathrm{pair}}$, at a given coupling these curves are bound to cross each other at a characteristic temperature $T^{*}$.  
This temperature, which can thus be obtained at any coupling throughout the BCS-BEC crossover, has the meaning of a ``crossover temperature'' below which inter-pair correlations are built up 
out of intra-pair correlations that are already present above this temperature.
[The values of $\xi_{\mathrm{pair}}$ in Fig.~\ref{Figure-10} for $T \rightarrow T_{c}^{-}$ slightly differ from those in Fig.~\ref{Figure-11} for $T \rightarrow T_{c}^{+}$,
because Fig.~\ref{Figure-10} is obtained at the mean-field level while Fig.~\ref{Figure-11} includes the effect of pairing fluctuations.]

The coupling dependence of the crossover temperature $T^{*}$ obtained
in this way is shown in Fig.~\ref{Figure-12} (full line), where it is
compared with the ``pair dissociation temperature'' (dotted line)
obtained in Section~\ref{sec:BCS_finite_T} at the mean-field level
(the latter was already reported as the dashed line in
Fig.~\ref{Figure-9}).  It is rather remarkable that the overall
coupling dependences of these two \emph{crossover} temperatures about
coincide with each other, even though they have been obtained through
different physical approaches.
With the present analysis based on the range of the correlation
functions, however, this crossover temperature acquires the more
physical meaning for the building up of inter-pair correlations out of
intra-pair correlations that develop above this temperature.

For this reason, and as already remarked in Section~\ref{sec:nsr},
precursor pairing phenomena (like the pseudo-gap energy to be
discussed in Section~\ref{sec:pseudo-gap}) are expected to occur in a
range of intermediate temperatures, which are above the critical
temperature $T_c$ but below the crossover temperature $T^{*}$.
    
\subsection{Single-particle spectral function and pseudo-gap}
\label{sec:pseudo-gap}

The term pseudo-gap was originally introduced to indicate a minimum in
the density of states at the Fermi level \cite{Mott-1969}.  As such,
it can be directly related to the shape of the \emph{single-particle
  spectral function} $A(\vek{k},\omega)$, which embodies the essential
information on the single-particle Green's function
$G(\vek{k},\omega_{n})$ through the spectral representation
\begin{equation}
G(\vek{k},\omega_{n}) = \int_{- \infty}^{+ \infty} \!\! d \omega \, \frac{A(\vek{k},\omega)}{i\omega_{n} - \omega} 
\label{spectral-representation}
\end{equation}
\noindent
and to which the following sum rule is associated:
\begin{equation}
\int_{- \infty}^{+ \infty} \!\! d \omega \, A(\vek{k},\omega) = 1 \, .
\label{sum-rule}
\end{equation}

\begin{figure}[t]
\begin{center}
\includegraphics[width=16cm,angle=0]{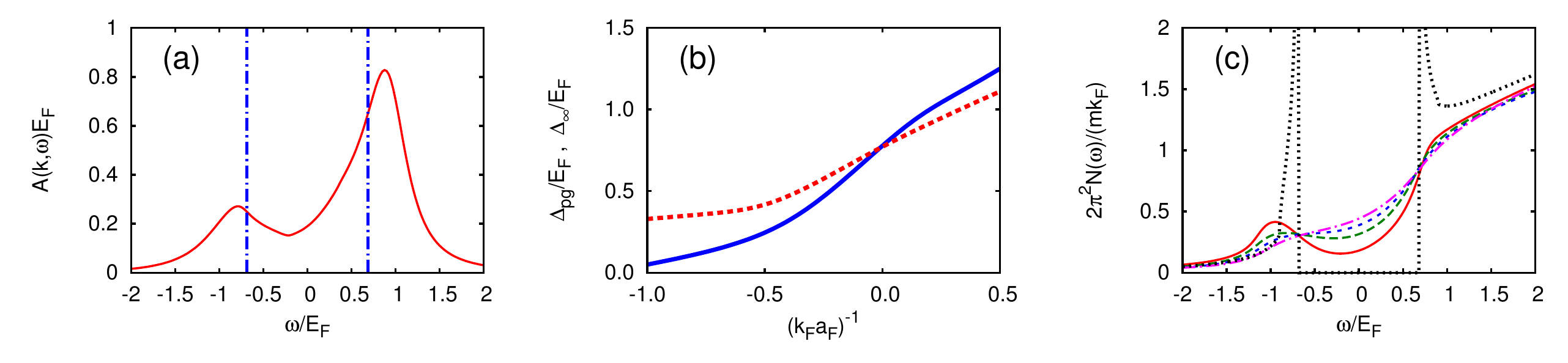}
\vspace{-0.5cm}
\caption{(a) Single-particle spectral function $A(k,\omega)$ vs $\omega$ at unitarity, as obtained at $T = 0$ within mean field (dashed-dotted lines) and at $T_c$ with the inclusion of pairing
fluctuations (full line).  
The corresponding wave vectors are $k = 0.76 \, k_F$ and $k = \, 0.91 k_F$, at which the maximum of the lower branch of the dispersion relation occurs in the two cases.
(b) Pseudo-gap $\Delta_{\rm pg}$  (full line) and parameter $\Delta_{\infty}$ related to short-distance
physics (dashed-line) vs.~the coupling $(k_Fa_F)^{-1}$ at $T_c$. 
(c) Density of states $N(\omega)$ vs $\omega$ calculated within the $t$-matrix approximation at unitarity and various temperatures:
$T_c$ (full line), $1.2 \, T_c$ (long-dashed line), $1.4 \, T_c$ (short-dashed line), and $1.65 \, T_c$ (dot-dashed line).
The corresponding mean-field result at $T = 0$ is also reported (dotted line).  
The non-interacting value $N_0=m k_F /( 2 \pi^2)$ of the density of states at the Fermi surface is used for normalization.
[Adapted from Refs.~\cite{Palestini-2012} and \cite{Palestini-2010}.]}
\label{Figure-13}
\end{center}
\end{figure}

Analogous considerations apply to the superfluid phase below $T_c$
for a Fermi gas with an attractive inter-particle interaction.  In
particular, within BCS mean-field one gets:
\begin{equation}
\mathcal{G}_{11}(\vek{k},\omega_{n}) = \frac{\varv_{\vek{k}}^2}{i\omega_{n} + E_{\vek{k}}} \, + \, \frac{u_{\vek{k}}^2}{i\omega_{n} - E_{\vek{k}}} 
\label{G-11-BCS}
\end{equation}
\noindent
with reference to the upper diagonal element of
Eq.(\ref{Green's-functions-BdG}) (once specified to a homogeneous
system and using the notation introduced after
Eq.(\ref{BdG-equations-homogeneous})).  The corresponding
single-particle spectral function
\begin{equation}
A(\vek{k},\omega) = \varv_{\vek{k}}^2 \, \delta(\omega + E_{\vek{k}}) \, + u_{\vek{k}}^2 \, \delta(\omega - E_{\vek{k}}) 
\label{BCS-single-particle-spectral-function}
\end{equation}
\noindent
contains two delta-like peaks symmetrically located about $\omega = 0$
(although with different weights), which correspond to the upper ($+
E_{\vek{k}}$) and lower ($- E_{\vek{k}}$) branches of the BCS
dispersion.  For values of $k = |\vek{k}|$ that satisfy $\xi_{\vek{k}}
= k^2/(2m) - \mu = 0$ such that $E_{\vek{k}} = |\Delta|$, the spacing
between these two peaks provides a measure of $2 |\Delta|$.  This
situation is represented by the dashed-dotted lines in
Fig.~\ref{Figure-13}(a) for the case of unitarity and $T = 0$.  Within
the BCS mean-field, when $T$ approaches $T_c$ and $|\Delta|$
vanishes these two peaks merge eventually into a single
(free-particle) peak centered at $\omega = 0$.

The situation changes when including pairing fluctuations, which give
rise to precursor pairing effects above $T_c$.  Extensive studies
along these lines \cite{Palestini-2012} have, in particular, been
performed within the $t$-matrix approximation discussed in
Section~\ref{sec:nsr}, whereby the single-particle Green's function
$G(\vek{k},\omega)$ contains the self-energy
(\ref{Sigma-NSR}). 
The formation of a pseudo-gap was studied also in nuclear physics and calculated for nuclear matter in Ref.~\cite{Schnell-1999} (cf. Fig.~\ref{fig:SchnellPRL83pseudo-gap} in
Section~\ref{sec:pn_finite_T}).
As it was already
mentioned in Section \ref{sec:nsr}, to get the single-particle
spectral function $A(\vek{k},\omega)$ it is essential for the
self-energy (\ref{Sigma-NSR}) to appear in the denominator of
$G(\vek{k},\omega)$ like in Eq.(\ref{Dyson}) and not in the expanded
form (\ref{Dyson1storder}).  Otherwise, the poles of
$G(\vek{k},\omega)$ in the complex $\omega$-plane cannot be located in
a physically meaningful way.  
Figure~\ref{Figure-13}(a) shows the
corresponding calculation at unitarity and $T_c$, where two broad
and overlapping peaks now appear in $A(\vek{k},\omega)$ in the place
of the two delta-like peaks obtained at $T=0$ within mean field.  This
result evidences how a real gap at $T = 0$ transforms into a
pseudo-gap at $T_c$, through a partial filling of the spectral
function in between the two peaks.

As an example, Fig.~\ref{Figure-13}(b) shows the coupling dependence of the pseudo-gap $\Delta_{\rm pg}$
at $T_c$ (full line), which can be extracted from the two-peak structure of $A({\bf k},\omega)$ shown in Fig.~\ref{Figure-13}(a) \cite{Palestini-2010}.
For comparison, Fig.~\ref{Figure-13}(b) also reports the coupling dependence of the parameter $\Delta_{\infty}$, related to short-distance physics,
 introduced in Section 
\ref{sec:intra-inter-pair} (dashed line). Note that the two quantities $\Delta_{\rm pg}$ and $\Delta_{\infty}$ differ considerably from each other
over most of the coupling range. Physically, this difference is because  $\Delta_{\rm pg}$ is defined at
small wave vectors while $\Delta_\infty$ is defined at large wave vectors, such that one does not expect them to be correlated with each other. In this respect, the crossing of the two curves near unitarity is to be considered as accidental.
This coincidence has sometimes led in the literature to identify the expression (\ref{Delta-infinity}) for $\Delta_{\infty}$ with the pseudo-gap $\Delta_{\rm pg}$ \cite{Chen-2005,Tsuchiya-2011,Kashimura-2014}.

Reference to a specific wave vector can be avoided by performing an
average of $A(\vek{k},\omega)$ over all $\vek{k}$.  In particular,
this procedure may support to the presence of a pseudo-gap in cases
when the two peaks of $A(\vek{k},\omega)$ strongly overlap each other,
at least in some ranges of $\vek{k}$.  In this way one is led to
define the single-particle density of states:
\begin{equation}
N(\omega) \, = \, \int \!\! \frac{d\vek{k}}{(2 \pi)^3} \, A(\vek{k},\omega) \, .
\label{density-of-states}
\end{equation}
\noindent
Figure \ref{Figure-13}(c) shows plots of $N(\omega)$ at unitarity for
several temperatures at and above $T_c$ \cite{Palestini-2012}.
Here, the depression of $N(\omega)$ about $\omega = 0$ confirms the
existence of a pseudo-gap, which is seen to progressively disappear at
temperatures somewhat below the crossover temperature $T^{*}$ (in the
present case, $T^{*}$ is approximately $2 \, T_c$).  For comparison,
also reported in Fig.~\ref{Figure-13}(c) is the density of states
obtained within mean field at $T=0$, which shows two sharp peaks
located at $\pm |\Delta|$.  Similar results have also been obtained in
two dimensions \cite{Marsiglio-2015}, which are qualitatively similar
to the ones shown here in three dimensions.  However, the temperature
window above $T_c$ in which precursor pairing occurs is wider in two
than in three dimensions, owing to the increased importance of
fluctuations. 

When spanning the BCS-BEC crossover so as to approach the BEC limit,
the question naturally arises about \emph{the boundary} between the
pseudo-gap (many-body) regime and the molecular (two-body) regime.
This is because in the BEC limit when all fermions are bound in pairs,
the depression of the single-particle density of states would just
reflect the presence of the molecular (two-body) bound state.  It is
then clear that this molecular regime should not be confused with the
pseudo-gap regime of interest, which originates from genuine many-body
effects and is not simply a manifestation of two-body binding.  This
is even more so in two dimensions, where a two-body bound state occurs
for any value of the inter-particle coupling.  To resolve this issue,
a careful analysis is required on the shape of the lower branch of the
dispersion relation for given coupling and temperature above $T_c$.
Accordingly, in Ref.~\cite{Perali-2011} the position of the wave
vector was determined, at which the maximum of the lower branch occurs
and past which the branch eventually backbends, identifying in this
way a characteristic wave vector $k_{L}$ called the Luttinger wave
vector.  By this analysis, it was found that the value of $k_{L}$
remains finite (near $k_F$) even past unitarity where the gas is
strongly interacting, thereby signaling the presence of a remnant
Fermi surface and the importance of the Fermi statistics to pairing.
Afterwards once entering the molecular regime, $k_{L}$ rapidly
decreases toward zero and vanishes at about $(k_F a_F)^{-1}\simeq 0.5$
\cite{Perali-2011}.  The conclusion was that the boundary between the
pseudo-gap and molecular regimes resides just where $k_{L}$ vanishes
and the underlying Fermi surface disappears. 
This criterion generalizes to finite temperature and beyond mean field the
zero-temperature mean-field criterion, which identifies the boundary between the BCS and BEC sides of the crossover
with the change of the sign of the chemical potential $\mu$ \cite{Leggett-1980} (see \cite{Perali-2011,Marsiglio-2015} for a discussion of 
the connections and differences between these two criteria). 

\subsection{Gor'kov-Melik-Barkhudarov (screening) corrections}
\label{sec:gmb}

In the weak-coupling (BCS) limit, the exponential dependence of
$\Delta_0$ and $T_c$ on $(k_Fa_F)^{-1}$ (given, respectively, by
Eq.~(\ref{Delta-exact}) and by Eq.~(\ref{T-c-w-c}), whereby
$\Delta_0/T_c = \pi/e^{\gamma}$) leaves open the possibility of the
presence of sub-leading contributions in the exponent beyond the
leading contribution $(k_F a_F)^{-1}$, which may thus affect the
pre-factors in the expressions of $\Delta_0$ and $T_c$.  That this
is actually the case was proved by Gor'kov and Melik-Barkhudarov (GMB)
\cite{Gorkov-1961} not long after the BCS result, by introducing
corrections to the BCS mean field.  We provide here a slightly
different derivation of their original result, by focusing
specifically on the correction to $T_c$ for which one can exploit
the pairing fluctuations approach of Section~\ref{sec:nsr} valid for
the normal phase.

We recall that the expression (\ref{T-c-w-c}) for the superfluid
critical temperature $T_c$ can be alternatively obtained by
approaching $T_c$ from above in terms of the Thouless criterion
\cite{Thouless-1960}.  According to this criterion, the instability of
the normal phase when approaching $T_c$ is signaled by the
divergence at $Q=0$ of the particle-particle propagator (or
$t$-matrix) $\Gamma_0(Q)$ given by Eq.~(\ref{Gamma-0}).  This is
equivalent to the condition $m/(4\pi a_F) + R_{\mathrm{pp}}(Q=0) = 0$
in terms of the regularized particle-particle bubble
$R_{\mathrm{pp}}(Q)$ of Eq.~(\ref{bubble-pp-exact}).  In weak coupling
for this quantity one obtains the value:
\begin{equation}
R_{\mathrm{pp}}(Q=0) \simeq N_0 \ln\left(\frac{8 e^{\gamma} E_F} {\pi e^2 T} \right)
\label{approximate-pp-bubble}
\end{equation}
\noindent
where $N_0= m k_F/(2 \pi^2)$, so that the BCS result (\ref{T-c-w-c})
for $T_c$ is readily recovered.

\begin{figure}[h]
\begin{center}
\includegraphics[width=11.5cm,angle=0]{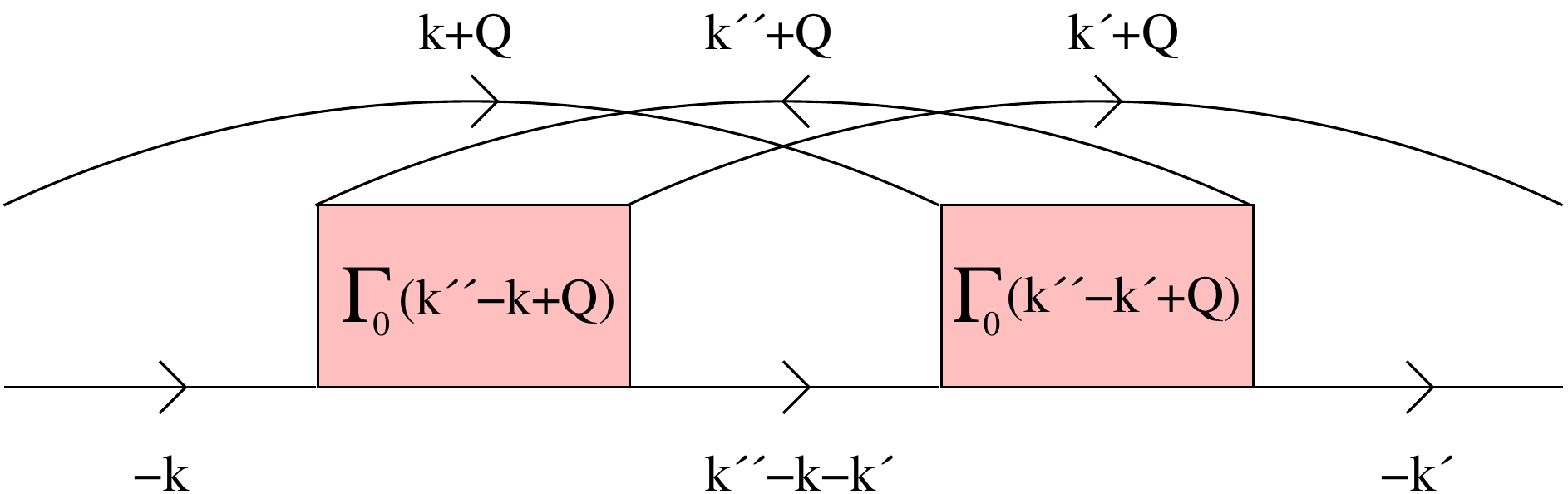}
\caption{Graphical representation of the bosonic-like self-energy $\Sigma_{\mathrm{GMB}}^{\rm pp}(Q)$ that enters the condition (\ref{GMB-condition}) for the critical temperature, 
according to the GMB approach.  
Boxes stand for the particle-particle propagator $\Gamma_0$ and thin lines for the bare fermionic single-particle propagator $G_0(k)$.}
\label{Figure-14}
\end{center}
\end{figure}

To go beyond the BCS result for $T_c$, the above form of the Thouless criterion needs to be modified by including particle-particle (${\rm pp}$) ``self-energy''
effects in the particle-particle propagator $\Gamma_0(Q)$.  
In their simplest form, these effects are represented by the crossed-ladder
diagram $\Sigma_{\mathrm{GMB}}^{\rm pp}(Q)$ depicted in Fig.~\ref{Figure-14},
which was originally identified in Ref.~\cite{Gorkov-1961} at the
lowest significant order in the small parameter $k_F a_F$.
In this way, the modified criterion for the critical temperature reads:
\begin{equation}
\Gamma_0(Q=0)^{-1} - \, \Sigma_{\mathrm{GMB}}^{\rm pp}(Q=0) = 0 
\label{GMB-condition}
\end{equation}
\noindent
with $- \Gamma_0(Q=0)^{-1} = m/(4 \pi a_F) + R_{\mathrm{pp}}(Q=0)$
according to Eq.(\ref{Gamma-0}).  Note how the criterion
(\ref{GMB-condition}) resembles the Hugenholtz-Pines condition for the
critical temperature of a system of interacting bosons
\cite{Hugenholtz-1959}.

The quantity $\Sigma_{\mathrm{GMB}}^{\rm pp}(Q=0)$ can be readily calculated in
the weak-coupling limit ($k_Fa_F)^{-1} \ll -1$ through the following
steps \cite{Gorkov-1961}: (i) Both particle-particle propagators
occurring in Fig.~\ref{Figure-14} are approximated by the constant
value $\Gamma_0 \simeq - 4 \pi a_F/m$, according to a standard result
of the Galitskii approach \cite{Galitskii-1958}; (ii) Owing to this
approximation, the particle-hole (or polarization) bubble
$\chi_{\mathrm{ph}}(k+k')$
\begin{equation}
\chi_{\mathrm{ph}}(k+k') = \int \! \frac{d\vek{k''}}{(2 \pi)^3} \frac{f(\xi_{\vek{k}+\vek{k}'+\vek{k''}}) - f(\xi_{\vek{k''}})} 
{\xi_{\vek{k}+\vek{k}'+\vek{k''}}-\xi_{\vek{k''}}-i(\omega_{n}+\omega_{n'})} \,,
\label{p-h-polarization}
\end{equation}
\noindent
which is nested in the central part of the diagram of
Fig.~\ref{Figure-14}, gets effectively disentangled from the two
particle-particle bubbles $\chi_{\mathrm{pp}}(Q=0)$ which are present
on the sides of the diagram; (iii) To avoid an ensuing ultraviolet
divergence in the particle-particle channel, $\chi_{\mathrm{pp}}$ is
then replaced by its regularized version $R_{\mathrm{pp}}$ according
to Eq.~(\ref{bubble-pp-exact}) (in practice, this procedure is
equivalent to introducing a cutoff of the order $k_F$ in the
integration over $|\vek{k}|$); (iv) Correspondingly, in the
particle-hole bubble (\ref{p-h-polarization}) one takes the
zero-temperature limit (whereby $\omega_{n}+\omega_{n'}=0$), sets
$\mu=E_F$, and performs an average over the Fermi sphere, to yield:
\begin{equation}
\bar{\chi}_{\mathrm{ph}}(0) = - N_0 \ln(4e)^{1/3} \, ;
\label{p-h-polarization-approximate}
\end{equation}
\noindent
(v) Grouping all the above approximate results, one obtains eventually
\begin{equation}
\Sigma_{\mathrm{GMB}}^{\rm pp}(Q=0) \simeq \left(\frac{4\pi a_F}{m}\right)^2 R_{\mathrm{pp}}(Q=0)^2 \, \bar{\chi}_{\mathrm{ph}}(0) 
\label{Sigma_GMB-approximate}
\end{equation}
\noindent
in such a way that the condition (\ref{GMB-condition}) is approximated
by the following expression:
\begin{equation}
\frac{m}{4 \pi a_F} + R_{\mathrm{pp}}(Q=0) + \left(\frac{4 \pi a_F}{m} \right)^2 R_{\mathrm{pp}}(Q=0)^2 \, \bar{\chi}_{\mathrm{ph}}(0) = 0 \, ;
\label{GMB-condition-approximate}
\end{equation}
\noindent
(vi) In weak coupling, one can further approximate $(4 \pi a_F/m) \,
R_{\mathrm{pp}}(Q=0) \simeq -1$ in the last term on the left-hand side
of Eq.~(\ref{GMB-condition-approximate}).  In this way, the equation
for $T_c$ reduces to the simple form:
\begin{equation}
\frac{m}{4 \pi a_F} + R_{\mathrm{pp}}(Q=0) + \bar{\chi}_{\mathrm{ph}}(0) = 0 \, .
\label{GMB-condition-approximate-final}
\end{equation}
\noindent
With the expressions (\ref{approximate-pp-bubble}) for
$R_{\mathrm{pp}}(Q=0)$ and (\ref{p-h-polarization-approximate}) for
$\bar{\chi}_{\mathrm{ph}}(0)$,
Eq.~(\ref{GMB-condition-approximate-final}) yields eventually the
result:
\begin{equation}
T_c = \frac{1}{(4e)^{1/3}} \, \frac{8 e^{\gamma} E_F}{\pi e^2} \, \exp [\pi/(2 k_F a_F)]                      
\label{Tc-GMB}
\end{equation}
\noindent
which is smaller by the factor $(4e)^{1/3} \simeq 2.2$ with respect to
the BCS result (\ref{T-c-w-c}).  Note that it is for the presence in
Eq.~(\ref{GMB-condition-approximate-final}) of the particle-hole (or
polarization) bubble $\chi_{\mathrm{ph}}$ (which gives the leading
contribution to the screening in an electron gas \cite{Fetter-1971})
that the GMB result is often referred to as a ``screening correction''
to $T_c$.

By a similar token, particle-hole corrections affect also the value of
the BCS gap, by replacing the term $m/(4 \pi a_F)$ on the right-hand
side of Eq.(\ref{gap-homogeneous}) with $m/(4 \pi a_F) +
\bar{\chi}_{\mathrm{ph}}(0)$ just like in
Eq.~(\ref{GMB-condition-approximate-final}).  At zero-temperature,
this yields a reduction of the BCS result $\Delta_0 = (8 E_F/e^2)
\exp[\pi/(2k_Fa_F)]$ by the same factor of $2.2$ occurring in
Eq.~(\ref{Tc-GMB}), thereby maintaining the value of the ratio
$\Delta_0 / T_c = \pi / e^{\gamma}$ of the BCS theory.

In the context of the BCS-BEC crossover, it would be of interest
\cite{Bulgac-2006} to assess whether corrections of the GMB type (as
embodied by the diagram of Fig.~\ref{Figure-14}) can still yield
significant effects on $T_c$ and $\Delta$, when carried over to the
unitary limit $(k_Fa_F)^{-1}=0$ where one knows that a remnant Fermi
surface is still active (cf. Section~\ref{sec:pseudo-gap}). A partial
answer to this question was provided in
Refs.~\cite{Yu-2009a,Ruan-2013}, where the expression
(\ref{GMB-condition-approximate-final}) was taken to hold for all
couplings across the BCS-BEC crossover, with the only provision of
using the appropriate value of the chemical potential (and not simply
$E_F$) when calculating $\bar{\chi}_{\mathrm{ph}}(0) = 0$.  
 [A completely different approach was instead followed in
  Refs.~\cite{Floerchinger-2008,Floerchinger-2010,Tanizaki-2014}, where particle-particle and
  particle-hole fluctuations were included simultaneously in the framework
  of the functional renormalization-group approach.]  
 To get a complete answer to this question, however, the typical approximations of the
weak-coupling (BCS) limit that led to the expression
(\ref{GMB-condition-approximate-final}) should be abandoned and the
full wave-vector and frequency dependence of the particle-particle
propagator $\Gamma_0(Q)$ should be retained in the calculation of
$\Sigma_{\mathrm{GMB}}^{\rm pp}$ in Fig.~\ref{Figure-14}.  
Work along these lines has recently appeared \cite{Pisani-2018}.

In any case, the fact that
experimental \cite{Nascimbene-2010,Ku-2012} and Quantum Monte Carlo
(QMC) \cite{Burovski-2006a,Burovski-2008,Goulko-2010} results for
$T_c$ at unitarity are only by about $25\%$ lower than the
NSR result indicates that screening effects on top of the $t$-matrix
are less important at unitarity than in the weak-coupling (BCS)
regime.

In nuclear physics, the GMB corrections were instead considered for the gap parameter $\Delta$ \cite{Schulze-2001}.  
With more than two species of fermions, the reduction of the gap can turn over into an enhancement \cite{Heiselberg-2000}.  
In nuclear and neutron matters, screening of the pairing force has also been considered at higher densities \cite{Cao-2006}.  
Not only the particle-hole bubble exchange has been considered, but these bubbles have been summed up to the full RPA. 
Once polarization effects are no longer treated to the lowest order as within the GMB approach, the problem arises that the screening of the force and the coupling of 
the single-particle motion to the two-particle excitations have to be treated on the same footing.  
This introduces a single-particle self-energy energy $\Sigma(\vek{k},\omega)$ that depends on wave vector and frequency, a feature that has to be taken into account because 
the Fermi step gets reduced by the wave-function renormalization factor
\begin{equation}
Z^{-1}_{\vek{k}} = 1-\left. \frac{\partial\Sigma(\vek{k},\omega)} {\partial \omega} \right|_{\omega = \xi_{\vek{k}}} \, .
\label{Z-factor}
\end{equation}
The gap equation modified in this way reads \cite{Baldo-2000}:
\begin{equation}
\Delta_{\vek{k}} = \int \frac {d\vek{k}'}{(2\pi)^3} V_{\mathrm{eff}}(\vek{k},\vek{k}') \frac{Z_{\vek{k}}Z_{\vek{k}'}\Delta_{\vek{k}'}} {2\sqrt{\xi_{\vek{k}'}^2+\Delta_{\vek{k}'}^2}}
\label{screened-gap}
\end{equation}
where the effective potential $V_{\mathrm{eff}}$ contains the RPA
screening correction [cf. diagrams (b) and (c) of Fig.~\ref{fig:screened-force} in Section \ref{sec:nn_cold}] in addition to the Born term [cf. diagram (a) of Fig.~\ref{fig:screened-force}
in Section \ref{sec:nn_cold}]. 
In neutron matter this kind of
  screening reduces the gap as a function of $k_F$ by about 30$\%$,
  bringing the results in quite close agreement with QMC calculations
  \cite{Gezerlis-2010} (see Sect.~\ref{sec:nn_cold}).  In symmetric
  nuclear matter an anti-screening effect also exists, because in this
  case attractive density fluctuations become more important than spin
  fluctuations.  This enhancement effect, however, is over-compensated
  by the particle-phonon coupling in the self-energy, so that in the
  end the gap becomes even smaller than in the BCS approximation.  In
  the case of symmetric nuclear matter no QMC calculations exist to
  compare with.

\subsection{NSR theory in nuclear physics and the approach by Zimmermann and Stolz}
\label{sec:ZS}
In nuclear physics, too, the $t$-matrix (sometimes referred to also as $T$-matrix) approach has a long history, to treat bound state formation and correlations at finite temperatures \cite{Roepke-1982,Roepke-1983}. 
Based on the work by Zimmermann and Stolz (ZS) \cite{Zimmermann-1985}, an improved Beth-Uhlenbeck equation for hot nuclear matter was treated with the $t$-matrix \cite{Schmidt-1990}. 
In 1995 the NSR approach was applied to nuclear matter for the treatment of the BEC-BCS transition in the case of proton-neutron pairing, which leads to the bound state
of the deuteron in strong coupling \cite{Stein-1995}.

In the ZS approach of Ref.\cite{Zimmermann-1985}, the self-energy $\Sigma$ is given by the diagram of Fig.~\ref{Figure-8}(a) similar to the NSR approach, albeit with an important difference.  
The ZS approach is based on a perturbative treatment, where the Dyson equation is truncated at first order in the self-energy like in Eq.(\ref{Dyson1storder}) for the NSR approach,
that is, $G \approx G_0 + G_0^2 \, \Sigma$ where $\Sigma$ denotes the self-energy calculated in ladder approximation with single-particle propagators $G_0$.  
The difference is that now $G_0$ is not the bare propagator $1/(i \omega_n -\xi_{\vek{k}})$ of the NSR approach, but is a quasi-particle propagator that takes into account a mean-field-like
quasi-particle energy shift due to $\Sigma$, that is, $G_0 = 1/(i\omega_n-\xi^*_{\vek{k}})$ where $\xi^*_{\vek{k}} = \xi_{\vek{k}} + \Sigma^R(\vek{k},\xi^*_{\vek{k}})$ and $\Sigma^R$ is the retarded
self-energy.  
This is because in nuclear physics it is crucial to include the (Hartree-Fock) mean-field shift in the single-particle energies \cite{Jin-2010}.  
Also in the case of a contact interaction, however, the importance of including a mean-field shift was pointed out in Refs.~\cite{Perali-2002,Pieri-2004b} for the BCS side of the
crossover and in Ref.~\cite{Pantel-2014} for the whole BCS-BEC crossover.  
With the above replacement $\xi \rightarrow \xi^*$, one ends up with an expression for the correlated density slightly different from that of the NSR approach, of the form
\begin{equation}
n_{\text{corr}} = - \! \int \frac{d\vek{Q}}{(2\pi)^3} \! \int \frac{d\Omega}{\pi} \, b'(\Omega) \, \bigg(\delta(Q,\Omega)- \frac{1}{2}\sin[2\delta(Q,\Omega)]\bigg) 
\label{rhocorrzs}
\end{equation}
where $b'(\Omega)$ is the derivative of the Bose function $b(\Omega)
= 1/(e^{\Omega/T}-1)$ with respect to $\Omega$ and $\delta(Q,\Omega)=
\Imag \ln[R_{\mathrm{pp}}(Q,\Omega) + m/4\pi a_F]$ is the in-medium
scattering phase shift with $R_{\rm pp}$ given by
Eq.(\ref{bubble-pp-exact}).  In the presence of a bound state (for
which $a_F > 0$), $\delta = \pi$ in the energy range between the
bound-state energy and the continuum threshold.  The expression
(\ref{rhocorrzs}) for the correlated density was first used to
describe the BCS-BEC crossover in nuclear matter
\cite{Stein-1995,Jin-2010}, and was later used for a contact
interaction \cite{Pantel-2014}.

\begin{figure}[t]
\begin{center}
\parbox{6.5cm}{\includegraphics[width=8.0cm]{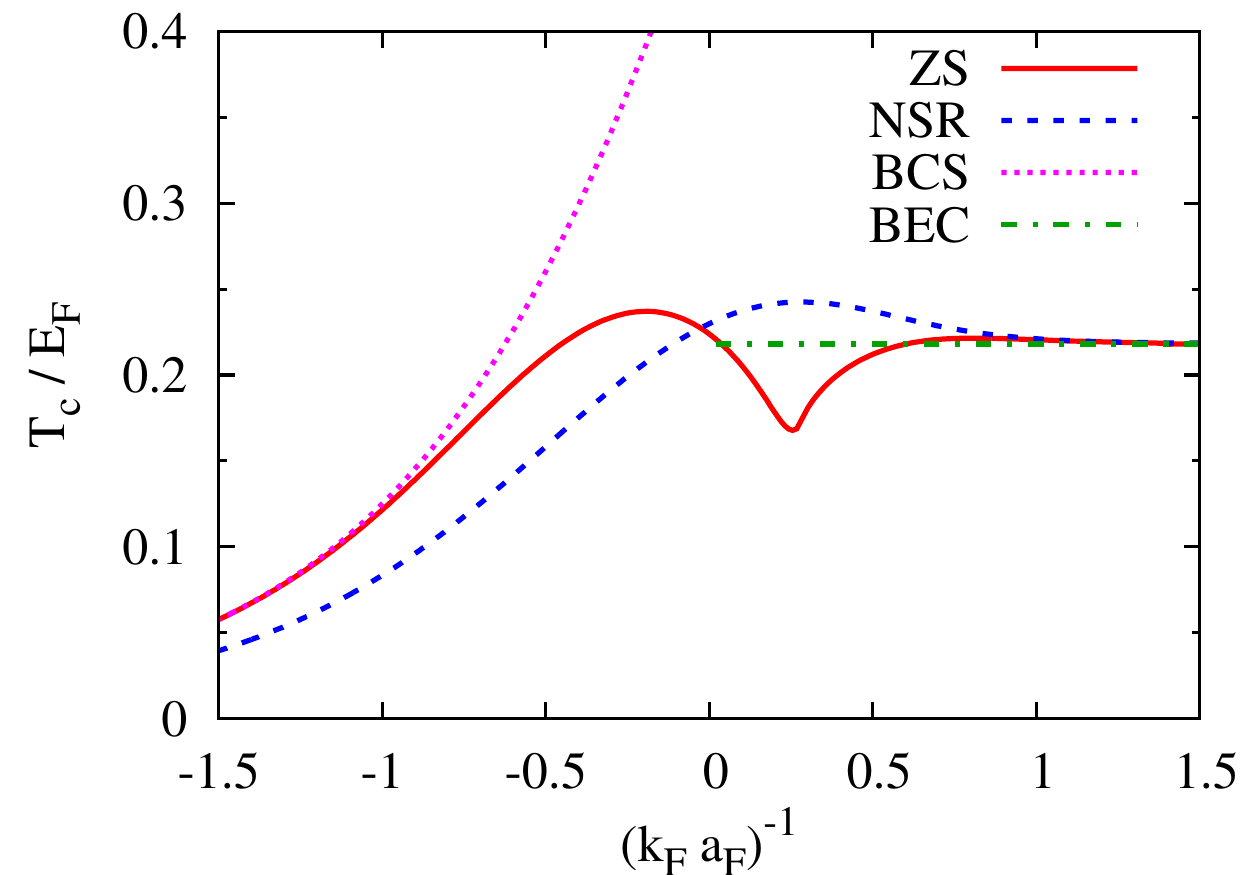}}
\end{center}
\caption{The dependence of the critical temperature $T_c$ (in units of $E_F$) on the interaction parameter $(k_Fa_F)^{-1}$ is compared for several approximations: ZS approach (solid line); 
NSR approach (dashed line); BCS result (dotted line); BEC limit (dashed-dotted line).  
The NSR result of Fig.~\ref{Figure-9} has been also reported for convenience.
[Adapted from Ref.~\cite{Pantel-2014}.]}
\label{Figure-15}
\end{figure}

Figure \ref{Figure-15} shows the coupling dependence of the critical
temperature obtained by the ZS approach, which interpolates between
the BCS and BEC limits, and compares it with that obtained by other
approaches.  Note that, while in the BCS limit the ZS approach
converges to the BCS result (\ref{T-c-w-c}), the NSR approach (which
neglects the mean-field shift) converges to the BCS result divided by
$e^{1/3}$, as already mentioned in Section~\ref{sec:nsr}.  In Section
\ref{sec:nuclearsystems} the ZS approach will be employed to describe
the deuteron case in nuclear physics.

\subsection{Residual interaction among composite bosons}
\label{sec:b-b-interaction}

In the (extreme) BEC limit of the BCS-BEC crossover, all fermions get
paired into tight composite bosons (dimers) which interact with each
other through a residual (albeit small) interaction.  To the extent
that this gas of composite bosons is dilute, at low energy the
residual interaction can be characterized by the boson-boson
scattering length $a_{B}$, in a similar fashion as the inter-particle
interaction of the original fermionic system was characterized by the
fermionic scattering length $a_F$.  It is also clear that $a_{B}$ is
bound to be proportional to $a_F$ with a constant coefficient, in such
a way that $a_{B}$ vanishes simultaneously with $a_F$ in the extreme
BEC limit when $(k_F a_F)^{-1} \gg +1$.

From a physical point of view, the progressive loss of fermionic
character of the system when approaching the BEC limit becomes
apparent by looking at the shape of the single-particle spectral
function discussed in Section~\ref{sec:pseudo-gap}.  When approaching
the BEC limit, the left (hole) peak at negative frequency in the
single-particle spectral function looses progressively its weight because all particles are bound in pairs
\cite{Pieri-2004a,Pieri-2004b,Palestini-2012}.

\begin{figure}[t]
\begin{center}
\includegraphics[width=15cm,angle=0]{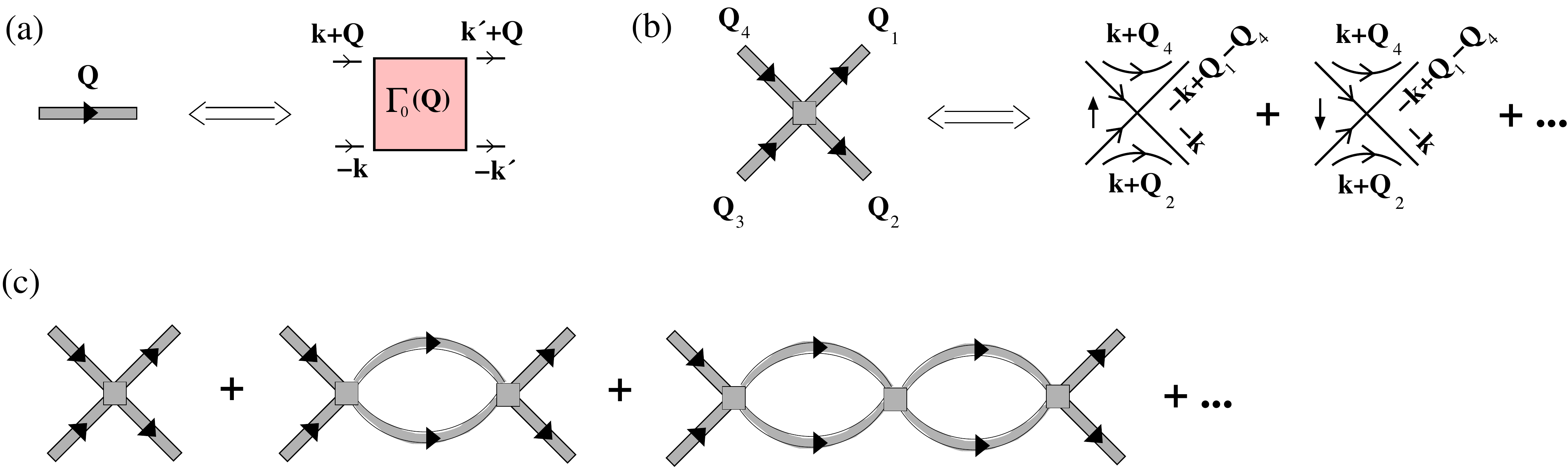}
\caption{(a) Correspondence between the bare bosonic propagator (left) and the particle-particle propagator (right).  
(b) The dimer-dimer residual interaction is represented graphically by a block (where $Q_{1}+Q_{2}=Q_{3}+Q_{4}$ in four-vector notation).  
The first two terms of the dimer-dimer residual interaction are explicitly shown (right), where thin lines represent bare fermionic single-particle propagators $G_0$ 
and fermionic spin labels are indicated in the internal lines.  
The dimer-dimer residual interaction contains, in principle, an additional infinite number of terms (dots), where in the intermediate states one of the dimers is broken 
into its fermionic constituents.  
(c) $T$-matrix for the dimer-dimer scattering problem, where successive blocks are connected through a sequence of dimer-dimer bubbles.}
\label{Figure-16}
\end{center}
\end{figure}

In terms of the diagrammatic structure, the above features are
reflected in the facts that: (i) The only diagrams that survive the
regularization procedure (\ref{regularization}) are those which are
built directly on the particle-particle propagator $\Gamma_0$ given by
the expression (\ref{Gamma-0}); (ii) In the BEC limit, the
particle-particle propagator $\Gamma_0$ is proportional to the bare
bosonic (dimer) propagator [cf. Eq.~(\ref{Gamma-0-BEC_limit})].  The
correspondence between these two propagators is shown graphically in
Fig.~\ref{Figure-16}(a).

To construct the full diagrammatic structure in the BEC limit, one has
then to connect different particle-particle
propagators $\Gamma_0$ with each other through four-leg structures, which, when taken
together, correspond to an effective residual interaction between the
composite bosons.  These structures (which are infinite in number)
correspond to all different kinds of virtual processes by which the
fermionic character of the constituent particles manifests itself when
two composite bosons undergo isolated scattering events.  Accordingly,
the associated diagrams are calculated in the limit of vanishing
density.  The collection of all these diagrammatic structures is
schematically represented in Fig.~\ref{Figure-16}(b) by a block which
connects two incoming and two outgoing composite bosons.

The first two terms contributing to the residual interaction among
composite bosons are shown in Fig.~\ref{Figure-16}(b).  When evaluated
in the limit of vanishing four-vectors, these terms yield the value $8
\pi a_F/(2 m)$ of the coefficient of the cubic term in the GP equation
(\ref{GP-equation}), thereby implying the result $a_{B} = 2 a_F$
between the fermionic and bosonic scattering lengths.  This result
corresponds to the Born approximation for the dimer-dimer scattering
problem \cite{Sa-de-Melo-1993,Haussmann-1993,Haussmann-1994}.  It
turns out that this result can be improved by introducing the
dimer-dimer $T$-matrix which considers a sequence of
dimer-dimer scatterings instead of a single scattering, as shown
schematically in Fig.~\ref{Figure-16}(c).  [Note that here the label
 $T$-matrix is meant to distinguish this
  dimer-dimer sequence from the $t$-matrix ($\Gamma_0$) of the
  constituent fermions.]  The calculation of the dimer-dimer
$T$-matrix (and thus of $a_{B}$) is then reduced to
the calculation of two infinite perturbation series, namely, the
series that builds up the dimer-dimer residual interaction
(cf. Fig.~\ref{Figure-16}(b)) and the series that obtains the complete
dimer-dimer $T$-matrix through an infinite sequence
of this dimer-dimer residual interaction
(cf. Fig.~\ref{Figure-16}(c)).  By this scheme, the Born approximation
corresponds to the first-order term of both perturbation series.

The dimer-dimer $T$-matrix beyond the Born
approximation was first considered in Ref.~\cite{Pieri-2000}.  There,
only the first two terms of the dimer-dimer residual interaction shown
explicitly in Fig.~\ref{Figure-16}(b) were retained and the
corresponding diagrammatic series for the dimer-dimer
$T$-matrix of Fig.~\ref{Figure-16}(c) was summed up,
obtaining the value $a_{B} \simeq 0.75a_F$ for the bosonic scattering
length.  Later on, a complete treatment was given in
Refs.~\cite{Petrov-2004,Petrov-2005} for the numerical solution of the
corresponding four-body Schr\"{o}dinger equation, yielding $a_{B}
\simeq 0.60a_F$.  This value of $a_{B}$ was subsequently confirmed in
Refs.~\cite{Brodsky-2005,Brodsky-2006}, where a full diagrammatic
treatment was given by summing up not only the series for the
dimer-dimer $T$-matrix of Fig.~\ref{Figure-16}(c)
but \emph{also} the infinite series for the residual dimer-dimer
interaction of Fig.~\ref{Figure-16}(b), and by lattice-effective-field theory-methods~\cite{Elhatisari-2017} which yielded in addition 
the estimate $r_e\simeq -0.43 a_F$ for the dimer-dimer effective range parameter. This value has recently been corrected to $r_e \simeq 0.13 a_F$  by the four-body calculation of Ref.~\cite{Deltuva-2017}), which has confirmed the previous results of Refs.~\cite{vonStecher-2008,D'Incao-2009}.
The problem was recently
reconsidered in Ref.~\cite{Alzetto-2013} for a two-component Fermi
mixture in which the two different fermionic species have different
masses.  It was found that, in the large mass ratio domain, the
dominant processes are just the repeated dimer-dimer scatterings
considered originally in Ref.~\cite{Pieri-2000} (that is, those
explicitly depicted in Figs.\ref{Figure-16}(b) and (c)), with the
conclusion that the ensuing approximation is asymptotically correct
for large mass ratio (and thus provides, in practice, a good
approximation for any mass ratio). 

It is worth emphasizing that the presence of the branch cut, which occurs in the BEC
limit at sufficiently high energy in the spectral representation
(\ref{spectral-representation-Gamma-0}) of the particle-particle
propagator $\Gamma_0$, is bound to give an important contribution to
the \emph{ab-initio} calculation of the boson-boson scattering length
$a_{B}$ in terms of the scattering length $a_F$ of the constituent
fermions.  On physical grounds, this feature is expected to occur
because the fermionc nature of the constituent particles of the
composite bosons has to show up just when these bosons collide against
each other, such that energies that are much higher than the kinetic
energy of the colliding bosons come into play in the intermediate
(virtual) states of the collision.  To some extent the logic here is
similar to Frank Wilczek's account for the origin of mass
\cite{Wilczek-2012}, when this quantity is calculated from first
principles and not merely regarded as a phenomenological parameter by
remaining in the low-energy sector where quantum mechanics effectively
suppresses the complexity of composite systems.

For all these reasons, the above results for the dimer-dimer
scattering obtained in the BEC limit of the BCS-BEC crossover can be
considered almost unique, to the extent that it has been possible to obtain from first principles the relevant
scattering properties of composite (bosonic) objects in terms of the
scattering properties of the constituent (fermionic) particles.  Only very recently, it has become possible
to describe also deuteron-deuteron scattering by solving the four-body Schr\"odinger equation for two neutrons and two protons \cite{Deltuva-2015}.
A corresponding information is not available, for instance, for a gas of
bosons like $^{4}\mathrm{He}$ atoms, whose inter-particle potential is
commonly modeled by a simplified phenomenological form (like the
hard-sphere model) and the bosons are considered to be effectively
point-like objects.

The above results for the dimer-dimer scattering were obtained in the limit of vanishing density, which is relevant to an extremely dilute gas of composite bosons in the BEC limit.  
At higher densities, where the Pauli principle starts to become important and the Fermi functions are no longer negligible, an in-medium corrected four-fermion equation
of the following form has to be considered (see Ref.~\cite{Roepke-1998}): 
\begin{align}
 [E-(\varepsilon_{k_1}+\varepsilon_{k_2}+\varepsilon_{k_3}+\varepsilon_{k_4})] \, \Psi_{k_1k_2k_3k_4} 
& = [1-f(\xi_{k_1})-f(\xi_{k_2})] \sum_{k'_1k'_2} \bar{\varv}_{k_1k_2k'_1k'_2} \Psi_{k'_1k'_2k_3k_4} 
\nonumber \\
& + [1-f(\xi_{k_1})-f(\xi_{k_3})] \sum_{k'_1k'_3} \bar{\varv}_{k_1k_3k'_1k'_3} \Psi_{k'_1k_2k'_3k_4}  + \mathrm{permutations} \, .
\label{4-body}
\end{align}
Here, $\bar \varv_{k_1k_2k_3k_4} = \langle k_1k_2|\varv|k_3k_4\rangle - \langle k_1k_2|\varv|k_4k_3\rangle$ is the anti-symmetrized matrix element of the two-body force 
and the indices $k_i = (\vek{k}_i, \sigma_i)$ contain momenta and spin (and possibly isospin, like in nuclear physics).  
The eigenvalue of this Schr\"odinger type of equation is denoted by $E$.
The phase-space factors $[1-f(\xi_{k_1})-f(\xi_{k_2})]$ are the same as in the two particle Bethe-Salpeter equation and take care of the fact that particles cannot scatter into points 
of phase space which are already occupied.  
The above in-medium four-fermion equation could be used to study the density dependence of $a_B$.  
It will be explicitly employed in Secion~\ref{sec:nuclearsystems} below to reveal quartet condensation (but it could as well be used to investigate bi-exciton versus exciton condensation in semiconductors).

\subsection{Pairing fluctuations in polarized Fermi gases}
\label{sec:nsr-polarized}

The NSR approach discussed in Section \ref{sec:nsr} is readily
extended to the polarized case with different spin populations, by
introducing two different chemical potentials $\mu_{\sigma}$ for the
two spin species $\sigma$ in the bare fermionic single-particle
propagators $G_{0 \sigma}$ that enter the particle-particle propagator
$\Gamma_0$ and the self-energy (\ref{Sigma-NSR}) (which thus depends
also on $\sigma$).  This approach, however, breaks down near
unitarity, because if one calculates the densities for $\mu_\uparrow >
\mu_\downarrow$ one finds $n_\uparrow < n_\downarrow$ in some regions of the phase
diagram \cite{Liu-2006,Parish-2007}.  A warning about this problem
exists already in the unpolarized case, where the spin susceptibility
becomes negative at temperatures slightly above $T_c$ in a region
about unitarity \cite{Kashimura-2012,Kashimura-2013} (see also Ref.~\cite{Maly-1996}).  The size of
this region can be significantly reduced by avoiding the expansion
(\ref{Dyson1storder}) of the Dyson equation and using the $t$-matrix
approximation discussed in Section \ref{sec:nsr}, although even in
this case the spin susceptibility at $T_c$ remains negative in a
small region on the BEC side of unitarity~\cite{Kashimura-2012}.  

To overcome this problem of a negative spin susceptibility for all temperatures above $T_c$ throughout the whole BCS-BEC crossover, it turns out that it is sufficient to 
introduce some degree of self-consistency in the self-energy of Fig.~\ref{Figure-8}(a).
This has been done either by using the ZS approach of Section~\ref{sec:ZS} where a quasi-particle energy shift is introduced in all particle lines of the self energy of Fig.~\ref{Figure-8}(a) \cite{Pantel-2014}, or by replacing the bare single-particle propagator that closes the loop in Fig.~\ref{Figure-8}(a) by a dressed one \cite{Kashimura-2012,Kashimura-2013}.  
The latter approximation (referred to as the Extended T-Matrix Approximation (ETMA)) was applied to calculate the critical temperature of the polarized gas throughout the BCS-BEC crossover \cite{Kashimura-2012}.  
Specifically, assuming that the transition to the superfluid phase is of second order, in Ref.~\cite{Kashimura-2012} the Thouless criterion $\Gamma_0(Q=0)^{-1}=0$ was solved simultaneously with the number equations $n_{\sigma}= \int \! \frac{d\vek{k}}{(2 \pi)^3} \, T \, \sum_{n} \, e^{i\omega_{n}\eta} \, G_{\sigma}(k)$, where $G_{\sigma}$ is dressed by the ETMA self-energy $\Sigma_{\sigma}$ via the Dyson equation and the particle-particle propagator $\Gamma_0(Q)$ is determined by Eq.~(\ref{bubble-pp-exact}) for $R_{pp}$ with the replacement 
$\xi_{\vek{k}+\vek{Q}} \to \xi_ {\vek{k}+\vek{Q}\uparrow}$ and $\xi_{\vek{k}} \to \xi_ {\vek{k} \downarrow}$.  
This form of the Thouless criterion corresponds to the vanishing of the gap $\Delta$ in the mean-field equation (\ref{Sarma-gap}) for a polarized superfluid.

\begin{figure}[t]
\begin{center}
\includegraphics[width=10.0cm,angle=0]{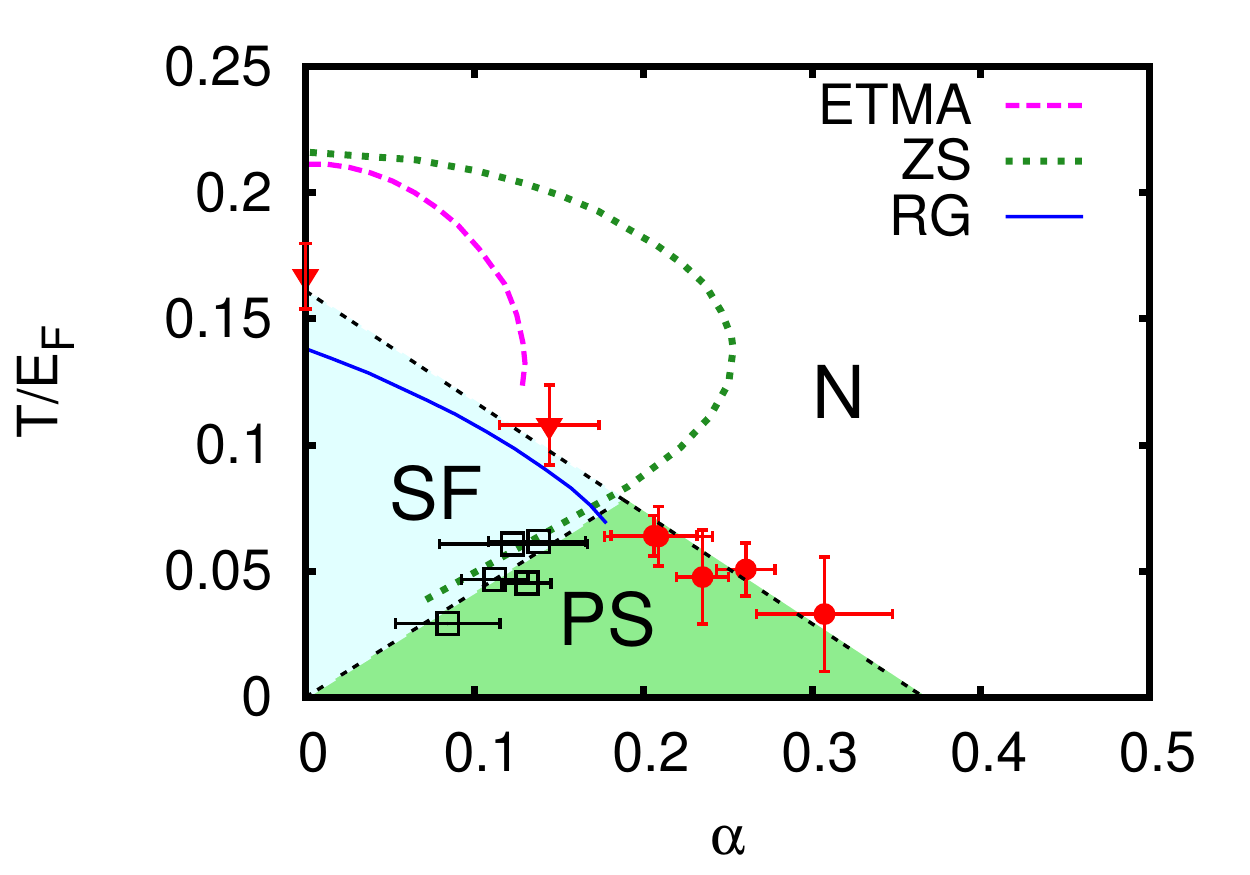}
\end{center}
\caption{Phase diagram for a unitary Fermi gas in the plane temperature (in units of $T_F$) vs polarization $\alpha$.  
Symbols correspond to the experimental data from Ref.~\cite{Shin-2008b} for $\alpha \neq 0$ and from Ref.~\cite{Ku-2012} for $\alpha=0$.
Circles: boundary between the normal phase (N) and the phase separation (PS) region; Triangles: boundary between the normal phase
and the superfluid phase (SF); Squares: boundary between the superfluid phase and the phase separation region.  
Following Ref.~\cite{Shin-2008b}, data are interpolated by straight lines to identify the three regions in the phase diagram. 
The second-order transition lines obtained by the theoretical ETMA \cite{Kashimura-2012}, ZS \cite{Pantel-2014}, and renormalization group (RG) \cite{Gubbels-2013} 
approaches are also reported for comparison.}
\label{Figure-17}
\end{figure}

At unitarity, the critical temperature obtained in this way decreases
for increasing polarization, from the value $T_c/E_F=0.21$ for
balanced populations to the value $T_c/E_F \simeq 0.12$ at a
critical polarization $\alpha_{c} = 0.13$ where the curve develops a
re-entrant behaviour (cf. the dashed line in Fig.~\ref{Figure-17}), in
analogy to what occurs within mean-field (cf.~Fig.~\ref{Figure-7} of
Section \ref{sec:polarized}).  This analogy would suggest one that the
re-entrant behaviour could be eliminated by taking into account the
FFLO instability or phase separation.  However, a problem arises at finite temperature with
the FFLO instability within the present scheme, because a second-order
phase transition to the FFLO phase would require the Thouless
criterion to be satisfied for finite value $\vek{Q}_0$ of the pair
momentum $\vek{Q}$ (that is, $\Gamma_0(\vek{Q}_0,
\Omega_{\nu}=0,)^{-1}=0$) rather than for $\vek{Q}=0$ as usually.
This, in turn, implies that $\Gamma_0(\vek{Q}, \Omega_{\nu}=0)$
diverges like $(|\vek{Q}|-|\vek{Q}_0|)^{-2}$ when $|\vek{Q}| \to
|\vek{Q}_0|$ (in the present rotationally invariant case $\Gamma_0$
depends only on the magnitude $|\vek{Q}|$ of the wave vector).  This
divergence is non-integrable in the expression of the self-energy, in
contrast to the standard case where the divergence of $\Gamma_0$ when
$\vek{Q}\to 0$ is compensated by the factor $Q^2$ resulting from the
spherical integration over $\vek{Q}$.  As a consequence, when $T \to
T_c$ the self-energy $\Sigma_{\sigma}(k)$ would diverge for all wave
vectors and frequencies, a result which does not allow for a solution
of the number equations.  
This failure of having a second-order phase transition associated with the FFLO phase within the NSR theory was pointed out in Ref.~\cite{Ohashi-2002b}, following a similar
conclusion drawn in Refs.~\cite{Shimahara-1998,Shimahara-1999} within a Ginzburg-Landau approach.  
The above argument appears to be quite general  and is thus applicable, at finite temperature, to any version of the $t$-matrix approximation.

The above argument, on the other hand, will not be applicable if the
transition from the normal to the FFLO phase is of \emph{first order},
since in this case the transition occurs before the divergence of the
particle-particle propagator will show up.  In this respect, it was
already mentioned in Section \ref{sec:polarized} that the detailed
analysis of Ref.~\cite{Mora-2005} in the weak-coupling limit has shown
that the transition from the normal to the FFLO phase is indeed of
first order even at the mean-field level.  For ultra-cold Fermi gases,
for which the two spin populations are separately conserved, this
first-order transition corresponds to a phase separation between the
polarized normal gas and a polarized FFLO superfluid.  In general,
this kind of phase separation will compete with the phase separation
between the normal phase and a standard BCS superfluid (where Cooper
pairs condense with zero center-of-mass wave vector). 
Finally, we mention some recent work on this matter where an analysis of the stability of the FFLO phase with respect to fluctuations was considered for the superfluid phase at finite temperature. 
While Refs.~\cite{Radzihovsky-2009,Radzihovsky-2011} conclude that low-energy fluctuations act to disorder the FFLO phase at long distances, yielding a phase with an algebraic decay of correlations, 
the renormalization-group analysis of Ref.~\cite{Pawel-2017} points to an instability of the FFLO phase towards phase separation between the normal and the standard BCS superfluid phase.  

The experimental phase diagram for the polarized unitary Fermi gas was obtained in Ref.~\cite{Shin-2008b} and shows that indeed, beyond a critical polarization $\alpha_{tc}\approx 0.20$, 
the transition from the normal to the superfluid phase turns from second- to first-order, with a phase separation occurring between a polarized normal gas and a polarized superfluid (with no indication, however, about the BCS or FFLO nature of the superfluid component).  
This experimental phase diagram is shown in Fig.~\ref{Figure-17}, where the second-order transition lines obtained by three different theoretical approaches are also reported: 
the ETMA of Ref.\cite{Kashimura-2012}, the ZS approach of Ref.~\cite{Pantel-2014}, and the renormalization group (RG) approach of Refs.~\cite{Gubbels-2008,Gubbels-2013}. 
The theoretical results are
physically meaningful only for the part of the curves that precedes
the re-entrant behaviour, and for this reason the re-entrant branch is
only partially reported.  The physical branch can instead be compared
with the experimental data for the second-order transition to the
superfluid phase (triangles).  One sees that, although the ETMA and ZS
curves are somehow off the experimental data, they still represent an
improvement over the mean-field curve of Fig.~\ref{Figure-7}(b).  The
RG approach curve approximates instead quite well the experimental
data for the second-order transition, including also the position of
the tricritical point (that corresponds to the end point of the RG
curve and to the upper vertex of the experimental PS region).  
This agreement appears quite remarkable, even though one should warn that there is some 
arbitrariness in the choice of couplings to be kept in the RG equations and in the way the RG flow 
is implemented, as pointed out in Ref.~\cite{Gubbels-2013}.
On the other hand, theoretical calculations are still missing for the
first-order (N-PS and SF-PS) transition lines of Fig.~\ref{Figure-17},
which would eventually allow for a quantitative comparison with the
experimental data.  [In this respect, it should be mentioned that the
  ``pseudo-gap" approach of Ref.~\cite{He-2007}, which at zero
  temperature reduces to the mean-field approximation discussed in
  Section \ref{sec:polarized}, cannot be used for such a quantitative
  comparison,  while the functional renormalization group results of Ref.~\cite{Boettcher-2015} are expressed in terms of 
  grand-canonical variables and for this reason they cannot be directly compared with the phase diagram of Fig.~\ref{Figure-17}].

\begin{figure}[t]
\begin{center}
\includegraphics[width=7.0cm,angle=0]{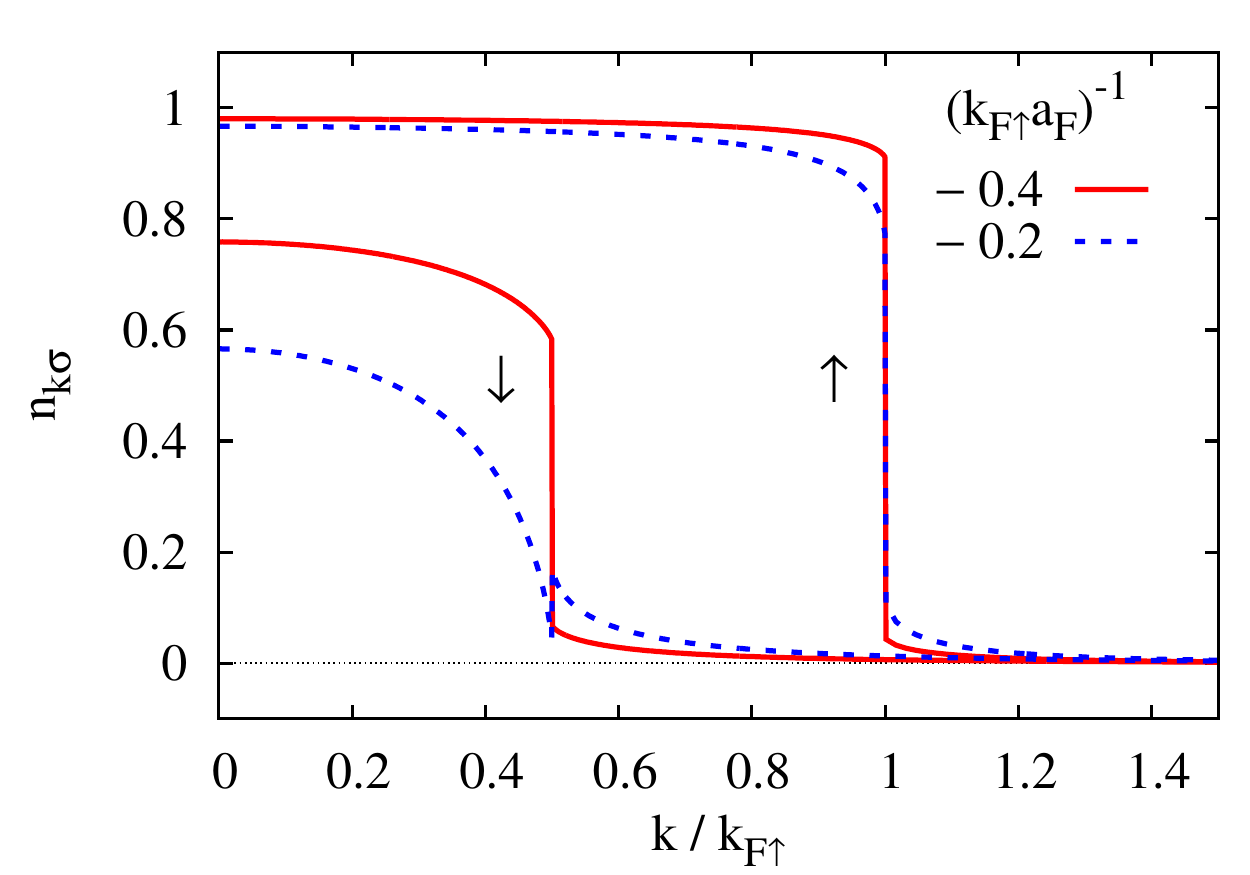}
\end{center}
\caption{Occupation numbers for majority ($\uparrow$) and minority ($\downarrow$) fermions vs the wave vector $k$ in the normal phase at $T=0$ when $k_{F\downarrow} =
k_{F\uparrow}/2$, obtained within the pp-RPA approach for two values of the interaction strength.
[Adapted from Ref.~\cite{Urban-2014}.]}
\label{Figure-18}
\end{figure}

As a final remark, we discuss the relevance of the Luttinger theorem in the context of imbalanced systems.                                                                                                                                                                                              
According to this theorem \cite{Luttinger-1960}, interaction effects in the normal phase do not modify the position of the Fermi step in wave-vector space at $T=0$. 
In the polarized case, one has to impose a strong polarization such that the system remains normal even at $T=0$, in such a way that a jump is expected to occur
in the occupation numbers $n_{\vek{k}\sigma}$ at the respective Fermi wave vectors $k_{F\sigma} = (6\pi^2n_\sigma)^{1/3}$. 
As an illustration, Fig.~\ref{Figure-18} shows $n_{\vek{k}\sigma}$ obtained for $k_{F\downarrow} = k_{F\uparrow}/2$ within the pp-RPA approach at $T=0$ for two different couplings.
It turns out that the depletion of $n_{\vek{k}\sigma}$ below $k_{F\sigma}$ due to interaction effects is compensated by the filling of $n_{\vek{k}\sigma}$ above $k_{F\sigma}$, 
in such a way that the Luttinger theorem is satisfied for each $\sigma$ as shown analytically in Ref.~\cite{Urban-2014}.
However, one sees from Fig.~\ref{Figure-18} that for the stronger coupling $(k_{F\uparrow} a_{F})^{-1} = -0.2$ the jump of $n_{\vek{k}\downarrow}$ at $k_{F\downarrow}$ is positive instead of being negative, resulting in a negative quasi-particle residue $Z_{\vek{k}\downarrow}$.
This shortcoming signals the breakdown of the perturbative treatment (like in Eq.(\ref{Dyson1storder})) of the self-energies $\Sigma_{\sigma}$ in the pp-RPA approach.
It is relevant to mention that the $T=0$ formalism of the pp-RPA approach, where the bare particle propagators are defined in terms of the Fermi wave vectors $k_{F\sigma}$ instead of the chemical potentials $\mu_\sigma$ \cite{Fetter-1971} and in terms of which the results of Fig.~\ref{Figure-18} have been obtained, has been shown to be equivalent to the ZS approach in the $T\to 0$ limit 
\cite{Urban-2014}, due to the inclusion of the mean-field-like quasi-particle energy shift (cf. Section~\ref{sec:ZS}).
For this reason, the results obtained above for the Luttinger theorem at $T=0$ within the pp-RPA approach apply also to the ZS approach reported in the phase diagram of Fig.~\ref{Figure-17}. 
 
\section{BCS-BEC crossover in ultra-cold Fermi gases}
\label{sec:fermigases}

Ultra-cold Fermi gases have provided a unique opportunity for
realizing experimentally the BCS-BEC crossover.  This is because a
method was found to vary the scattering length $a_F$ of the
two-fermion problem, from negative to positive values across the
resonance where $a_F = \pm \infty$, while keeping  the density of
the system (and thus the corresponding Fermi wave vector $k_F$) fixed.  In
this way, the experimental data have been associated with and labeled
by the value of the dimensionless coupling parameter $(k_F a_F)^{-1}$
which is also used theoretically to drive the BCS-BEC crossover,
thereby allowing for a direct and unambiguous comparison between
experiments and theory.  Below we shall describe the main aspects and
achievements of the BCS-BEC crossover that have emerged with
ultra-cold Fermi gases during the last several years.  We refer to
Refs.~\cite{Bloch-2008,Giorgini-2008} for further reviews on the
BCS-BEC crossover with ultra-cold fermionic atoms.

\subsection{Fano-Feshbach resonances and the interaction-induced crossover}
\label{sec:Feshbach}

In the context of ultra-cold Fermi gases, the Fano-Feshbach resonances
\cite{Fano-1935,Fano-1961,Feshbach-1962} occur via the coupling of
alternative scattering channels, when a bound level in a
\emph{closed} molecular channel can be displaced continuously with
respect to the threshold of an \emph{open} channel.  This occurs by
exploiting the different Zeeman coupling felt by the two (closed and
open) channels in an external magnetic field $B$ \cite{Chin-2010}.

\begin{figure}[h]
\begin{center}
\includegraphics[width=9.5cm,angle=0]{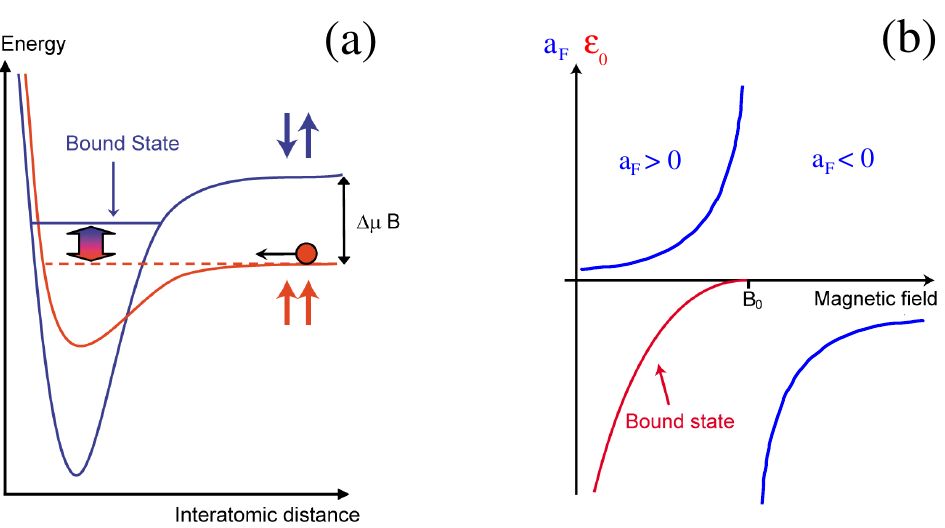}
\caption{(a) Coupling of closed and open dimer scattering channels, which can be displaced relative to each other by the coupling to a magnetic field $B$. 
(b) The corresponding scattering length vs the magnetic field. When $a_F > 0$, a bound state with binding energy $\varepsilon_0= (m a_F^2)^{-1}$ sets in. 
[Adapted from http://cua.mit.edu/ketterle\_group.] }
\label{Figure-19}
\end{center}
\end{figure}

The atom-atom interaction can thus be varied, from a condition when
the resonant bound state is embedded in the continuum above threshold,
to a condition when a true bound state exists below threshold
(cf. Fig.~\ref{Figure-19}(a)).  Correspondingly, the scattering length
$a_F$ for the low-energy scattering changes from negative to positive
values, being associated with attractive or repulsive interactions in
the two cases, and diverges when the bound state sets in
(cf. Fig.~\ref{Figure-19}(b)).

\begin{figure}[h]
\begin{center}
\includegraphics[width=12.0cm,angle=0]{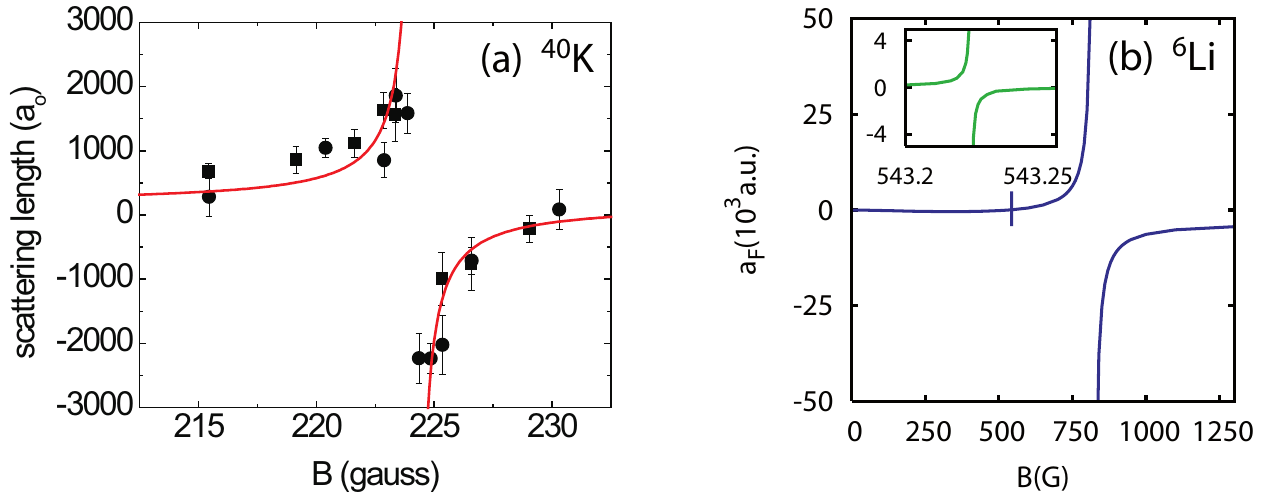}
\caption{(a) Measurements of the scattering length (in units of the Bohr radius $a_0$) vs the magnetic field $B$ near the broad Fano-Feshbach resonance of $^{40}\mathrm{K}$ Fermi atoms.
(b) Calculated scattering length $a_F$ (in atomic units) for the collision of two fermionic $^{6}\mathrm{Li}$ atoms vs the magnetic field $\mathrm{B}$ (in Gauss), which detunes the scattering 
from the closed channel.
The inset show the behaviour of $a_F$ for the narrow resonance at about $543 G$. 
[Panel (a) adapted from Ref.~\cite{Regal-2003a}; panel (b) adapted from Ref.~\cite{Simonucci-2005}.]}
\label{Figure-20}
\end{center}
\end{figure}

In particular, for a ``broad'' Fano-Feshbach resonance (with respect
to the variations of $B$), such that the effective range can be
neglected in the expression of the scattering amplitude, the
interaction between atoms with different spin labels can be modeled by
a contact interaction with a negative strength $\varv_0$, where, in
turn, $\varv_0$ can be eliminated in terms of $a_F$ like in
Eq.(\ref{regularization}) \cite{Simonucci-2005}.  What is here meant by ``spin'' is the
quantum number associated with the atomic hyperfine levels that are
split apart by the magnetic field itself.  The two hyperfine levels of
lowest energy are then conventionally referred to as spin $\uparrow$
and spin $\downarrow$, and can be populated independently with
$N_{\uparrow}$ and $N_{\downarrow}$ atoms (in practice, $N =
N_{\uparrow} + N_{\downarrow}$ is at most of the order of $10^{6}$).
For instance, in $^{6}\mathrm{Li}$ for $B$ large enough the states
$|\downarrow \rangle$ and $|\uparrow \rangle$ correspond to
$|m_{I}=1,m_{J}=-\frac{1}{2}\rangle$ and
$|m_{I}=0,m_{J}=-\frac{1}{2}\rangle$, respectively, where $I=1$ and
$J=\frac{1}{2}$ are the nuclear and electron spin quantum numbers with
projections $m_{I}$ and $m_{J}$.  In addition, the dependence of $a_F$
on $B$ can be inferred either directly from the experiment
\cite{Regal-2003a} (cf. Fig.~\ref{Figure-20}~(a)) or from a molecular
calculation \cite{Simonucci-2005,Chin-2010}
(cf. Fig.~\ref{Figure-20}~(b)).

For a Fano-Feshbach resonance, the resonant condition $a_F = \pm
\infty$, which occurs when the bound state crosses the threshold of
the continuum, corresponds to the ``unitarity limit" whereby $(k_F a_F
)^{-1} =0$.  While in principle the BCS condition of strongly
overlapping Cooper pairs corresponds to $(k_Fa_F)^{-1} \ll -1$ and the
BEC condition for molecular bosons corresponds to $(k_F a_F)^{-1} \gg
+1$, it turns out that the evolution from the BCS to BEC limits is
best studied in the limited range $-1 \lesssim (k_F a_F )^{-1}
\lesssim +1$ for the following reasons.  On the one hand, it is found
by theoretical calculations that this limited range is sufficient in
practice for realizing the BCS-BEC crossover in most physical
quantities.  On the other hand, the range $-1 \lesssim (k_F a_F )^{-1}
\lesssim +1$ is just where the experiments can be realized because:
(i) when $(k_Fa_F)^{-1} \ll -1$ the experimental signal is usually too
weak to be detected; (ii) when $(k_F a_F)^{-1} \gg +1$ the molecules
(which are in the highest vibrational state of the interatomic
potential) undergo rapid vibrational relaxation via three-body
collisions, leading to heating and trap losses which set an upper
limit on the value of $(k_F a_F)^{-1}$ up to which the experimental
data can be trusted.

\subsection{Main experimental results}
\label{sec:experiments}

The distinctive and rather unique feature of the BCS-BEC crossover
with trapped ultra-cold Fermi gases is that a variety of targeted
experiments have been realized with these systems over the last
decade, which have specifically allowed to follow the continuous
evolution of the system between the two (BCS and BEC) limits.  The
large majority of these experiments have been realized with $^6$Li
atoms, using the broad Fano-Feshbach resonance occurring at 832 G to
tune the interaction, while the remaining experiments used $^{40}$K
atoms and the broad Fano-Feshbach resonance at 202.2 G.  In both
cases, the interaction between fermions can be accurately modeled by
the contact potential described in detail in the previous sections.
It should be recalled in this context that, initially, most of the
theoretical studies of the BCS-BEC crossover with ultra-cold atoms
were instead based on a ``two-channel" model, which includes
explicitly the closed-channel molecules in the Hamiltonian
\cite{Holland-2001,Ohashi-2002a,Stoof-2004,Bruun-2004,Stajic-2004,Diehl-2006}.  
In Ref.~\cite{Simonucci-2005} it was eventually clarified, however, that
for a broad resonance a ``single-channel" contact potential is
adequate to describe the system in an extended range of magnetic field
that spans the whole BCS-BEC crossover.  This property of broad
Fano-Feshbach resonances is fundamental for using ultra-cold atoms to
simulate the BCS-BEC crossover in a universal fashion, independent of
the details of the inter-atomic interaction.

An additional aspect to be considered for the use of ultra-cold gases
as effective ``quantum simulators", is that ultra-cold gases require a
trap to hold the neutral atoms together.  Most useful in this sense
prove to be the \emph{optical traps}, which allow for an independent
application of a magnetic field to tune a Fano-Feshbach resonance.  In
practice, these traps act on the atoms as an external potential
proportional to the laser intensity, which has usually the shape of an
inverted Gaussian.  For most purposes, it is sufficient to expand this
inverted Gaussian up to second order around its minimum, leading to an
anisotropic harmonic potential of the form
\begin{equation}
V_{\mathrm{ext}}(\vek{r}) = \frac{1}{2} m \left( \omega_{x}^2 x^2 + \omega_{y}^2 y^2 + \omega_{z}^2 z^2 \right) 
\label{external_potential}
\end{equation}
\noindent
where $\vek{r} = (x,y,z)$, with typical values of the frequencies ($\omega_{x},\omega_{y},\omega_{z}$) ranging from tens to thousands of Hz.

When comparing theory and experiments, one thus needs, in general, to
consider the effects of the trapping potential
(\ref{external_potential}).  This is usually done by relying on a
local-density approximation (LDA), whose conditions of validity are
normally satisfied in the experiments.  By this approximation, the
trapped gas is regarded as a (continuous) collection of independent
homogeneous systems characterized by a ``local" chemical potential
$\mu(\vek{r}) = \mu_{\rm trap} -V_{\mathrm{ext}}(\vek{r})$, with a
common temperature $T$ and scattering length $a_F$.  [To avoid
  ambiguities, we have here introduced the chemical potential
  $\mu_{{\rm trap}}$ of the trapped system to distinguish it from that
  of the homogeneous system].  Physical quantities then result by
averaging (that is, by integrating) over the contributions at
different $\vek{r}$.  In this context, it is also useful to introduce
the trap Fermi energy $E_{F \, {\rm trap}} = \omega_0 \, (3 N)^{1/3}$
(associated with an ideal balanced Fermi gas at $T=0$), where
$\omega_0 = (\omega_{x} \omega_{y} \omega_{z})^{1/3}$ is the
geometrically averaged trap frequency \cite{Pethick-2008}.  In terms
of $E_{F \, {\rm trap}}$, one can also define the Fermi wave vector
$k_{F \, {\rm trap}} = \sqrt{2 m E_{F \, {\rm trap}}}$ for the trap
and form the dimensionless coupling parameter $(k_{F \, {\rm trap}}
a_F )^{-1}$.  [Note, however, that these numbers do \emph{not}
  coincide with the local values of $k_F$ and $(k_Fa_F)^{-1}$
  associated with the density of the system at the trap center.]
Thermodynamic quantities can be made dimensionless by using
appropriate powers of $E_{F \, {\rm trap}}$, $k_{F \, {\rm trap}}$ and
$N$, in such a way that they depend solely on the interaction
parameter $(k_{F \, {\rm trap}} a_F )^{-1}$ and the dimensionless
temperature $T/E_{F \, {\rm trap}}$ (see, e.g,.~Refs.~\cite{Perali-2003,Perali-2004b,Haussmann-2008}).  
Recently, novel experimental
methodologies have become available which apply tomographic
reconstruction techniques to the bare experimental data by still
relying on the above LDA approximation \cite{Shin-2007,Ho-2010}, thus
allowing for a direct extraction of quantities characteristic of the
homogeneous gas from the experiments made on the trapped gas.

An impressive series of experiments has been realized with ultra-cold
Fermi gases over the last decade, covering a broad number of physical
(both thermodynamic and dynamical) quantities.  These experiments
include:

\begin{enumerate-roman}
\item The initial production of long-lived composite bosons
  \cite{Regal-2003b,Strecker-2003,Cubizolles-2003,Jochim-2003b};
\item The condensation of composite bosons
  \cite{Greiner-2003,Jochim-2003a,Zwierlein-2003};
\item The evidence of a condensate extending also to the BCS side of
  the crossover \cite{Regal-2004,Zwierlein-2004} with the
  ``projection" technique introduced in Ref.~\cite{Regal-2004} (see
  Refs.~\cite{Perali-2005} and \cite{Zwierlein-2005a} for a
  theoretical and experimental analysis of this technique,
  respectively);
\item The evidence for a pairing gap with radio-frequency spectroscopy
  \cite{Chin-2004,Greiner-2005a,Shin-2007} and its quantitative
  measurement using a small population imbalance
  \cite{Schirotzek-2008};
\item The study of collective modes in a trap, in particular, the
  measurement of frequencies and damping rates of breathing
  \cite{Kinast-2004a,Bartenstein-2004b,Kinast-2004b,Altmeyer-2007a,Riedl-2008},
  radial quadrupole \cite{Altmeyer-2007b,Wright-2007}, scissors
  \cite{Wright-2007,Riedl-2008}, and higher-nodal axial modes
  \cite{Tey-2013};
\item The measurement of several thermodynamic properties of the
  unitary Fermi gas across the superfluid transition, including the
  critical temperature
  \cite{Kinast-2005,Luo-2007,Nascimbene-2010,Ku-2012}, the specific
  heat and entropy \cite{Kinast-2005,Luo-2007}, the equation of state,
  chemical potential and pressure
  \cite{Nascimbene-2010,Navon-2010,Horikoshi-2010,Ku-2012};
\item The precise measurement of the closed-channel contribution in
  $^6$Li, that has confirmed the validity of the single-channel model
  \cite{Partridge-2005};
\item The detection of pairing correlations with atom shot noise
  measurements \cite{Greiner-2005b}, following the proposal of
  Ref.~\cite{Altman-2004};
\item The observation of vortices throughout the whole crossover
  \cite{Zwierlein-2005b};
\item The detection of phase separation
  \cite{Zwierlein-2006a,Partridge-2006,Shin-2006} and the
  measurement of the critical temperature
  \cite{Zwierlein-2006b,Shin-2008b} and polarization
  \cite{Zwierlein-2006b,Shin-2008a,Olsen-2015} in imbalanced Fermi
  gases;
\item The measurement of first- \cite{Joseph-2007} and second-sound
  \cite{Sidorenkov-2013} velocities;
\item The measurement of the critical velocity for the superfluid flow
  \cite{Miller-2007,Moritz-2015};
\item The estimate of the Cooper pair size $\xi_{\rm pair}$ \cite{Schunck-2008};
\item The measurement of single-particle spectral features (with
  evidence of a pseudo-gap) with wave-vector resolved radio-frequency
  spectroscopy \cite{Stewart-2008,Gaebler-2010};
\item The measurement of the Tan's contact
  \cite{Stewart-2010,Kuhnle-2010,Kuhnle-2011,Sagi-2012,Hoinka-2013}
  and the verification of universal relations connected with this
  quantity \cite{Stewart-2010,Kuhnle-2010};
\item The measurement of the shear viscosity and of its ratio with the
  entropy density \cite{Cao-2011,Elliott-2014,Joseph-2015,Bluhm-2017}, to test
  the ``universal minimum" conjectured for this quantity by
  string-theory methods \cite{Kovtun-2005};
\item The measurement of static \cite{Sanner-2011} and dynamic spin
  \cite{Hoinka-2012} and density \cite{Hoinka-2013,Lingham-2014}
  response functions;
\item The evidence for a superfluid quenching of the moment of inertia
  \cite{Riedl-2011} and the measurement of the superfluid fraction
  \cite{Sidorenkov-2013};
\item The measurement of ``current-voltage" characteristic curves for
  transport through a quantum point contact in the unitary Fermi gas
  across the superfluid transition \cite{Husmann-2015};
\item The detection of the Josephson's oscillations throughout the
  BCS-BEC crossover \cite{Valtolina-2015}.
\end{enumerate-roman}

The above experiments have explored rather completely the BCS-BEC
crossover in three dimensions.  More recently, experimental studies
have been conducted for the BCS-BEC crossover also in systems with
reduced dimensionality. In particular, for a two-dimensional gas
experiments have analyzed pairing effects
\cite{Froehlich-2011,Sommer-2012} and single-particle spectral
features \cite{Feld-2011}, collective modes and the shear viscosity
\cite{Vogt-2012}, the ground-state pressure \cite{Makhalov-2014}, the
effects of spin imbalance \cite{Ong-2015}, the superfluid critical
temperature \cite{Ries-2015,Murthy-2015}, and the equation of state
\cite{Fenech-2016,Boettcher-2016}.  Finally, Ref.~\cite{Liao-2010} has
analyzed the phase-separation in a one-dimensional spin-imbalanced
Fermi gas, confirming the predictions based on the Bethe-ansatz
\cite{Orso-2007} and providing indirect evidence for the occurrence of FFLO ordering in the system.
 
\subsection{The unitary limit}  
\label{sec:unitarylimit}

In the context of ultra-cold Fermi gases, special attention has been devoted to the ``unitary limit" of the inter-particle interaction, when the scattering length is infinite.  
At the two-body level, in this limit the scattering cross section approaches $\sigma(k)= 4\pi/k^2$ at low relative wave vector $k$, which corresponds to the unitarity bound 
for the ($s$-wave) scattering cross section and stems directly from the unitarity condition of the S-matrix (see, e.g., Ref.~\cite{Taylor-1972}).

At the many-body level, interest in this limit of the interaction was stimulated by G. Bertsch at the 10th conference on ``Recent progress in many-body theories" 
(see Refs.~\cite{Baker-2000} and \cite{Baker-1999}).  
As a challenge to the participants, Bertsch proposed the problem of determining the ground-state properties of a (homogeneous) system of fermions interacting via a contact interaction
tuned at infinite scattering length (then referred to as the ``unitary Fermi gas"). 
This case is particularly interesting also for a dilute neutron gas, where the scattering length is much larger than the inter-particle distance (cf. Section \ref{sec:nuclearsystems}).
What is intriguing of the unitary limit is that there are no length scales associated with the interaction, paradoxically somehow just like for the non-interacting Fermi gas.  
The only length scale that remains at zero temperature is the average inter-particle spacing $k_F^{-1}$ which is fixed by the density.  
More generally, the thermodynamic properties of the unitary Fermi gas depend only on density and temperature.  
A few universal functions of the dimensionless temperature $T/E_F$ (or, at zero temperature, a few universal numbers) then suffice to characterize the properties of the
unitary Fermi gas and for this reason the unitary gas is said to be \emph{universal}. In addition, several nontrivial relationships between thermodynamic observables can be derived \cite{Ho-2004}, which are particularly useful to extract from experiments thermodynamic quantities that cannot directly be measured (or are especially difficult to measure).
 
For the unitary Fermi gas, the absence of a length scale associated
with the interaction has interesting consequences also for trapped
systems.  For instance, it can be shown \cite{Perali-2004} that at
zero temperature the spatial profile $n(\vek{r})$ of the density
within LDA is completely specified in terms of the value of a single
parameter $\xi$.  This parameter (referred to as the Bertsch
parameter) is defined as the ratio between the ground-state energy of
the unitary {\em homogeneous} Fermi gas and that of the
non-interacting system (this ratio corresponds also to $\mu/E_{\rm F}$
as a consequence of scale invariance).  The argument goes as follows.
For a contact potential at $T=0$, the chemical potential in units of
the Fermi energy depends only on the dimensionless coupling $(k_F
a_F)^{-1}$, such that for the homogenous gas $\mu/E_{\rm F}=
F\left((k_F a_F)^{-1}\right)$ where $F$ is a universal function.  The
LDA then implies that for the trapped system:
\begin{equation}
\mu(\vek{r})/E_F(\vek{r}) = F\left((k_F(\vek{r}) a_F)^{-1}\right)
\end{equation}
with the local quantities $k_F(\vek{r})=(3\pi^2 n(\vek{r}) )^{1/3}$,
$E_F(\vek{r})= k_F(\vek{r})^2/ 2 m$, and $\mu(\vek{r})=\mu_{{\rm
    trap}} - V_{{\rm ext}}(\vek{r})$.  A remarkable feature of the
unitary limit is that the ``local'' dimensionless coupling
$(k_F(\vek{r}) a_F)^{-1}$ does not depend on the position $\vek{r}$ in
the trap, such that $\mu(\vek{r})/E_F(\vek{r}) = F(0) \equiv \xi$ as
for the homogeneous gas.  This relation can be inverted to get
$E_F(\vek{r}) = \mu(\vek{r})/\xi$, thus yielding for the density
profile the expression:
\begin{equation}
n(\vek{r}) = \xi^{-3/2} [2m (\mu_{{\rm trap}} - V_{{\rm ext}}(\vek{r}))]^{3/2}/(3\pi^2)
\label{profile}
\end{equation}
which has the same form of the non-interacting Fermi gas apart from
the pre-factor $\xi^{-3/2}$.  In addition, the expression $N = \int
d\vek{r} \, n (\vek{r})$, once applied alternatively to
Eq.(\ref{profile}) and to the corresponding equation for the
non-interacting gas, yields the relation $\mu_{\mathrm{trap}}/E_{F \,
  \mathrm{trap}} = \xi^{1/2}$, which connects the ratio between the
chemical potential $\mu_{\mathrm{trap}}$ of the unitary Fermi gas and the
Fermi energy $E_{F \, \mathrm{trap}} = \omega_0 \, (3 N)^{1/3}$ of the
non-interacting Fermi system in the trap, to the corresponding ratio
$\xi$ for the homogeneous gas.  Equation (\ref{profile}) also implies
that the axial and radial radii of the cloud at unitarity are reduced
by a factor $\xi^{1/4}$ with respect to the corresponding values of
the non-interacting Fermi gas, a result that was used to obtain some
of the first experimental estimates of the Bertsch parameter
\cite{Gehm-2003,Bartenstein-2004a}.  The spatial invariance of the
local coupling strength for the trapped unitary Fermi gas leads also
to the relation
\begin{equation}
\Delta({\vek{r}})= \frac{\eta}{\xi}  \, \mu(\vek{r})
\label{eta-xi}
\end{equation}
for the local gap parameter $\Delta(\vek{r})$, where $\eta$ is the
ratio $\Delta/E_F$ for the homogenous unitary Fermi gas at $T=0$.
Also the gap profile of the trapped unitary Fermi gas has thus the
form of an inverted parabola.

 \subsection{Theoretical approaches to the  unitary limit and comparison with experiments}  
\label{sec:theoretical}
The universality of the unitary Fermi gas and the possibility to realize it in experiments with ultra-cold atoms has motivated a number of theoretical studies of this system, and in particular of its ground-state energy characterized by the Bertsch parameter $\xi$. 
The first estimate of the Bertsch parameter was given in
 
 Ref.~\cite{Baker-1999} by using a Pad\'e-approximant
extrapolation of the perturbative expansion of the ground-state energy
of a dilute Fermi gas, with the result $\xi= 0.43 \pm 0.13$.  
Since then, a number of theoretical works have calculated the value of the
Bertsch parameter. Analytic or semi-analytic works  have improved the
simple BCS mean-field estimate $\xi=0.59$ by including pairing fluctuations in the superfluid phase 
with diagrammatic~\cite{Pieri-2004b,Perali-2004,Haussmann-2007} or functional integral approaches~\cite{Hu-2006,Diener-2008}, by using
$\epsilon$~\cite{Nishida-2006,Nishida-2007a,Nishida-2009} and large-$N$ expansions \cite{Veillette-2007}, 
or by the functional renormalization group approach~\cite{Diehl-2007,Bartosch-2009}. 

A few words about these approaches are required. 

The $\epsilon$ expansion exploits the fact that, in $d=2$ dimensions the unitary Fermi gas is a non-interacting Fermi gas, while in $d=4$ it is a gas of non-interacting dimers \cite{Nussinov-2004}. 
By calculating the loop integrals using dimensional regularization, one can perform calculations in arbitrary dimensions near $d=2$ or $d=4$ and make expansions in $\bar{\epsilon}=d-2$ or in $\epsilon = 4-d$,
respectively. 
To get predictions for the physical case with $d=3$, one then tries to match these two expansions.
Depending on the details of the approximations used for this matching, one obtains $\xi = 0.364$ to $ 0.391$ \cite{Nishida-2007a}. 
The $\epsilon$ expansion can also be used for finite temperature calculations \cite{Nishida-2007b}. 
For instance, for the critical temperature this approach gives $T_{c}/E_{F} = 0.183\pm 0.014$.

The large-$N$ expansion provides another way to circumvent the absence of a small expansion parameter in the unitary Fermi gas. 
Here, one introduces $N$ species (``flavors'') of fermions distinguished by an index $i$ ($i = 1, \cdots, N$), each having two spin projections ($\sigma =\uparrow,\downarrow$) and mutually interacting via \cite{Veillette-2007}
\begin{equation}
H_{\text{int}} = \frac{g}{N} \sum_{i,j=1}^{N} \int \! d\vek{r} \, \psi^\dagger_{i\uparrow}(\vek{r}) \, \psi^\dagger_{i\downarrow}(\vek{r}) \, \psi_{j\downarrow}(\vek{r}) \, \psi_{j\uparrow}(\vek{r})\,,
\end{equation}
$\psi_{i\sigma}(\vek{r})$ being the field operator for the fermion species $i$ and spin $\sigma$ and $g$ the coupling constant. 
Note that the interaction has a $\mathrm{Sp}(2N)$ symmetry and can transform a pair of one species $j$ into a pair of another species $i$.
For this reason, the case $N=2$ is different from the isospin in nuclear physics, where the numbers of protons and neutrons are instead separately conserved. 
Note in addition that the number of species $N$ is a purely formal parameter (like the quantity $\epsilon$ above) and the physical results correspond to $N=1$. 
What is relevant here is that, in the limit $N\to\infty$, the BCS mean-field approximation becomes exact, and corrections in $1/N$ can be calculated by including the RPA on top of it. 
Including only the leading term and the $1/N$ correction, and then extrapolating the result to $N=1$, one finds for the Bertsch parameter the value $\xi = 0.28$.

The same $\epsilon$ and large-$N$ expansions were used also in Ref.~\cite{Nikolic-2007}, where they have been framed in the standard approach for critical phenomena.
In this case, the starting point was the observation that the unitary Fermi gas in the zero-density limit can be interpreted as a quantum multi-critical point at $\mu=(k_{F}a_{F})^{-1}=h=T=0$, 
which can be used as the fixed point of a renormalization-group analysis.

Finally, the functional renormalization group starts from a flow equation for the effective-action generating functional. 
Approximate solutions of the flow equation can be found by using truncations in the space of possible functionals (see Refs.~\cite{Diehl-2010,Metzner-2012} for recent reviews of this method).

Before we turn to the fully numerical works, let us mention that at
high temperature, the fugacity $e^{\beta\mu}$ can be used as a small
expansion parameter, leading to the virial expansion. Of course it
cannot be used to compute the universal parameters $\xi$ or $\eta$,
but it allows for the calculation of the equation of state of the
unitary Fermi at high temperatures which has been measured to high
precision. While the second-order virial coefficient for a contact
interaction  was calculated long time
ago~\cite{Beth-1937}, the virial expansion was pushed to higher order
only recently.  In particular, the third-order virial coefficient for
the unitary Fermi gas was obtained in
Refs.~\cite{Liu-2009,Kaplan-2011,Leyronas-2011} with different
techniques (see Ref.~\cite{Liu-2013} for a review).  For the
fourth-order virial coefficient, on the other hand, a discrepancy that
was found between the theoretical estimates
\cite{Rakshit-2012,Ngampruetikorn-2015} and the experimental value
\cite{Nascimbene-2010,Ku-2012} was resolved only recently by a fully
numerical calculation \cite{Yan-2016} that agrees with the
experimental value.

A completely different way of approaching the unitary Fermi gas
consists in trying to solve the problem using  fully numerical
techniques. 

One can distinguish between studies of particles in a harmonic potential
and studies of particles in a box with periodic boundary conditions.
Calculations for particles in a harmonic potential are especially
interesting because of their similarity with nuclear systems and
because they allow one to extract the coefficients of the leading
gradient terms (i.e., beyond LDA) in the energy density functional
\cite{Carlson-2014}. Such calculations have been done using the direct
diagonalization of the Schr\"odinger equation in the basis of
correlated Gaussians (CG) for few particles (up to six in Ref.~\cite{Blume-2007} and up to ten in Ref.~\cite{Yin-2015}),
and for larger particle numbers using Green's function
\cite{Chang-2007} and fixed-node diffusion Monte-Carlo (FN-DMC)
\cite{Blume-2007}, lattice Monte-Carlo \cite{Endres-2011}, and
auxiliary-field Monte-Carlo (AFMC) \cite{Carlson-2014}. The pairing
gap, defined as the energy difference between odd and even numbers of
particles (corresponding to the lowest quasiparticle energy, which is
generally smaller than the local gap at the trap center), was discussed
in Refs.~\cite{Chang-2007,Blume-2007}. Unfortunately, results of
different calculations do not agree with each other (see Fig.~10 of
Ref. \cite{Carlson-2012}).  It should be stressed,
however, that while FN-DMC is just a variational method
with an unknown error bar a priori,  AFMC and lattice Monte-Carlo 
are in principle exact (and free from the fermionic sign problem for the
imbalanced gas). In practice, it turns out to be quite difficult to reach the zero-range interaction limit in this kind of calculations.
It is also not clear why the lattice calculations
\cite{Endres-2011} show pronounced shell effects (that is, strongly reduced
energies per particle at shell closures $N=2,8,20,40$) in contrast to
all other approaches, where the shell effects seem to be washed out by
the strong pairing.

Calculations for particles in a box with periodic boundary conditions
are better suited for the extrapolation to a uniform gas, although it remains difficult to
control finite size effects. These
calculations, too, were performed using different numerical methods. Like
in the trapped case, the results show strong variations. The
predictions for the Bertsch parameter $\xi$ vary between $\xi =
0.25$ in the the first lattice calculation \cite{Lee-2006} and $\xi =
0.44$ in the first DMC calculation \cite{Carlson-2003}; in later
lattice \cite{Lee-2008} and FN-DMC calculations
\cite{Astrakharchik-2004,Carlson-2005,Morris-2010,Forbes-2011,Li-2011} the value of $\xi$ slowly
converged over the the years (see table VI of Ref. \cite{Endres-2013})
towards the value $\xi = 0.37$ that was obtained in the latest AFMC
\cite{Carlson-2011} and lattice \cite{Endres-2013} calculations
simulating up to 66 particles. (The AFMC technique was also used for
neutron matter, see Section~\ref{sec:nn_cold}.)
 
Correspondingly, a number of experimental works has determined $\xi$
with increasing precision
\cite{OHara-2002,Gehm-2003,Bartenstein-2004a,Bourdel-2004,Stewart-2006,Partridge-2006,Luo-2009,Navon-2010,Ku-2012},
culminating with the value $\xi=0.370 \pm 0.005$ obtained in
Ref.~\cite{Ku-2012} (this value takes into account a small
systematic-error correction associated with a subsequent more precise
determination of the position of the broad Fano-Feshbach resonance of
$^6$Li~\cite{Zuern-2013}).  This experimental value agrees well with
the state-of-the art theoretical results $\xi=0.372 \pm 0.005$ and $\xi= 0.366^{+0.016}_{-0.011}$
obtained in Refs.~\cite{Carlson-2011} and \cite{Endres-2013}, respectively. 
In addition, the value $\eta = 0.44 \pm
0.03$ for the parameter $\eta$ of Eq.(\ref{eta-xi}) was determined
experimentally in Ref.~\cite{Schirotzek-2008}, in good agreement with
the QMC result $\eta = 0.45 \pm 0.05$ of Ref.~\cite{Carlson-2008}.
These comparisons exemplify the high degree of accuracy that is
nowadays possible to reach in experiments with ultra-cold Fermi gases,
thereby pointing out the effectiveness of these gases as ``analog"
quantum simulators of many-body problems.

The unitary Fermi gas has been analyzed in detail also at finite
temperature.  The main experiments that have contributed to the
characterization of the thermodynamic and dynamical properties of the
unitary Fermi gas have already been mentioned in Section
\ref{sec:experiments}.  On the theoretical side, the (homogeneous)
unitary Fermi gas has represented a central issue of
finite-temperature QMC calculations, and actually most of the
numerical efforts have been focused on this specific coupling value of
the BCS-BEC crossover.  The auxiliary-field QMC method has extensively
been applied to the calculation of thermodynamic quantities, such as
the temperature dependence of the chemical potential, energy, and
entropy \cite{Bulgac-2006,Bulgac-2008a,Drut-2012} as well as of the
wave-vector distribution and ``contact" constant \cite{Drut-2011}.  In
principle, this method is unbiased for thermodynamic quantities,
although the effects due to space and time discretization, the
finite-size of the sample, and the use of a finite-range interaction
introduce significant sources of uncertainty (which are discussed in
detail in Ref.~\cite{Drut-2012}).  The same method has also been
applied to the calculation of dynamical quantities, such as the
spectral-weight function \cite{Bulgac-2011} and the (frequency
dependent) spin susceptibility \cite{Bulgac-2013a} and shear
viscosity \cite{Wlazlowski-2012,Bulgac-2013b,Bulgac-2015}.  It should
be recalled, however, that the extraction of dynamical quantities from
QMC simulations relies on a numerical solution of the problem of the
analytic continuation of a function from the imaginary to the real
axis.  Such a numerical solution is notoriously an ill-posed problem,
which unavoidably requires uncontrolled assumptions on the properties
and the overall structure of the function to be continued
analytically.  For these reasons, the results for dynamical quantities
extracted from QMC simulations should be considered with caution.

Alternative Monte Carlo methods based on diagrammatic expansions of
the partition function or of the single-particle self-energy have also
been applied to the (homogeneous) unitary Fermi gas.  The determinant
diagrammatic Monte Carlo method has been used in
Refs.~\cite{Burovski-2006a,Burovski-2006b,Burovski-2008} and in
Ref.~\cite{Goulko-2010} for the calculation of the critical
temperature, chemical potential, and energy of the unitary Fermi gas.
Their results for the critical temperature $T_c/E_F= 0.152 \pm
0.007$ \cite{Burovski-2006a,Burovski-2008} and $T_c/E_F=0.173\pm
0.006$~\cite{Goulko-2010} slightly disagree with each other, probably
because of the different extrapolations to the continuum limit from
the lattice model used in the calculations.  These results, however,
are both consistent with the experimental values $T_c/E_F= 0.157 \pm
0.015$ \cite{Nascimbene-2010}, $0.17 \pm 0.01$ \cite{Horikoshi-2010},
and $0.167 \pm 0.013$ \cite{Ku-2012}. 
In this respect, it should be noted that a crucial advance of Ref.~\cite{Ku-2012} was the determination of the 
temperature directly from the density profiles without the use of a fitting procedure or an external thermometer.

 In addition, the bold
diagrammatic Monte Carlo method \cite{Prokofev-2008} was used in
Ref.~\cite{Van-Houcke-2012} to obtain the equation of state of the
unitary Fermi gas in the normal phase.  This method is formulated
directly in the thermodynamic limit and can thus be applied to the
continuum system with a genuine contact interaction.  It is based on
the idea of sampling stochastically Feynman diagrams for the
(imaginary time) self-energy using ``bold" (that is, self-consistent)
fermionic single-particle propagators $G$ and particle-particle
propagators $\Gamma$, summing up in this way a huge number of
diagrams.  The method relies on the assumption that the Feynman series
of self-consistent (skeleton) diagrams, possibly regularized with
re-summation techniques for diverging series, converges to the
physical self-energy $\Sigma$.  This assumption, however, was challenged in Ref.~\cite{Kozik-2015} (see also
Ref.~\cite{Rossi-2015}), where explicit counter-examples were given.
The reported excellent agreement with the experimental data for the
unitary Fermi gas \cite{Van-Houcke-2012}, on the other hand, seems to
indicate {\em a posteriori} that, at least for the unitary Fermi gas
in the normal phase, a good convergence to the physical self-energy
can be reached.  Clearly, an {\em a priori} knowledge about the
expected range of validity of the method would be desirable.
Progress in this direction was recently made in Ref.~\cite{Rossi-2016}, where
a sufficient condition was established for the convergence to the correct result by the self-consistent skeleton scheme.


We conclude by pointing out that, although the unitary Fermi gas has
been at the forefront of both theory and experiments in the field of
ultra-cold Fermi gases, under the perspective the BCS-BEC crossover
the unitary limit $(k_F a_F)^{-1}=0$ represents just one point (and
not necessarily the most significant one) within the
intermediate-coupling region.  For instance, as discussed in Section
\ref{sec:pseudo-gap}, the boundary between the fermionic (pseudo-gap)
and bosonic (molecular) regimes of the BCS-BEC crossover is found to
occur past unitarity at about $(k_F a_F)^{-1}\simeq 0.5$
\cite{Perali-2011}.  In addition, the maximum value of the superfluid
critical velocity (which signals where superfluidity is most robust)
does not occur exactly at unitarity but slightly past it, as obtained
by theoretical calculations \cite{Spuntarelli-2007} and measured by
experiments \cite{Miller-2007,Moritz-2015} (see Section
\ref{sec:Josephson} below).  Finally, in two dimensions, where the
unitary gas (that is, a scale-invariant interacting Fermi gas) cannot
be defined, the intermediate region of the BCS-BEC crossover is still
perfectly meaningful, and the unitary limit can be replaced with
alternative concepts characterizing the crossover region
\cite{Marsiglio-2015}.

\subsection{Tan contact}
\label{sec:Tancontact}

A major advance in the understanding of the physical properties of many-body systems with short-range interaction (like ultra-cold Fermi gases) has recently come from 
the introduction of a set of universal relations, expressed in terms of a quantity called the \emph{contact} $C$, which connect the strength of short-range two-body correlations
to the thermodynamics and provide powerful constraints on the behaviour of the system irrespective of its state
\cite{Tan-2008a,Tan-2008b,Tan-2008c}.

Originally, Tan derived the universal relations for the contact $C$ by
a suitable enforcement of the boundary conditions at short
inter-particle distance on the many-body Schr\"{o}dinger wave-function
for otherwise non-interacting particles.  Later on, these relations
were re-derived within quantum field theory methods involving the
operator product expansion \cite{Braaten-2008a,Braaten-2008b}, as well as using the
Schr\"odinger formalism in the coordinate representation
\cite{Combescot-2009}.  

From a physical and intuitive point of view, the contact $C$ describes
how the two-body problem locally merges into the surrounding many-body
problem, by considering the short-distance behaviour of the pair
correlation function for opposite-spin fermions
\cite{Tan-2008a,Braaten-2012}:
\begin{equation}
g_{\uparrow \downarrow}(\boldsymbol{\rho}) \underset{(\boldsymbol{\rho} \rightarrow 0)}{\longrightarrow} \frac{C}{(4 \, \pi)^2} 
\left( \frac{1}{\rho^2} - \frac{2}{a_F \rho} + \cdots \right) 
\label{pair-correlation-function-short-range-contact}
\end{equation}
\noindent
which generalizes the expression
(\ref{pair-correlation-function-short-range}) obtained in Section
\ref{sec:intra-inter-pair} at the level of the $t$-matrix.

Alternatively, the contact can be defined via the large
wave-vector tail of the fermionic distributions $n_\sigma(k)$ with
$k=|\vek{k}|$, which behaves asymptotically as \cite{Tan-2008a} 
\begin{equation}
  n_\sigma(k)\approx \frac{C}{k^4}\,.
\end{equation}

With a zero-range inter-particle interaction, both the kinetic and interaction energies diverge but their sum is well-defined, such that the internal
energy density $E/V$ and the pressure $P$ can be expressed in terms of the contact $C$ as \cite{Tan-2008a,Tan-2008b}: 
\begin{equation}
  \frac{E}{V} = \frac{C}{4\pi a_Fm} + \int\frac{d\vek{k}}{(2\pi)^3}
  \frac{k^2}{2m}\left(n_\sigma(k)-\frac{C}{k^4}\right)
\end{equation}
\begin{equation}  
  P = \frac{2}{3}\frac{E}{V} + \frac{C}{12\pi a_Fm}
\end{equation}
\noindent
where $V$ is the volume occupied by the system.
One can also show that the
derivative of the energy (at constant entropy) and of the free energy (at constant temperature) with respect to the scattering length can be related to the contact through the following
``adiabatic relations" \cite{Tan-2008b,Braaten-2012}:
\begin{equation}
\left(\frac{\partial E}{\partial a_F^{-1}}\right)_{S} = - \frac{C V}{4 \pi m}
\end{equation}
\begin{equation}
\left(\frac{\partial F}{\partial a_F^{-1}}\right)_{T} = - \frac{C V}{4 \pi m} \, .
\end{equation}

Additional quantities depend only on the short-range behaviour of the pair correlation function and can therefore be expressed in terms
of the contact $C$. In particular, the high-frequency tail of the
radio-frequency (RF) spectrum $I_{\mathrm{RF}}(\omega)$
\cite{Pieri-2009}  approaches
$I_{\mathrm{RF}}(\omega) \approx C \omega^{-3/2}/(2^{3/2} \pi^2)$ (where $\omega$ is in units of $E_F$ and $\int_{-\infty}^{+\infty}
\! d\omega I_{\mathrm{RF}}(\omega) = 1/2$), which holds provided final-state effects in the RF transition can be
neglected (see Ref.~\cite{Braaten-2010} for a derivation of the high-frequency tail based on a short-time operator product expansion).
Another example is the large
wave-vector tail of the static structure factor $S(Q) \approx C
[1/(8Q) - 1/(2\pi a_F Q^2)]$ where $Q = |\vek{Q}|$ \cite{Hu-2010}. Also the number of closed-channel molecules near a
Fano-Feshbach resonance is proportional to the contact
\cite{Werner-2008,Zhang-2009}.
For a comprehensive review of the universal relations involving the
contact, cf.~Ref.~\cite{Braaten-2012}.

Albeit useful at a formal level, the above relations do not allow one to actually compute the value of the contact $C$, as a function of coupling and temperature.
We now turn to this point, restricting ourselves to the case of a balanced gas.

The simplest way to introduce (and calculate) the
contact is to consider the large wave-vector behaviour of the fermionic
distribution $n(k)$ (per spin component).  At the mean-field level,
$n(k)$ is given by the factor $|\varv_{\vek{k}}|^2 = (1 -
\xi_{\vek{k}} / E_{\vek{k}})/2$ that enters the BCS wave function
(\ref{BCS-wave-function}).  Upon expanding this factor for large $k$,
one obtains:
\begin{equation}
n(k)=|\varv_{\vek{k}}|^2 \simeq \frac{\Delta^2}{4 \, \xi_{\vek{k}}^2} \simeq \frac{(m \Delta)^2}{k^{4}} 
\label{expansion-v_k-mean_field}
\end{equation}
\noindent
which allows one to approximate the contact $C$ with the quantity $(m \Delta)^2$  
at this level of approximation. 
In particular, in the BEC limit at
$T=0$, one gets $(m \Delta)^2 = 4 \pi n_0 / a_F$ from the relation
between the gap $\Delta$ and the wave function $\Phi$ of the
Gross-Pitaevskii equation (\ref{GP-equation}), where $n_0$ is the
condensate density within the present approximation.  In the BCS
regime, on the other hand,  $\Delta$
is exponentially small in the coupling parameter $(k_F a_F)^{-1}$ and thus the value $(m\Delta)^2$ cannot properly represent the contact in this limit.
However, one knows on general grounds that in weak coupling $C$ must reduce to the expression $(2\pi a_F n)^2$ as it can be readily obtained from the adiabatic 
relation reported above, where $\pi a_F n^2/m$ represents the leading coupling dependence of the free energy at $T=0$.

Inclusion of pairing fluctuations somewhat  modifies the mean-field results and recovers, in particular, the correct value for the contact in weak coupling at $T=0$. The simplest way to assess the effect of pairing fluctuations is to consider the normal phase above $T_c$  in terms of the non-self-consistent $t$-matrix approximation.
 For large $|\vek{k}|$ the
approximate expressions (\ref{approximate-Sigma-NSR}) and
(\ref{approximate-density-BEC_limit}) hold at any coupling, such that
one obtains
\begin{equation}
n(k) \simeq - \, \Delta_{\infty}^2 \,\, T \sum_{n} \, G_0(k)^2 \, G_0(-k)
\label{expansion-n_k-pairing_fluctuations}
\end{equation} 
\noindent
for each spin component, where the quantity $\Delta_{\infty}$ was
defined in Eq.(\ref{Delta-infinity}).  [Strictly speaking, in the
  limit of large $|\vek{k}|$ to the result
  (\ref{approximate-Sigma-NSR}) one should add the term $- \Gamma_0(k)
  \, n_F$ where $n_F$ is the particle density due to free
  fermions. However, this additional term turns out to be irrelevant
  for the calculation of $n(k)$.]  With the further result which holds
for large $|\vek{k}|$
\begin{equation}
T \! \sum_{n} \,G_0(k)^2 \, G_0(-k) \simeq - \frac{1}{4 \, \xi_{\vek{k}}^2} \simeq - \frac{m^2}{|\vek{k}|^{4}} \, ,                        
\label{triple-G-0-partial}
\end{equation}
\noindent
Eq.(\ref{expansion-n_k-pairing_fluctuations}) becomes $n(k) \simeq (m
\Delta_{\infty})^2/|\vek{k}|^{4}$ thus identifying $(m
\Delta_{\infty})^2$ with the contact $C$ within the present approximation, as anticipated in Section
\ref{sec:intra-inter-pair}.  The identification $C =(m
\Delta_{\infty})^2$ where $\Delta_{\infty}$ is given by
Eq.(\ref{Delta-infinity}) was originally introduced in
Ref.~\cite{Pieri-2009} within the non-self-consistent $t$-matrix, and
later generalized in Ref.~\cite{Van-Houcke-2013} to include
all possible higher-order effects in the particle-particle propagator.

\begin{figure}[t]
\begin{center}
\includegraphics[width=14.0cm,angle=0]{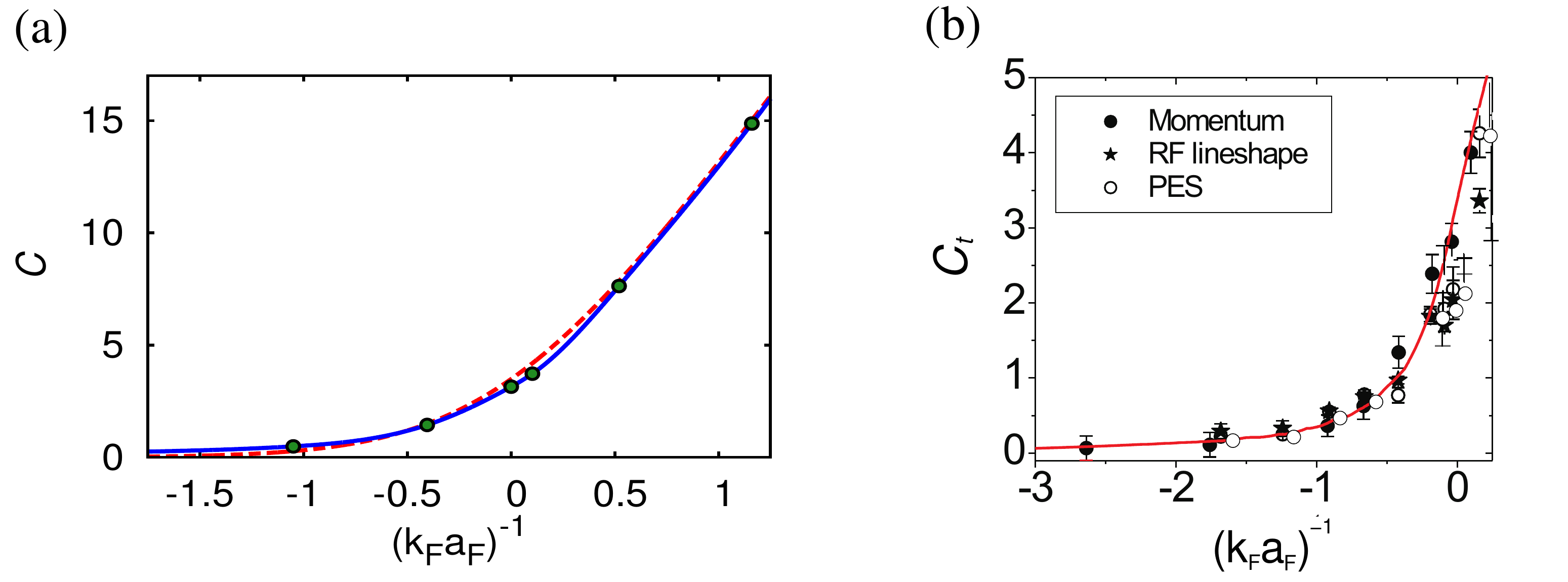}
\vspace{-0.3cm}
\caption{(a) Coupling dependence of the Tan's contact $C$ at $T = 0$ (in units of $k_F^4/3\pi^2$), obtained in terms of the gap $\Delta$ at the mean-field level (dashed line) 
and with the inclusion of pairing fluctuations in terms of the parameter $\Delta_{\infty}$ given by Eq.(\ref{Delta-infinity}) (circles and full line).  
(b) Experimental data of the (trap-averaged) contact $C_t$ (in units of $N k_F$, where $N$ is the total number of atoms and $k_F$ is the trap Fermi wave vector) for a gas 
of $^{40}\mathrm{K}$ atoms vs $(k_F a_F)^{-1}$.  
Data were obtained alternatively from the tail of the wave-vector (momentum) distribution (full circles), photo-emission spectrometry PES (empty circles), and the high-frequency 
tail of the RF line-shape (stars).  
[Panel (a) reproduced from Ref.~\cite{Palestini-2014}; panel (b) adapted from Ref.~\cite{Stewart-2010}.]}
\label{Figure-21}
\end{center}
\end{figure}

The quantity $\Delta_{\infty}^2$ can be calculated analytically in the
BCS and BEC limits.  Starting from the definition
(\ref{Delta-infinity}), in the BCS limit one expands $\Gamma_0(Q)$
given by Eq.(\ref{Gamma-0}) to second order in $4 \pi a_F/m$, so that
$\Gamma_0(Q) \simeq - \frac{4 \pi a_F}{m} \left( 1 - \frac{4 \pi
  a_F}{m} \, R_{pp}(Q) \right)$ with $R_{pp}(Q)$ given by
Eq.(\ref{bubble-pp-exact}).  This yields:
\begin{eqnarray}
(m \, \Delta_{\infty})^2 &\simeq& \left( 4 \pi a_F \right)^2 \int \!\! \frac{d\vek{Q}}{(2 \pi)^3} \, T \! \sum_{\nu} \, e^{i \Omega_{\nu} \eta} \, \chi_{pp}(\vek{Q},\Omega_{\nu})
\nonumber \\
& = & \left( 4 \pi a_F \right)^2 \!\! \int \!\! \frac{d\vek{Q}}{(2 \pi)^3} \, T \! \sum_{\nu} \, e^{i \Omega_{\nu} \eta} 
                                                          \int \!\! \frac{d\vek{k}}{(2 \pi)^3} \, T \! \sum_{n} \, G_0(\vek{Q}-\vek{k},\Omega_{\nu}-\omega_{n}) \, G_0(\vek{k},\omega_{n}) 
\,\,\, = \,\,\, \left( 2 \pi a_F n \right)^2
\label{Delta-infinity-BCS}
\end{eqnarray}
\noindent
where $\chi_{pp}$ is the particle-particle bubble given by
Eq.(\ref{particle-particle-bubble}) and $n$ is the (full) particle
density.  The value (\ref{Delta-infinity-BCS}) vanishes quadratically
with $|a_F|$ and is thus dominant with respect to the mean-field value
of the contact in the BCS limit, as discussed previously.  In the BEC
limit, on the other hand, using the approximate form
(\ref{Gamma-0-BEC_limit}) for $\Gamma_0$ one obtains:
\begin{equation}
 (m \, \Delta_{\infty})^2 \simeq - \frac{8 \pi}{a_F} \, \int \!\! \frac{d\vek{Q}}{(2 \pi)^3} \, T \! \sum_{\nu} \, 
\frac{e^{i \Omega_{\nu} \eta}}{i \Omega_{\nu} - \frac{\vek{Q}^2} {4 m} + \mu_B} = \frac{4 \pi n}{a_F} \, .
\label{Delta-infinity-BEC}
\end{equation}
\noindent
The result (\ref{Delta-infinity-BEC}) formally coincides with that
obtained within mean field, with the difference that here $n$ refers
to the full and not to the condensate density.  This kind of analysis
can be extended also to the superfluid phase below $T_c$. The
numerical results for the contact obtained at $T=0$ throughout the
BCS-BEC crossover with the inclusion of pairing fluctuations (circles and full line) and at the mean-field level (dashed line) are shown
in Fig.~\ref{Figure-21}(a).

Experimentally, the expected universality of the contact $C$ was
demonstrated in Ref.~\cite{Stewart-2010} for an ultra-cold gas of
trapped fermionic ($^{40}\mathrm{K}$) atoms, where the (trap-averaged)
contact $C_t$ was measured at a temperature of about $0.1 T_F$ by
three independent methods.  Specifically, through (i) the large
wave-vector tail of the fermionic distribution $n(k)$,
(ii) the high frequency tail of the RF spectrum
$I_{\mathrm{RF}}(\omega)$, as well as (iii) using wave-vector-resolved
RF (photo-emission) spectroscopy.  The data obtained by these three
independent experimental methods are reported in
Fig.~\ref{Figure-21}(b).  Here, the mutual agreement (obtained within
the experimental errors) of the different results can be considered as
an empirical proof of the universality of Tan's relations.

\begin{figure}[h]
\begin{center}
\includegraphics[width=9.0cm,angle=0]{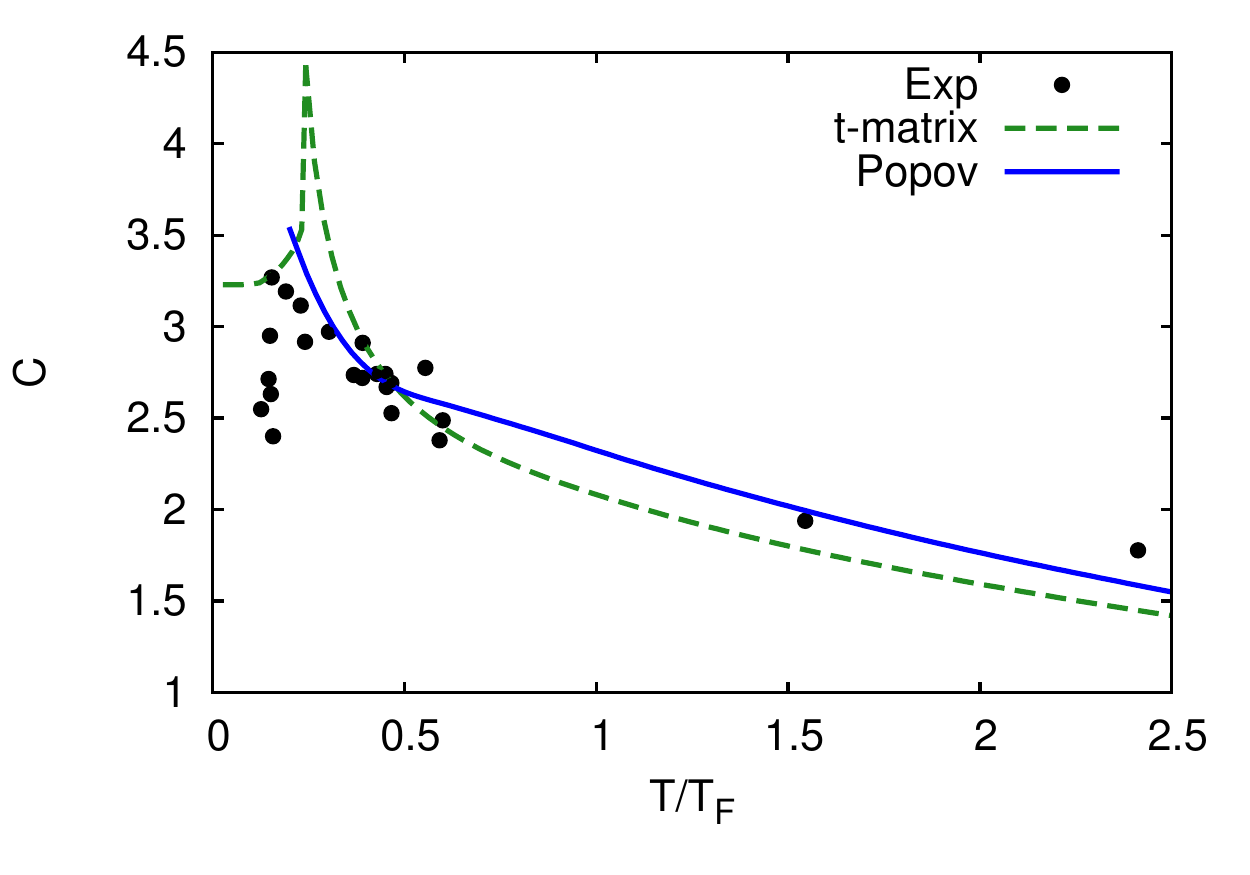}
\caption{Temperature dependence of the contact $C$ (in units of $k_F^4/3\pi^2$) for the homogeneous gas at unitarity.  
The experimental values from Ref.~\cite{Sagi-2012} (circles) are compared with the theoretical calculations of Ref.~\cite{Palestini-2010} based on the non-self-consistent 
$t$-matrix approximation (dashed line) and with an improved approximation based on the work of Ref.~\cite{Pieri-2005a} that also includes correlations among pre-formed pairs (full line).
[Courtesy of F. Palestini.]
}
\label{Figure-22}
\end{center}
\end{figure}

Besides the dependence of the contact $C$ on coupling at a given
temperature, its dependence on temperature at a given coupling is also
of interest.  This dependence was first determined analytically in the
low- and high-temperature limits in Ref.~\cite{Yu-2009b}.  It was found
that in the low-temperature limit the contact increases like $T^{4}$
due to phonons, while in the high-temperature limit the contact
decreases like $1/T$, thereby implying that the contact must achieve a
maximum at an intermediate temperature (which is expected to be of the
order of the Fermi temperature $T_F$).  [Higher orders in a high-temperature expansion were obtained in Ref.~\cite{Sun-2015}.] 
The full temperature
dependence of the contact for the homogeneous gas at unitarity was
then calculated in Ref.~\cite{Palestini-2010} from the
large wave-vector tail of the fermionic distribution $n(k)$, both
below and above $T_c$, and is reported Fig.~\ref{Figure-22} (dashed
line).  This calculation is based on the non-self-consistent
$t$-matrix approximation of Section \ref{sec:nsr}, which is known to
become exact in the high-temperature limit \cite{Combescot-2006a}, and
presents an enhancement of the contact $C$ when approaching $T_c$
from above that was attributed to pseudo-gap effects (a broad maximum
above $T_c$ was also found by QMC calculations~\cite{Drut-2011}).
Above $T_c$, Fig.~\ref{Figure-22} shows also an improved calculation
(full line) based on an extended version of the non-self-consistent
$t$-matrix \cite{Pieri-2005a}, which includes correlations among
pre-formed pairs that are missing in the $t$-matrix and which improves
on the value of $T_c$ and on the high-temperature behaviour of $C$.
The temperature dependence of the contact for the unitary Fermi gas has further been studied 
with the self-consistent t-matrix~\cite{Enss-2011},  Gaussian pairing-fluctuations~\cite{Hu-2011}, and 
functional renormalization group~\cite{Boettcher-2013} approaches.

Figure \ref{Figure-22} further reports the experimental measurements
of the contact for the homogeneous gas (circles), based on a technique
that allows the local properties of the trapped cloud to be probed
\cite{Sagi-2012}.  These measurements show a gradual decrease of the
contact with increasing temperature as predicted by theory, while at
the edge of the experimentally attainable temperatures (which is about
where the experimental value of $T_c$ is expected to occur) a sharp
decrease of the contact is observed.  Later on, based on the evidence
that the contact gets enhanced near $T_c$, the question was raised
about whether the contact displays a critical behaviour near a
continuous transition, in spite of the fact that it remains finite
even for strongly interacting fermions both in the normal and
superfluid phases \cite{Chen-2014}.

In nuclear physics, the range of the interaction is not short enough to allow finding a $1/k^4$ tail in the occupation numbers. 
Nevertheless, the concept of the contact has been generalized to the nuclear physics context as well (see Section \ref{sec:nuclear-contact} below).

\subsection{Josephson effect}
\label{sec:Josephson}

The Josephson effect is probably the most striking consequence of the
spontaneous symmetry breaking of the phase $\varphi(\vek{r})$ of the
complex gap (order) parameter $\Delta(\vek{r}) = |\Delta(\vek{r})|
e^{i \varphi(\vek{r})}$, which is at the essence of the macroscopic
quantum coherence of superconductivity.  In its original formulation
\cite{Josephson-1962}, the Josephson effect is the phenomenon for
which a super-current flows indefinitely without any voltage applied
across a physical constriction or a junction (which can be either a
thin insulating barrier or a point of contact) that weakens the
superconductivity in between two superconductors.  Such Josephson
junctions have important applications in quantum-mechanical circuits
\cite{Barone-1982}.  At a microscopic level, the phenomenon amounts to
the coherent tunneling of Cooper pairs across a potential barrier
placed at the interface between two superconductors.  It results in a
\emph{characteristic relation} $J(\delta \varphi)$ between the
super-current $J$ and the asymptotic phase difference $\delta \varphi$
across the potential barrier, which in its simplest version has the
form $J(\delta \phi) = J_0 \sin(\delta \varphi)$ where $J_0$ is the
maximum value of the current attainable with the barrier
\cite{Barone-1982}.

The BdG equations (\ref{BdG-equations}) can be used to calculate the
relation $J(\delta \phi)$ for an arbitrary shape of the potential
barrier.  And not only in the conventional situations of BCS weak
coupling when Cooper pairs tunnel across the barrier, but also under
more general conditions when the inter-particle interaction is spanned
up to the BEC strong-coupling limit where composite bosons are instead
responsible for tunneling.  In a recent study of the Josephson effect
throughout the BCS-BEC crossover at zero temperature
\cite{Spuntarelli-2007,Spuntarelli-2010}, a slab geometry was adopted
where a potential barrier $V_{\mathrm{ext}}(x)$ (either
rectangular or Gaussian) is embedded in a homogeneous superfluid that
extends to infinity on both sides of the barrier.  Accordingly, the
order parameter has the form $\Delta(x) = |\Delta(x)| e^{2i[q x +
    \phi(x)]}$, where $q = J m / n_0$ is the wave vector associated
with the uniform super-current ($n_0$ being the bulk fermionic number
density) and $2 \phi(x)$ is the local phase accumulated by the order
parameter over and above the reference value $2 q x$ in the absence of
the barrier, such that $\delta \varphi = 2 [\phi(x = + \infty) -
  \phi(x = - \infty)]$ across the barrier.  It turns out that, while a
full implementation of self-consistency for the solution of the BdG
equations is not required on the BCS side, self-consistency is
essential on the BEC side to account for the non-linearity of the GP
equation (\ref{GP-equation}) to which the BdG equations reduce in this
limit (cf. Section~\ref{sec:GLandGP}).  
\begin{figure}[h]
\begin{center}
\includegraphics[width=13.5cm,angle=0]{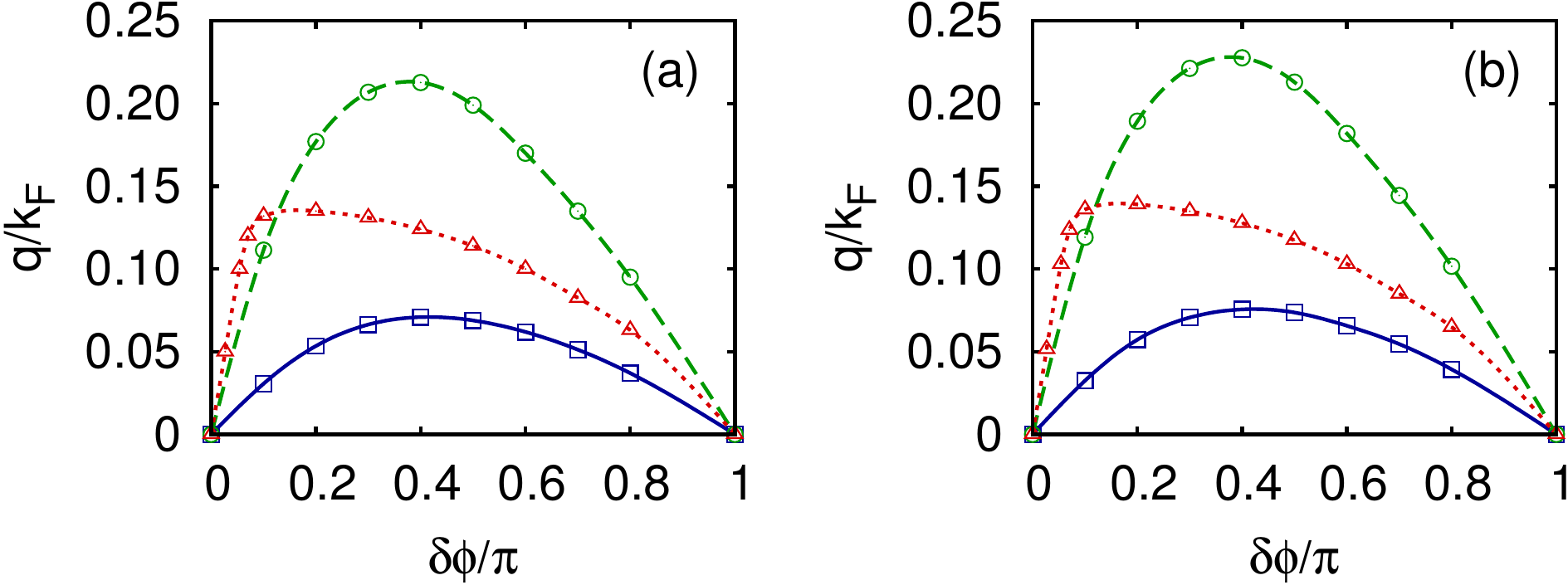}
\caption{Josephson characteristics for: (a) A rectangular barrier centered at $x=0$ of width $L k_F = 4.0$ and height $V_0/E_F = 0.1$;
(b) A Gaussian barrier $V_{\mathrm{eff}}(x) = V_0 e^{-x^2/(2 \sigma^2)}$ with $\sigma k_F = 1.6$ and the same value of $V_0$.
The three curves correspond to $(k_F a_F)^{-1} = -1.0$ (dotted lines), $0.0$ (dashed lines) and $+1.5$ (full lines).  
[Reproduced from Ref.~\cite{Spuntarelli-2010}.]}
\label{Figure-23}
\end{center}
\end{figure}
Typical Josephson
characteristics obtained in this way throughout the BCS-BEC crossover
are shown in Fig.~\ref{Figure-23}, where they are seen to depend
markedly on the inter-particle coupling $(k_F a_F)^{-1}$ for a given
barrier.

\begin{figure}[t]
\begin{center}
\includegraphics[width=8.0cm,angle=0]{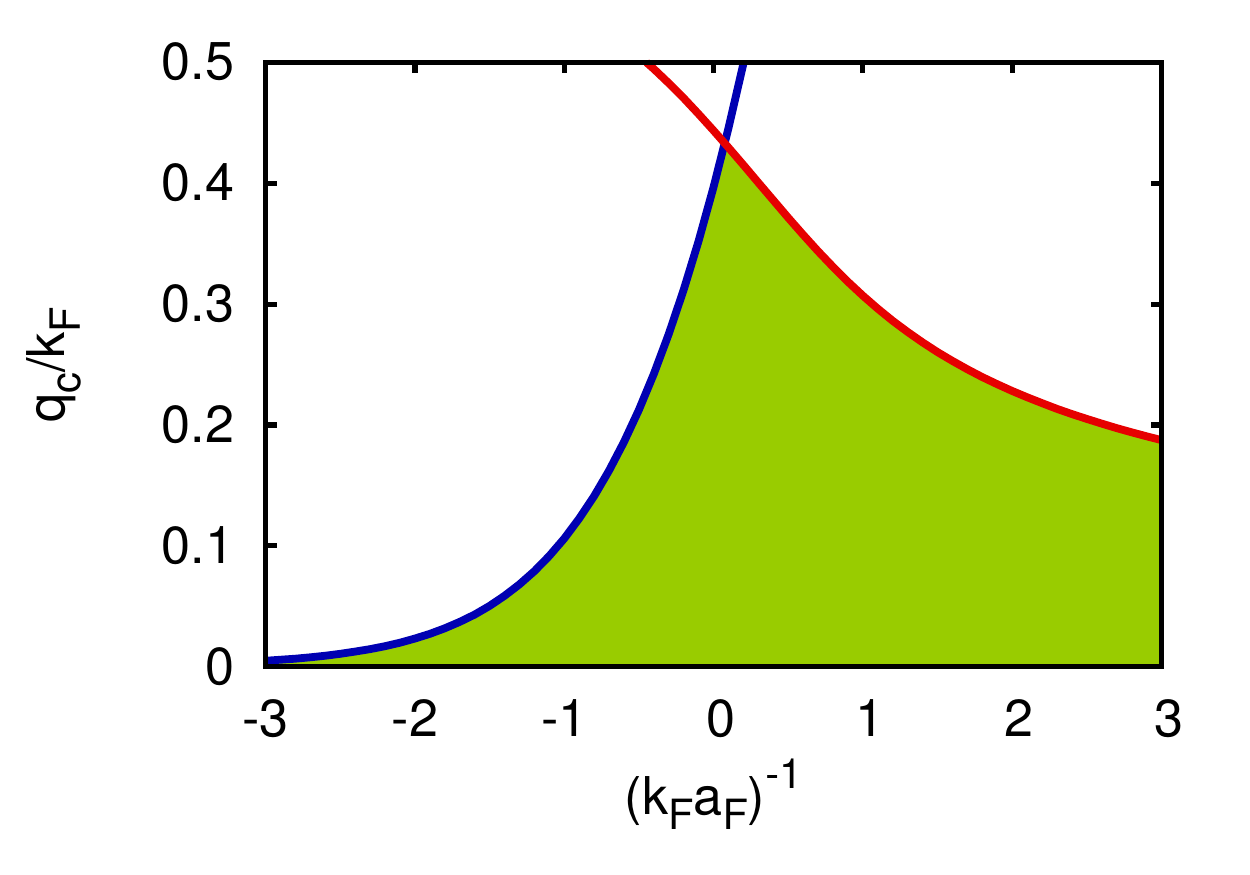}
\caption{Critical velocity vs coupling. The shaded region identifies the allowed region outside which the superfluid flow becomes unstable.  
The left (blue) and right (red) boundary curves are associated with pair-breaking and sound-mode excitations. 
[Reproduced from Ref.~\cite{Spuntarelli-2010}.]}
\label{Figure-24}
\end{center}
\end{figure}

\begin{figure}[h]
\begin{center}
\includegraphics[width=14cm,angle=0]{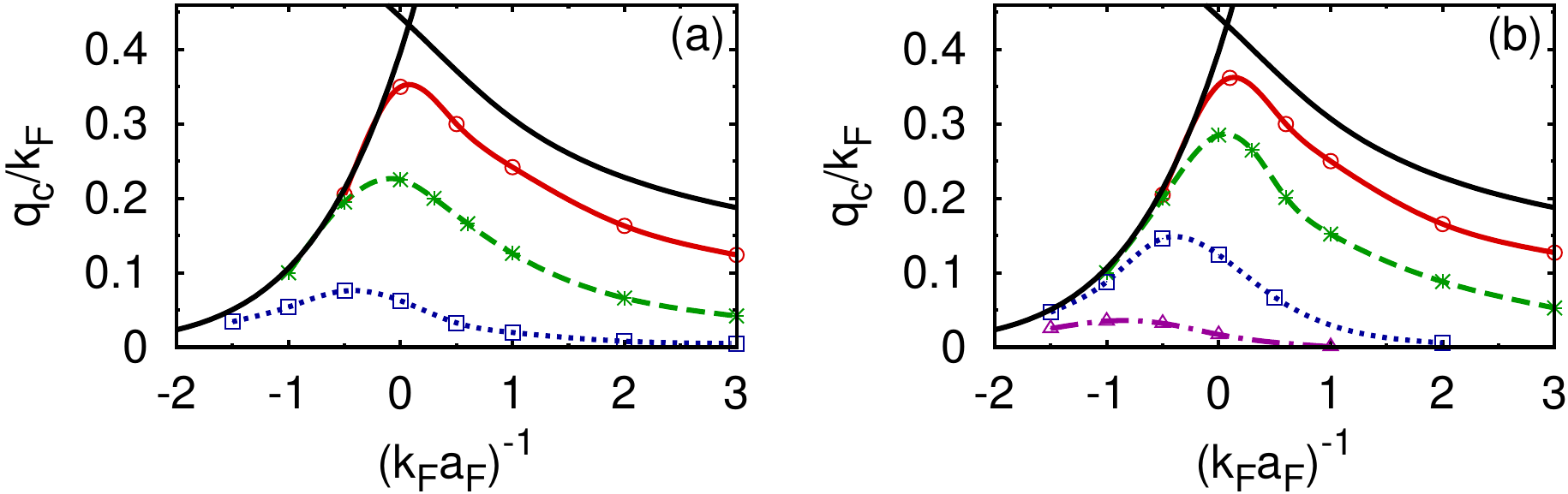}
\caption{Maximum velocity $q_{c}/m$ obtained from the maximum of the Josephson characteristics vs $(k_F a_F)^{-1}$, for several rectangular barriers with: 
(a) $L k_F = 2.65$ and $V_0/E_F = (0.02,0.10,0.40)$ from top to bottom; (b) $L k_F = 5.30$ and $V_0/E_F = (0.01,0.05,0.20,0.50)$ from top to bottom.  
The Landau critical velocity is also reported (full lines without symbols). 
[Reproduced from Ref.~\cite{Spuntarelli-2010}.]}
\label{Figure-25}
\end{center}
\end{figure}

For given coupling, the shape of the Josephson characteristics varies also
with the width and height of the barrier.  In particular, these
variations affect the value of the maximum Josephson current $J_0$
which increases as the ratio $V_0/E_F$ decreases.  For any given
coupling, it is thus of interest to determine the maximum allowed
value of the Josephson current obtained by progressively lowering the
barrier height, which corresponds to an \emph{intrinsic} upper value
$q_{c}/m$ for the velocity of the superfluid flow.  On physical
grounds, a vanishing small (albeit non zero) barrier acts as an
impurity that probes the stability of the homogeneous superfluid flow
and plays a similar role to the walls of a container in the context of
the \emph{Landau criterion for superconductivity}
\cite{Abrikosov-1975}.  In this respect, alternative dissipation
mechanisms are active on the two sides of unitarity and are associated
with the onset of quasi-particle excitations of a different nature,
namely, pair-breaking excitations on the BCS side and sound-mode
quanta on the BEC side.  In the presence of a barrier, the allowed
values of the superfluid flow are thus expected to lie inside the
shaded region of Fig.~\ref{Figure-24}, which is bounded by the
critical velocities associated with the above two kinds of
excitations.

Detailed numerical calculations with the BdG equations have confirmed this expectation, leading one to conclude that the Josephson effect is most robust in the unitary regime 
where $q_{c}$ attains its maximum \cite{Spuntarelli-2007,Spuntarelli-2010}.  
Examples of calculations of the maximum Josephson current for given barrier are reported in Fig.~\ref{Figure-25}.  
A similar conclusion has also been confirmed experimentally, through a direct measurement of the critical velocity of an ultra-cold superfluid Fermi gas in the BCS-BEC crossover
(cf. Fig.2 of Ref.~\cite{Miller-2007} and Fig.2 of Ref.~\cite{Moritz-2015}).

In terms of the above analysis, by spanning the BCS-BEC crossover it
was possible to smoothly connect the Josephson effect conceived
originally for fermions \cite{Josephson-1962,Barone-1982} to the
corresponding effect for bosons (as measured, e.g., in
Ref.~\cite{Engels-2007}).  This connection, in turn, has allowed for a
deeper understanding of the Josephson effect
\cite{Spuntarelli-2007,Spuntarelli-2010}.

\subsection{Collective modes and anisotropic expansion}\label{sec:collective}

Collective modes and the expansion from an anisotropic trap allow one
to obtain interesting information on the dynamical regime of a system.
A simple example is a non-interacting Fermi gas in an anisotropic trap
with $\omega_x > \omega_y$.  When the trap is switched off at $t = 0$,
each atom follows its ballistic trajectory $\vek{r}(t) = \vek{r}(0) +
t \, \vek{p}(0)/m$, so that after a long time of flight the cloud
profile reflects the spherical shape of the wave-vector distribution
at $t=0$.  Accordingly, the aspect ratio $R = \sqrt{\langle x^2\rangle
  / \langle y^2\rangle}$, which was initially given by $R =
\omega_y/\omega_x < 1$, approaches $R=1$ for long times.  The
situation is completely different in the hydrodynamic regime.  If the
cloud is initially in hydrostatic equilibrium, the pressure gradient
has to compensate the force of the trap.  For the above example, this
means that the pressure gradient is stronger in the $x$ than in the
$y$ direction, so that when the gas is released from the trap it will
be accelerated more strongly in the $x$ than in $y$ direction.
Accordingly, during the expansion the aspect ratio will turn from
$R<1$ to $R>1$.

Differences between the collisionless and hydrodynamic regimes can be
observed also in small-amplitude collective oscillations.  Among the
different modes that have been studied, the radial quadrupole mode
exhibits a particularly drastic change.  This mode is an oscillation
of the gas in a cylindrical trap (with $\omega_x = \omega_y \equiv
\omega_r$), which can be excited by squeezing the gas in $x$ direction
and simultaneously expanding it in $y$ direction for a short time (so
as to keep the volume constant).  In this way, the cloud shape
oscillates with frequency $\approx 2\omega_r$ in the collisionless
regime \cite{Menotti-2002} and $\sqrt{2}\omega_r$ in the hydrodynamic
regime \cite{Stringari-1996,Baranov-2000,Cozzini-2003}.

The hydrodynamic behaviour can be a consequence of superfluidity (the
so-called collisionless hydrodynamics).  For instance, in the BEC
limit the dynamics of the gas at $T=0$ is described by the
time-dependent Gross-Pitaevskii equation for dimers, which can be
reduced to the continuity equation
\begin{equation}
\dot{n} + \vek{\nabla}\cdot (n\vek{v}) = 0
\label{eq:continuity}
\end{equation}
and the Euler equation
\begin{equation}
\dot{\vek{v}} + \vek{\nabla}\left(\frac{\vek{v}^2}{2} +
  \frac{V_{\mathrm{ext}}}{m} + \frac{\mu_{\mathrm{loc}}}{m} \right) = 0
\label{eq:Euler}
\end{equation}
for large particle numbers \cite{Dalfovo-1999}.  Here,
$\vek{v}(\vek{r},t)$ is the velocity field and
$\mu_{\mathrm{loc}}(\vek{r},t)$ the local chemical potential
calculated from the density $n(\vek{r},t)$.  [Note that one can
  rewrite $\vek{\nabla}\mu_{\mathrm{loc}} = (1/n) \vek{\nabla}P$ where
  $P$ is the pressure.]  On the BCS side, it can be shown that at
$T=0$ the time-dependent BdG equations, too, can be reduced to the
continuity (\ref{eq:continuity}) and Euler (\ref{eq:Euler}) equations,
provided the dynamics is sufficiently slow to avoid pair breaking and
the Cooper pair size is sufficiently small compared to the system size
\cite{Urban-2006,Tonini-2006}.

The system behaves hydrodynamically also in the normal phase when
there occur enough collisions between atoms to maintain local
equilibrium (the so-called collisional hydrodynamics).  In this case,
Eqs.~(\ref{eq:continuity}) and (\ref{eq:Euler}) can be derived from
the Boltzmann equation \cite{LandauLifshitz10}.  If the system is very
dilute (like at high temperature) or weakly interacting, the normal
phase will be in the collisionless regime.  In this case, the
collision term of the Boltzmann equation (see Eq.~(\ref{eq:Boltzmann})
below) can be neglected and the system dynamics described by the
Vlasov equation:
\begin{equation}
\dot{f} + \frac{\vek{p}}{m}\cdot \vek{\nabla}_{\vek{r}} f -
(\vek{\nabla} V)\cdot\vek{\nabla}_{\vek{p}} f = 0\,,
\label{eq:Vlasov}
\end{equation}
where $f(\vek{r},\vek{p},t)$ is the distribution function per spin
state (we assume $f_\uparrow = f_\downarrow = f$) and $V =
V_{\mathrm{ext}}+U$ is the potential that takes into account the trap
$V_{\mathrm{ext}}$ and a mean-field-like potential $U$.
Theoretically, most challenging are the situations when the system
cannot be described by the above limiting cases.  We discuss two
examples where the intermediate regimes have explicitly been studied.

We first consider the $T=0$ case and discuss how well hydrodynamics
reproduces the results of the time-dependent BdG theory.  For
small-amplitude collective oscillations, the time-dependent BdG
equations can be linearized around equilibrium.  [In nuclear physics
  this procedure is referred to as the Quasiparticle Random-Phase
  Approximation (QRPA), while in condensed matter is known as RPA.]
QRPA calculations were performed in the BCS regime for the collective
modes of a gas in a spherical trap with $\omega_z = \omega_r$
\cite{Bruun-2001,Grasso-2005}.  It was found that hydrodynamics
is only valid if the gap $\Delta$ (at the trap center) is much larger
than the level spacing $\omega_r$ in the trap; otherwise, the
frequency of the collective mode can be significantly smaller than
that predicted by hydrodynamics, and the mode gets damped due to its
coupling to two-particle excitations (pair breaking).  In
Ref.~\cite{Combescot-2004} this pair-breaking effect was
suggested as an explanation for the frequency shift and damping of the
radial breathing mode observed experimentally in
Ref.~\cite{Bartenstein-2004b}.

The QRPA equations were also solved for the collective modes in a uniform gas \cite{Marini-1998,Combescot-2006b}.  
At small wave vector $q$, the QRPA dispersion $\omega(q)$ of the collective mode follows the linear behaviour $\omega = cq$ of the Bogoliubov-Anderson sound
\cite{Bogoliubov-1959,Anderson-1958}, where $c= \sqrt{(1/n)\partial P/\partial n}$ is the hydrodynamic sound velocity.  
In the BCS regime (but also at unitarity), for increasing $q$ the dispersion relation $\omega(q)$ approaches the threshold for pair breaking at about $2\Delta$,
and the QRPA dispersion $\omega(q)$ becomes progressively flat with a negative curvature $d^2\omega/dq^2 < 0$.  
Analogous results were obtained in Ref.~\cite{Forbes-2013}, where equations similar to the time-dependent BdG equations (albeit adapted to the unitary limit)
were solved, as well as in Ref.~\cite{Martin-2014} where the QRPA was solved for neutron matter. 
On the BEC side where $\mu < 0$, the situation is different since the pair-breaking threshold is no longer flat as a function of $q$ \cite{Combescot-2006b}. 
In this case, the QRPA dispersion $\omega(q)$ has a positive curvature but does not enter the continuum.  
Deep in the BEC limit, the QRPA reproduces the dispersion $\omega = \sqrt{(cq)^2+(q^2/2m_B)^2}$ of the Bogoliubov mode which is obtained from the linearized 
time-dependent GP equation for the condensate of composite bosons with mass $m_B = 2m$ \cite{Dalfovo-1999}.  
In this case, the sound velocity $c = \sqrt{8\pi a_F/m_B^2}$ corresponds to the (Born approximation) value $a_{B} = 2 a_F$ of the boson-boson scattering length, as discussed in Section
\ref{sec:b-b-interaction}.

We next consider the regime of sufficiently strong pairing and long
wavelength, such that hydrodynamics can be applied at $T=0$.  The
question then arises about what happens at finite temperature, when
thermal quasi-particle excitations form a normal-fluid component in
the superfluid.  If the collision rate of the thermal quasi-particles
is high enough (that is, much higher than the frequency of the
collective oscillation), also this normal component can be treated
hydrodynamically and the collective modes can be described in the
framework of Landau two-fluid hydrodynamics \cite{Taylor-2005}.
In particular, the two-fluid hydrodynamics predicts the existence of
second sound (for which normal and superfluid components oscillate out
of phase).  In a recent experiment with $^6$Li atoms at unitarity
\cite{Sidorenkov-2013}, second sound was indeed observed and the
temperature dependence of the normal-fluid density was extracted.

In the BCS regime, on the other hand, the normal component is more likely to be in the collisionless regime.  
In this case, one can again apply the time-dependent BdG theory (or its linearized form QRPA).  
A simplification, which is computationally less demanding, is provided by a semiclassical quasi-particle transport theory developed initially for clean superconductors 
in Ref.~\cite{Betbeder-Matibet-1969}. 
It can be derived by a formal expansion of the time-dependent BdG equations in powers of the Planck constant $\hbar$ \cite{Urban-2006}, in analogy to
the derivation of the Vlasov equation (\ref{eq:Vlasov}) from the time-dependent Hartree-Fock theory \cite{Ring-1980}.  
In this quasi-particle transport theory, the normal component is described in terms of a quasi-particle distribution function $\nu(\vek{r},\vek{p},t)$, which satisfies a Vlasov-like 
equation of motion
\begin{equation}
\dot{\nu} + (\vek{\nabla}_{\vek{p}}E)\cdot\vek{\nabla}_{\vek{r}} \nu -(\vek{\nabla}_{\vek{r}}E)\cdot\vek{\nabla}_{\vek{p}} \nu = 0 \, ,
\end{equation}
where $E$ is the quasiparticle energy in the local rest frame of the superfluid (see Refs.~\cite{Urban-2006,Urban-2007} for the explicit expressions).  
The superfluid velocity is obtained by solving the continuity equation for the total (superfluid and normal) density and current at the same time.

\begin{figure}[h]
\begin{center}
\includegraphics[width=15cm]{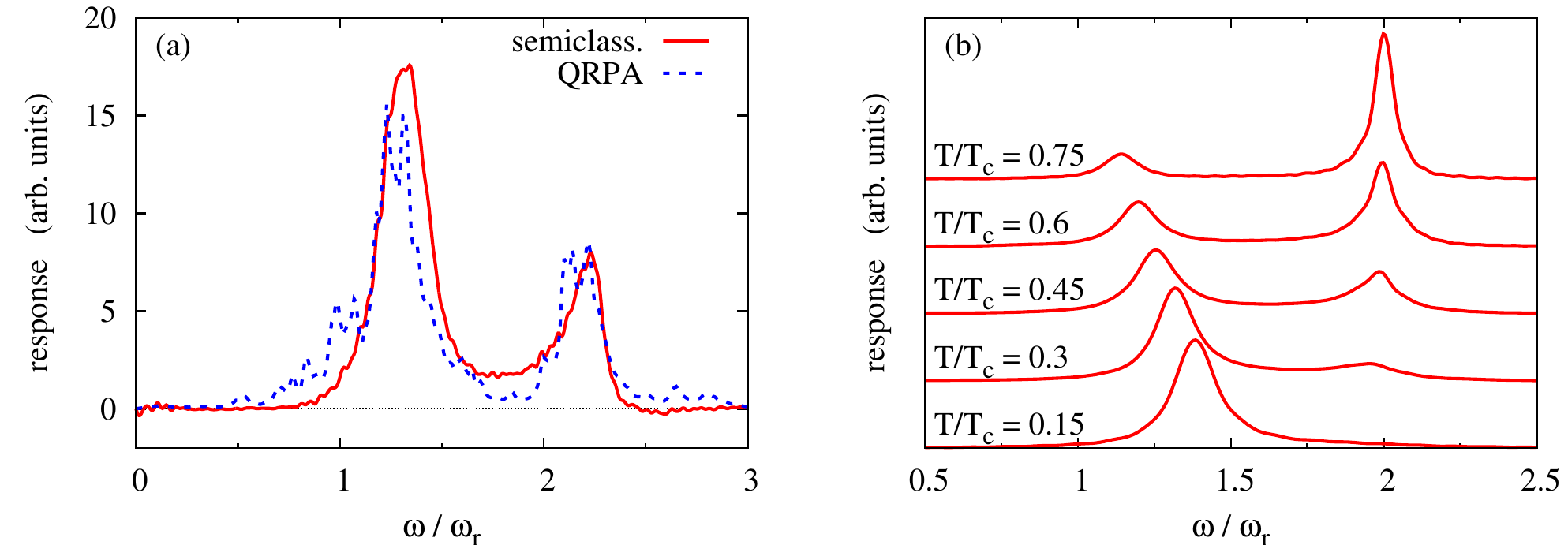}
\caption{(a) Quadrupole response function for a gas of $17000$ atoms in a spherical trap with $(k_F a_F)^{-1} = -1.5$ and $T = 0.5 T_c$. 
The results of the semiclassical quasi-particle transport theory of Ref.~\cite{Urban-2007} (solid line) are compared to the QRPA results of Ref.~\cite{Grasso-2005} (dotted line). 
(b) Semiclassical results for the quadrupole response of a gas of $400000$ atoms in a cylindrical trap with $(k_F a_F)^{-1} = -1.5$ and various temperatures.
[Panel (a) adapted from Ref.~\cite{Urban-2007}; panel (b) adapted from Ref.~\cite{Urban-2008b}.]}
\label{Figure-26}
\end{center}
\end{figure}

Figure~\ref{Figure-26}(a) shows a comparison of the quadrupole
response function (that is, the Fourier transform of the quadrupole
moment $\langle x^2-y^2\rangle$ after its excitation at $t=0$) for $T
\approx T_c/2$, as obtained within the semiclassical theory and the
QRPA.  With this set of parameters, hydrodynamics works well at $T=0$
since in this case the gap at the center of the trap is $\approx
6\omega_r$.  At $T = 0.5 T_c$, one sees from this figure that the
agreement between QRPA and the semiclassical theory remains
satisfactory.  The presence of the normal component leads to a strong
damping (which is reflected in the width of the peak) of the
hydrodynamic mode at $\sqrt{2}\omega_r$, as well as to the presence of
a second damped mode with frequency $2\omega_r$ corresponding to the
collisionless normal gas.

While the QRPA is limited for computational reasons to spherical traps
with small particle numbers, the semiclassical theory can be applied
also to realistic geometries and particle numbers \cite{Urban-2008b},
as shown in Fig.~\ref{Figure-26}(b). One distinguishes again two
peaks, whose strength is shifted for increasing temperature from the
hydrodynamic mode at left to the ballistic mode at right. In addition
to being broadened, the hydrodynamic mode is also shifted to lower
frequencies, which allows for a qualitative explanation of the
experimental result of Ref.~\cite{Altmeyer-2007a} as shown in
Ref.~\cite{Urban-2008b}.

\begin{figure}[t]
\begin{center}
\includegraphics[width=15cm]{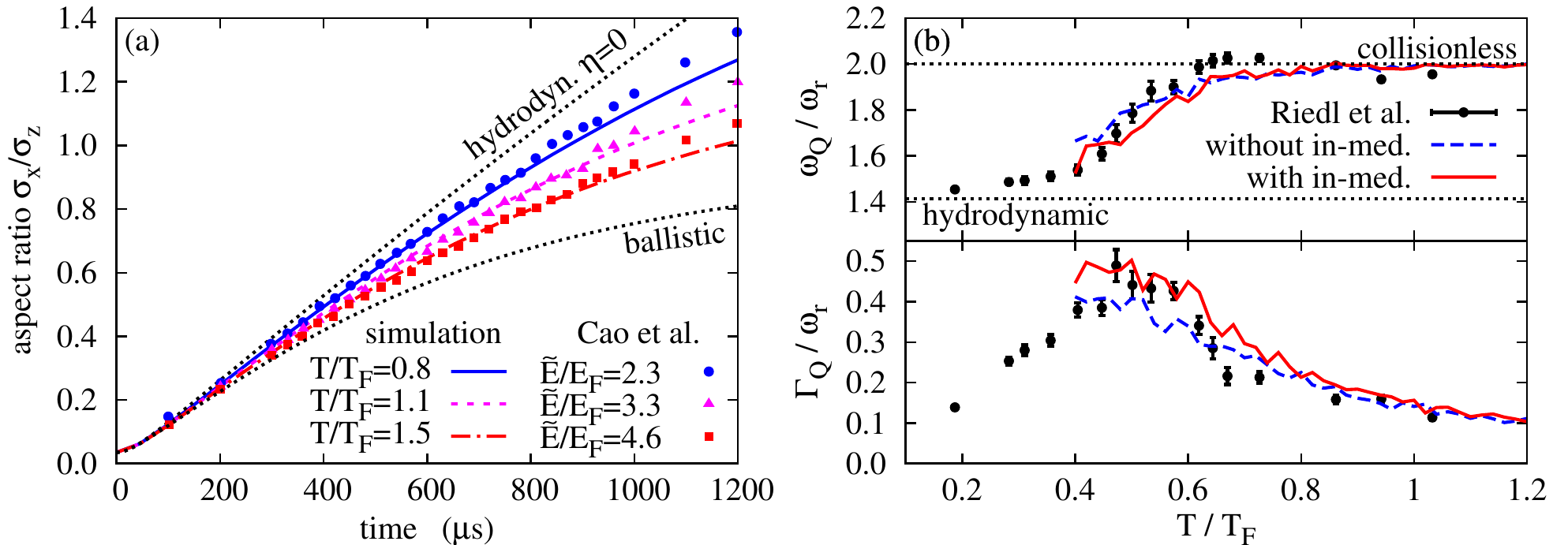}
\caption{Comparison of the numerical solutions of the Boltzmann equation for the unitary Fermi gas in the normal phase (lines) with the experimental data (symbols). 
(a) Time dependence of the aspect ratio during the anisotropic expansion of the cloud for different initial temperatures (experimental data from Ref.~\cite{Cao-2011}). 
(b) Temperature dependence of the frequency (top) and damping rate (bottom) of the quadrupole mode (experimental data from Ref.~\cite{Riedl-2008}).
[Adapted from Ref.~\cite{Pantel-2015}.]}
\label{Figure-27} 
\end{center}
\end{figure}

Finally, it may happen that the effects of collisions in the normal
component are not negligible but insufficient to guarantee
hydrodynamic behaviour.  In this case, neither the time-dependent BdG
equations nor the Landau two-fluid description are applicable (to our
knowledge, this theoretically difficult case has not been studied thus
far).  The situation becomes much simpler for temperatures $T > T_c$
where the entire system is in the normal phase.  In this case, the
transition from collisional hydrodynamic to collisionless behaviour,
which is observed experimentally as function of temperature or
coupling \cite{Wright-2007,Riedl-2008}, has been theoretically
described in the BCS regime up to unitarity in terms of the Boltzmann
equation
\cite{Toschi-2003,Massignan-2005,Riedl-2008,Chiacchiera-2009,Lepers-2010,Chiacchiera-2011,Pantel-2015}:
\begin{equation}
\dot{f} + \frac{\vek{p}}{m}\cdot \vek{\nabla}_{\vek{r}} f -
(\vek{\nabla} V)\cdot\vek{\nabla}_{\vek{p}} f \\ = -\int
\frac{d \vek{p}_1}{(2\pi)^3} \int d\Omega \frac{d\sigma}{d\Omega}
\frac{|\vek{p}-\vek{p}_1|}{m} [ff_1(1-f_2)(1-f_3)-f_2f_3(1-f)(1-f_1)] \, .
\label{eq:Boltzmann}
\end{equation}
[This equation is sometimes referred to as the Boltzmann-Nordheim or Boltzmann-Uehling-Uhlenbeck equation.]  
In the collision integral on the right-hand side, $f_i$ is a short-hand notation for  $f(\vek{r},\vek{p}_i,t)$, $\Omega$ is the solid angle defining the direction of $\vek{p}_2-\vek{p}_3\equiv 2\vek{q}$, 
while the total wave vector and the absolute value of the relative wave vector are determined by the conservation requirements 
$\vek{p}_2+\vek{p}_3 = \vek{p}+\vek{p}_1 \equiv \vek{k}$ and $|\vek{p}_2-\vek{p}_3| = |\vek{p}-\vek{p}_1| \equiv 2q$.  
The factors $(1-f)$ and $(1-f_i)$ in the collision integral underline the occurrence of Pauli blocking, which is absent in the classical Boltzmann equation.

To a first approximation, the cross section $d\sigma/d\Omega =
a_F^2/(1+q^2a_F^2)$ for the scattering of two atoms in vacuum can be
used in Eq.~(\ref{eq:Boltzmann}).  However, the cross section and thus
the collision rate may be strongly affected by medium effects.  In the
medium, the cross section is given by $d\sigma/d\Omega =
|m\Gamma^R(\vek{k},\omega)/4\pi|^2$, with $\omega =
k^2/4m+q^2/m-2\mu$.  As $\Gamma^R$ develops a pole at $T=T_c$,
$d\sigma/d\Omega$ can be strongly enhanced already at temperatures
above but close to $T_c$.  The effect of this enhancement on the
viscosity of the gas was studied in
Refs.~\cite{Bruun-2005,Bluhm-2014}.  The in-medium $t$-matrix enters
also the left-hand side of Eq.(\ref{eq:Boltzmann}), through the
mean-field-like contribution to $V$ calculated from the self-energy
\cite{Chiacchiera-2009,Pantel-2015}.

As an example of the quantitative precision that can be reached with
the Boltzmann equation, Fig.~\ref{Figure-27} compares the solution of
the Boltzmann equation, obtained in Ref.~\cite{Pantel-2015} for the
unitary Fermi gas in the normal phase, with the experimental data.
One sees that both (a) the expansion of the cloud from an anisotropic
trap and (b) the quadrupole mode are well described by the theory.
Unfortunately, these observables are not very sensitive to medium
effects, since they give strong weight to parts of the cloud far away
from the cloud center (where the local $T/T_F$ is large and medium
effects are weak). In addition, the enhanced collision rate and the
mean-field-like potential act into opposite directions
\cite{Pantel-2015}, making the observation of medium effects even more
difficult.

\subsection{Quantum vortices and moment of inertia}
\label{sec:vortices}

In contrast to superconducting materials, ultra-cold trapped Fermi
gases are neutral systems which are well isolated from their
surrounding.  In addition, all physical properties of these systems
are expected to vary smoothly throughout the BCS-BEC crossover, also
owing to the presence of the trap.  For instance, the fact that the
normal and condensed gas clouds have similar size and shape makes it
difficult to detect condensation on the BCS side.  For this reason, a
stringent proof of the superfluid behaviour of ultra-cold trapped Fermi
gases has long been elusive.  In this context, two experiments were
performed specifically to reveal the presence of the superfluid phase
\cite{Zwierlein-2005b,Riedl-2011}, by setting an ultra-cold trapped
Fermi gas into rotation.  In both experiments, an equal mixture of
$^{6}\mathrm{Li}$ atoms in the lowest two atomic (hyperfine) states
was prepared in a cigar-shaped trap.

\begin{figure}[h]
\begin{center}
\includegraphics[width=11.0cm,angle=0]{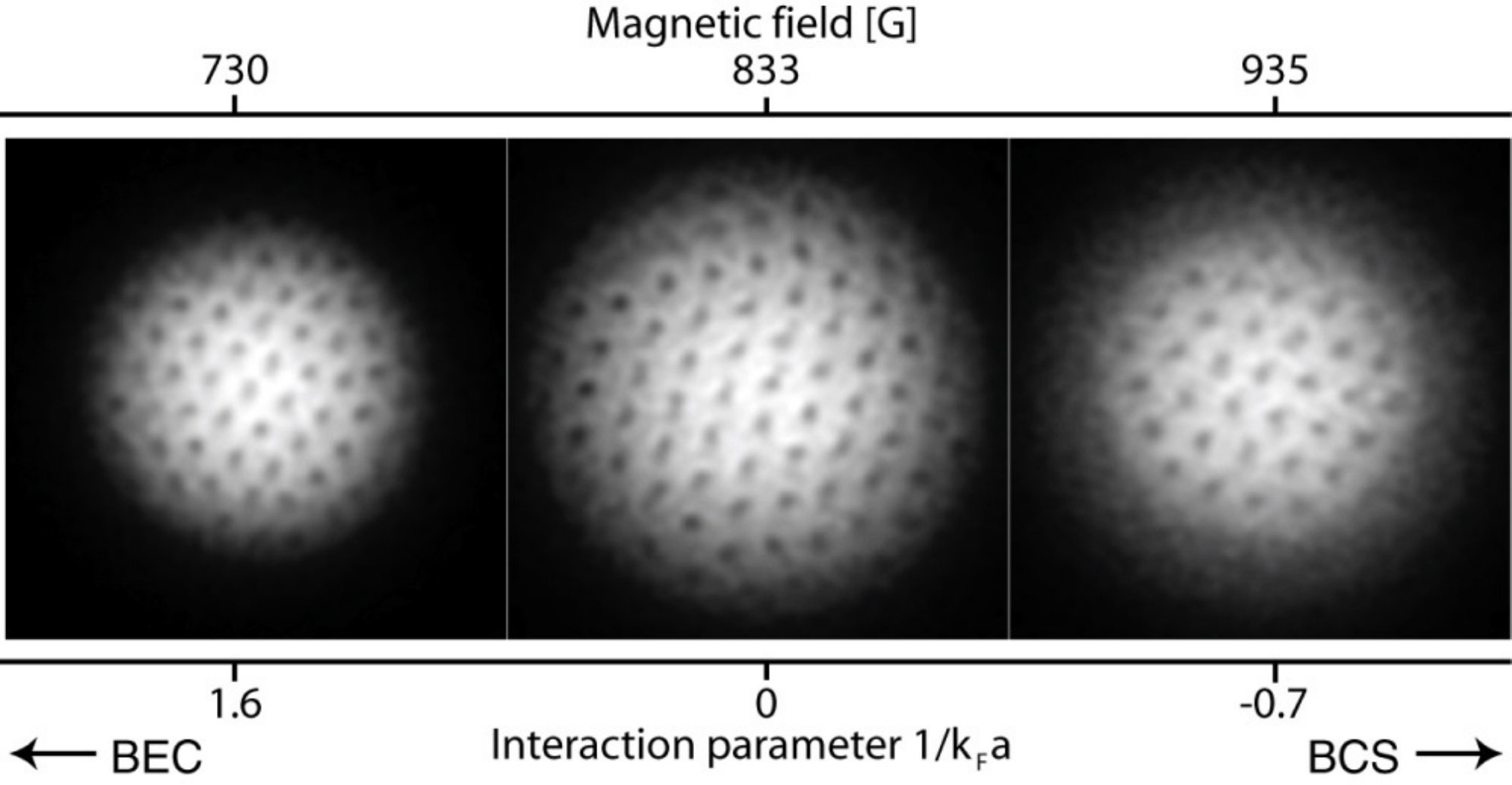}
\caption{Observation of vortex lattices across the BCS-BEC crossover. [Reproduced from Ref.~\cite{Zwierlein-2005b}.]}
\label{Figure-28}
\end{center}
\end{figure}

In the first experiment \cite{Zwierlein-2005b}, large arrays of
vortices were generated at low temperature throughout the BCS-BEC
crossover by sweeping a Fano-Feshbach resonance, thereby providing the
first direct evidence for the occurrence of the superfluid phase in a
\emph{fast} rotating trapped Fermi gas and establishing an analogy
with type-II superconductors and neutron stars.  This is because the
occurrence of a quantized vortex lattice is a direct consequence of
the existence of a macroscopic wave function that describes the
superfluid, with the gradient of the phase of the wave function being
proportional to the velocity field of the superfluid.  Figure
\ref{Figure-28} shows examples of these vortex lattices, that were
detected in Ref.~\cite{Zwierlein-2005b} by imaging the cloud after
ballistic expansion in order to increase the contrast.

In the second experiment \cite{Riedl-2011}, the superfluid behaviour
was revealed in a \emph{slowly} rotating trapped Fermi gas by the
quenching of the moment of inertia $\Theta$, which was measured at
unitarity for increasing temperature until the classical rigid-body
value $\Theta_{\mathrm{cl}}$ of the normal state was reached.  The
basic idea for a quenched moment of inertia as a signature of
superfluidity dates back to more than fifty years ago in nuclear
physics, where a moment of inertia below the classical rigid-body
value was attributed to superfluidity \cite{Ring-1980}.  In the
experiment of Ref.~\cite{Riedl-2011}, a slow rotation was required to
transfer a finite angular momentum $L$ to the system while avoiding at
the same time the presence of vortices, which are energetically
favored only above a critical rotation frequency $\Omega_{c_{1}}$ (in
this respect, the experiment of Ref.~\cite{Riedl-2011} is
complementary to the experiment of Ref.~\cite{Zwierlein-2005b}).

Theoretically, the slow rotation of a trapped Fermi gas at $T=0$ was studied in Ref.~\cite{Farine-2000}, where it was shown that the moment of inertia of a superfluid gas 
rotating around the $z$ axis reaches the irrotational value $\Theta_{\mathrm{irrot}} = \Theta_{\mathrm{cl}} (\omega_x^2 - \omega_y^2) / (\omega_x^2 + \omega_y^2)$ 
provided the condition $\Delta\gg (\omega_x,\omega_y)$ is satisfied (which coincides with the condition for the validity of superfluid hydrodynamics discussed in Section \ref{sec:collective}).
If $\Delta$ and $(\omega_x,\omega_y)$ are instead of the same order of magnitude, the moment of inertia is larger than $\Theta_{\mathrm{irrot}}$ and lies between $\Theta_{\mathrm{irrot}}$
and $\Theta_{\mathrm{cl}}$, as it is the case of atomic nuclei \cite{Durand-1985}. 
An early work on this subject was published by Migdal \cite{Migdal-1959}.

\begin{figure}[h]
\begin{center}
\includegraphics[width=15.0cm]{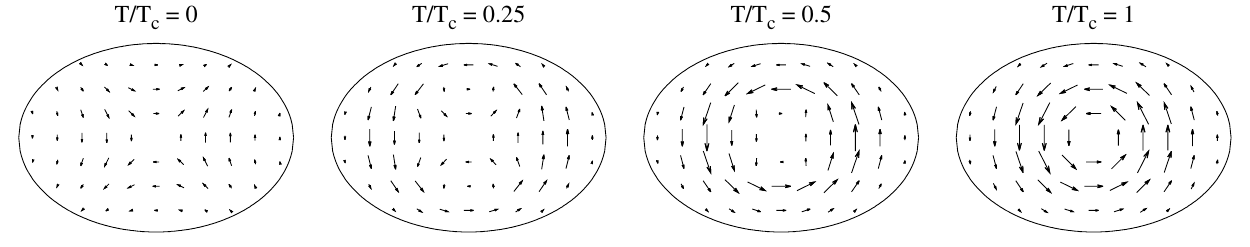}
\caption{Current densities in the $x-y$ plane for a gas in the BCS regime at temperatures $T/T_c = 0.0$, $0.25$, $0.5$, and $1.0$ (from left to right), in a trap with $\omega_x/\omega_y = 0.7$
slowly rotating about the $z$ axis. 
[Adapted from Ref.~\cite{Urban-2005}.]}
\label{Figure-29}
\end{center}  
\end{figure}

The temperature dependence of the moment of inertia was studied in
Refs.~\cite{Urban-2003,Urban-2005}, where it was shown that in the
limit $\Delta \gg (\omega_x,\omega_y)$ the current in the rotating
trap can be decomposed as in Landau two-fluid hydrodynamics into a
normal component rotating like a rigid body and a superfluid component
with an irrotational velocity field.  As an example, the resulting
current densities in a trap with $\omega_x/\omega_y = 0.7$ are
displayed in Fig.~\ref{Figure-29} at different temperatures, and show
how the irrotational velocity field at $T=0$ turns continuously into a
rigid-body rotation at $T=T_c$.  The rotational component, which is
proportional to the normal-fluid fraction, is concentrated at the
surface at low temperature and penetrates more and more into the trap
center upon approaching $T_c$.  As a consequence, the moment of
inertia smoothly reaches the rigid-body value at $T = T_c$.

Similar effects as in a slowly rotating gas at finite temperature may
appear in a rapidly rotating gas already at zero temperature.  In
particular, it was suggested in Ref.~\cite{Bausmerth-2008} that
it might be possible to put the system into fast rotation without
generating vortices by adiabatically increasing the angular velocity.
In this case, in the limit of axial symmetry $\omega_x = \omega_y$ it
is expected that the system separates into a superfluid phase at rest
near the center surrounded by a rotating normal phase.  Since in the
strongly-coupled regime the normal and superfluid phases have
different equations of state, this would have an observable effect on
the density profile, which would enable one to determine the
difference in energies of the normal and superfluid phases.  As
pointed out in Ref.~\cite{Urban-2008a}, however, the rotation
itself destroys a certain fraction of Cooper pairs, in analogy to what
happens at finite temperature in the two-fluid model.  As a
consequence, an intermediate layer appears in between the superfluid
and normal phases, where the rotational current gradually builds up
from the inner superfluid phase to the outer normal phase, and the
discontinuity in the density profile is eventually washed out.

\begin{figure}[h]
\begin{center}
\includegraphics[width=9.0cm,angle=0]{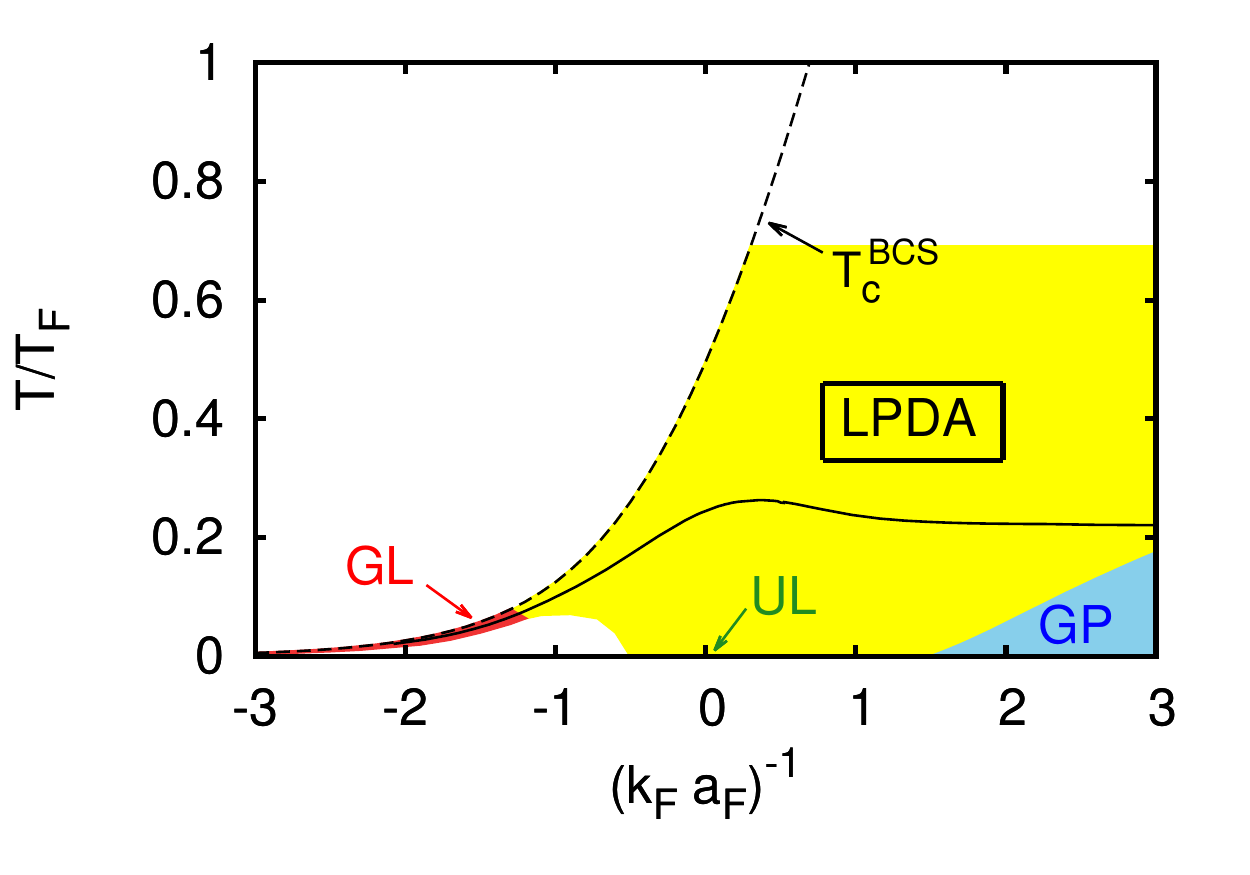}
\caption{Temperature vs coupling phase diagram across the unitary limit (UL), in which the regions of validity of the GL, GP, and LPDA
equations are schematically indicated by the colored sectors. 
The critical temperatures of the homogeneous system at the mean-field level (dashed line) and within the $t$-matrix approximation (full line) are also reported 
(cf. Fig.~\ref{Figure-9}) to set the relevant boundaries.}
\label{Figure-30}
\end{center}
\end{figure}

Regarding the case of a rapidly rotating system with vortices, as it
was observed in the experiment \cite{Zwierlein-2005b}, it turns out
that this case is theoretically challenging since it requires the
description of complex arrays of vortices in a trapped system across the
BCS-BEC crossover.  Approaches based on the BdG equations discussed in
Section \ref{sec:bdg-hfb} have so far been limited to a single vortex
line \cite{Nygaard-2003,Sensarma-2006,Simonucci-2013} or to vortex
lattices in small 2D systems \cite{Feder-2004,Tonini-2006}.  Extending
these calculations to vortex lattices in realistic 3D systems appears
to be computationally prohibitive, owing to exceeding computation time
and memory space.  Nevertheless, finding solutions with large vortex
patterns has recently become possible in terms of the LPDA equation
introduced in Ref.~\cite{Simonucci-2014}, where a ``Local Phase
Density Approximation'' to the BdG equations was obtained by a
suitable \emph{double coarse graining} of those equations throughout
the BCS-BEC crossover, which deals with the magnitude and phase of the
order parameter on a different footing.  In this way, the BdG
equations were replaced by a single differential equation for
$\Delta(\vek{r})$ in close analogy with the GL and GP equations
discussed in Section \ref{sec:GLandGP}.  With the important
difference, however, that the LPDA equation holds over an extended
region of the temperature-coupling phase diagram of the BCS-BEC
crossover, and reduces to the GL and GP equations in appropriate
ranges of coupling and temperature as shown schematically in
Fig.~\ref{Figure-30}.  A full theoretical analysis based on the LPDA
equation for the two experiments with rotating traps made in
Refs.~\cite{Zwierlein-2005b,Riedl-2011} was reported in
Ref.~\cite{Simonucci-2015}, where a vector potential $\vek{A}(\vek{r})
= m \, \vek{\Omega} \times \vek{r}$ was associated to the trap
rotating with frequency $\vek{\Omega}$.

\begin{figure}[h]
\begin{center}
\includegraphics[width=16cm,angle=0]{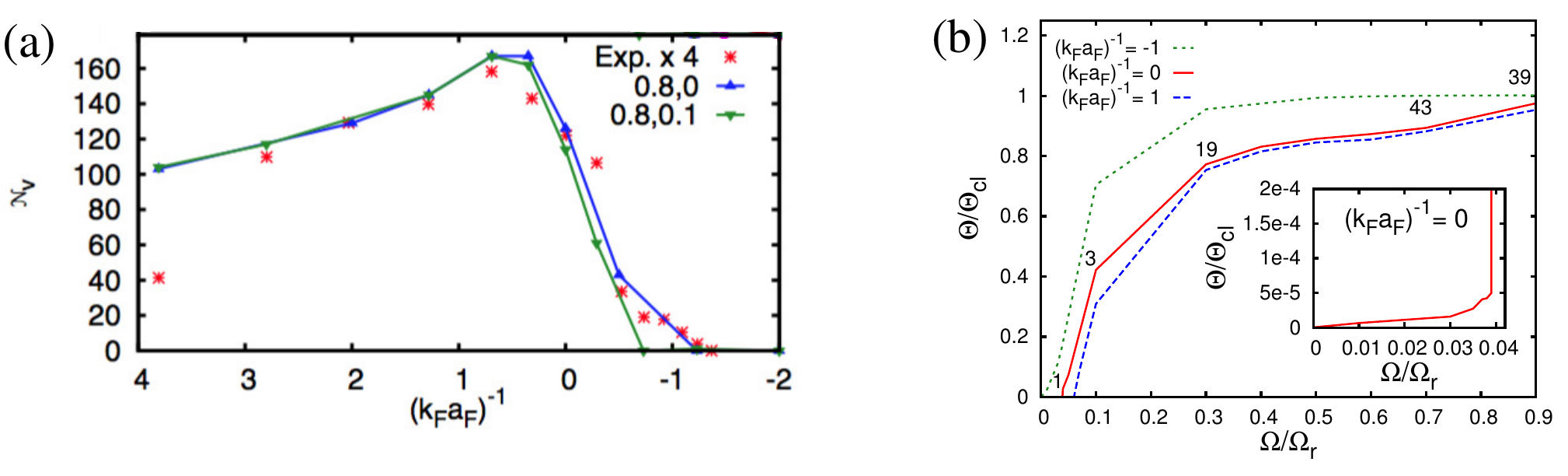}
\caption{(a) The experimental values for the number of vortices $\mathcal{N}_{v}$ from Ref.~\cite{Zwierlein-2005b} multiplied by a factor of four (stars) are compared with the results 
of the calculations of Ref.~\cite{Simonucci-2015} for $\Omega = 0.8$ (in units of the radial trap frequency $\Omega_{\mathrm{r}}$) and two different temperatures (in units of $T_F$).  
The calculations use the same trap parameters of the experiment.  
(b) Moment of inertia $\Theta$ (in units of its classical value $\Theta_{\mathrm{cl}}$) vs the angular frequency $\Omega$ (in units of $\Omega_{\mathrm{r}}$), for three couplings at zero temperature. 
The inset shows the tiny increase of $\Theta$ at unitarity before its sharp rise at the critical frequency $\Omega_{c_{1}}$.  
[Adapted from Ref.~\cite{Simonucci-2015}.]}
\label{Figure-31}
\end{center}
\end{figure}

Figure \ref{Figure-31}(a) shows a comparison of the LPDA calculation of Ref.~\cite{Simonucci-2015} with the experimental data of Ref.~\cite{Zwierlein-2005b}.  
In this figure, the experimental values for the number of vortices $\mathcal{N}_{v}$ have been multiplied by a factor of $4$ common to all couplings, because it was also found in
Ref.~\cite{Simonucci-2015} that the Feynman's theorem for the number of vortices for unit area (which is known to apply to an infinite array of vortices \cite{Nozieres-1990}) is 
satisfied here only in (about) $1/4$ of the cloud area.  
Similarly to the Josephson effect discussed in Section \ref{sec:Josephson}, also in this case by looking at the number of vortices one concludes that superfluidity is most robust in the unitary regime.

Figure \ref{Figure-31}(b) shows instead the frequency dependence of the moment of inertia $\Theta = L / \Omega$ obtained in terms of the total angular momentum $L$ 
at zero temperature for three different couplings, using the trap parameters of Ref.~\cite{Riedl-2011}.  
At unitarity, the number of vortices that enter the cloud are also indicated for a choice of angular frequencies, from which one concludes that not too many vortices are required to stabilise 
the moment of inertia to its classical value $\Theta_{\mathrm{cl}}$.  
The inset of Fig.~\ref{Figure-31}(b) provides a closer view at the behaviour of $\Theta$ at unitarity in a narrow frequency range about the critical frequency $\Omega_{c_{1}}$, 
at which the first vortex nucleates in the cloud.  
A smooth increase in $\Theta$ is found to occur before the sharp rise at $\Omega_{c_{1}}$, owing to the finite particle number.
In addition, this increase of $\Theta$ before $\Omega_{c_{1}}$ has a linear dependence, in line with a general argument discussed in Ref.~\cite{Bertsch-1999}.

\section{BCS-BEC crossover in nuclear systems}
\label{sec:nuclearsystems}

Nuclear systems (nuclei, neutron stars, proto-neutron stars, etc.) contain two types of nucleons: protons and neutrons. 
Consequently, three types of pairing can occur: neutron-neutron ($nn$), proton-proton ($pp$), and proton-neutron ($pn$). 
Only the latter channel shows a bound state, the deuteron ($d$ or $^2$H). 
As a consequence, one can imagine a crossover from a BEC of deuterons to a BCS state of $pn$ Cooper pairs.

In contrast to the cold-atom case, where the interaction strength can be varied (leading to the interaction-induced crossover of Section \ref{sec:Feshbach}), 
the nucleon-nucleon interaction is of course fixed. 
In nuclear matter the crossover is thus realised by varying the density.
At low density, the deuteron is bound, but the binding energy decreases with increasing density. 
At some stage, the deuteron becomes unbound and turns into a $pn$ Cooper pair. 
The possibility of such a density-induced crossover is actually not limited to nuclear systems.
It relies only on the fact that the interaction has a finite range~\cite{Andrenacci-1999}, since then the relevant matrix elements of the interaction get small as the momentum transfer, which is of the order of $k_F$ and then increases with density, exceeds the wave-vector range of the interaction.

In ordinary nuclei, $nn$ and $pp$ pairing win against $pn$ pairing because of $n$-$p$ asymmetry, i.e., there are usually more neutrons than protons. 
Since there is no $pp$ or $nn$ bound state, the BEC regime does not exist for these channels. 
However, since the $nn$ system is ``almost'' bound, the $nn$ scattering length is very large and a situation similar to the unitary limit can be realised in low-density neutron matter.

There is also a completely different way how one can pass from a BEC to a BCS state in nuclear systems. 
The $\alpha$ particle ($^4$He nucleus made of two protons and two neutrons) is much more strongly bound than the deuteron mentioned above. 
Therefore, one may think that at low density, a BEC of $\alpha$ particles is energetically more favourable than a BEC of deuterons. 
But as we will discuss below, an $\alpha$ particle is much more sensitive to the presence of a finite density than a deuteron, and therefore, at some density, the BEC of $\alpha$ particles 
disappears in favour of a BCS state of $pn$ (or $nn$ and $pp$) Cooper pairs mentioned above.

Nuclear pairing is, of course, a quite old subject dating back to the even-odd staggering of nuclear binding and the strong reduction of the moment of inertia of deformed nuclei. 
Those issues, together with their history, are well described in textbooks (see, e.g., Refs.~\cite{Bohr-1969,Ring-1980}).
More modern reviews can be found in Refs.~\cite{Brink-2005,Dean-2003,Gezerlis-2014}. 
Since our aim here is to concentrate on the BCS-BEC crossover and since this phenomenon is so far elusive in finite nuclei (see later), 
this Section will mainly concentrate on infinite nuclear or infinite neutron matter with only short glimpses on what happens in finite nuclei.

\subsection{Deuteron in symmetric nuclear matter and proton-neutron
  pairing}
\label{sec:deuteron}

The deuteron is the only bound state of two nucleons. 
Its spin is $S=1$ (spin triplet). 
To have an antisymmetric wave function, it must be isospin singlet (i.e., $T=0$), and the orbital angular momentum can take the values $L=0,2$ 
since the total spin of the deuteron is $J^P=1^+$. 
It is, indeed, well known that the deuteron has a quadrupole ($L=2$) component in its wave function as a consequence of the non-central (tensor) force. 
With $E_B/A=1.1$ MeV (where $E_B = 2.2$ MeV is the binding energy and $A=2$ is the number of nucleons), its binding energy is very weak compared 
to the most bound nucleus $^{56}$Fe with $E_B/A= 8.8$ MeV or the $\alpha$ particle with $E_B/A=7.1$ MeV. 
Nonetheless, $pn$ pairing can exhibit the BCS-BEC crossover phenomenon \cite{Alm-1993,Stein-1995,Baldo-1995}. 
This can be of a certain importance, for instance, to explain the deuteron production in heavy-ion collisions where the expansion of the system after the collision results 
in low-density nuclear matter \cite{Baldo-1995}.
One may also imagine that in collapsing stars or in not completely cooled neutron stars, where a certain fraction of protons still exists, $pn$ pairing can play a role.

The BCS-BEC crossover in nuclear matter has been investigated within the BCS framework in Ref.~\cite{Baldo-1995}. 
At finite temperature, the BCS equation for the wave-vector dependent gap $\Delta_{\vek{k}}$ has the usual form 
\begin{equation}
\Delta_{\vek{k}} = - \sum_{\vek{k'}} V(\vek{k},\vek{k'}) \frac{\Delta_{\vek{k'}}}{2E_{\vek{k'}}}[1-2f(E_{\vek{k'}})] 
\label{gapeqn}
\end{equation}
which extends Eq.(\ref{gap-eq}) to finite temperature.
Here, $V(\vek{k},\vek{k'})$ denotes the matrix element of the nucleon-nucleon force, $E_{\vek{k}} = \sqrt{\xi_{\vek{k}}^2+\Delta_{\vek{k}}^2}$ the quasiparticle energy,
with $\xi_{\vek{k}} = \varepsilon_{\vek{k}}-\mu$ and $f(E) = 1/(e^{E/T}+1)$ the Fermi function. 
In Ref.~\cite{Baldo-1995}, a separable form of the Paris force \cite{Haidenbauer-1984} was used for $V(\vek{k},\vek{k'})$, reproducing nucleon-nucleon phase shifts and the
deuteron binding energy. 
The density-dependent nucleon single-particle energies $\varepsilon_{\vek{k}}$ were obtained within the Br\"uckner-Hartree-Fock (BHF) approach.

We recall from Section~\ref{sec:BCS_homogeneous} that the BCS occupation numbers at finite temperature are given by the expression: 
\begin{equation}
n_{\vek{k}} = \frac{1}{2}\left(1-\frac{\xi_{\vek{k}}}{E_{\vek{k}}}[1-2f(E_{\vek{k}})] \right ).
\label{BCSoccupation}
\end{equation}
Similarly to what was done in Section~\ref{sec:BCSwavefunction}, multiplying both sides of Eq.~(\ref{gapeqn}) by $\xi_{\vek{k}}[1-2f(E_{\vek{k}})]/2E_{\vek{k}} = 1-2n_{\vek{k}}$ 
and defining $\kappa_{\vek{k}} = \Delta_{\vek{k}}[1-2f(E_{\vek{k}})]/2E_{\vek{k}}$, one recovers Eq.~(\ref{alternative-gap-equation}) also at finite temperature, in the form:
\begin{equation}
2\varepsilon_{\vek{k}}\, \kappa_{\vek{k}} + (1-2n_{\vek{k}}) \sum_{\vek{k'}} V(\vek{k},\vek{k'})\, \kappa_{\vek{k'}} = 2 \mu\,\kappa_{\vek{k}} \, .
\label{schroedingereqn}
\end{equation}
As discussed in Section~\ref{sec:BCSwavefunction} for the zero-temperature case, this shows that in the low-density limit, where $\varepsilon_{\vek{k}} \to k^2/(2m)$ and $n_{\vek{k}}\to 0$, 
the gap equation reduces to the Schr\"odinger equation, and the pairing tensor $\kappa_{\vek{k}}$ coincides with the deuteron wave function, whose eigenvalue $2\mu$ tends towards 
(minus) the deuteron binding energy $-2.2$ MeV. 
Note, however, that this argument holds only at zero temperature.
For any fixed non-zero temperature, the low-density limit corresponds to the classical limit whereby $\mu\to -\infty$, such that the gap equation has only the trivial solution $\Delta_{\vek{k}} = 0$
[i.e., $\kappa_{\vek{k}} = 0$]. 

\begin{figure}[t]
\begin{center}
\includegraphics[width=15cm]{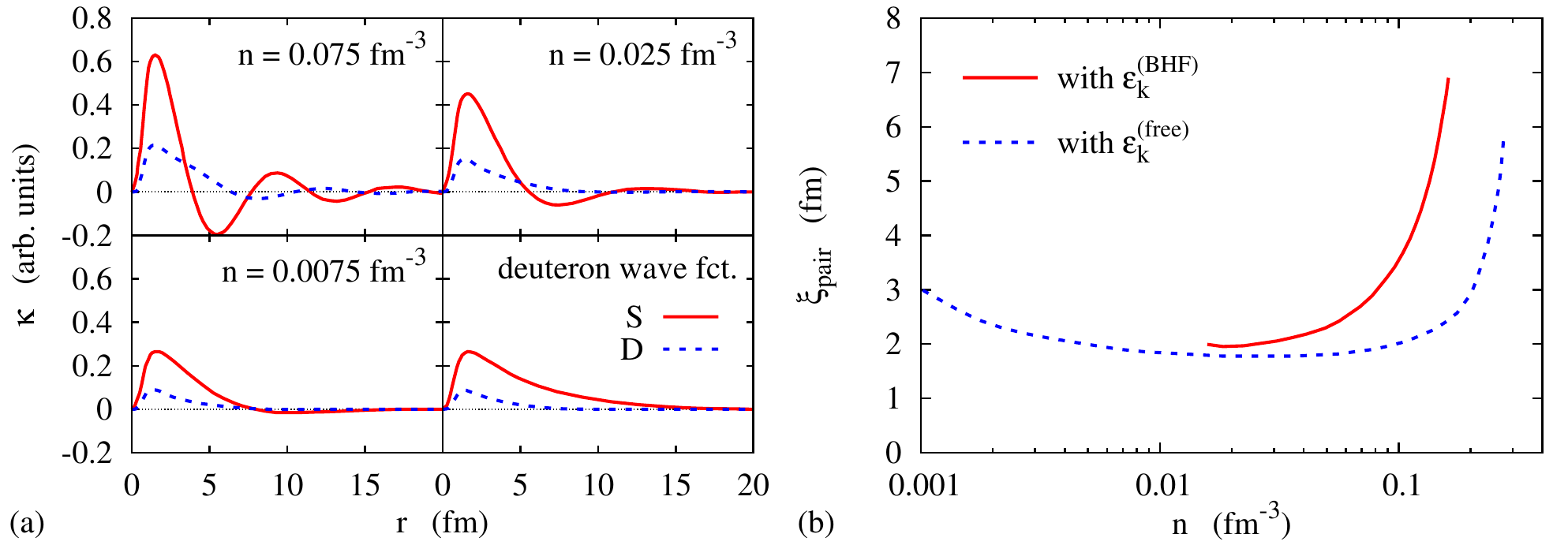}
\end{center}
\caption{\label{fig:deuteronbcs} (a) $S$ and $D$ components of the pairing tensor in coordinate space at $T=0$ for three different values of the density. 
The panel at the lower right displays the free deuteron wave function for comparison. 
(b) Deuteron size as a function of density.
[Panel (a) adapted from Ref.~\cite{Baldo-1995}; panel (b) adapted from Ref.~\cite{Lombardo-2001a}]}
\end{figure}

Figure \ref{fig:deuteronbcs}(a) shows the $S$-wave and $D$-wave components of the pairing tensor $\kappa(r)$ (defined as the Fourier transform of $\kappa_{\vek{p}}$) as a function of $r$ at zero temperature for four
different densities $n$. 	
The typical oscillatory behaviour of the Cooper pair wave function results at higher densities. 
As the density decreases, on the other hand, the Cooper pair wave function reduces to the bound-state wave function of the free deuteron.

These results can be used to determine the size $\xi_{\mathrm{pair}}$ of the deuteron in matter [cf. Eq.~(\ref{CL}) of Section~\ref{sec:pairingcorrelations}], according to the expression
\cite{Lombardo-2001a,Pistolesi-1994}:
\begin{equation}
\xi_{\mathrm{pair}}^2 = \frac{\int d\vek{r}\, r^2 |\kappa(r)|^2}{\int d\vek{r}\, |\kappa(r)|^2} = \frac{\sum_{\vek{k}} |\nabla_{\vek{k}} \kappa_{\vek{k}}|^2}{\sum_{\vek{k}} |\kappa_{\vek{k}}|^2} \, .
\end{equation}
Results for the deuteron size as function of the density are shown in Fig.~\ref{fig:deuteronbcs}(b). 
In the lower curve (dashed line), free single-particle energies $p^2/(2m)$ were used, while the upper curve (solid line) was obtained with BHF single-particle energies $\varepsilon_{\vek{k}}$. 
Note how the deuteron size first shrinks with increasing density until it reaches a minimum at $n \sim 0.036$ fm$^{-3}$, after which it starts to increase. 
This initial size shrinking was also noticed in a more general context in Ref.~\cite{Andrenacci-2000}.  A qualitatively similar behavior was already 
revealed in 1970, when the deuteron size was calculated as a function of the distance to the
$\alpha$-particle in $^6$Li \cite{Itonaga-1970}.
Note further that, although the deuteron size initially shrinks, it gets monotonously less bound with increasing density.
Neglecting the BHF mean field and the wave vector dependence of $V(\vek{k},\vek{k'})$ at low wave vectors, one can show from Eq.~(\ref{schroedingereqn}) that, to linear order in the density, 
the chemical potential is given by $\mu = \mu_D + \pi n/(2mp_D)$ where $\mu_D = -1.1$ MeV is one half of (minus) the free deuteron binding energy and $p_D = \sqrt{2m|\mu_D|}$.

\subsection{Asymmetric nuclear matter and the BCS-BEC crossover}
\label{sec:deuteron-asy}

Thus far we have considered only symmetric nuclear matter, with equal densities of protons and neutrons. 
However, in astrophysical situations such as proto-neutron stars, there is usually a strong neutron excess. 
It is thus important to consider also the asymmetric (i.e., isospin imbalanced) situation when there are more neutrons than protons. 
Formally, this is completely analogous to the situation of pairing between spin $\uparrow$ and $\downarrow$ in a spin-imbalanced (polarized) system discussed in Section~\ref{sec:polarized}.
Here, for the sake of illustration, we concentrate on the blocking effect (breached-pair solution) and leave aside the FFLO solution and phase separation.
[Recall, however, from Section~\ref{sec:polarized} that, at least for the contact potential, the breached-pair solution corresponds to a physically stable phase only in the BEC limit.]

A first study of this type, which considers the blocking effect due to unpaired neutrons, was undertaken in Ref.~\cite{Lombardo-2001b}, where the $pn$ gap was studied as a
function of asymmetry and density. 
It was found that asymmetry quenches the gap in the weak-coupling BCS regime, but that it does not play a role in the strong-coupling (low-density) regime. 
This is understandable since in this regime the system will form a BEC of deuterons and a Fermi sea of the remaining neutrons. 
At sufficiently low density, the surrounding neutrons will not perturb the deuteron bound state.
\begin{figure}[t]
\begin{center}
\includegraphics[height=7cm,angle=-90]{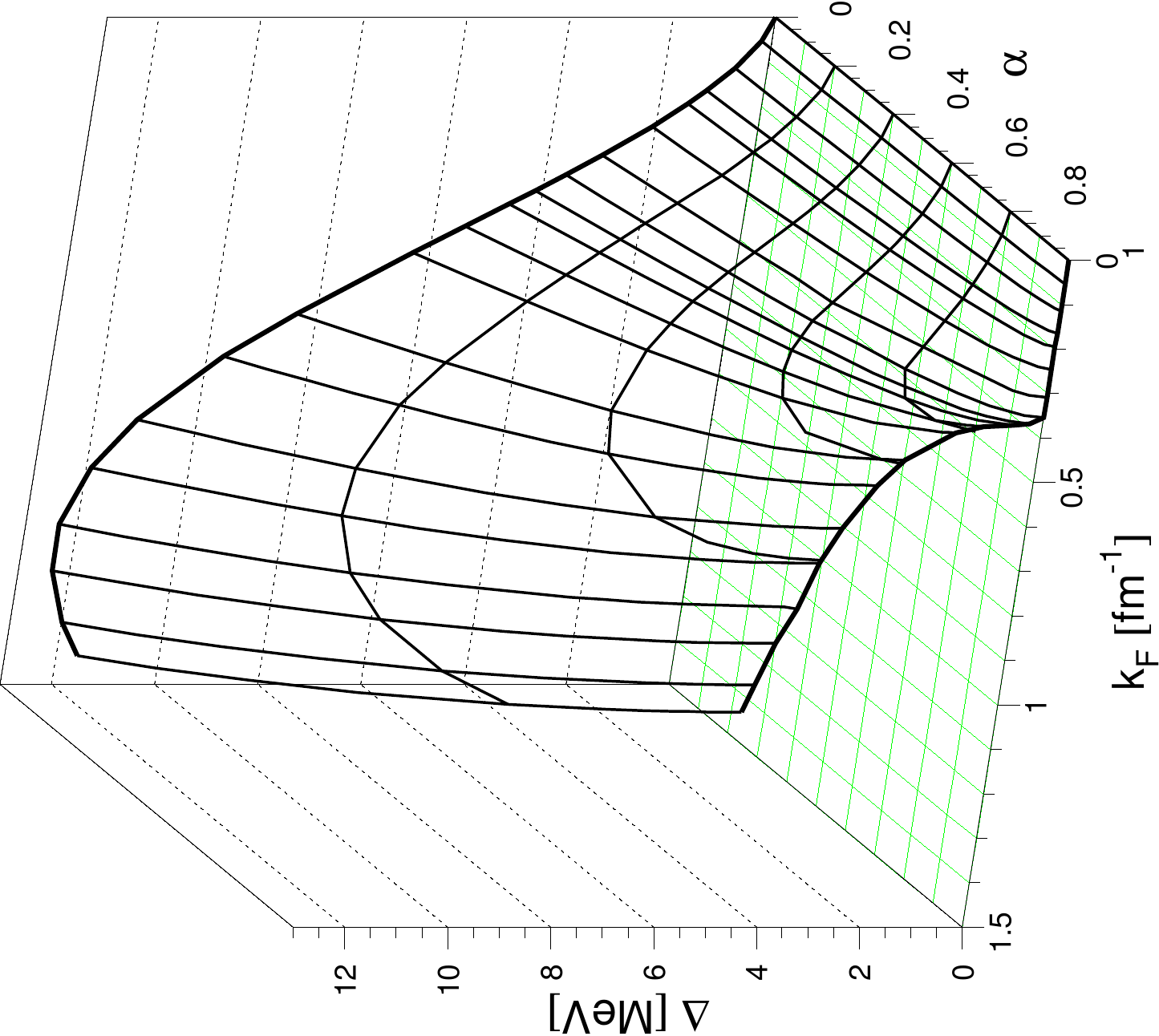}
\end{center}
\caption{\label{Fig:LombardoPRC64-Fig4} Gap as a function of $k_F = (3\pi^2 n/2)^{1/3}$ (where $n=n_n+n_p$) and asymmetry $\alpha = (n_n-n_p)/(n_n+n_p)$. 
[Reproduced from Ref.~\cite{Lombardo-2001b}.]}
\end{figure}
Figure \ref{Fig:LombardoPRC64-Fig4} shows the gap as a function of $k_F=(3\pi^2 n/2)^{1/3}$, where $n = n_p+n_n$ is the total density, and of the asymmetry $\alpha = (n_n -n_p)/(n_n + n_p)$. 
In these calculations, the Argonne $V_{14}$ interaction \cite{Wiringa-1984} and free single-particle energies were used. 
One sees that even at very large asymmetries (i.e., $n_n \gg n_p$) there remains a finite gap in the $k_F \rightarrow 0$ region,
while in the weak-coupling regime (larger $k_F$) pairing does not survive past  some critical asymmetry.

Analytically, this can be shown as follows. 
In the strong-coupling limit, where the chemical potential is negative and for very low densities tends to $\mu_D = -E_B/2 \sim -1.1$ MeV, the gap and particle number equations 
somehow exchange their roles as already noted in Section~\ref{sec:BCS_finite_T}. 
The gap equation goes over into the Schr{\"o}dinger equation for the deuteron, while the gap can be determined from the number equation in the following way. 
Consider, for example, the density of the protons given by
\begin{equation}
n_p = \frac{(2m)^{3/2}}{2\pi^2}\int_{\varepsilon_{\mathrm{min}}}^{\infty} d\varepsilon\sqrt{\varepsilon} \frac{1}{2}\left[1-\frac{\varepsilon-\mu}{\sqrt{(\varepsilon-\mu)^2+\Delta^2}} \right] \, ,
\end{equation}
where the integral over $\vek{k}$ has been transformed in an integral over $\varepsilon = \vek{k}^2/(2m)$ and $\varepsilon_{\mathrm{min}} = \mu+\delta\varepsilon$ is the energy 
below which the the neutron states are fully occupied while the proton states are empty (at zero temperature), cf. Fig.~\ref{Figure-4}(b). 
In the $n\to 0$ limit, the width of this region tends towards zero. 
In addition, since $\mu\to \mu_D$ and $\Delta\ll |\mu_D|$, one can expand the expression within brackets up to order $(\Delta/\mu_D)^2$. 
The resulting integral can be performed analytically and one obtains $n_p \sim (n_\mu/2) (\Delta/\mu_D)^2$, where $n_\mu = (2m|\mu_D|)^{3/2}/(8\pi) \sim 0.00049$ fm$^{-3}$. 
Writing the asymmetry as $\alpha = 1-2n_p/n$, one ends up with the following low-density result for the gap \cite{Lombardo-2001b}
\begin{equation}
\frac{\Delta}{|\mu_D|} = \sqrt{\frac{n(1-\alpha)}{n_{\mu}}}\,,
\label{gap-asym}
\end{equation}
which confirms that in strong coupling the gap persists for any asymmetry, as seen in Fig.~\ref{Fig:LombardoPRC64-Fig4}.

In weak coupling the situation is different. 
As shown in Fig.~\ref{Figure-4}(a), the proton occupation numbers are now zero in a window $\mu\pm\delta\varepsilon$ with $\delta\varepsilon = \sqrt{\delta\mu^2-\Delta^2}$, 
where $\delta\mu = (\mu_p-\mu_n)/2$ is half the difference between the chemical potentials (analogous to $h = (\mu_\uparrow-\mu_\downarrow)/2$ of Section~\ref{sec:polarized}). 
For the gap equation, using a contact interaction with a coupling constant $\varv_0$ and a cutoff $\varepsilon_c$ instead of the regularization of Eq.(\ref{Sarma-gap}),
one obtains:
\begin{equation}
1=-\frac{(2m)^{3/2}\varv_0}{8\pi^2}\int_{-\mu}^{\varepsilon_c-\mu} d\xi \Theta(|\xi|-\delta \varepsilon)\sqrt{\frac{\xi+\mu}{\xi^2 + \Delta^2}} 
\label{blocked-gap}
\end{equation}
where $\xi = \varepsilon-\mu$.
This integral can be done with the usual approximations valid in the weak-coupling limit. 
Comparing the result with the corresponding weak-coupling result in the symmetric ($\delta\mu = 0$) case, one finds:
\begin{equation}
\sqrt{\delta\varepsilon^2+\Delta^2} + \delta \varepsilon = \Delta_0\,,
\end{equation}
where $\Delta_0$ is the gap in symmetric matter with the same value of $\mu$. 
One obtains eventually:
\begin{equation}
\Delta = \Delta_0\sqrt{1 - \frac{2\delta \varepsilon}{\Delta_0}} \, .
\label{result-gap-asy}
\end{equation}
From this result one sees that the gap decreases very rapidly and disappears when $2\delta\varepsilon=\Delta_0$, i.e., when the width of the window reaches the size of the gap in symmetric matter. 
Since $\Delta_0\ll\mu$ in weak coupling, one can expand the asymmetry $\alpha$ to leading order in $\delta\varepsilon/\mu$, with the result $\alpha = 3\delta\varepsilon/(2\mu)$. 
Inserting this into Eq.~(\ref{result-gap-asy}), one concludes that the gap vanishes continuously, although with infinite slope, when \cite{Lombardo-2001b}
\begin{equation}
\alpha_{\mathrm{max}} = \frac{3\Delta_0}{4\mu} \, ,
\end{equation}
as it can be seen in Fig.~\ref{Fig:LombardoPRC64-Fig4} for large values of $k_F$.
Concerning further the dependence of $\Delta$ on $\delta\mu$, the asymmetry sets in and the gap starts to decrease only after the window $\delta\varepsilon$ opens up, that is, 
when $\delta\mu$ reaches $\Delta_0$. 
But then, while $\delta\varepsilon$ and the asymmetry $\alpha$ increase, $\delta\mu$ \emph{decreases} as one can show with the help of above expressions, and the gap eventually vanishes when $\delta\mu = \delta\varepsilon = \Delta_0/2$. \cite{Lombardo-2001b}. This corresponds to the re-entrant behavior already found by Sarma \cite{Sarma-1963}   for condensed matter systems long time ago.  As it was argued in Section~\ref{sec:polarized}, however, this unphysical behavior is eliminated when the FFLO phase or phase separation are considered.  For nuclear matter this was done, for instance, in Ref.~\cite{Stein-2014}.

\subsection{Proton-neutron correlations at finite temperature}
\label{sec:pn_finite_T}
Finite-temperature effects are very important in heavy-ion collisions and supernova matter, when low-density nuclear matter can be produced.
As discussed in Section~\ref{sec:pairingfluctuations}, the BCS theory is not suitable for the calculation of the critical temperature $T_c$ in the BEC and crossover regimes, 
because of the existence of pair correlations (that give rise to pre-formed non-condensed pairs in the pseudo-gap phase above $T_c$). 
In nuclear matter, pair fluctuations were first studied in Ref.~\cite{Schmidt-1990} within the ZS approach discussed in Section~\ref{sec:ZS}.

To reduce the complexity of the calculation, we follow Refs.~\cite{Schmidt-1990,Stein-1995,Jin-2010} and consider a separable potential. 
In its simplest form, the potential in a given interaction channel $\alpha$ can be written in the form:
\begin{equation}
V_\alpha(k,k') = -\lambda_\alpha \varv(k)\varv(k')\,,
\label{VYamaguchi}
\end{equation}
where $k$ and $k'$ are in- and out-going wave vectors in the center-of-mass frame of the two nucleons, $-\lambda_\alpha$ is the coupling constant, and $\varv(k)$ is a form factor 
(which, for instance, in the case of the Yamaguchi potential \cite{Yamaguchi-1954}, has the form $\varv(k) = 1/(k^2+\beta^2)$). 
In this way, the expression for the $t$-matrix (with the diagrams shown in Fig.~\ref{Figure-8}(b)) becomes analytic up to a quadrature like in the case of a contact interaction
(cf. Eqs.~(\ref{Gamma-0}) and (\ref{bubble-pp-exact})),  leading to the following expressions:
\begin{equation}
\Gamma_\alpha(k,k',Q,\omega) = \frac{V_\alpha(k,k')}{1-J_\alpha(Q,\omega)}
\label{Gammaseparable}
\end{equation}
where $Q$ and $\omega$ are the total wave vector and energy, respectively, and
\begin{equation}
J_\alpha(Q,\omega) = \int \frac{d\vek{k}}{(2\pi)^3} V_\alpha(k,k) \frac{1-f(\xi_{\vek{Q}/2+\vek{k}})-f(\xi_{\vek{Q}/2-\vek{k}})} {\omega-\xi_{\vek{Q}/2+\vek{k}}-\xi_{\vek{Q}/2-\vek{k}}+i0} \, .
\label{Jalpha}
\end{equation}
The form factor makes this integral finite and no regularization procedure is needed. 
Concerning the single-particle energies $\xi_{\vek{k}}$  that enter the expression (\ref{Jalpha}), different approximations have been used in the literature. 
In Ref.~\cite{Stein-1995}, the wave-vector dependence of the mean field was neglected and the constant shift was absorbed in an effective chemical potential $\mu^*$,
yielding $\xi_{\vek{k}} = k^2/2m-\mu^*$. 
In Ref.~\cite{Jin-2010}, the wave-vector dependent mean field was instead calculated with the D1 Gogny force \cite{Decharge-1980} in Hartree-Fock (HF) approximation and then 
expanded up to second order, yielding $\xi_{\vek{k}} = k^2/2m^*-\mu^*$ (effective mass approximation).

\begin{figure}[t]
\begin{center}
\includegraphics[width=16cm]{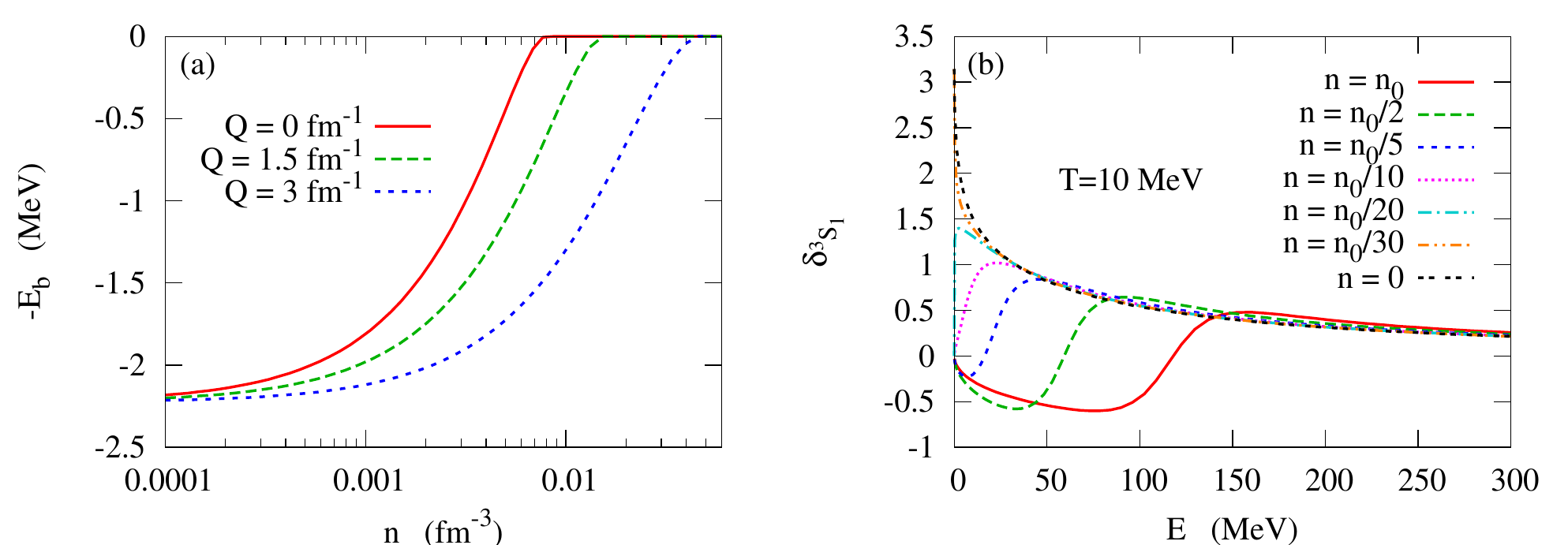}
 \end{center}
\caption{\label{fig:Jin-2010Fig23} (a) In-medium deuteron binding energy $E_b$ as a function of the density $n$ for different deuteron wave vectors $Q$ 
at a fixed temperature of $T=10$ MeV. 
(b) In-medium scattering phase shift in the $^3S_1$ channel for $Q=0$ as a function of $E = k^2/m^*$ for different densities ($n_0=0.17$ fm$^{-3}$ 
is the saturation density of nuclear matter) and $T = 10$ MeV. 
[Adapted from Ref.~\cite{Jin-2010}.]}
\end{figure}

Note that in the $^3S_1$ channel, the $t$-matrix (\ref{Gammaseparable}) can have a pole at an energy $\omega_b(Q)$ below the two-particle threshold $\omega_0(Q) = Q^2/4m^*-2\mu^*$. 
This pole corresponds to the deuteron bound state, and one can define the in-medium binding energy of the deuteron as $E_b = \omega_0-\omega_b$. 
Owing to Pauli blocking, the deuteron gets less bound with increasing density, but this effect is weaker for a deuteron that moves with respect to the medium,
as it can be seen from Fig.~\ref{fig:Jin-2010Fig23}(a).
The wave vector above which the deuteron is bound is called the Mott wave vector $Q_{\mathrm{Mott}}$ \cite{Schmidt-1990}. 
At very low density, the deuteron is always bound and $Q_{\mathrm{Mott}} = 0$. 
The density at which the bound state at $Q=0$ disappears at a given temperature is called the Mott density.  
The in-medium nucleon-nucleon phase shifts $\delta_\alpha$ can readily be obtained from $1/(1-J_\alpha) = e^{i\delta_\alpha} / |1-J_\alpha|$. 
As an example, Fig.~\ref{fig:Jin-2010Fig23}(b) shows the phase shift in the $^3S_1$ channel for $Q=0$ at different densities, as function of the energy $E = \omega+2\mu^* = k^2/m^*$. 
One sees that at higher densities, for instance at $n \geq n_0/5$ ($n_0 = 0.17$ fm$^{-3}$ being the saturation density of nuclear matter), the phase shift is negative in the
low-energy region and then becomes positive as the energy increases. 
The energy where the phase shift crosses zero is $\omega = 0$, i.e., $E = 2\mu^*$. 
At lower densities, when $\mu^*$ is negative, the phase shift is positive at low energy. 
At some very low density, the value of the phase shift at $E = 0$ changes from 0 to $\pi$.
This happens precisely at the density below which the deuteron is bound, in analogy with what is required in vacuum by Levinson's theorem (see, e.g., Ref.~\cite{Taylor-1972b}).

Like in the NSR approach, the self-energy $\Sigma$ is given by the diagram shown of Fig.~\ref{Figure-8}(a), with an important difference, however,
as explained in Section~\ref{sec:ZS}. 
In the NSR approach, the thin lines of Fig.~\ref{Figure-8} represent the bare single-particle propagator $G_0$,
while here they represent the HF propagator $G_{\mathrm{HF}}$ which contains the HF single-particle energies $\xi_{\vek{k}}$ instead of the free ones $k^2/2m-\mu$.
The Dyson equation now reads $G^{-1} = G_{\mathrm{HF}}^{-1}-\tilde{\Sigma}$, where $\tilde{\Sigma}$ is the self-energy  with the HF mean field subtracted. 
A caution is in order at this point.
On the one hand, the density-dependent Gogny force is supposed to take already into account the correlation effects on the single-particle energies $\xi_{\vek{k}}$. 
On the other hand, when calculating $\Sigma$ with the $t$-matrix (\ref{Gammaseparable}), including only the
channels $\alpha= {^3S_1}$ (deuteron) and $\alpha={^1S_0}$ which are most relevant for pairing correlations, one cannot expect $\Sigma$ to give the correct quasi-particle energies. 
For this reason, it is convenient to subtract the full energy shift generated by the self-energy, by setting
\begin{equation}
\tilde{\Sigma}(\vek{k},\omega) = \Sigma(\vek{k},\omega) - \Sigma(k,\xi_{\vek{k}}) \, ,
\end{equation}
which allows for a clear separation of dynamic correlation effects encoded in the energy dependence of $\tilde{\Sigma}$ and effects from
the mean-field shift contained in $\xi_{\vek{k}}$.

To discuss the correlation correction to the density as a function of $\mu$, one starts from the expression
\begin{equation}
n(T,\mu) = -4 T\sum_{\nu,\vek{k}}G(k,i\omega_\nu)
\label{n-from-G}
\end{equation}
where the factor of 4 comes from the sum over spin and isospin. 
If the Dyson equation is again truncated at first order, such that $G \approx G_{\mathrm{HF}} + G_{\mathrm{HF}}^2\tilde{\Sigma}$, one can write the total density as in Eq.~(\ref{rhocorrzs}) 
as a sum of the HF density $n_{\mathrm{HF}}$ plus corrections in the form \cite{Schmidt-1990,Stein-1995,Jin-2010}:
\begin{equation}
n = n_{\mathrm{HF}} + n_{\mathrm{corr}} = n_{\mathrm{HF}}+n_{\mathrm{bound}}+n_{\mathrm{scatt}} \, .
\label{n-with-corr}
\end{equation}
Here, the bound-state contribution reads
\begin{equation}
n_{\mathrm{bound}} = 6 \!\!  \int \limits_{|\vek{Q}| > Q_{\mathrm{Mott}}} \!\!\!\!\frac{d \vek{Q}}{(2 \pi)^3}  b(\omega_b(Q))
\label{n-bound}
\end{equation}
where $b(\omega) = 1/(e^{\omega/T}-1)$ is the Bose function. 
This term gives the nucleon density corresponding to a Bose gas of deuterons. 
The factor of 6 takes into account that the deuteron has 3 spin projections and each deuteron contains 2 nucleons. 
In addition, the scattering-state contribution reads:
\begin{equation}
n_{\mathrm{scatt}} = -6\!\! \int \limits_{|\vek{Q}| > Q_{\mathrm{Mott}}} \!\!\!\!\frac{d \vek{Q}}{(2 \pi)^3}   b(\omega_0(Q)) -6 \sum_{\alpha=\,^3S_1,\,^1S_0}
  \int \!\!\frac{d \vek{Q}}{(2 \pi)^3} \int_{\omega_0(Q)}^\infty \frac{d\omega}{\pi} \left(\frac{d}{d\omega}b(\omega)\right) \left(\delta_\alpha-\tfrac{1}{2}\sin 2\delta_\alpha\right) \, .
\label{n-scatt}
\end{equation}

\begin{figure}[t]
\begin{center}
\includegraphics[width=15cm]{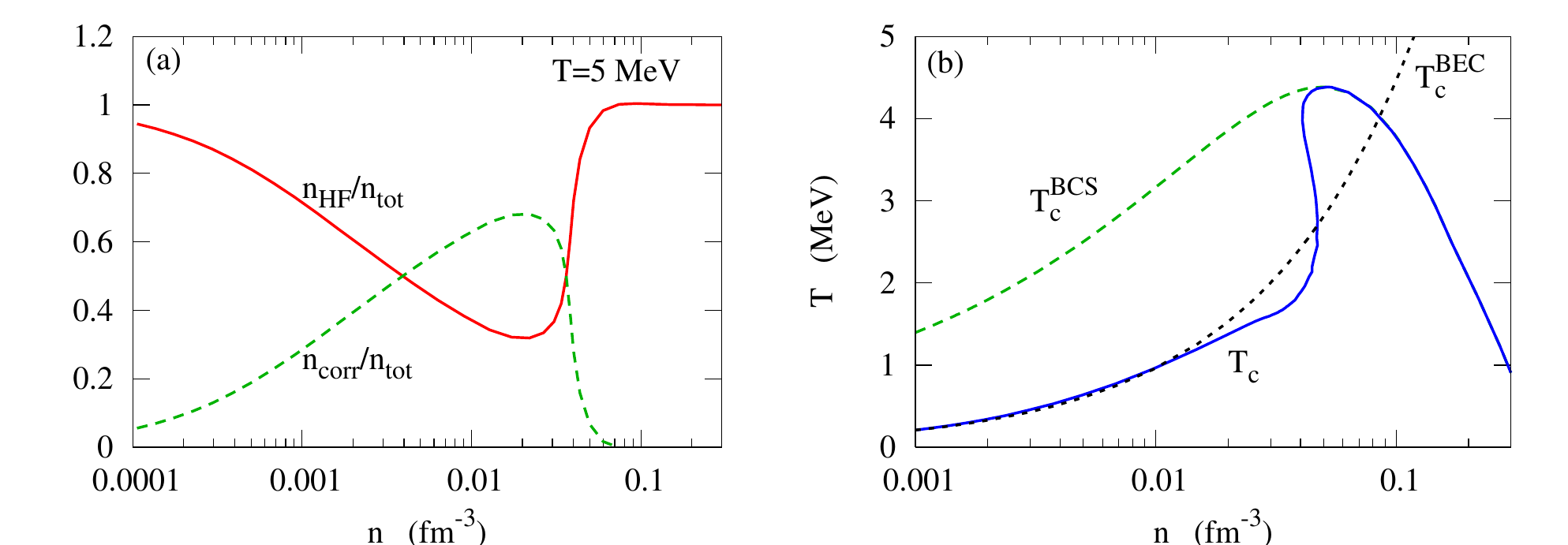}
\end{center}
\caption{(a) HF (solid line) and correlation (dashed line) contributions to the total density $n_{\mathrm{tot}} = n_{\mathrm{HF}}+n_{\mathrm{corr}}$ for $T = 5$ MeV 
as a function of the total density. 
(b) Critical temperature (solid line) as a function of density interpolating between BCS (long-dashed line) and BEC (short-dashed line) critical temperatures in symmetric nuclear matter.
[Adapted from Ref.~\cite{Jin-2010}.]
\label{fig:composition5_Tc}}
\end{figure}

As an example of the importance of the correlation contribution to the density, Fig.~\ref{fig:composition5_Tc}(a) shows the composition 
of the system at temperature $T = 5$ MeV.  
One sees that the correlation contribution to the total density is important at low density ($n<n_0/4$). 
When $n=0.02$ fm$^{-3}$, the correlated part is even larger than the HF part, meaning that most of the nucleons are in correlated pairs in this density region. With increasing temperature, the ratio of the correlated density to the total density decreases, but the density region with sizeable nucleon correlations is enlarged. 
Note that the correlation contribution is not separated into bound and scattering state contributions since, taken separately, 
they are not very meaningful, as discussed in Ref.~\cite{Schmidt-1990}. 
For instance, if the temperature is much larger than the deuteron binding energy, the first term of the scattering-state contribution (\ref{n-scatt}) almost exactly 
cancels the bound-state contribution (\ref{n-bound}).

For given chemical potential $\mu$, the critical temperature $T_c$ is obtained from the Thouless criterion, which in the present case reads $J_{^3S_1}(Q=0,\omega=0) = 1$. 
When one calculates $T_c$ as a function of the density, the corrections $n_{\mathrm{bound}}$ and $n_{\mathrm{scatt}}$ are crucial in the strong-coupling regime, 
since only with these corrections one recovers the critical temperature $T_c^{BEC} = \pi/m (n/6\zeta(3/2))^{2/3}$ for the Bose-Einstein
condensation of deuterons in the low-density limit. 
At higher densities, the result interpolates between $T_c^{\mathrm{BEC}}$ and the BCS critical temperature $T_c^{\mathrm{BCS}}$, as shown in Fig.~\ref{fig:composition5_Tc}(b).
Note  further that, in some range of densities, $T_c$ as a function of $n$ is double-valued. 
This is a consequence of the first-order liquid-gas phase transition of nuclear matter. 
We will discuss this phenomenon in detail in Section~\ref{sec:liquidgas}.

As discussed in Section~\ref{sec:pseudo-gap}, the pre-critical pair fluctuations lead to the formation of a pseudo-gap. 
For the case of low-density nuclear matter, the single-particle spectral function and the pseudo-gap in the level density
 (which corresponds to the density of states in condensed matter)
were studied in Ref.~\cite{Schnell-1999}. 
In this work, the authors consider again the $t$-matrix approximation to the self-energy of Fig.~\ref{Figure-8}(a). 
After analytic continuation, the imaginary part of the retarded self-energy reads:
\begin{equation}
\Imag \Sigma(1, \omega) = \sum_2 \Imag \Gamma(12, 12, \omega + \xi_2+i0) [f(\xi_2) + b(\omega + \xi_2)]
\label{ImM}
\end{equation}
where the short-hand notation ``1'' stands for wave vector  $\vek{k}_1$, spin $\sigma_1$, and isospin $\tau_1$ quantum numbers (and analogously
for ``2''). 
The real part is obtained from the imaginary part according to the Kramers-Kronig relation. 
Here, $\Gamma$ is again calculated with the separable Yamaguchi potential (\ref{VYamaguchi}), however, with the single-particle
energies $\xi_k = k^2/2m+U(k)-\mu$ obtained by a self-consistent Galitskii-Feynman calculation from the real part of the self-energy, i.e., $U(k) = \Real\Sigma(k,\xi_k)$. 
Note that this is different from the Br\"uckner-Hartree-Fock (BHF) approach \cite{Fetter-1971}, where the ``backward'' (hole-hole) contributions are discarded 
[i.e., the factor $(1-f_1-f_2)$ in the numerator of Eq.~(\ref{Jalpha}) is replaced by $(1-f_1)(1-f_2)$]. 
As we will see, this is important since the BHF approximation entails unphysical results for the level density.

\begin{figure}[h]
\begin{center}
\includegraphics[width=7.5cm]{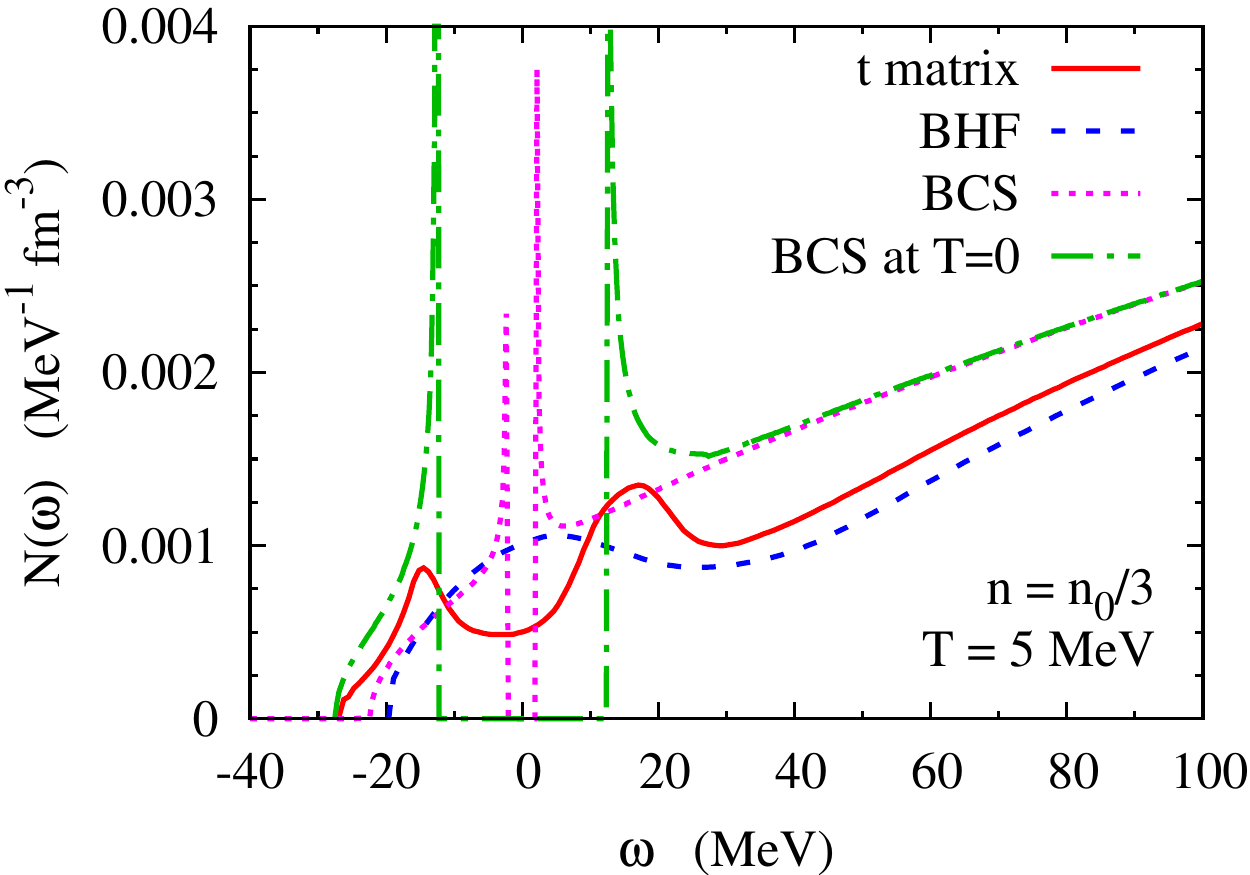}
\end{center}
\caption{\label{fig:SchnellPRL83pseudo-gap} Level density for $T=5$ MeV and $n = n_0/3$ within the quasiparticle approximation, using $U(k)$ from the $t$-matrix approach 
(solid line) and from the BHF approximation (dashed line).
For comparison, BCS results for $T=5$ MeV (dotted line) and $T=0$ (dot-dashed line) are also shown. 
[Adapted from Ref.~\cite{Schnell-1999}.]}
\end{figure}
Using the quasi-particle approximation, whereby  the full spectral function in Eq.~(\ref{density-of-states}) is replaced by $A(k,\omega) = \delta(\omega-\xi_k)$, one 
obtains the level density from the expression:
\begin{equation}
N(\omega) = \bigg[ \frac{1}{k^2}\frac{\partial \xi_k}{\partial k}\bigg]^{-1} \qquad \text{calculated at} \; \xi_{k}=\omega\, .
\label{leveldensity}
\end{equation}
Figure \ref{fig:SchnellPRL83pseudo-gap} shows the formation of the pseudo-gap in the level density approaching the critical temperature from above, for  
low particle density $n= n_0/3$. 
Note that when $T=5$ MeV the BHF calculation yields an opposite behaviour compared to the $t$-matrix approach, underlining the necessity
to incorporate the hole-hole contribution. 
Figure \ref{fig:SchnellPRL83pseudo-gap} also shows the standard BCS gap (dotted line) in the level density, which is quite small owing to the closeness of $T=5$ MeV 
to the critical value  $T_c^{\mathrm{BCS}} =5.08$ MeV of BCS theory. 
With the full theory, the critical temperature is instead $T_c = 4.34$ MeV and the calculation is performed in the normal-fluid regime. 
The full line shows that the pair fluctuations  lead to a pseudo-gap, with a strong depletion of the level density around the Fermi energy.
However, the pseudo-gap effect is overestimated in the quasi-particle approximation. 
If one calculates the level density from the full spectral function $A(k,\omega) = -1/\pi \Imag [1/(\omega-k^2/2m-\Sigma(k,\omega)+\mu)]$, the depletion in the level density is still present but it is somewhat  weaker than that  shown in
Fig.~\ref{fig:SchnellPRL83pseudo-gap} (cf.~Fig.~2 of Ref.~\cite{Schnell-1999}).
Note the overall similarity with the results presented in Fig.~\ref{Figure-13}(c) for a contact potential. 

\subsection{Effect of pairing on the liquid-gas transition in symmetric nuclear matter} 
\label{sec:liquidgas}

In symmetric nuclear matter one cannot pass continuously from the BEC to the BCS side, because, at intermediate densities, homogeneous nuclear matter is thermodynamically 
unstable and transforms into an inhomogeneous mixture of ``gas'' (dilute) and ``liquid'' (dense) phases (for a review, see, e.g., Ref.~\cite{Chomaz-2004}). 
This occurs already at the HF level, but the inclusion of the correlated density in Eq.~(\ref{n-with-corr}) has sizeable effects on the phase diagram.

\begin{figure}[h]
\begin{center}
\includegraphics[width=16cm]{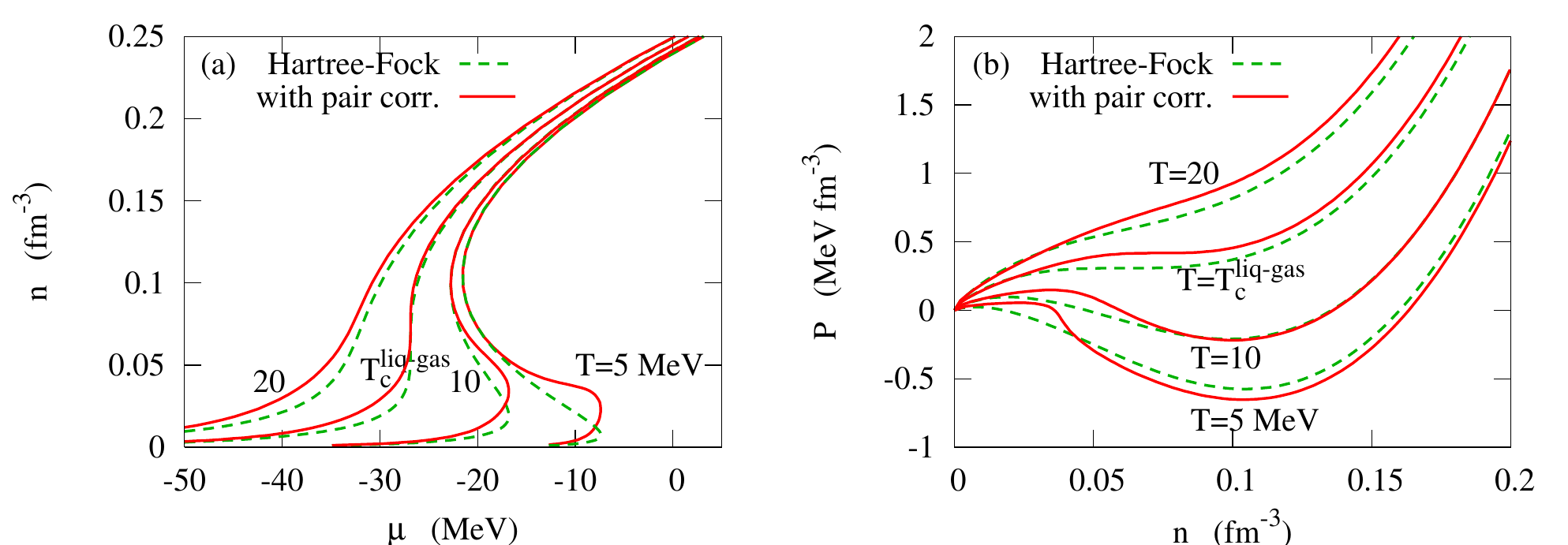}
\end{center}
\caption{(a) Density of symmetric nuclear matter at $T = 20$, $15.9$, $10$, and $5$ MeV (from left to right) as a function of the chemical potential. 
(b) Corresponding pressure as function of the density. 
The dashed lines are Gogny HF results, while the solid lines include the pair correlations in the normal phase.
[Adapted from Ref.~\cite{Jin-2010}.]
\label{fig:nvsmu_pressure}}
\end{figure}
We begin by discussing the relation between the density $n$ and the chemical potential $\mu$. 
Figure~\ref{fig:nvsmu_pressure}(a) shows $n$ as a function of $\mu$ for different temperatures, where the dashed lines represent the HF results $n_{\mathrm{HF}}$ 
(obtained with the Gogny force \cite{Decharge-1980}) and the solid lines include the correction $n_{\mathrm{corr}}$ due to pair correlations (cf.~Eq.~(\ref{n-with-corr})). 
This comparison shows that, for given chemical potential, the correlations increase the density. 
In the high-density region, the results with and without correlations converge to the same value, implying that the correlations fade away at high density as expected. 
For example, at $T=5$ MeV, the two results coincide starting from $n=0.07$ fm$^{-3}$. 
This is a consequence of the Mott mechanism, as discussed at length in Ref.~\cite{Schmidt-1990}. 
As mentioned in Section~\ref{sec:pn_finite_T}, the critical number density where the bound state (at $Q=0$) disappears is called Mott density. 
When the temperature changes from 5 MeV to 10, 15.9, and 20 MeV, the Mott density changes from 0.07 fm$^{-3}$ to 0.12, 0.18, and 0.22 fm$^{-3}$,
implying that the mean-field approximation is valid in the high-density region. 
Below this region, the contribution of the nucleon-nucleon correlations is important.

From Fig.~\ref{fig:nvsmu_pressure}(a) one also sees that, below a certain critical temperature $T_c^{\mathrm{liq-gas}}=15.9$ MeV 
(which has nothing to do with the superfluid critical temperature), $n$ is no longer a single-valued function of $\mu$.
Therefore, the curves in Fig.~\ref{fig:nvsmu_pressure}(a) are generated in practice by making a loop over the HF density and not over $\mu$. 
The presence of a region of densities where the chemical potential decreases with increasing density is a typical feature of a liquid-gas phase transition. 
In the present framework, the liquid-gas critical temperature coincides with the mean-field result \cite{Song-1990,Ventura-1992}. 
This is an artifact of the perturbative treatment of correlation effects, as explained in Section~\ref{sec:pn_finite_T}. 
In a more complete (self-consistent) treatment, $T_c^{\mathrm{liq-gas}}$ should be reduced when deuteron (and heavier)
clusters are included, as shown in Ref.~\cite{Roepke-1983}.

To determine the boundary of this first-order phase transition, one needs to consider the pressure. 
In principle, one can get the pressure $P(T,\mu)$, as a function of temperature and chemical potential, from the number density $n(T,\mu)$ by integrating the
thermodynamic relation $n = (\partial P/\partial \mu)_T$ over $\mu$:
\begin{equation}
P(T,\mu)=\int_{-\infty}^{\mu}n(T,\mu^\prime)d\mu^\prime \, .
\label{Pofmu}
\end{equation}
However, this procedure has to be modified in the temperature range below $T_c^{\mathrm{liq-gas}}$, where $n$ is not a single-valued function of $\mu$. 
In this region, one can obtain the pressure by transforming the integral over $\mu$ into an integral over $n_{\mathrm{HF}}$ as follows:
\begin{equation}
P(T,n_{\mathrm{HF}}) = \int_0^{n_{\mathrm{HF}}} \!\!\!n(T,n'_{\mathrm{HF}})
 \left(\frac{\partial\mu}{\partial n'_{\mathrm{HF}}}\right)_T \!dn'_{\mathrm{HF}} \, .
\label{Pofn}
\end{equation}
Since $\mu$ is a single-valued function of $n_{\mathrm{HF}}$ (cf. dashed line in Fig.~\ref{fig:nvsmu_pressure}(a)), this integral is well defined.

The resulting pressure is shown in Fig.~\ref{fig:nvsmu_pressure}(b) as a function of the density, where again the solid lines are the full results
including pair correlations and the dashed lines are HF results. 
The main effect of the nucleon-nucleon correlations is to increase the pressure at very low densities, with the exception of the case
$T = 5$ MeV for which the pressure at high densities is lower than the HF result.

\begin{figure}[h]
\begin{center}
\includegraphics[width=15cm]{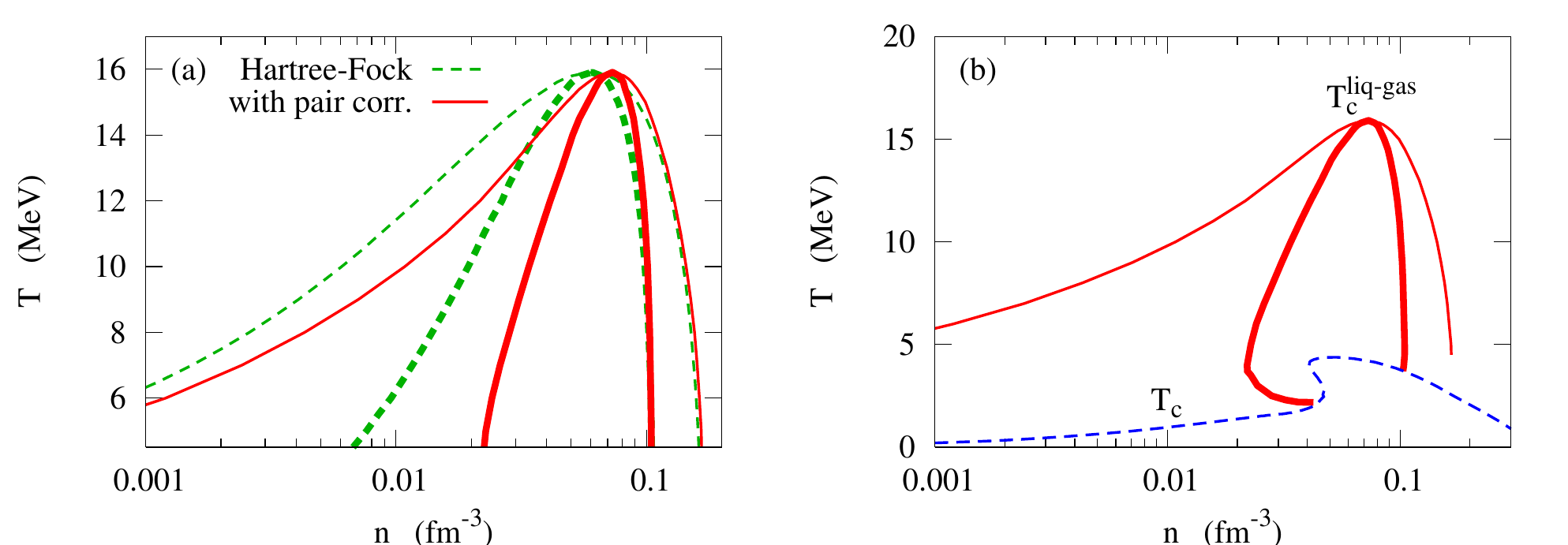}
\end{center}
\caption{(a) The liquid-gas phase diagram of symmetric nuclear matter as a function of density and temperature (for $T \geq 4.5$ MeV). 
The thin lines are the boundary of the coexistence region, while the thick lines are the boundary of the spinodal region. 
Solid lines: with correlations; dashed lines: mean field results. 
(b) Combination of the liquid-gas phase diagram of (a) and the superfluid phase diagram shown in Fig.~\ref{fig:composition5_Tc}(b). 
Upper thin line: limits of the coexistence region; thick line: spinodal region in the normal phase, dashed line: transition to the superfluid phase.
[Adapted from Ref.~\cite{Jin-2010}.]
\label{fig:liquidgas}}
\end{figure}

From Fig.~\ref{fig:nvsmu_pressure}(b) one sees that there is a density region where $\partial P/\partial n < 0$. 
In this so-called spinodal region, small density fluctuations grow exponentially, resulting in phase separation (visible as multi-fragmentation in heavy-ion collisions). 
The boundaries of the spinodal region can be obtained from the zeros of $\partial P/\partial n$ (or, equivalently, of $\partial\mu/\partial n$) and are shown as
the thick solid line in Fig.~\ref{fig:liquidgas}.
Outside the spinodal region, there is a region in the phase diagram where the system is metastable and it is energetically favourable to phase separate. 
The coexistence region of the liquid and gas phases of nuclear matter are determined by the conditions for chemical and mechanical equilibrium:
\begin{equation}
P(T,n_1) = P(T,n_2)\qquad\mbox{and}\qquad \mu(T,n_1)=\mu(T,n_2) \, .
\end{equation}
The result is shown in Fig.~\ref{fig:liquidgas} as the thin solid line. 
Note that, within the present approach, the high-density boundary of the coexistence region cannot be computed for $T<4.5$ MeV,
because the calculation of the pressure at a given density $n$ from Eq.~(\ref{Pofn}) necessitates the calculation of all densities $n'<n$.
This would, in fact, include the density at $n = 0.05$ fm$^{-3}$ where the superfluid $T_c$ reaches its maximum of about $4.5$ MeV (cf. Fig.~\ref{fig:composition5_Tc}(b)).

For comparison, the corresponding mean-field results for the phase boundaries are presented in Fig.~\ref{fig:liquidgas}(a)  (dashed lines), 
which coincide with Fig.~6 of Ref.~\cite{Ventura-1992}. 
Comparing the results with and without correlations, one sees that the correlations decrease the phase-transition temperature in the low-density region and reduce the
unstable region of the liquid-gas phase transition considerably. 
This is an expected result, since the presence of deuterons in the gas can be considered as a first step towards the formation of liquid droplets. 
In the high-density region, the effect of the correlations is almost negligible, which is also plausible because there the deuteron is unbound.
The stabilisation effect is probably even stronger for $\alpha$-particle condensation (to be discussed in Section \ref{subsec:quartetcondensation}) than for deuterons, 
since $\alpha$ particles are about seven times more strongly bound than deuterons.

In Fig.~\ref{fig:liquidgas}(b), the superfluid critical temperature $T_c$ (dashed line), the liquid-gas coexistence region (thin line), and the spinodal instability region (thick line) have been
combined in a single phase diagram. 
As explained above, the present approach does not allow one to calculate the liquid-gas coexistence curve for $T < 4.5$ MeV.
However, extrapolating the thin solid curve to lower temperatures and recalling that at $T=0$ the liquid phase becomes 
stable at saturation density, it is clear that the coexistence curve will cross the superfluid $T_c$ curve at $n\sim n_0$, implying that homogeneous nuclear matter with 
pairing is stable above this density, as one would expect. 
From the results of Ref.~\cite{Stein-1998} one can presume that the liquid-gas coexistence region will be slightly reduced below the superfluid critical temperature, but this effect
should be almost negligible in the case of symmetric nuclear matter considered here \cite{Stein-1998,Su-1987}. 
At low densities, superfluid matter is never stable, because the superfluid $T_c$ curve remains always below the coexistence curve.

The spinodal curve, given by the thick line of Fig.~\ref{fig:liquidgas}, can be calculated until it reaches the superfluid region. 
One thus concludes that superfluid nuclear matter is metastable below $n\sim 0.045$ fm$^{-3}$ and above $n\sim 0.1$ fm$^{-3}$.
Note that, at low density, the density region where the gas phase is metastable is strongly enlarged by correlations,
especially when approaching the superfluid transition temperature. 
This confirms that the correlations have a stabilizing effect with respect to phase separation, even though the intermediate region of the BCS-BEC crossover still lies in the unstable region of the liquid-gas phase transition.
Note that also at $T=0$ symmetric nuclear matter is not stable at sub-saturation densities, but the exact boundaries of the spinodal region are not yet known.

\subsection{Neutron-neutron pairing at zero temperature}
\label{sec:nn_cold}
In the $nn$ case, the BEC side of the BCS-BEC crossover cannot be realized because there is no $nn$ bound state.
As a matter of fact, it is generally believed that two neutrons are nearly bound (although experimentally the possibility of an extremely weakly bound state is not excluded). 
As a consequence, the $S$-wave scattering length $a_{nn} = -18.5$ fm \cite{Gardestig-2009} is quite large compared to the range of the $nn$ interaction, 
which is only of the order of $R\sim 1$ fm. 
Hence, at low density the situation $R < 1/k_F < |a_{nn}|$  can be realized, similarly to trapped atoms near the unitary limit.
 
With increasing density, however, the finite
range of the interaction plays an increasingly important role. The
effect of the effective range on the equation of state was studied
within a $t$-matrix theory in Ref. \cite{Schwenk-2005}. The finite range of the interaction leads to a reduction
of pairing correlations at higher density, so that one passes from the
strongly interacting to the BCS regime.

It is assumed that in the inner crust of neutron stars there is a dilute gas of neutrons coexisting with a Coulomb lattice of dense nuclear clusters (for a review, see, e.g., Ref.~\cite{Chamel-2008}). 
The density of the neutron gas varies continuously from $0$ to $\sim 0.08$ fm$^{-1}$ as one goes deeper into the star. 
The distance between the heavy clusters can be very large (up to more than 100 fm) and the neutron gas can be approximately regarded as uniform. 
The $nn$ pairing in this gas can have important observable consequences on the cooling of the star \cite{Fortin-2010}.

Several studies in the low-density limit of $nn$ pairing have been performed.
In Ref.~\cite{Matsuo-2006}, the BCS gap equation (\ref{gapeqn}) was solved for the case of pure neutron matter. 
For the interaction $V(\vek{k},\vek{k}')$, the D1 Gogny force \cite{Decharge-1980} and the so-called G3RS force \cite{Tamagaki-1968} were used. 
While the former is a density-dependent effective in-medium interaction, the latter is a bare nucleon-nucleon force, which is relatively simple but reproduces the $nn$ scattering 
phase shifts in the $^1S_0$ channel. 
The single-particle energies $\xi_k$ were calculated with the Gogny HF mean field. 
It should, however, be mentioned that the Gogny force in the $^1S_0$ channel is also density independent and relatively close to the bare force 
(with scattering length $-13.5$ fm).

\begin{figure}[t]
\begin{center}
\includegraphics[width=14cm]{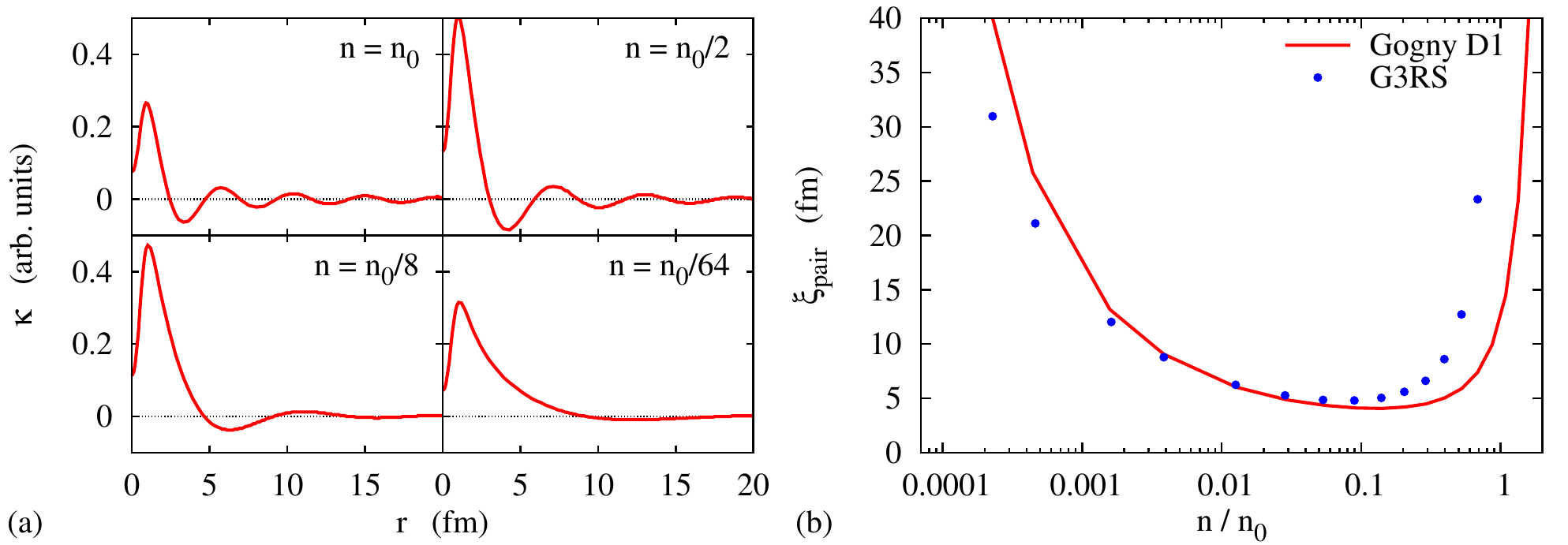}
\end{center}
\caption{\label{Matsuofigures} (a) Pairing tensor in coordinate space as a function of $r$ in neutron matter for various densities ($n_0 = 0.085$ fm$^{-3}$ is half the saturation density 
of symmetric nuclear matter). 
(b) Coherence length of the $nn$ Cooper pair in neutron matter as a function of the density.
[Adapted from Ref.~\cite{Matsuo-2006}.]
}
\end{figure}

Figure \ref{Matsuofigures}(a) shows the $nn$ pairing tensor as a function of position $r$ for various densities. 
In the zero-density limit, the wave function looks like that of a bound state although there is only a resonance very close to zero energy. 
In Fig.~\ref{Matsuofigures}(b), the $nn$ coherence length (pair size) is shown as a function of density. 
Although qualitatively it looks similar to the $pn$ case of  Fig.~\ref{fig:deuteronbcs}(b), the important difference is that,
in the zero-density limit, the coherence length of the $nn$ pairs diverges (since there is no $nn$ bound state) while the coherence length of $pn$ pairs approaches 
the deuteron rms radius. 
Note also that the minimal size of the $nn$ Cooper pair stays slightly above the $pn$ one.
Similar results were found using relativistic mean-field (RMF) single-particle energies and the Bonn-B potential in the gap equation \cite{Sun-2010}.

The nuclear pairing problem is further complicated by the fact that the nuclear force contains a three-body piece, which is not very well known. 
Recently, a $nn$ pairing calculation was performed in Ref.~\cite{Maurizio-2014} including two-body and three-body forces obtained from chiral perturbation theory. 
According to this study, the (repulsive) three-body force does not give rise to a very strong reduction of the gap in neutron matter. 
As with other well tested phenomenological pairing forces like the Gogny D1S force \cite{Decharge-1980}, one obtains a gap which culminates at $\sim 2.5$ MeV at $k_F \sim 0.8 $ fm$^{-1}$. 
In a recent study of the $nn$ gap \cite{Drischler-2017}, extensive numerical investigations of the $^1S_0$ gap (and the ${^3P_2} - {^3F_2}$ one) were performed in the BCS approach 
using chiral effective field theory also including three body forces, with the aim of assessing the uncertainty of the $nn$ gap in BCS approximation. 
These results confirm the earlier findings, with the conclusion that at the BCS level the uncertainties are not  important
(at least, as far as the $^1S_0$ $nn$ gaps are concerned).

\begin{figure}[h]
\begin{center}
\includegraphics[width=15cm]{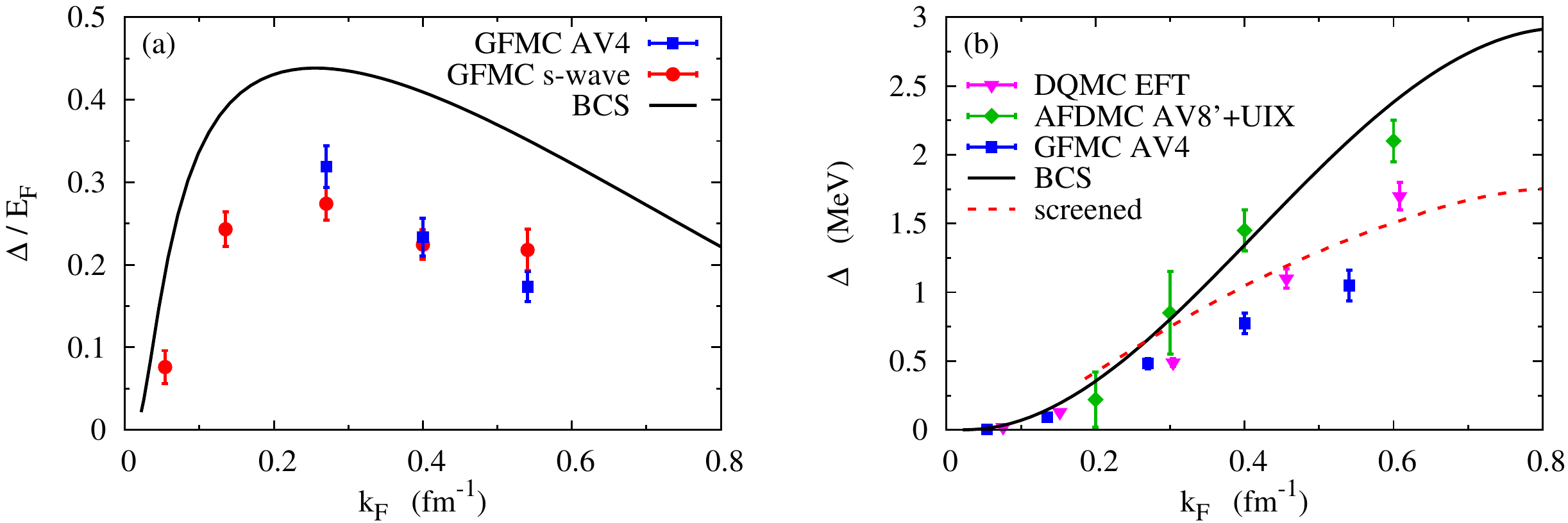}
\end{center}
\caption{\label{fig:neutronDeltaEF-Carlson} (a) Ratio $\Delta/E_F$ and (b) gap $\Delta$ in neutron matter as a
function of the Fermi wave vector $k_F$, as obtained in different QMC
calculations (symbols): variational and Green's function Monte Carlo (GFMC)
\cite{Gezerlis-2010}; determinantal quantum Monte Carlo (DQMC)
\cite{Abe-2009}; auxiliary field diffusion Monte Carlo (AFDMC)
\cite{Gandolfi-2008}. For comparison, the BCS results (solid line) and
results of Ref.~\cite{Cao-2006} that include screening corrections (dashed
line) are also shown. [Adapted from Ref. \cite{Gezerlis-2010}.]}\
\end{figure}
Low-density neutron matter was also studied with more sophisticated Quantum-Monte-Carlo (QMC) approaches. 
These are ab-initio calculations that adopt a realistic nucleon-nucleon interaction. 
As an example, Fig. \ref{fig:neutronDeltaEF-Carlson}(a) shows the ratio
$\Delta/E_F$ of the gap to the Fermi energy $E_F = k_F^2/(2m)$
obtained by the variational and Green's function Monte-Carlo (GFMC)
method of Ref. \cite{Gezerlis-2010}. 
The blue squares were obtained with the AV4 interaction, a simplified version of the Argonne $V_{18}$
interaction keeping $s$ and $p$ wave contributions, while the red circles were obtained keeping only the $s$-wave interaction.
 As expected, the gap is entirely determined by the $s$-wave interaction at these low densities. 
It appears that the QMC results for $\Delta/E_F$ are suppressed compared to the BCS ones (solid line). Even
with this suppression, the maximum value of $\Delta/E_F$ of about 0.3
is reached at $k_F \sim 0.27$ fm$^{-1}$, corresponding to a very low
density of $\sim 0.0007$ fm$^{-3}$. 
This ratio corresponds to a strong coupling situation, with $(k_F a_{nn})^{-1} \sim -0.2$, which is close to, but not 
precisely at, the unitary limit. At higher densities, $(k_F a_{nn})^{-1}$ gets even closer to zero, 
but the finite range of the interaction leads to a strong reduction of the gap. As a consequence,  high-density neutron
matter is again in the weak-coupling (BCS) regime. In
Fig. \ref{fig:neutronDeltaEF-Carlson}(b), the gap $\Delta$ of the GFMC calculations of Ref. \cite{Gezerlis-2010} (squares) is compared with 
the results of other QMC calculations. 
The determinantal quantum Monte-Carlo (DQMC) results
of Ref. \cite{Abe-2009} (triangles), obtained with a low-momentum
interaction derived from pionless effective-field theory (EFT), agrees
at low density very well with the GFMC results of
Ref. \cite{Gezerlis-2010} and also shows a suppression with respect
to the BCS gap (solid line). This suppression is mainly a consequence of the
particle-hole fluctuations, analogous to the Gor'kov-Melik-Barkhudarov
(GMB) corrections discussed in Section~\ref{sec:gmb}
for a contact interaction. These screening effects, which are
missing in the BCS approach, are in principle automatically included in the exact
QMC calculations. Surprisingly, the auxiliary
field diffusion Monte-Carlo (AFDMC) calculation of
Ref. \cite{Gandolfi-2008} (diamonds), which uses a more complete
interaction (AV8' two-body plus Urbana IX three-body force), shows
almost no suppression of the gap with respect to the BCS result. As pointed out in Ref.~\cite{Gezerlis-2010}, this discrepancy between the
AFDMC results of Ref.~\cite{Gandolfi-2008} and the other QMC results may be due to
the trial wave function used in Ref.~\cite{Gandolfi-2008}. 

\begin{figure}[h]
\begin{center}
\includegraphics[width=8.0cm]{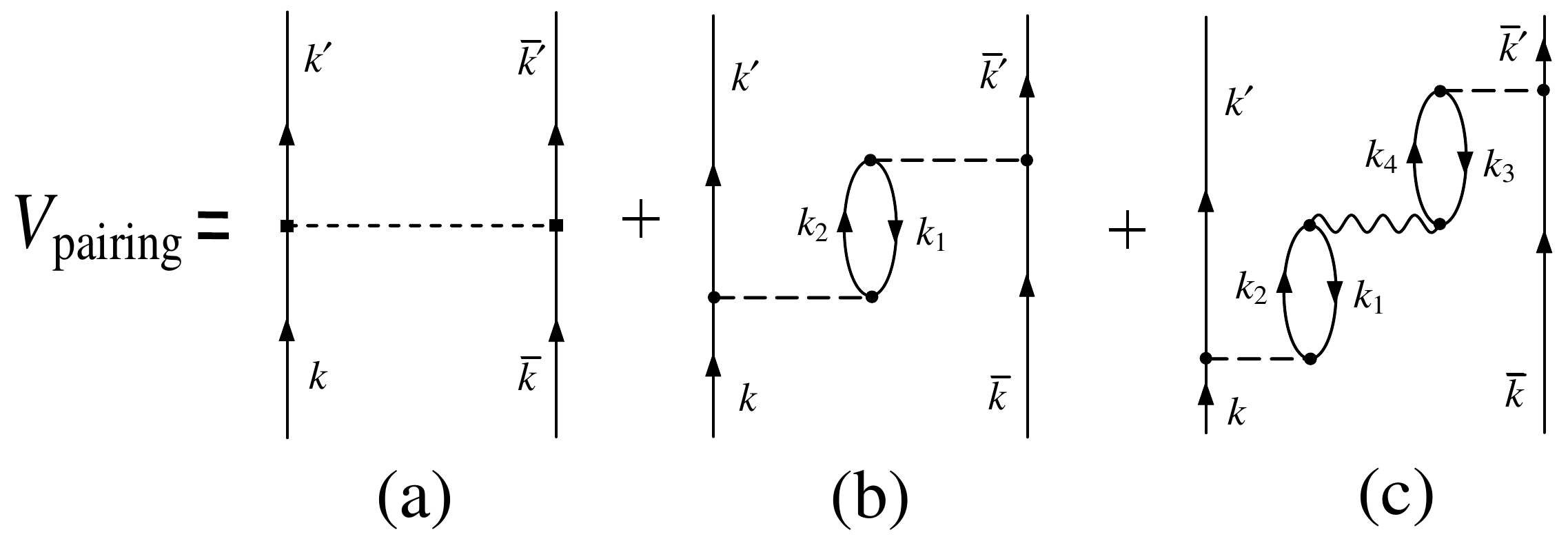}
\end{center}
\caption{Diagrams for the pairing interaction with screening from bubble exchanges.}
\label{fig:screened-force}
\end{figure}

The inclusion of screening (or anti-screening) corrections is a delicate task. 
Only a few theoretical studies in infinite nuclear and neutron matters exist (see, e.g., Ref.~\cite{Schwenk-2004} and references therein). 
The reason is that nuclear systems are strongly interacting, while medium polarisation can only be tackled perturbatively for technical reasons. 

A step towards a
non-perturbative treatment of screening, based on a renormalization
group approach, was made in Ref. \cite{Schwenk-2003}, leading to a very strong suppression of the gap. The fact
that the gap, like in nuclear matter, depends exponentially on any
uncertainty of the effective pairing force makes a reliable prediction
very difficult. 

In any case, one has to go beyond the lowest-order GMB approach (see Section~\ref{sec:gmb}). 
Nonetheless, the results obtained in Ref.~\cite{Cao-2006} with
screening corrections are quite close to the Monte-Carlo results in
neutron matter (see dashed line in
Fig. \ref{fig:neutronDeltaEF-Carlson}(b)).  It is therefore worth explaining how pairing is affected by medium polarization in neutron matter and symmetric nuclear matter. 
One of the subtle points is that one has to treat polarization effects consistently in the pairing force and nucleon
self-energies. 
This is because quite often strong cancellations of both contributions occur. 
The screening terms which we refer to are depicted in Fig.~\ref{fig:screened-force}.
\begin{figure}[h]
\begin{center}
\includegraphics[width=16cm]{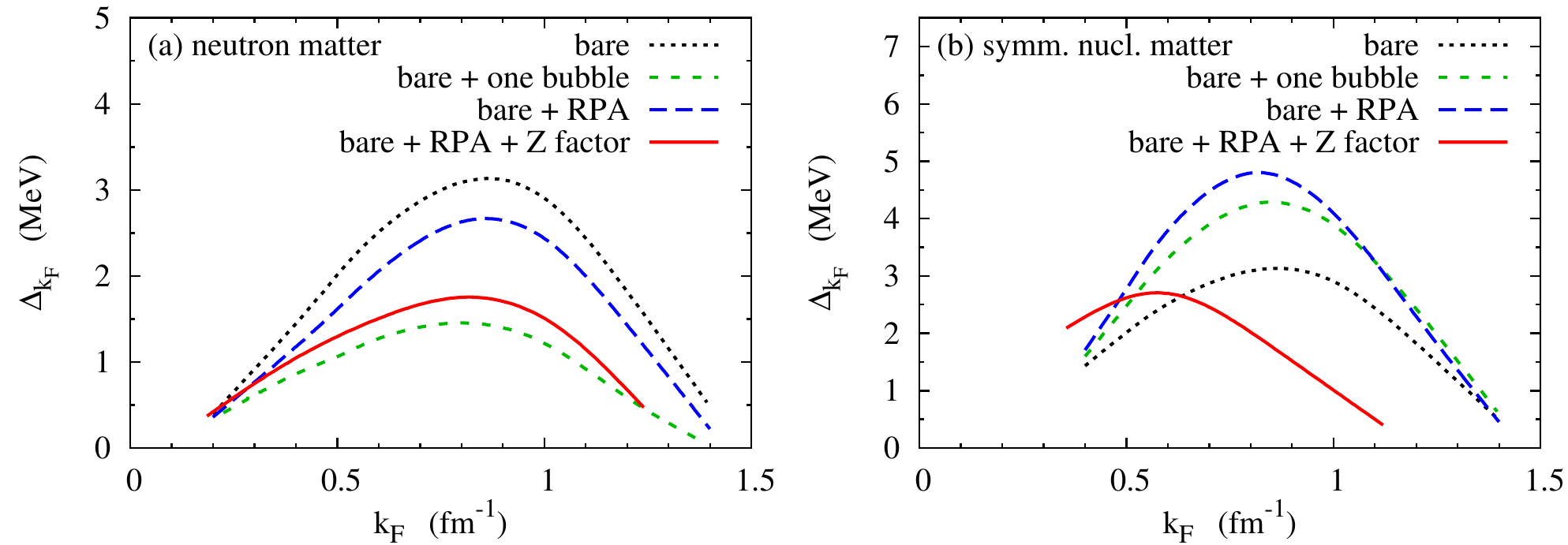}
\end{center}
\caption{Neutron-neutron pairing gaps in the $^1S_0$ channel for (a) neutron matter and (b) symmetric nuclear matter obtained with different approximations. 
Dashed line: using only the bare interaction of Fig.~\ref{fig:screened-force}(a);
short-dashed line: including the one bubble exchange of Figs.~\ref{fig:screened-force}(a) and (b); 
long-dashed line: including the full RPA in the screened interaction of Figs.~\ref{fig:screened-force}(a), (b), and (c); 
solid line: including the $Z$ factors in addition to the fully screened interaction.
[Data taken from Ref.~\cite{Cao-2006}.]}
\label{fig:screened-gaps}
\end{figure}
Here, diagram (a) corresponds to the bare force (Born term), while diagrams (b) and (c) stand for the whole bubble series (RPA). 
The bubbles are attached to the particle-particle lines via the G-matrix \cite{Brueckner-1955,Day-1967}. 
The corresponding gap equation is given in Eq.~(\ref{screened-gap}). 
There is, however, a problem with the calculation of the multi-bubble contributions. 
The particle-hole interaction in the bubble series can be approximated by the Landau parameters. 
However, as a consequence of the liquid-gas instability, there appears the well-known low-density singularity of the RPA in nuclear matter 
(with Landau parameter $F_0=-1$). 
This problem, discussed in Ref.~\cite{Lombardo-2001c} (see also references therein) is remedied by dressing the vertex insertions according to the Babu-Brown
induced-interaction theory \cite{Babu-1973}. 
The results are shown in Fig.~\ref{fig:screened-gaps} for (a) neutron matter and (b) symmetric nuclear matter. 
The red solid lines should be considered as the final results.

As already mentioned, the gap in neutron matter compares quite well with the QMC results of Ref.~\cite{Gezerlis-2010}. 
One may extrapolate from this that also the gap in symmetric nuclear matter is quite reliable. 
However, the strong shift of the peak position towards lower density is slightly suspicious, also because this feature is not seen in any of the more phenomenological
approaches. 
If true, this feature would imply that, at saturation, there is practically a vanishing neutron-neutron gap in symmetric nuclear
matter. 
Nevertheless, this does not necessarily mean that in finite nuclei the superfluid properties are not well reproduced, since there an average over whole volume and surface is performed.

It is also interesting to see from Fig.~\ref{fig:screened-gaps} that, concerning the induced force, in symmetric nuclear matter there is actually anti-screening while in neutron matter there is screening. 
This stems from the fact that, contrary to neutron matter, also proton bubbles can be exchanged which have an attractive effect. 
Only when in addition the dynamic mean field corrections (i.e., the $Z$-factors defined in Eq.~(\ref{Z-factor})) are included, the final gap becomes quenched.

More recently \cite{Zhang-2016}, the induced pairing interaction was calculated at various asymmetries. 
The screening in neutron matter goes smoothly over to anti-screening in symmetric nuclear matter. 
About half way in between there is practically total cancellation between screening and anti-screening and only the mean field remains active. 
However, so far the corresponding gaps have not yet been calculated.

\subsection{Neutron matter at finite temperature}
\label{sec:nn_finite_T}
At finite temperature, the correlated density of Eq.(\ref{n-with-corr}) has to be included in the calculation of $T_c$ as a function of $n$, 
as discussed in Sections~\ref{sec:nsr} and \ref{sec:pn_finite_T}. 
This was done, e.g., in Ref.~\cite{Ramanan-2013}, where an effective low-momentum interaction $V_{\mathrm{low-}k}$ was used \cite{Bogner-2007}. 
Since this interaction is not separable, the calculation of the $t$-matrix, the critical temperature, and the correlated density is slightly more involved than with a separable potential.

The technique used in Ref.~\cite{Ramanan-2013} is based on the so-called Weinberg eigenvalues \cite{Weinberg-1963}. 
The starting point is the integral equation for the $nn$ $t$-matrix, of the form: 
\begin{equation}
\Gamma(\vek{k},\vek{k}',\vek{Q}, \omega) = V(\vek{k},\vek{k}') + \int\frac{d\vek{k}''}{(2\pi)^3} V(\vek{k},\vek{k}'') G^{(2)}_0(\vek{k}'',\vek{Q},\omega) \Gamma(\vek{k}'',\vek{k}',\vek{Q},\omega) \, .
\label{eq:nnTmatrix}
\end{equation}
Here,  $\vek{Q}$ is the total wave vector of the pair, $\vek{k}$ and $\vek{k}'$ are the initial and final wave vectors in the center-of-mass frame, and
\begin{equation}
G^{(2)}_0(\vek{k},\vek{Q},\omega) = \frac{1-f(\xi_{\vek{Q}/2+\vek{k}})-f(\xi_{\vek{Q}/2-\vek{k}})} {\omega-\xi_{\vek{Q}/2+\vek{k}}-\xi_{\vek{Q}/2-\vek{k}}+i0}
\end{equation}
is the (retarded) non-interacting two-particle propagator in the medium with $\xi_{\vek{k}} = k^2/2m-\mu$. 
Equation~(\ref{eq:nnTmatrix}) can be readily  solved in the basis where the operator $VG^{(2)}_0$ is diagonal, which implies  finding 
the (generally complex) eigenvalues $\eta_\nu$ that correspond  to the eigenfunctions $\psi_\nu$, such that:
\begin{equation}
\int\frac{d\vek{k}'}{(2\pi)^3} V(\vek{k},\vek{k}') G^{(2)}_0(\vek{k}',\vek{Q},\omega) \psi_\nu(\vek{k}',\vek{Q},\omega) = \eta_\nu(Q,\omega)\psi_\nu(\vek{k},\vek{Q},\omega) \, .
\end{equation}
In the $S$-wave case here considered, the eigenvectors depend only on $Q = |\vek{Q}|$ and $k = |\vek{k}|$, but not on the angle between $\vek{Q}$ and $\vek{k}$, such that 
$G^{(2)}_0$ can be averaged over this  angle. 
The critical temperature can then be obtained from the condition $\eta_\nu(Q=0,\omega=0) = 1$, since this corresponds to a pole in the $t$ matrix. 

To obtain the correction to the density, the Dyson series of the single-particle Green's function can be truncated to first order like in 
the original NSR  approach, by writing  $G \approx G_0 + G_0^2(\Sigma-\Sigma_1)$ where the self-energy $\Sigma$ is again calculated within  the $t$-matrix approximation (cf. Fig.~\ref{Figure-8}(a)). 
Note that the first-order (HF) contribution $\Sigma_1$ has been subtracted, assuming that it is already included in the single-particle energies 
(cf. discussion in Section~\ref{sec:pn_finite_T}). 
In this case, one can write the density as $n = n_{\mathrm{free}}+n_{\mathrm{corr}}-n_1$, where the correlation contribution can be expressed in terms of the Weinberg eigenvalues 
$\eta_\nu$, in the form
\begin{equation}
n_{\mathrm{corr}} = -\frac{\partial}{\partial\mu}\int\frac{Q^2 dQ}{2\pi^2} \int\frac{d\omega}{\pi} b(\omega)\Imag\sum_\nu \log(1-\eta_\nu(Q,\omega))\,,
\end{equation}
while the correction from the HF term is given by:
\begin{equation}
n_1 = 2\int \frac{d\vek{k}}{(2\pi)^3} \frac{\partial f(\xi_{\vek{k}})}{\partial\xi_{\vek{k}}} \Sigma_1(k) \, .
\end{equation}

\begin{figure}[t]
\begin{center}
\includegraphics[width=7cm]{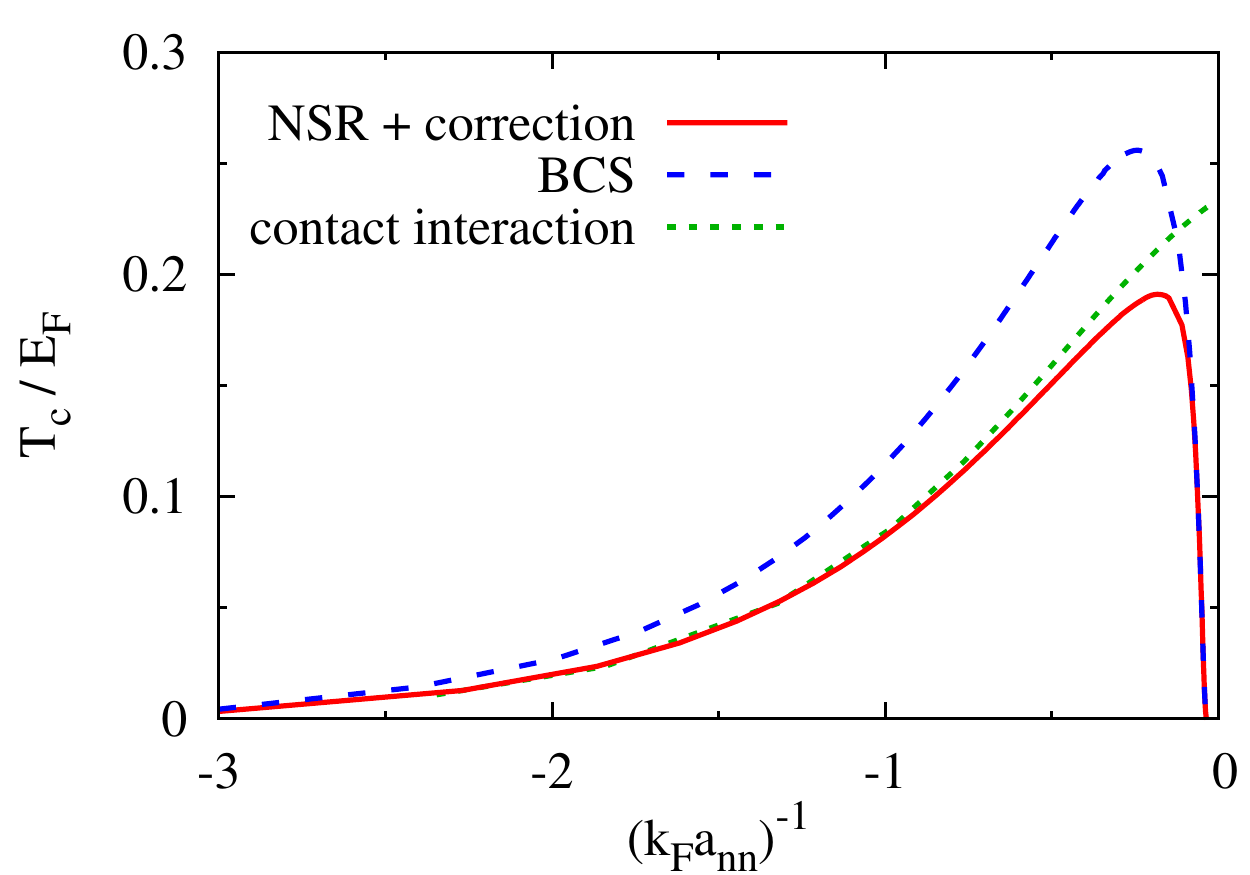}
\end{center}
\caption{\label{RamananFig8} Solid line: critical temperature of neutron matter in units of the Fermi energy, as a function of the dimensionless parameter 
$(k_Fa_{nn})^{-1}$, obtained within the NSR approach  adapted to neutron matter \cite{Ramanan-2013}. 
For comparison, the dashed line shows the BCS result and the dotted line the NSR result for a contact interaction.
[Adapted from Ref.~\cite{Ramanan-2013}.]
}
\end{figure}
Figure~\ref{RamananFig8} shows the resulting dependence of $T_c$ on $n$ via the dimensionless parameter $(k_Fa_{nn})^{-1} $(with $k_F = (3\pi^2 n)^{1/3}$).
The results obtained by the BCS theory (dashed line) and the original NSR approach with a contact interaction 
(dotted line) are also shown for comparison.
In the figure, the limits $(k_Fa_{nn})^{-1} \to -\infty$ and $(k_Fa_{nn})^{-1} \to 0$ correspond to low and high density, respectively. 
In both limits, neutron matter is in the BCS phase: at low density, because $a_{nn} < 0$ is finite;  at high density, because of the finite range of the interaction~\cite{Andrenacci-1999}. 
The maximum of $T_c/E_F$ is reached for $(k_Fa_{nn})^{-1} \sim -0.2$, which corresponds to a density of 
$\sim 0.0007\, \mathrm{fm}^{-3} \sim 0.004\,n_0$ (which is approximately where also the ratio $\Delta/E_F$ is maximum, cf. Fig~\ref{fig:neutronDeltaEF-Carlson}(a)).
Such a dilute neutron gas exists probably between the nuclear clusters in certain regions of the inner crust of neutron stars. 
One can say that, at this density, the neutron gas is very close to the unitary limit. 
Note however that, as it was the case for the gap (cf. Figs.~\ref{fig:neutronDeltaEF-Carlson} and \ref{fig:screened-gaps}), also the critical temperature is expected to be lowered by 
GMB-like screening corrections \cite{Gorkov-1961,Cao-2006} which  were not included here.

\subsection{Short-range correlations and generalized nuclear contact}
\label{sec:nuclear-contact}
In contrast to ultra-cold atoms, the inter-particle distance and the
interaction range in nuclei are not separated by orders of magnitude
and, therefore, one cannot expect to find a clean $1/k^4$ tail in the
occupation numbers \cite{Weiss-2015a}. As a
consequence, there is no nuclear analog of the Tan's relations of
Section~\ref{sec:Tancontact}. Nevertheless, the contact formalism has
been generalized to the nuclear physics context.

As pointed out in Ref.~\cite{Weiss-2015a}, a kind
of universality of short-range proton-neutron correlations was already
observed more than 60 years ago. Namely, it was found that the cross
section for the nuclear photoeffect at photon energies above $150$ MeV
is approximately given by \cite{Levinger-1951} 
\begin{equation}
\sigma = L \frac{NZ}{A}\sigma_d\,,
\end{equation}
where $\sigma_d$ is the photodisintegration cross section of the
deuteron. This suggests that the photon is absorbed by a correlated
proton-neutron pair whose wave function is similar to that of the
deuteron at short distances. The so-called Levinger constant $L\approx
6$ is universal in the sense that it is roughly the same for all
nuclei, and it has been recently related to the contact
\cite{Weiss-2015a,Weiss-2016}.

Since in nuclear physics there is not only spin but also isospin, one 
has now two contacts $C_s$ and $C_t$ depending on whether the pair is
in a spin singlet ($S=0, T=1$) or triplet ($S=1, T=0$ as the deuteron)
state. In terms of the contacts $C_s$ and $C_t$, the Levinger constant
$L$ can be written as \cite{Weiss-2015a}
\begin{equation}
  L = \frac{a_t}{8\pi}\frac{A}{NZ} (C_s+C_t)\,,
\end{equation}
where $a_t$ is the $pn$ scattering length in the triplet channel.

Although the $s$ wave is dominant at short distances, a generalization
of the contact formalism to higher partial waves was presented in
Ref. \cite{Weiss-2015b}. In that work, the
momentum distributions of light nuclei, obtained in variational
Quantum Monte-Carlo calculations, were analyzed. It was found that the large
wave-vector tails become universal, although not proportional to
$1/k^4$, for $k\gtrsim 4$ fm$^{-1}$, in contrast to
Ref. \cite{Hen-2015} where a universal $1/k^4$
behavior was found in the range $1.6$ fm$^{-1} \lesssim k\lesssim 3.2$
fm$^{-1}$.

Also inelastic electron scattering experiments were interpreted in
terms of the contact \cite{Hen-2014}. It was observed
that for each high-momentum proton knocked out of the nucleus, most of
the time another nucleon is knocked out with momentum back-to-back to
the first one, confirming the picture of short-range two-body
correlations. Interestingly, $np$ correlations are about 20 times
stronger than $pp$ correlations. Hence, in a neutron-rich nucleus, the
probability to be in the large wave-vector tail of the distribution is higher for protons than for neutrons.

\subsection{Quartet BEC with applications to nuclear systems: alpha
  condensation in infinite nuclear matter}
\label{subsec:quartetcondensation}
The possibility of quartet (that is, $\alpha$-particle) condensation in nuclear systems has come to the forefront only in recent years. 
This may be due to the fact that the problem of quartet condensation (the condensation of four tightly correlated fermion) is technically far more difficult than pairing. 
In addition, the BCS-BEC transition for quartets is very different (as we shall  see) from the pair case, to the extent that  the weak-coupling BCS-like regime of long coherence length does not exist for quartets. 
Rather, at high density quartets dissolve into two Cooper pairs.

Quartets are  present in nuclear systems, while they are much rarer in other fields of physics. 
One knows that two excitons in semiconductors can form a bound state and the question has been raised  in the past whether bi-excitons can condense \cite{Nozieres-1982}. 
In future cold-atom devices, it might be possible to trap four different species of fermions which, with the help of Fano-Feshbach resonances, could form quartets (note that four different fermions are necessary to form quartets owing to  Pauli principle).
 Theoretical models have already been treated and a quartet phase predicted in this context \cite{Capponi-2007,Capponi-2008}.

We begin the theoretical description of quartet condensation, by briefly recalling what is done in the standard $S$-wave pairing. 
In the equation for the pairing tensor (anomalous average) $\kappa_{\vek{k}_1 \vek{k}_2} = \langle c_{\vek{k}_1} c_{\vek{k}_2}\rangle$ (with spin and isospin dependence suppressed)
\begin{equation}
\kappa_{\vek{k}_1 \vek{k}_2} = \frac{1 - n_{\vek{k}_1} - n_{\vek{k}_2}} {\varepsilon_{\vek{k}_1} + \varepsilon_{\vek{k}_2} - 2 \mu } 
\sum_{\vek{k}'_1 \vek{k}_2'} \langle \vek{k}_1\vek{k}_2|V|\vek{k}_1'\vek{k}_2'\rangle \kappa_{\vek{k}_1' \vek{k}_2'} \, ,
\end{equation}
where $\varepsilon_{\vek{k}}$ is the kinetic energy (possibly, with a HF shift) and 
$\langle \vek{k}_1 \vek{k}_2|V|\vek{k}_1'\vek{k}_2'\rangle = \delta(\vek{Q}-\vek{Q}') V(\vek{k} - \vek{k}')$ is the matrix element of the force with $\vek{Q}=\vek{k}_1 + \vek{k}_2 $ and $\vek{k}=(\vek{k}_1 - \vek{k}_2)/2$ the center-of-mass and relative wave vectors, respectively, one recognises the two-particle Bethe-Salpeter equation, taken at the eigenvalue $E= 2\mu$. 
For pairs at rest (that is, with $\vek{Q}=0$),  inserting the standard BCS expression (\ref{BCSoccupation}) for the occupation numbers 
 leads to the gap equation (\ref{gapeqn}). 

\begin{figure}[t]
\begin{center}
\includegraphics[width=10cm]{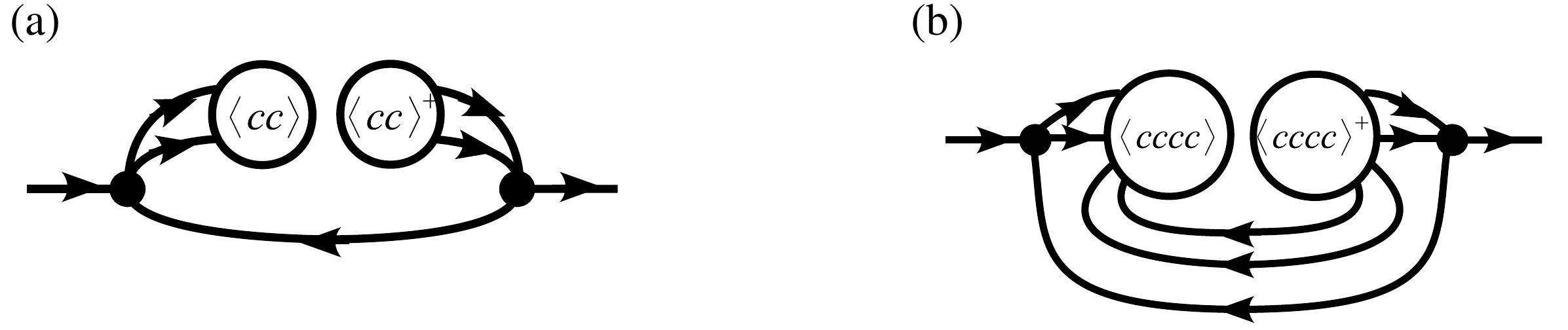}
\end{center}
\caption{\label{fig:Sogomassoperator} Single-particle mass operator in case of pairing (a) and quartetting (b).}
\end{figure}

The idea is to proceed in an analogous way for quartets. 
With the short-hand notation introduced after Eq.~(\ref{ImM}), the in-medium four-fermion Bethe-Salpeter equation for the quartet order parameter 
$K(1234) = \langle c_1c_2c_3c_4\rangle$ reads \cite{Sogo-2010a}:
\begin{equation}
(\varepsilon_1 + \varepsilon_2 + \varepsilon_3 + \varepsilon_4 - 4\mu)K(1234) = (1 -n_1-n_2)\sum_{1'2'}\langle 12|V|1'2'\rangle K(1'2'34) + \text{permutations} \, .
\label{quartetorderparametereqn}
\end{equation}
Although the above equation is a rather straightforward extension of the pairing to the quartet one, the crux lies in the problem of finding  
the single-particle occupation numbers $n_{\vek{k}}$ in the quartet case. 
To this end, it is again convenient to proceed in analogy with the pairing case. 
Eliminating there the anomalous Green's function from the $2\times 2$ set of Gor'kov equations \cite{Fetter-1971}, the retarded mass operator in the Dyson equation for the normal 
Green's function takes the form:
\begin{equation}
M_{1,1'} = \frac{|\Delta_1|^2}{\omega + \xi_1}\delta_{1,1'}
\label{M-pair}
\end{equation}
where $\xi_1 = \varepsilon_1-\mu$ and the gap  is defined by
\begin{equation}
\Delta_1 = \sum_2 \langle 1\bar{1}|V|2\bar{2}\rangle \langle c_2c_{\bar{2}}\rangle \, ,
\end{equation}
``$\bar{1}$'' being  the time reversed state of ``$1$''. 
The graphical representation of Eq.(\ref{M-pair}) is given in Fig.~\ref{fig:Sogomassoperator}(a).

In the case of quartets, the derivation of a single-particle mass operator is more involved, and we give here only the final expression
(for a detailed derivation, see Refs.~\cite{Sogo-2010a,Schuck-2014}):
\begin{equation}
M^{\mathrm{quartet}}_{1,1}(\omega ) = \sum_{234}\frac{\Delta_{1234}[\bar f_2\bar f_3\bar f_4 + f_2f_3f_4]\Delta^*_{1234}}{\omega +\xi_2+\xi_3+\xi_4}
\label{M-quartet}
\end{equation}
where $f_i = f(\xi_i)$ is the Fermi function and $\bar f = 1-f$. 
In addition, the quartet gap matrix is given by:
\begin{equation}
\Delta_{1234} = \sum_{1'2'}\langle 12|V|1'2'\rangle \langle
c_{1'}c_{2'}c_3c_4\rangle
\label{eq:quartetmassoperator}
\end{equation}
The quartet mass operator (\ref{M-quartet}) is also depicted in Fig.~\ref{fig:Sogomassoperator}(b).

Although the derivation of Eqs.(\ref{M-quartet}) and (\ref{eq:quartetmassoperator}) is slightly intricate, the final result looks rather plausible
on physical grounds. 
For instance, the three backward going fermion lines in Fig.~\ref{fig:Sogomassoperator}(b) give rise to the Fermi occupation factors in the numerator of
Eq.~(\ref{eq:quartetmassoperator}). 
This makes a marked  difference with pairing (cf.~Fig.~\ref{fig:Sogomassoperator}(a)), where  only a single fermion
line with $\bar{f} + f = 1$ appears. 
Once the mass operator has been obtained, the occupation number can be calculated via the standard procedure and the system of equations for the quartet
order parameter is closed.

\begin{figure}[t]
\begin{center}
\includegraphics[width=13cm]{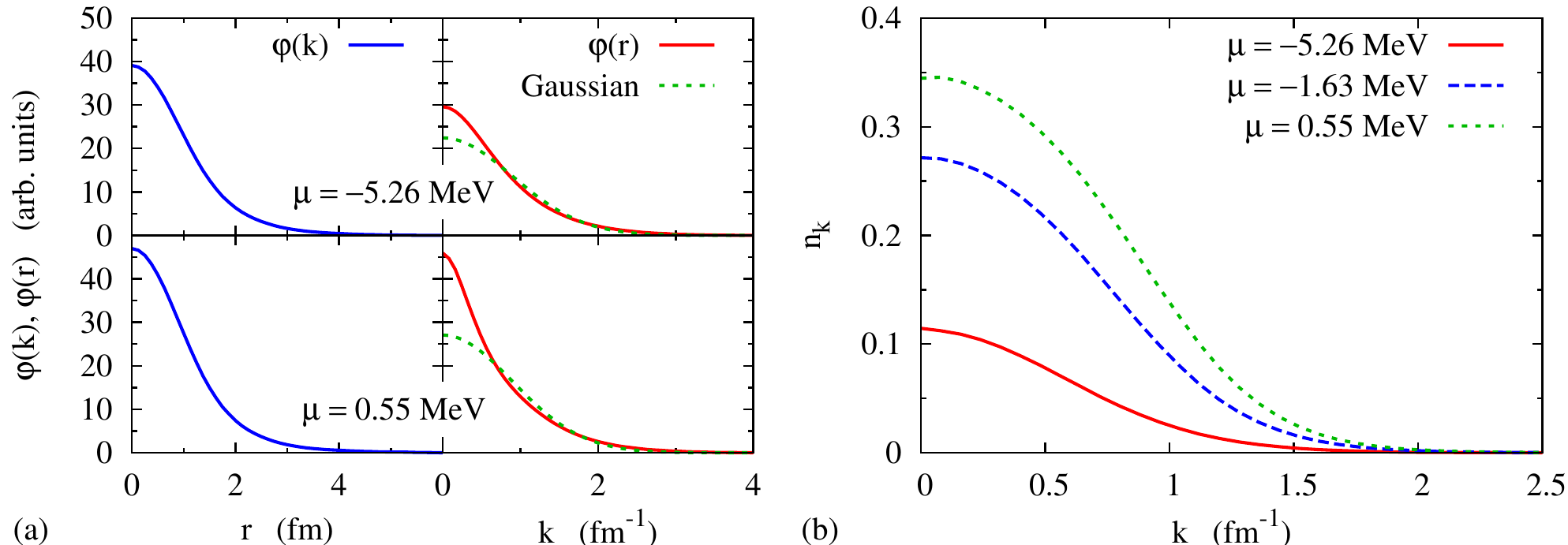}
\end{center}
\caption{\label{Sogofig6} (a) Single-particle wave functions $\varphi$ in position (left) and wave-vector (right) space for two different chemical potentials at zero temperature. 
A Gaussian function is also shown for comparison (dashed line). 
(b) Single-particle occupation numbers $n_{k}$ vs $k$.
[Data taken from Ref.\cite{Sogo-2010a}.]
}
\end{figure}

For the moment we concentrate on the case $T=0$. 
Even in this case, solving  numerically this complicated non-linear set of four-body equations by brute force is out of question. 
Luckily, there exists a very efficient and simplifying approximation that comes to the rescue. 
In nuclear physics, it is known that, because of its strong binding, it is a good approximation to treat an $\alpha$ particle in mean field as long as
it is projected on good total momentum. 
In Eq.(\ref{eq:quartetmassoperator}), one can therefore make the ansatz (see also Ref.~\cite{Kamei-2005}):
\begin{equation}
\langle c_1c_2c_3c_4 \rangle \rightarrow \varphi(\vek{k}_1) \,\varphi(\vek{k}_2) \, \varphi(\vek{k}_3) \, \varphi(\vek{k}_4) \, \delta(\vek{k}_1+\vek{k}_2+\vek{k}_3+\vek{k}_4)\,,
\label{eq:alphaprojectedmf}
\end{equation}
where $\varphi$ is a $0S$ single-particle wave function in wave-vector space (the spin-isospin singlet wave function is again suppressed). 
This ansatz, which is an eigenstate of the total momentum operator with eigenvalue $\vek{K}=0$, makes the problem manageable, since it reduces the calculation to the self-consistent determination of $\varphi(\vek{k})$.
Below, an example will be given where the high efficiency of the product ansatz is demonstrated. 
Note, however, that the bare nucleon-nucleon force cannot be used with the mean-field ansatz (\ref{eq:alphaprojectedmf}).
It is thus convenient to take a separable potential  with two parameters (strength and range), which can be  adjusted to energy and radius of 
a free $\alpha$ particle. 
Figure \ref{Sogofig6}(a) shows the single-particle wave function in position and wave-vector space for two values of  the chemical potential. 
For the larger chemical potential, one sees the wave function deviates considerably from a Gaussian. 
It is also found that, for slightly positive $\mu$, the system of equations has no longer solution and self-consistency cannot be achieved.

The evolution of the occupation number $n_{k}$ with $\mu$ (density) shown in Fig.~\ref{Sogofig6}(b) is very interesting. 
At slightly positive $\mu$ where the system has no longer solution, the occupation number is still far from unity. 
The largest occupation number one obtains lies at around $n_{\vek{k}=0} \sim 0.35$. 
This result is completely different from the BCS-BEC crossover in the pairing case, when $\mu$ can vary from negative to positive values
and the occupation number saturates at unity for $\mu$ well inside  the positive region, see Section \ref{sec:BCSwavefunction}. 
This shows that,  in the case of quartetting, the system is still far from the regime of weak coupling and large coherence length, when the equations that describe it 
 stop having solution. 
From the extension of the wave functions one also sees that the size of the $\alpha$ particles has barely increased. 
Before giving  an explanation for this behaviour, we consider the critical temperature for which the breakdown of the solution 
is more clearly seen.

To determine the critical temperature for the onset of quartet condensation, the equation for the order parameter (\ref{quartetorderparametereqn}) has to be  linearised,
by replacing the correlated occupation number $n_{\vek{k}}$  therein by the free Fermi-Dirac distribution $f_{\vek{k}}$ at finite temperature. 
Determining the temperature where the equation is satisfied gives the desired critical temperature $T^{\alpha}_c$. 
This is the Thouless criterion for the critical temperature of pairing \cite{Thouless-1960}, transposed to the quartet case.
\begin{figure}[t]
\begin{center}
\includegraphics[width=13cm]{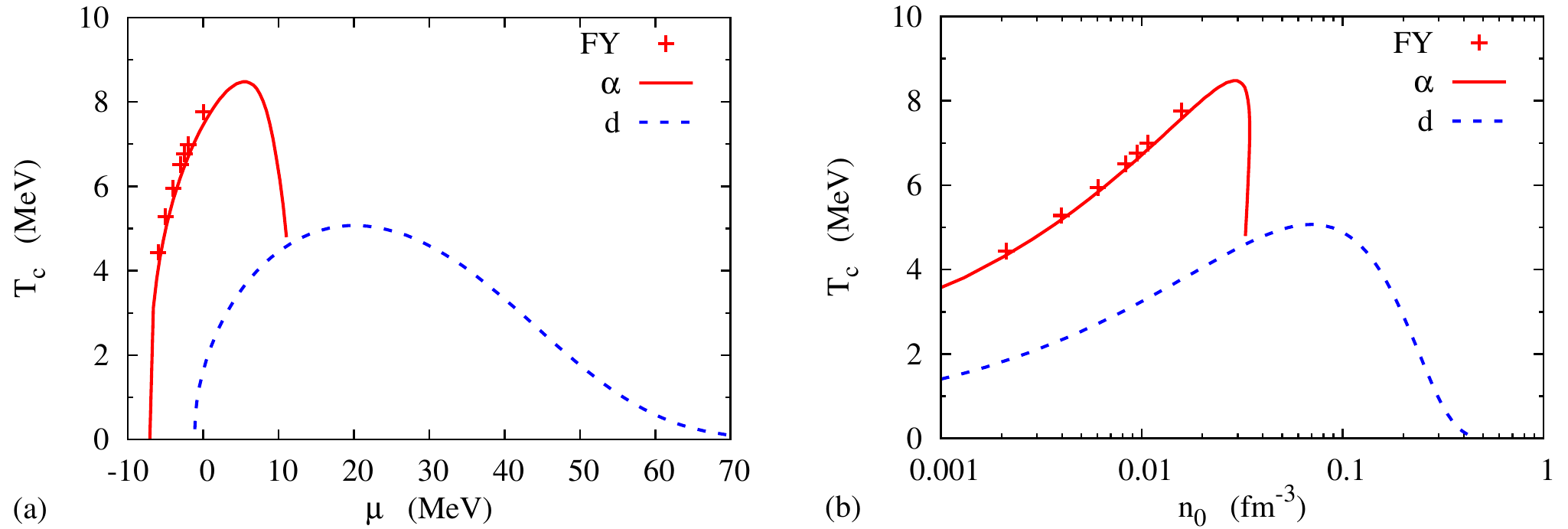}
\end{center}
\caption{\label{SogoTc} Critical temperatures for the condensation of $\alpha$ particle (with binding energy $E_B/A\sim 7.1$ MeV - solid lines) and of deuteron (with $E_B/A\sim 1.1$ MeV - dashed lines), as a function (a) of the chemical potential $\mu$ and (b) of the uncorrelated density $n_0$ (see also Ref.~\cite{Roepke-1998}). 
The crosses were obtained by solving the linearized version of Eq.~(\ref{quartetorderparametereqn}) with the Faddeev-Yakubosvki (FY) method.
[Adapted from Ref.~\cite{Sogo-2009}.]}
\end{figure}
Figure \ref{SogoTc} shows the evolution of $T^{\alpha}_c$ as a function of the chemical potential and density \cite{Sogo-2009}, and explicitly demonstrates the excellent 
performance of the momentum-projected mean-field ansatz for the quartet order parameter. 
The crosses correspond to the full solution of Eq.~(\ref{quartetorderparametereqn}) in the linearised finite-temperature regime with the rather realistic Malfliet-Tjohn bare nucleon-nucleon potential \cite{Malfliet-1969} using the Faddeev-Yakubovsky method, while the continuous line corresponds to the projected mean-field solution (the full solution is available
only for negative chemical potentials). 
One clearly sees the breakdown of quartetting at small positive $\mu$, while $pn$ pairing (in the deuteron channel) continues smoothly into the large $\mu$ region.
[The occurrence of the breakdown of $T^{\alpha}_c$ at a somewhat larger positive $\mu$ with respect to the full solution of the quartet gap equation with the ansatz (\ref{eq:alphaprojectedmf}) at $T=0$, may be due to the fact that here we are at finite temperature - see also the discussion below.]
Note that the density $n_0$ in Fig. \ref{SogoTc}(b) is
the uncorrelated one. Actually, a calculation {\em \`a la} Nozi\`eres and
Schmitt-Rink for the quartet case is still missing. Therefore, no plot can be shown for $T_c$,
as a function of the correlated density, as it was the case for
deuteron pairing, see Fig. \ref{fig:composition5_Tc}(b).

It is worth mentioning that in the isospin polarised case with more neutrons than protons, $pn$ pairing is much more affected than quartetting (due to the much stronger binding of
the $\alpha$ particle) and finally loses against $\alpha$ condensation \cite{Sogo-2010b}. 
As a consequence, in the quartetting case, and contrary to the pairing case, the dissolution of $\alpha$ particles seems to occur quite abruptly as a function of density, 
although it is still unknown whether this is an abrupt transition or a (rather sharp) crossover. 
It could be possible that $\alpha$ particles coexist with ($pn$, $nn$, or $pp$) Cooper pairing at somewhat positive values of $\mu$ (higher density) and disappear asymptotically much faster than deuteron pairing. 
Nevertheless, a phase transition from $\alpha$ condensation to Cooper pairing cannot be excluded and the breakdown of the solution at $T=0$ at small
positive $\mu$, discussed above, hints at a quantum
phase transition with the density as control parameter. 
To settle this question, the equation for the quartet order parameter should be formulated at the outset in terms of BCS quasi-particles and solved as a function of density. 
This is a complicated problem to be addressed in the future. 

The above difference between pairing and quartetting has to do with the different level densities involved in the two systems.
In the pairing case, the single-particle mass operator contains only a single particle (hole) line propagator and the level density is given by:
\begin{equation}
g_{1h}(\omega) = -\frac{1}{\pi}\Imag \int\!\!\frac{d\vek{k}}{(2 \pi)^3} \frac{\bar{f}_{\vek{k}} + f_{\vek{k}}}{\omega + \xi_{\vek{k}} + i\eta} =\int\!\!\frac{d\vek{k}}{(2 \pi)^3}\delta(\omega+ \xi_{\vek{k}}) \, .
\end{equation}
In the case of three fermions, as is the case of quartetting, the corresponding level density is instead given by (see also Ref.~\cite{Blin-1986}):
\begin{equation}
g_{3h}(\omega) = -\frac{1}{\pi}\Imag \Tr \frac{\bar{f}_1\bar{f}_2\bar{f}_3 + f_1f_2f_3} {\omega + \xi_1 + \xi_2 + \xi_3 + i\eta} = \Tr (\bar{f}_1\bar{f}_2\bar{f}_3 + f_1f_2f_3)
\,\delta(\omega + \xi_1 + \xi_2 + \xi_3)\,.
\end{equation}

\begin{figure}[t]
\begin{center}
\includegraphics[width=15cm]{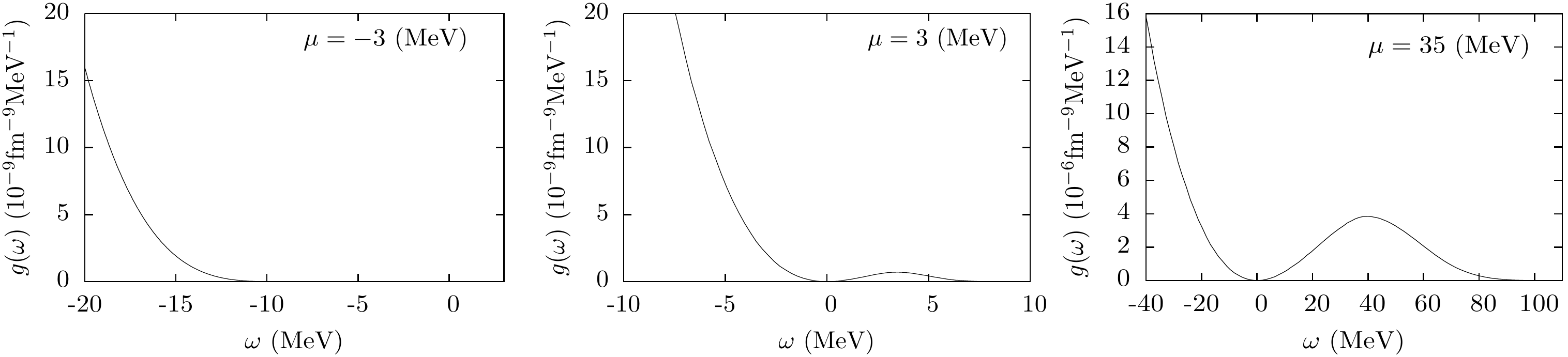}
\end{center}
\caption{\label{Sogoleveldensity} 3h-level density $g_{3h}(\omega)$ for three different values of the chemical potential $\mu$.
[Reproduced from Ref.~\cite{Sogo-2010a}.]}
\end{figure}
Figure~\ref{Sogoleveldensity} gives the results of $g_{3h}(\omega)$ for negative and positive values of $\mu$ at $T=0$. 
The interesting case is $\mu > 0$, for which  phase-space constraint and energy conservation cannot be fulfilled simultaneously at the Fermi 
level  (which corresponds to $\omega = 0$) and the level density is zero there. 
This is just the point where quartetting should build up, but if the level density is zero, there cannot be quartetting. 
In the case of pairing, on the other hand, there is no phase space restriction and the level density is finite at the Fermi level. 
For negative $\mu$, $f(\xi_{\vek{k}})$ vanishes at $T=0$ and is exponentially small at finite $T$, in such a way that there is no fundamental difference 
between $1h$ and $3h$ level densities. 
This explains the striking difference between pairing and quartetting in the weak-coupling regime. 
The same reasoning holds when considering the in-medium four-body equation (\ref{quartetorderparametereqn}). 
The relevant in-medium four-fermion level density is also zero at $4\mu$ for $\mu > 0$. 
Actually, the only case of an in-medium $n$-fermion level density which remains finite at the Fermi energy is (besides $n=1$) the $n=2$ case when the center-of-mass wave vector is zero, see Ref.~\cite{Xu-2016}.
That is why pairing is such a special case, different from condensation of all higher clusters. 
At finite temperature, on the other hand,  the level densities no longer pass through zero and only a strong depression may occur at the Fermi
energy. 
This is probably the reason why the breakdown of the critical temperature is slightly less abrupt than what happens at $T=0$.
\subsection{A glimpse at finite nuclei}
We end up this Section on nuclear systems by briefly considering the situation of the BCS-BEC crossover and alpha condensation in finite nuclei. 
The proton-neutron (deuteron like) pairing in heavier nuclei is very elusive, and a considerable activity is going on about this subject at present. 
For asymmetric nuclei, the suppression of $pn$-pairing can be understood in terms of what we have seen from the infinite-matter study. 
For $N\sim Z$ nuclei, on the other hand, one would expect to find $pn$ pairing because the bare proton-neutron attraction is stronger than the proton-proton or neutron-neutron one. However, these different pairing channels enter in competition with each other and it is not clear which one prevails. 
Recent publications of interest on this subject can be found in Refs.~\cite{Bulthuis-2016,Gezerlis-2011}.
\begin{figure}[h]
\begin{center}
\includegraphics[width=7.7cm,angle=90]{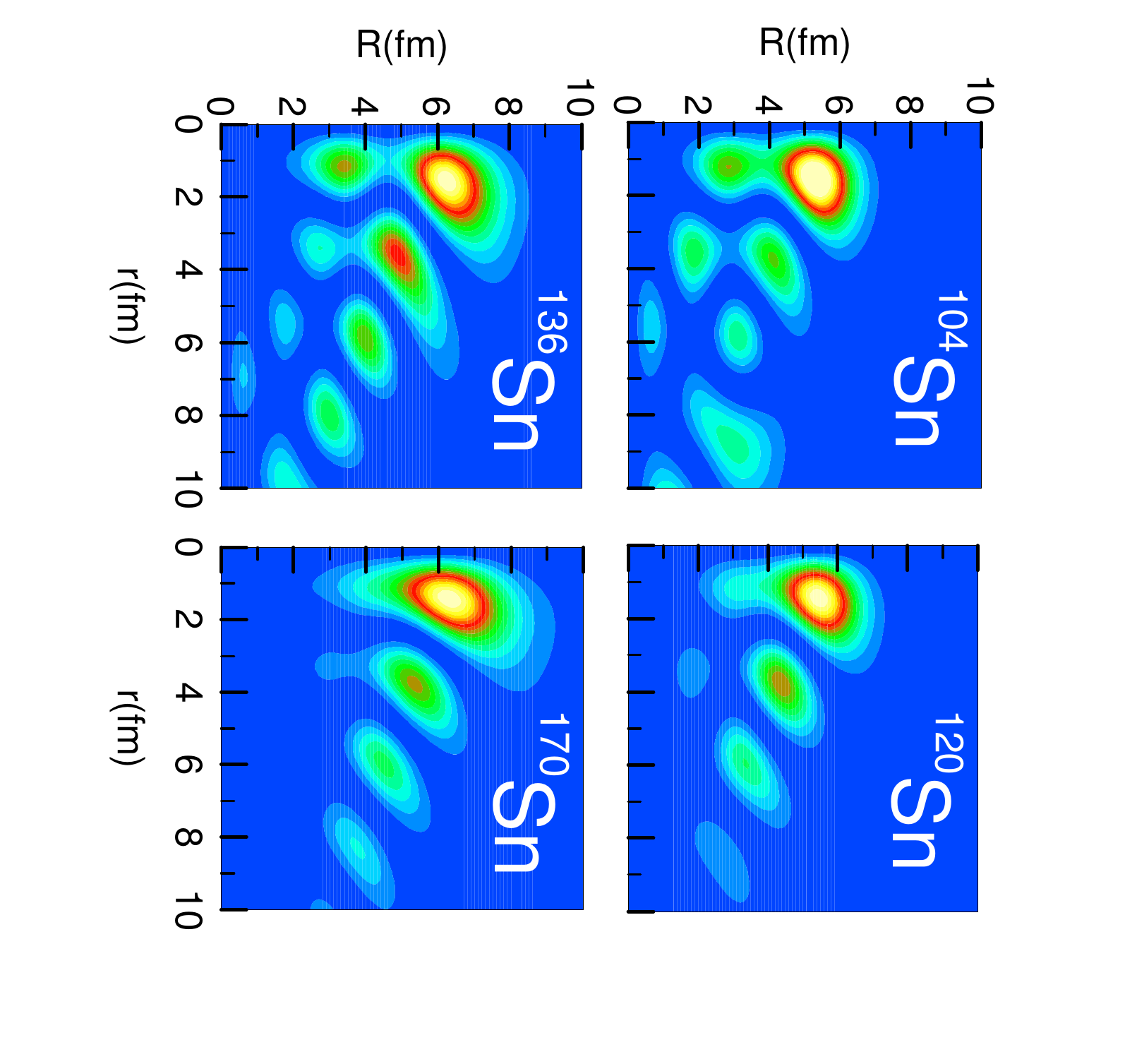}
\end{center}
\vspace{-1cm}
\caption{\label{fig:120Sn} Square of the pairing tensor, multiplied by $R^2r^2$, as a function of
relative ($r$) and center-of-mass ($R$) coordinates, in different Sn
isotopes. [Reproduced from Ref. \cite{Pillet-2007}.]}
\end{figure}

The situation is different if one considers single Cooper pairs in light nuclei (in this case, even though one cannot speak about BEC, yet 
the strong coupling limit may manifest itself). 
For example, the nucleus $^6$Li in its ground state has a strong $\alpha + d$ cluster structure. 
Because of the small number of nucleons, quite sophisticated Resonating Group Method (RGM) calculations could be performed in the past \cite{Fukushima-1978,Kamimura-1981}. 
It is interesting to note that, similar to Fig.~\ref{fig:deuteronbcs}, as a function of distance from the $\alpha$ particle the deuteron first shrinks before it enters the inner part of the $\alpha$. 
For this reason, the situation is qualitatively similar to the infinite-matter case. 
A similar situation may occur for $^{18}$F = $^{16}$O + $d$, where $^{18}$F has the quantum numbers of the deuteron in its ground state as is the case also 
for $^{42}$Sc = $^{40}$Ca + $d$. 
However, no studies concerning the strong coupling features of these systems are reported. 
Even though one may think that adding more deuterons to doubly magic nuclei would enhance the condensate aspect,
two deuterons form an $\alpha$ particle which is about seven times more strongly bound.
This fact may also obscure the $pn$-pairing in the $N\sim Z$ nuclei.

Important $nn$ correlations have been discussed in very neutron-rich nuclei, close to the neutron drip line, where a so-called neutron skin develops. 
The neutron Cooper pairs turn out to be spatially localized at  the surface of those nuclei, thus inducing a spatial asymmetry. 
What is  important for this effect to occur  is that the gap field is built up of states with even {\it and} odd parity. 
This can happen either when the value of the gap englobes different major shells or when an intruder single particle state invades a shell of opposite parity. 
Typical studies of this effect can be found in Refs.~\cite{Matsuo-2005,Pillet-2007,Pillet-2010,Hagino-2010}. 
This spatial concentration of $nn$ Cooper pairs in those nuclei (as shown in Fig.~\ref{fig:120Sn}) can be considered as a signature for the transition from weak to strong coupling,
as discussed in Section~\ref{sec:nn_cold}.

To summarize, in finite nuclei no clear evidence has been seen thus far of strong coupling and the BCS-BEC crossover in the
deuteron channel, but these effects  may play a certain role in compact stellar objects.
Concerning $nn$ pairing in neutron-rich nuclei, some transition from weak to strong coupling can actually be revealed.

\begin{figure}[h]
\begin{center}
\includegraphics[angle=-90,width=9cm]{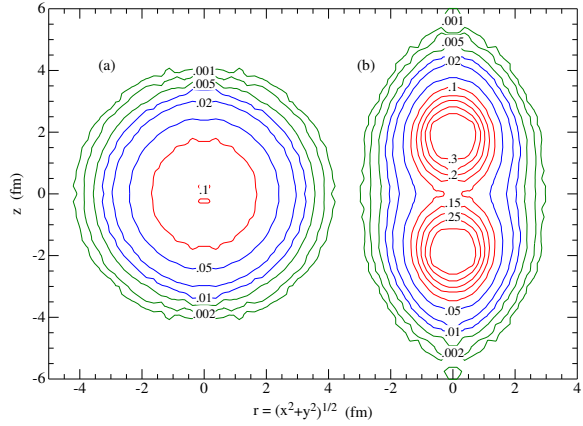}
\end{center}
\caption{\label{fig:8Be} QMC result for the density distribution of $^8$Be in its ground state, (a) in the laboratory frame and (b) in the intrinsic frame: 
The nucleus clearly consists of two $\alpha$ clusters.
[Adapted from Ref.~\cite{Wiringa-2000}.]}
\end{figure}
The BEC of $\alpha$ particles in nuclei has been widely  discussed over the last ten years. 
Usually, the ground states of nuclei are too compact to allow for $\alpha$ particles to form (the only exception being $^8$Be, see Fig.~\ref{fig:8Be}).
At least in lighter nuclei, however, there exist long lived states with a dilute density (about a factor 3-4 smaller than the ordinary density of nuclei), 
which allows for the appearance of $\alpha$ particles with little mutual overlap, forming an almost ideal Bose gas. 
One notable  state of this kind is the first excited $0^+$ state at 7.65 MeV in $^{12}$C, called the Hoyle state. 
Because of the abundance of $^{12}$C in the universe (giving rise to life on earth), the astrophysicist Fred Hoyle predicted in 1952 \cite{Hoyle54} that in stars the triple 
$\alpha$ reaction $\alpha + \alpha + \alpha \rightarrow ^{12}$C$^*$ must be accelerated due to the existence of an excited state in $^{12}$C at the right resonating energy, 
which he predicted to be very close to the value 7.65 MeV measured by Fowler a couple of years later \cite{Cook-1957}. 
It turns out that this state can be described to a good approximation as a state of low density, where all three $\alpha$ particles are ``condensed'' 
with their center-of-mass motion in the same $0S$ bosonic mean-field orbit \cite{Yamada-2012}. 
It seems that this is not the only $n\alpha$ nucleus exhibiting such an ``exotic'' state. 
A good candidate is the sixth $0^+$ state in $^{16}$O at 15.1 MeV \cite{Yamada-2012}. 
There may be $\alpha$ gas states in even heavier nuclei as well. 
Intense experimental activity is presently going on in this field. 
There also exist speculations that in heavier nuclei a surface layer of condensed $\alpha$ particles may exist \cite{Funaki-2009} and $\alpha$ clustering certainly also
exists in nuclei with spontaneous $\alpha$ decay \cite{Roepke-2014}.

\begin{figure}[t]
\begin{center}
\includegraphics[width=7cm]{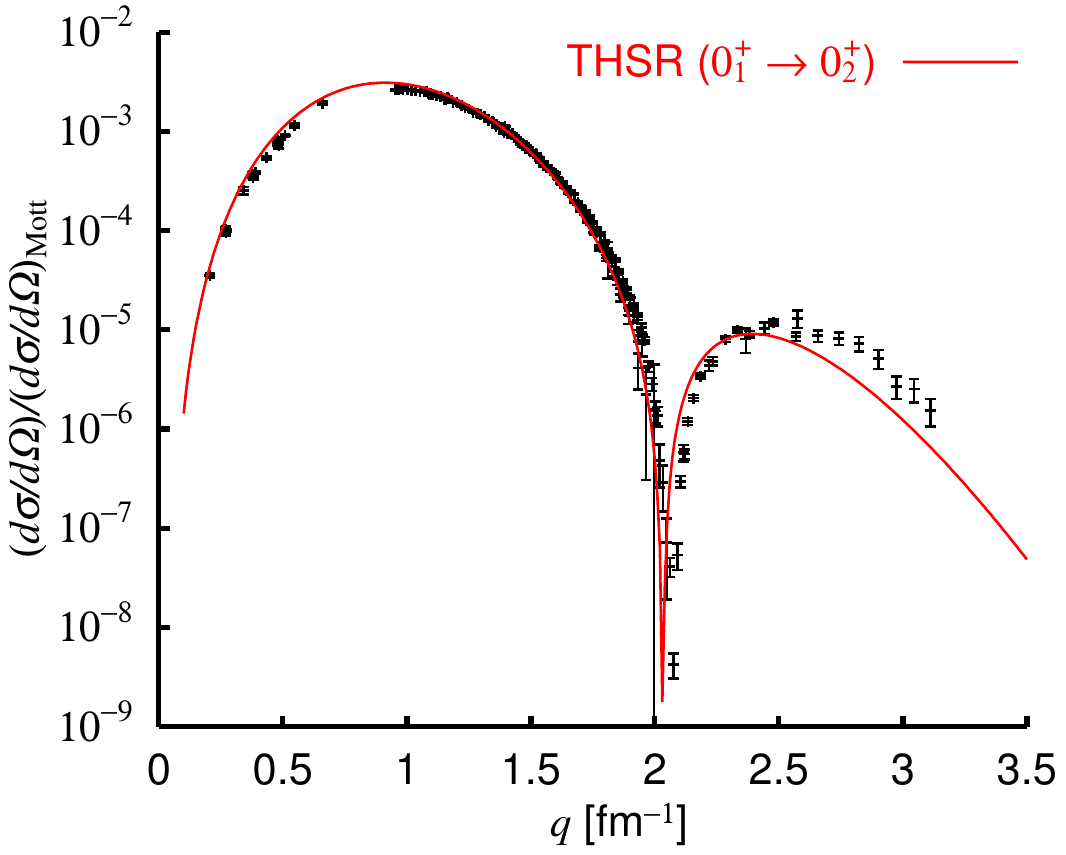}
\end{center}
\caption{Inelastic form factor calculated with the THSR wave function (BEC) and the one of Kamimura \textit{et al.} \cite{Fukushima-1978,Kamimura-1981}. 
The two results are on top of each other. 
Experimental data are from Ref.\cite{Chernykh-2007}.
[Courtesy of Y. Funaki.]
\label{fig:fmfct}}
\end{figure}

On the theoretical side, in small systems like nuclei one must deal with a number-conserving condensate ansatz. 
In Ref.~\cite{Tohsaki-2001}, the following variational ansatz was made for  the Hoyle state with, for instance, three $\alpha$ particles:
\begin{equation}
\Psi_{3\alpha} = {\mathcal A}[\phi(\alpha_1)\phi(\alpha_2)\phi(\alpha_3)]
\label{Psi_3}
\end{equation}
where $\phi(\alpha_i)$ the four-nucleon wave function of the i-th $\alpha$ particle and ${\mathcal A}$ the antisymmetrization operator   
with respect to all the $12$ nucleons.
Note the analogy of the expression (\ref{Psi_3}) with a number-projected BCS wave function.  
Generalization of this ansatz to  more than three $\alpha$ particles is straightforward. 
As mentioned in Section~\ref{subsec:quartetcondensation}, the $\alpha$ particle can be well described by a mean-field wave function as long as the center-of-mass motion is correctly separated. 
It is also well known that for light nuclei, harmonic oscillator wave functions with the oscillator length as parameter are excellent variational functions. 
The following ansatz was thus made in Ref.~\cite{Tohsaki-2001}:
\begin{equation}
\phi(\alpha_i) \propto \exp\bigg[-\frac{2}{B^2}\vek{R}_i^2\bigg]\varphi(\alpha_i)
\end{equation}
where $\vek{R}_i$ is the center-of-mass coordinate of the $i$-th $\alpha$ particle and
\begin{equation}
\varphi(\alpha_i) \propto \exp \bigg[-\frac{1}{2} \sum_{k,l=1}^4(\vek{r}_{i,k} - \vek{r}_{i,l})^2/(8b^2) \bigg]
\end{equation}
is a translationally invariant ``intrinsic'' $\alpha$-particle wave function. 
Here, the only variational parameters are then the widths $B$ and $b$. 
It generally turns out that, for $\alpha$-particle gas states, the width of the intrinsic $\alpha$-particle wave function stays close to its free value $b=1.35$ fm, 
whereas $B$ is much larger since the center-of-mass covers the whole nuclear volume. 
Performing the antisymmetrization and minimising the energy with respect to $B$ for fixed $b$, under the condition that the Hoyle state is orthogonal to the ground state, 
yields a {\emph single} $\alpha$-condensate $12$-nucleon wave function. 
Because of the Gaussian form of the center-of-mass motion of each $\alpha$, also the total center-of-mass can eventually be eliminated, so that the $12$-nucleon
3$\alpha$-cluster wave function is totally translationally invariant. 
The Hamiltonian used contains two-body and three-body effective forces of Gaussian type, whose parameters were adjusted many years before the description of 
the Hoyle state from $\alpha-\alpha$ scattering data and and the properties of a single free $\alpha$ particle \cite{Tohsaki-1994}. 
The Coulomb force was also included. 
In this way, all measured properties of the Hoyle state have been reproduced from a parameter-free theory with only a single variational parameter $B$. 
Among those properties we mention, for instance, the inelastic form factor from the ground to the Hoyle state obtained from electron scattering experiments \cite{Funaki-2006}, 
as shown in Fig.~\ref{fig:fmfct}.
This inelastic form factor is  quite sensitive to the size of the Hoyle state. 
Its rms radius was obtained to be 3.8 fm, which, when compared with that of the ground state of 2.4 fm, implies a volume increase by a factor 3-4,
thus confirming the low-density gas-like quartet structure of the Hoyle state which is similar to the one of $^8$Be shown in Fig.~\ref{fig:8Be} but with one additional $\alpha$ particle. 
Since  quartetting was found to exist only at low density also in nuclear matter, one can assume that the Hoyle state (and possibly more
$\alpha$ gas states in other nuclei) is a precursor to macroscopic quartet condensation. 
This should be considered in analogy to pairing, for which it is known that only a handful of Cooper pairs gives rise in superfluid nuclei 
to what can be considered as precursor to macroscopic neutron superfluidity in neutron stars. 

\begin{figure}[h]
\begin{center}
\includegraphics[width=7cm]{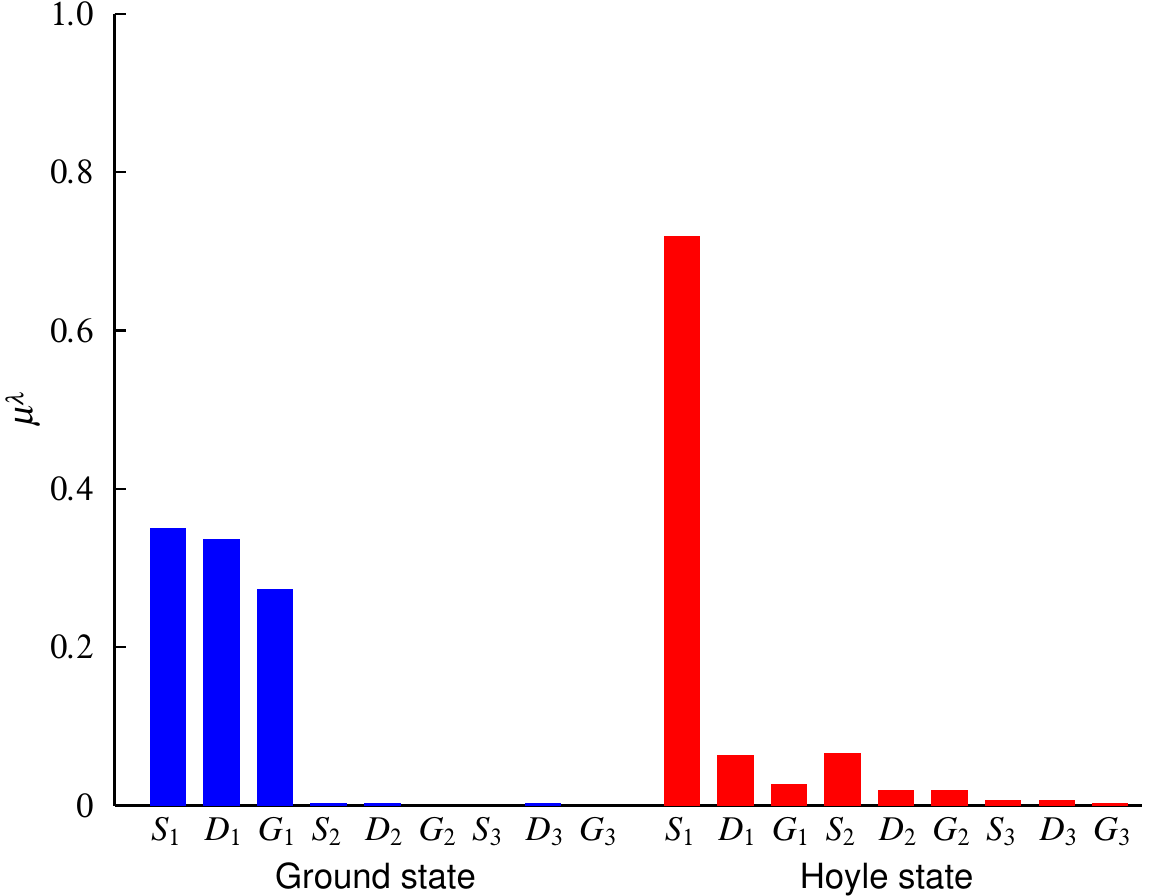}
\end{center}
\caption{\label{fig:3a-occ} 
Single $\alpha$ occupation probabilities for ground (left) and Hoyle state (right). 
The ground state complies with the pure shell model where the occupation is uniform, while in the Hoyle state the 0S occupancy is by over 70\% 
at least a factor of ten larger than in all other states. 
[Reproduced from Ref.~\cite{Yamada-2012}.]}
\end{figure}

A number-conserving quartet wave function does not necessarily imply that condensation occurs. 
One convenient way to theoretically investigate this question is to construct the single $\alpha$-particle density matrix $n(\vek{R}, \vek{R}')$, by integrating out from 
the density matrix of the total system all intrinsic coordinates and all center-of-mass coordinates but one. 
The diagonalisation of this matrix yields the occupation probabilities of the various $\alpha$ orbits. 
It was found that in the Hoyle state the three $\alpha$ particles are to over 70\% in the lowest 0S orbit, the remaining orbits having an occupancy reduced 
by at least a factor of ten. 
In contrast, the occupancies of the $^{12}$C ground states are equally distributed \cite{Yamada-2012}. 
At least theoretically, this result represents a clear signature that to a large extent the Hoyle state can be considered as an $\alpha$ condensate. 
However, the effect of anti-symmetrisation is to scatter the $\alpha$ particles out of the condensate 30\% of the time, which by itself 
is an interesting effect. 
Figure \ref{fig:3a-occ} shows the probability distributions of the $\alpha$ occupancies for the ground and Hoyle states.

Although more could be said about $\alpha$ clustering and $\alpha$ condensation in nuclei, we have given only a short overview of the present situation 
in finite nuclei because this report mainly concentrates on phenomena occurring in infinite matter. Recent more complete reviews on the subject can be found in
 Refs.~\cite{Schuck-2016,Tohsaki-2017}.
\section{Concluding remarks}
\label{sec:concl_rem}
Quite generally,  ``crossover" is a term that describes a situation when a system goes from being in one phase to being in another phase as a certain parameter is changed, 
without encountering a phase transition in between. 
For the BCS-BEC crossover, this means that by tuning the inter-particle interaction (or the density) the system goes from a BCS state where pairs of (opposite spin) fermions 
are described by Fermi statistics, to a BEC state where two-fermion dimers are described by Bose statistics. 
In this way, by freezing out the internal degrees of freedom of the fermion pairs, the statistics of fermions smoothly merges with the statistics of bosons. 
As a consequence, the BCS and BEC paradigms are not fully distinct from each other but rather are the two extrema of a continuum.

It appears fair to say that the BCS-BEC crossover should be considered one of the major scientific achievements that have occurred during the last several years. 
For this reason, our scope here has been to provide a detailed and as much as possible comprehensive review of the physics of the BCS-BEC crossover, and to apply it to the physical systems for which (the ultra-cold Fermi gases) this crossover has been explicitly realized experimentally, or for which (the nuclear systems) the crossover scenario is found consistent with various aspects of its phenomenology, or indirectly accessed from close analogy with ultra-cold gases.
For both systems, we have focused on the aspects of maximum cross interest. 
For this reason, less attention was paid to topics that are of more specific interest to the separate systems, such as the effects of the trapping potential 
in ultra-cold gases or the pairing in finite nuclei. 

Even within these restrictions, the topics covered by this paper are not exhaustive and important aspects have unavoidably been left out. 
In particular, special mention deserve the radio-frequency spectroscopy that was originally introduced to measure the pairing gap in the condensed state \cite{Chin-2004,Torma-2016}, and the thermodynamics of the gas cloud from which the equation of state was extracted \cite{Navon-2010,Ku-2012}. 
From the theoretical side, emphasis was not given to the results obtained via functional-integral approaches \cite{Hu-2006,Diener-2008,Salasnich-2016}
or quantum Monte Carlo methods \cite{Carlson-2012,Bulgac-2012}. 
Rather, a diagrammatic approach was preferred, because it can be more directly connected to one's physical intuition and can also be more readily subject to targeted improvements.

Additional interest in the BCS-BEC crossover comes from particle physics (and, in particular, from quark-gluon plasma, see, e.g., Refs.~\cite{Baym-1983,Yamamoto-2007}). 
In the ultimate analysis, this cross interest from different fields of physics to the BCS-BEC crossover is due to the fact that strongly-coupled quantum Fermi systems, even though
they are described by different microscopic theories, share the basic feature of having the same kind of spontaneously broken symmetry \cite{Nambu-1960,Yang-1962}.
Superconductors make thus no exception to the interest in the BCS-BEC crossover. 
As mentioned in Section~\ref{Historical-background}, actually the first motivation for introducing the BCS-BEC crossover originated in condensed matter.
Nowadays, there is growing direct evidence for the occurrence of this crossover in two-band superconductors with iron-based materials \cite{Lubashevsky-2012}, 
for which it was possible to detect the collapse of the small Fermi surface pocket, with the related single-particle dispersion in the condensed state becoming an inverted parabola 
(as discussed in Section~\ref{sec:pseudo-gap} in the context of ultra-cold gases).
Electron-hole bilayers appear also as promising candidates for the realization of the BCS-BEC crossover in condensed matter systems~\cite{Pieri-2007,Perali-2013,Li-2016,Lee-2016}. 
It is thus possible that in the near future the BCS-BEC crossover might be under the spotlight also in condensed matter, the field where it was originally envisaged.  On the other hand, condensation of quartets, very relevant
for nuclear systems, may in the future also be investigated  with ultra-cold atoms
when the trapping of four different fermion species will be achieved.
Definitely, the exploration of the BCS-BEC crossover in ultra-cold  Fermi gases and nuclear systems, which has been reviewed here, has enriched us with concepts, tools, and results that will be extremely precious for the investigations of this phenomenon in a broader range of contexts, in the years to come.

\vspace{0.15cm}
\noindent{\bf Acknowledgements}

\vspace{0.1cm}
The authors are grateful to M. Baldo and U. Lombardo for discussions on Section~\ref{sec:nuclearsystems}, and to A. Perali for bringing Ref.~\cite{Lubashevsky-2012} to their attention.


\bibliographystyle{elsarticle-num} 
\bibliography{phys_rep_BCS-BEC.bib}

\begin{thebibliography}{100}
\expandafter\ifx\csname url\endcsname\relax
  \def\url#1{\texttt{#1}}\fi
\expandafter\ifx\csname urlprefix\endcsname\relax\def\urlprefix{URL }\fi
\expandafter\ifx\csname href\endcsname\relax
  \def\href#1#2{#2} \def\path#1{#1}\fi

\bibitem{Bardeen-1957}
J.~Bardeen, L.~N. Cooper, J.~R. Schrieffer, {Theory of Superconductivity},
  Phys. Rev. 108 (1957) 1175--1204.

\bibitem{Schafroth-1957}
M.~R. Schafroth, S.~T. Butler, J.~M. Blatt, {Quasi-chemical equilibrium model
  for superconductivity}, Helv. Phys. Acta 30 (1957) 93--134.

\bibitem{Keldysh-1965}
L.~V. Keldysh, Y.~V. Kopaev, {Possible instability of the semimetallic state
  toward Coulomb interaction}, Sov. Phys. Solid State 6 (1965) 2219--2224,
  [Fiz. Tverd. Tela (Leningrad) 6 (1964) 2791].

\bibitem{Popov-1966}
V.~N. Popov, {Theory of a Bose gas produced by bound states of Fermi
  particles}, Sov. Phys. JETP 23 (1966) 1034--1039, [Zh. Eksp. Teor. Fiz. 36
  (1966) 1550-1558].

\bibitem{Eagles-1969}
D.~M. Eagles, Possible pairing without superconductivity at low carrier
  concentrations in bulk and thin-film superconducting semiconductors, Phys.
  Rev. 186 (1969) 456--463.

\bibitem{Leggett-1980}
A.~J. Leggett, Diatomic molecules and {C}ooper pairs, in: A.~Pekalski,
  R.~Przystawa (Eds.), {Modern trends in the theory of condensed matter}, Vol.
  115 of Lecture Notes in Physics, Springer-Verlag, Berlin, 1980, p.~13.

\bibitem{Nozieres-1985}
P.~{Nozi{\`e}res}, S.~{Schmitt-Rink}, {Bose condensation in an attractive
  fermion gas: From weak to strong coupling superconductivity}, J. Low Temp.
  Phys. 59 (1985) 195--211.

\bibitem{Randeria-1989}
M.~Randeria, J.-M. Duan, L.-Y. Shieh, {Bound states, Cooper pairing, and Bose
  condensation in two dimensions}, Phys. Rev. Lett. 62 (1989) 981--984.

\bibitem{Randeria-1990}
M.~Randeria, J.-M. Duan, L.-Y. Shieh, {Superconductivity in a two-dimensional
  Fermi gas: Evolution from Cooper pairing to Bose condensation}, Phys. Rev. B
  41 (1990) 327--343.

\bibitem{Micnas-1990}
R.~Micnas, J.~Ranninger, S.~Robaszkiewicz, Superconductivity in narrow-band
  systems with local nonretarded attractive interactions, Rev. Mod. Phys. 62
  (1990) 113--171.

\bibitem{Randeria-1992}
M.~Randeria, N.~Trivedi, A.~Moreo, R.~T. Scalettar, Pairing and spin gap in the
  normal state of short coherence length superconductors, Phys. Rev. Lett. 69
  (1992) 2001--2004.

\bibitem{Drechsler-1992}
M.~Drechsler, W.~Zwerger, {Crossover from BCS-superconductivity to
  Bose-condensation}, Annalen der Physik 504 (1992) 15--23.

\bibitem{Haussmann-1993}
R.~Haussmann, {Crossover from BCS superconductivity to Bose-Einstein
  condensation: A self-consistent theory}, Z. Phys. B 91 (1993) 291--308.

\bibitem{Pistolesi-1994}
F.~Pistolesi, G.~C. Strinati, {Evolution from BCS superconductivity to Bose
  condensation: Role of the parameter
  ${\mathit{k}}_{\mathit{F}}$\ensuremath{\xi}}, Phys. Rev. B 49 (1994)
  6356--6359.

\bibitem{Casas-1994}
M.~Casas, J.~M. Getino, M.~de~Llano, A.~Puente, R.~M. Quick, H.~Rubio, D.~M.
  van~der Walt, {BCS-Bose model of exotic superconductors: Generalized
  coherence length}, Phys. Rev. B 50 (1994) 15945--15952.

\bibitem{Baldo-1995}
M.~Baldo, U.~Lombardo, P.~Schuck, {Deuteron formation in expanding nuclear
  matter from a strong coupling BCS approach}, Phys. Rev. C 52 (1995) 975--985.

\bibitem{Inguscio-2007}
M.~Inguscio, W.~Ketterle, C.~Salomon (Eds.), {Ultra-Cold Fermi Gases}, Vol.
  CLXIV of Proceedings of the International School of Physics ``Enrico Fermi'',
  IOS Press, Amsterdam, 2007.

\bibitem{Zwerger-2012}
W.~Zwerger (Ed.), {The BCS-BEC Crossover and the Unitary Fermi Gas}, Vol. 863
  of Lecture Notes in Physics, Springer-Verlag, Berlin, 2012.

\bibitem{Baker-2000}
G.~A. Baker, {The MBX Challenge Competition: a neutron matter model}, in: R.~F.
  Bishop, K.~A. Gernoth, N.~R. Walet, Y.~Xian (Eds.), {The Proceedings of the
  10th International Conference Recent Progress in Many-body Theories}, Vol.~3
  of Series on Advances in Quantum Many-Body theory, World Scientific,
  Singapore, 2000, pp. 15--20.

\bibitem{Baker-1999}
G.~A. Baker, Neutron matter model, Phys. Rev. C 60 (1999) 054311.

\bibitem{Alm-1993}
T.~Alm, B.~L. Friman, G.~R{\"o}pke, H.~Schulz, Pairing instability in hot
  asymmetric nuclear matter, Nucl. Phys. A 551 (1993) 45--53.

\bibitem{Stein-1995}
H.~Stein, A.~Schnell, T.~Alm, G.~R{\"o}pke, {Correlations and pairing in
  nuclear matter within the Nozi{\`e}res-Schmitt-Rink approach}, Z. Phys. A 351
  (1995) 295--299.

\bibitem{Lombardo-2001b}
U.~Lombardo, P.~Nozi\`eres, P.~Schuck, H.-J. Schulze, A.~Sedrakian, {Transition
  from BCS pairing to Bose-Einstein condensation in low-density asymmetric
  nuclear matter}, Phys. Rev. C 64 (2001) 064314.

\bibitem{Andrenacci-1999}
N.~Andrenacci, A.~Perali, P.~Pieri, G.~C. Strinati, {Density-induced BCS to
  Bose-Einstein crossover}, Phys. Rev. B 60 (1999) 12410--12418.

\bibitem{Heckel-2009}
S.~Heckel, P.~P. Schneider, A.~Sedrakian, {Light nuclei in supernova envelopes:
  A quasiparticle gas model}, Phys. Rev. C 80 (2009) 015805.

\bibitem{Jin-2010}
M.~{Jin}, M.~{Urban}, P.~{Schuck}, {BEC-BCS crossover and the liquid-gas phase
  transition in hot and dense nuclear matter}, Phys. Rev. C 82 (2010) 024911.

\bibitem{Roepke-1998}
G.~R\"opke, A.~Schnell, P.~Schuck, P.~Nozi\`eres, Four-particle condensate in
  strongly coupled fermion systems, Phys. Rev. Lett. 80 (1998) 3177--3180.

\bibitem{Capponi-2007}
S.~Capponi, G.~Roux, P.~Azaria, E.~Boulat, P.~Lecheminant, Confinement versus
  deconfinement of {C}ooper pairs in one-dimensional spin-$3/2$ fermionic cold
  atoms, Phys. Rev. B 75 (2007) 100503(R).

\bibitem{Yamada-2012}
T.~{Yamada}, Y.~{Funaki}, H.~{Horiuchi}, G.~{R{\"o}pke}, P.~{Schuck},
  A.~{Tohsaki}, {Nuclear Alpha-Particle Condensates}, in: C.~{Beck} (Ed.),
  {Clusters in Nuclei, Vol. 2}, Vol. 848 of Lecture Notes in Physics,
  Springer-Verlag, Berlin, 2012, p. 229.

\bibitem{Schnell-1999}
A.~Schnell, G.~R\"opke, P.~Schuck, Precritical pair fluctuations and formation
  of a pseudogap in low-density nuclear matter, Phys. Rev. Lett. 83 (1999)
  1926--1929.

\bibitem{Stewart-2008}
J.~T. Stewart, J.~P. Gaebler, D.~S. Jin, {Using photoemission spectroscopy to
  probe a strongly interacting Fermi gas}, Nature 454 (2008) 744--747.

\bibitem{Gaebler-2010}
J.~P. Gaebler, J.~T. Stewart, T.~E. Drake, D.~S. Jin, A.~Perali, P.~Pieri,
  G.~C. Strinati, {Observation of pseudogap behaviour in a strongly interacting
  Fermi gas}, Nat. Phys. 6 (2010) 569--573.

\bibitem{Feld-2011}
M.~Feld, B.~Fr\"{o}hlich, E.~Vogt, M.~Koschorreck, M.~K\"{o}hl, {Observation of
  a pairing pseudogap in a two-dimensional Fermi gas}, Nature 480 (2011)
  75--78.

\bibitem{Matsuo-2006}
M.~Matsuo, {Spatial structure of neutron Cooper pair in low density uniform
  matter}, Phys. Rev. C 73 (2006) 044309.

\bibitem{Abe-2009}
T.~Abe, R.~Seki, {Lattice calculation of thermal properties of low-density
  neutron matter with pionless $\mathit{NN}$ effective field theory}, Phys.
  Rev. C 79 (2009) 054002.

\bibitem{Gezerlis-2010}
A.~Gezerlis, J.~Carlson, Low-density neutron matter, Phys. Rev. C 81 (2010)
  025803.

\bibitem{Ramanan-2013}
S.~Ramanan, M.~Urban, {BEC-BCS crossover in neutron matter with
  renormalization-group-based effective interactions}, Phys. Rev. C 88 (2013)
  054315.

\bibitem{Bulgac-2012}
A.~Bulgac, {The unitary Fermi gas: from Monte Carlo to density functionals},
  in: W.~Zwerger (Ed.), {The BCS-BEC Crossover and the Unitary Fermi Gas}, Vol.
  863 of Lecture Notes in Physics, Springer-Verlag, Berlin, 2012, pp. 305--374.

\bibitem{Carlson-2012}
J.~Carlson, S.~Gandolfi, A.~Gezerlis, Quantum {Monte Carlo} approaches to
  nuclear and atomic physics, Prog. Theor. Exp. Phys. 2012 (2012) 01A209.

\bibitem{Diehl-2010}
S.~Diehl, S.~Floerchinger, H.~Gies, J.~Pawlowkski, C.~Wetterich, {Functional
  renormalization group approach to the BCS-BEC crossover}, Annalen der Physik
  522 (2010) 615--656.

\bibitem{Nishida-2012}
Y.~Nishida, D.~Son, {Unitary Fermi gas, $\epsilon$ expansion, and
  nonrelativistic conformal field theories}, in: W.~Zwerger (Ed.), {The BCS-BEC
  Crossover and the Unitary Fermi Gas}, Vol. 863 of Lecture Notes in Physics,
  Springer-Verlag, Berlin, 2012, pp. 305--374.

\bibitem{Liu-2013}
X.-J. Liu, {Virial expansion for a strongly correlated Fermi system and its
  application to ultracold atomic Fermi gases}, Phys. Rep. 524 (2013) 37--83.

\bibitem{Pistolesi-1996}
F.~Pistolesi, G.~C. Strinati, {Evolution from BCS superconductivity to Bose
  condensation: Calculation of the zero-temperature phase coherence length},
  Phys. Rev. B 53 (1996) 15168--15192.

\bibitem{Hu-2006}
{H. Hu}, {X.-J. Liu}, {P. D. Drummond}, {Equation of state of a superfluid
  Fermi gas in the BCS-BEC crossover}, Europhys. Lett. 74 (2006) 574--580.

\bibitem{Diener-2008}
R.~B. Diener, R.~Sensarma, M.~Randeria, Quantum fluctuations in the superfluid
  state of the {BCS-BEC} crossover, Phys. Rev. A 77 (2008) 023626.

\bibitem{Randeria-2014}
M.~Randeria, E.~Taylor, {Crossover from Bardeen-Cooper-Schrieffer to
  Bose-Einstein Condensation and the Unitary Fermi Gas}, Ann. Rev. Cond. Matt.
  Phys. 5 (2014) 209--232.

\bibitem{Pitaevskii-2016}
L.~Pitaevskii, S.~Stringari, {Bose-Einstein Condensation and Superfluidity},
  Oxford University Press, Oxford, 2016.

\bibitem{Zwerger-2016}
W.~Zwerger, {Strongly Interacting Fermi Gases}, in: M.~Inguscio, W.~Ketterle,
  S.~Stringari, G.~Roati (Eds.), {Quantum Matter at Ultralow Temperatures},
  Vol. 191 of Proceedings of the International School of Physics ``Enrico
  Fermi'', IOS Press, Amsterdam, 2016, pp. 63--142.

\bibitem{Chin-2010}
C.~Chin, R.~Grimm, P.~Julienne, E.~Tiesinga, Feshbach resonances in ultracold
  gases, Rev. Mod. Phys. 82 (2010) 1225--1286.

\bibitem{Gubbels-2013}
K.~Gubbels, H.~Stoof, Imbalanced {Fermi} gases at unitarity, Phys. Rep. 525
  (2013) 255--313.

\bibitem{Giorgini-2008}
S.~Giorgini, L.~P. Pitaevskii, S.~Stringari, Theory of ultracold atomic {F}ermi
  gases, Rev. Mod. Phys. 80 (2008) 1215--1274.

\bibitem{Radzihovsky-2010}
L.~Radzihovsky, D.~E. Sheehy, {Imbalanced Feshbach-resonant Fermi gases}, Rep.
  Prog. Phys. 73 (2010) 076501.

\bibitem{Chevy-2010}
F.~Chevy, C.~Mora, {Ultra-cold polarized Fermi gases}, Rep. Prog. Phys. 73
  (2010) 112401.

\bibitem{Chen-2005}
Q.~Chen, J.~Stajic, S.~Tan, K.~Levin, {BCS-BEC crossover: From high temperature
  superconductors to ultracold superfluids}, Phys. Rep. 412 (2005) 1--88.

\bibitem{Brink-2005}
D.~Brink, R.~Broglia, Nuclear Superfluidity: Pairing in Finite Systems, Vol.~24
  of Cambridge Monographs on Particle Physics, Nuclear Physics and Cosmology,
  Cambridge University Press, Cambridge, 2005.

\bibitem{Dean-2003}
D.~J. Dean, M.~Hjorth-Jensen, Pairing in nuclear systems: from neutron stars to
  finite nuclei, Rev. Mod. Phys. 75 (2003) 607--656.

\bibitem{Gezerlis-2014}
A.~Gezerlis, C.~J. Pethick, A.~Schwenk, Pairing and superfluidity of nucleons
  in neutron stars, in: K.-H. Bennemann, J.~B. Ketterson (Eds.), Novel
  Superfluids: Vol. 2, Vol. 157 of International Series of Monographs on
  Physics, Oxford University Press, 2014, p. 580.

\bibitem{Schrieffer-1964}
{J. R. Schrieffer}, {Theory of Superconductivity}, Benjamin, New York, 1964.

\bibitem{Ring-1980}
P.~Ring, P.~Schuck, {The Nuclear Many Body Problem}, Springer-Verlag, Berlin,
  1980.

\bibitem{Spuntarelli-2010}
A.~Spuntarelli, P.~Pieri, G.~C. Strinati, {Solution of the Bogoliubov-de Gennes
  equations at zero temperature throughout the BCS-BEC crossover: Josephson and
  related effects}, Phys. Rep. 488 (2010) 111--167.

\bibitem{Mook-1973}
H.~A. Mook, J.~W. Lynn, R.~M. Nicklow, {Temperature dependence of the magnetic
  excitations in Nickel}, Phys. Rev. Lett. 30 (1973) 556--559.

\bibitem{Marsiglio-1990}
F.~Marsiglio, J.~E. Hirsch, {Hole superconductivity and the
  high-${\mathit{T}}_{\mathit{c}}$ oxides}, Phys. Rev. B 41 (1990) 6435--6456.

\bibitem{Palestini-2014}
F.~Palestini, G.~C. Strinati, {Temperature dependence of the pair coherence and
  healing lengths for a fermionic superfluid throughout the BCS-BEC crossover},
  Phys. Rev. B 89 (2014) 224508.

\bibitem{Cooper-1956}
L.~N. Cooper, {Bound electron pairs in a degenerate Fermi gas}, Phys. Rev. 104
  (1956) 1189--1190.

\bibitem{Garrido-2001}
E.~Garrido, P.~Sarriguren, E.~Moya~de Guerra, U.~Lombardo, P.~Schuck, H.~J.
  Schulze, Nuclear pairing in the $t=0$ channel reexamined, Phys. Rev. C 63
  (2001) 037304.

\bibitem{Bulgac-2002}
A.~Bulgac, Y.~Yu, {Renormalization of the Hartree-Fock-Bogoliubov equations in
  the case of a zero range pairing interaction}, Phys. Rev. Lett. 88 (2002)
  042504.

\bibitem{Yu-2003}
Y.~Yu, A.~Bulgac, Energy density functional approach to superfluid nuclei,
  Phys. Rev. Lett. 90 (2003) 222501.

\bibitem{Sa-de-Melo-1993}
C.~A.~R. S\'a~de Melo, M.~Randeria, J.~R. Engelbrecht, {Crossover from BCS to
  Bose superconductivity: Transition temperature and time-dependent
  Ginzburg-Landau theory}, Phys. Rev. Lett. 71 (1993) 3202--3205.

\bibitem{Regal-2003a}
C.~A. Regal, D.~S. Jin, {Measurement of positive and negative scattering
  lengths in a Fermi gas of atoms}, Phys. Rev. Lett. 90 (2003) 230404.

\bibitem{Carter-1995}
R.~M. Carter, M.~Casas, J.~M. Getino, M.~de~Llano, A.~Puente, H.~Rubio, D.~M.
  van~der Walt, {Coherence lengths for three-dimensional superconductors in the
  BCS-Bose picture}, Phys. Rev. B 52 (1995) 16149--16154.

\bibitem{Marini-1998}
M.~Marini, F.~Pistolesi, G.~C. Strinati, {Evolution from BCS superconductivity
  to Bose condensation: analytic results for the crossover in three
  dimensions}, Eur. Phys. J. B 1 (1998) 151--159.

\bibitem{Papenbrock-1999}
T.~Papenbrock, G.~F. Bertsch, {Pairing in low-density Fermi gases}, Phys. Rev.
  C 59 (1999) 2052--2055.

\bibitem{Salasnich-2005}
L.~Salasnich, N.~Manini, A.~Parola, {Condensate fraction of a Fermi gas in the
  BCS-BEC crossover}, Phys. Rev. A 72 (2005) 023621.

\bibitem{Strinati-2000}
G.~C. Strinati, {A survey on the crossover from BCS superconductivity to
  Bose-Einstein condensation}, Phys. Essays 13 (2000) 427--436.

\bibitem{de-Gennes-1966}
P.~G. {De Gennes}, {Superconductivity of Metals and Alloys}, Benjamin, New
  York, 1966, Ch.~5.

\bibitem{Bruun-1999}
G.~{Bruun}, Y.~{Castin}, R.~{Dum}, K.~{Burnett}, {BCS theory for trapped
  ultracold fermions}, Eur. Phys. J. D 7 (1999) 433--439.

\bibitem{Grasso-2003}
M.~Grasso, M.~Urban, {Hartree-Fock-Bogoliubov theory versus local-density
  approximation for superfluid trapped fermionic atoms}, Phys. Rev. A 68 (2003)
  033610.

\bibitem{Simonucci-2013}
S.~Simonucci, P.~Pieri, G.~C. Strinati, {Temperature dependence of a vortex in
  a superfluid Fermi gas}, Phys. Rev. B 87 (2013) 214507.

\bibitem{Pieri-2003}
P.~Pieri, G.~C. Strinati, Derivation of the {Gross-Pitaevskii} equation for
  condensed bosons from the {Bogoliubov-de Gennes} equations for superfluid
  fermions, Phys. Rev. Lett. 91 (2003) 030401.

\bibitem{Ohashi-2005}
Y.~Ohashi, A.~Griffin, {Single-particle excitations in a trapped gas of Fermi
  atoms in the BCS-BEC crossover region}, Phys. Rev. A 72 (2005) 013601.

\bibitem{Castorina-2005}
P.~Castorina, M.~Grasso, M.~Oertel, M.~Urban, D.~Zappal{\`a}, {Nonstandard
  pairing in asymmetric trapped Fermi gases}, Phys. Rev. A 72 (2005) 025601.

\bibitem{Jensen-2007}
L.~M. Jensen, J.~Kinnunen, P.~T\"orm\"a, {Non-BCS superfluidity in trapped
  ultracold Fermi gases}, Phys. Rev. A 76 (2007) 033620.

\bibitem{Nygaard-2003}
N.~Nygaard, G.~M. Bruun, C.~W. Clark, D.~L. Feder, {Microscopic structure of a
  vortex line in a dilute superfluid Fermi gas}, Phys. Rev. Lett. 90 (2003)
  210402.

\bibitem{Sensarma-2006}
R.~Sensarma, M.~Randeria, T.-L. Ho, {Vortices in superfluid Fermi gases through
  the BEC to BCS crossover}, Phys. Rev. Lett. 96 (2006) 090403.

\bibitem{Bulgac-2003}
A.~Bulgac, Y.~Yu, Vortex state in a strongly coupled dilute atomic fermionic
  superfluid, Phys. Rev. Lett. 91 (2003) 190404.

\bibitem{Bulgac-2002b}
A.~Bulgac, Local density approximation for systems with pairing correlations,
  Phys. Rev. C 65 (2002) 051305.

\bibitem{Bulgac-2007}
A.~Bulgac, {Local-density-functional theory for superfluid fermionic systems:
  The unitary gas}, Phys. Rev. A 76 (2007) 040502.

\bibitem{Bulgac-2008b}
A.~Bulgac, M.~McNeil~Forbes, Unitary {Fermi} supersolid: The
  {Larkin-Ovchinnikov} phase, Phys. Rev. Lett. 101 (2008) 215301.

\bibitem{Feder-2004}
D.~L. Feder, {Vortex arrays in a rotating superfluid Fermi gas}, Phys. Rev.
  Lett. 93 (2004) 200406.

\bibitem{Tonini-2006}
G.~{Tonini}, F.~{Werner}, Y.~{Castin}, {Formation of a vortex lattice in a
  rotating BCS Fermi gas}, Eur. Phys. J. D 39 (2006) 283--294.

\bibitem{Decharge-1980}
J.~Decharg\'e, D.~Gogny, {Hartree-Fock-Bogolyubov calculations with the D1
  effective interaction on spherical nuclei}, Phys. Rev. C 21 (1980)
  1568--1593.

\bibitem{Bender-2003}
M.~Bender, P.-H. Heenen, P.-G. Reinhard, Self-consistent mean-field models for
  nuclear structure, Rev. Mod. Phys. 75 (2003) 121--180.

\bibitem{Vretenar-2005}
D.~Vretenar, A.~V. Afanasjev, G.~A. Lalazissis, P.~Ring, {Relativistic
  Hartree-Bogoliubov theory: static and dynamic aspects of exotic nuclear
  structure}, Phys. Rep. 409 (2005) 101--259.

\bibitem{Erler-2011}
J.~Erler, P.~Kl\"upfel, P.-G. Reinhard, {Self-consistent nuclear mean-field
  models: example Skyrme-Hartree-Fock}, J. Phys. G 38 (2011) 033101.

\bibitem{Fetter-1971}
A.~L. Fetter, J.~D. Walecka, {Quantum Theory of Many-Particle Systems},
  McGraw-Hill, New York, 1971.

\bibitem{Gorkov-1959}
L.~P. Gor'kov, {Microscopic derivation of the Ginzburg-Landau equations in the
  theory of superconductivity}, Sov. Phys. JETP 9 (1959) 1364--1367, [Zh. Eksp.
  Teor. Fiz. 36 (1959) 1918-1923].

\bibitem{Baranov-1998}
M.~A. Baranov, D.~S. Petrov, {Critical temperature and Ginzburg-Landau equation
  for a trapped Fermi gas}, Phys. Rev. A 58 (1998) R801--R804.

\bibitem{Simonucci-2014}
S.~Simonucci, G.~C. Strinati, Equation for the superfluid gap obtained by
  coarse graining the {B}ogoliubov$-$de {G}ennes equations throughout the
  {BCS-BEC} crossover, Phys. Rev. B 89 (2014) 054511.

\bibitem{Clogston-1962}
A.~M. Clogston, Upper limit for the critical field in hard superconductors,
  Phys. Rev. Lett. 9 (1962) 266--267.

\bibitem{Chandrasekhar-1962}
B.~S. Chandrasekhar, A note on the maximum critical field of high-field
  superconductors, Appl. Phys. Lett. 1 (1962) 7--8.

\bibitem{Sarma-1963}
G.~Sarma, On the influence of a uniform exchange field acting on the spins of
  the conduction electrons in a superconductor, J. Phys. Chem. Solids 24 (1963)
  1029--1032.

\bibitem{Fulde-1964}
P.~Fulde, R.~A. Ferrell, Superconductivity in a strong spin-exchange field,
  Phys. Rev. 135 (1964) A550--A563.

\bibitem{Larkin-1964}
A.~I. Larkin, Y.~N. Ovchinnikov, Nonuniform state of superconductors, Sov.
  Phys. JETP 20 (1965) 762--769, [Zh. Eksp. Teor. Fiz. 47 (1964) 1136-1146].

\bibitem{Matsuda-2007}
Y.~Matsuda, H.~Shimahara, {Fulde-Ferrell-Larkin-Ovchinnikov State in heavy
  fermion superconductors}, J. Phys. Soc. Jpn. 76 (2007) 051005.

\bibitem{Sedrakian-1997}
A.~Sedrakian, T.~Alm, U.~Lombardo, Superfluidity in asymmetric nuclear matter,
  Phys. Rev. C 55 (1997) R582--R584.

\bibitem{Casalbuoni-2004}
R.~Casalbuoni, G.~Nardulli, {Inhomogeneous superconductivity in condensed
  matter and QCD}, Rev. Mod. Phys. 76 (2004) 263--320.

\bibitem{Alford-2008}
M.~G. Alford, A.~Schmitt, K.~Rajagopal, T.~Sch\"afer, Color superconductivity
  in dense quark matter, Rev. Mod. Phys. 80 (2008) 1455--1515.

\bibitem{Zwierlein-2006a}
M.~W. Zwierlein, A.~Schirotzek, C.~H. Schunck, W.~Ketterle, Fermionic
  superfluidity with imbalanced spin populations, Science 311 (2006) 492--496.

\bibitem{Partridge-2006}
G.~B. Partridge, W.~Li, R.~I. Kamar, Y.-a. Liao, R.~G. Hulet, Pairing and phase
  separation in a polarized {F}ermi gas, Science 311 (2006) 503--505.

\bibitem{Liu-2003}
W.~V. Liu, F.~Wilczek, Interior gap superfluidity, Phys. Rev. Lett. 90 (2003)
  047002.

\bibitem{Gubankova-2003}
E.~Gubankova, W.~V. Liu, F.~Wilczek, Breached pairing superfluidity: Possible
  realization in {QCD}, Phys. Rev. Lett. 91 (2003) 032001.

\bibitem{Forbes-2005}
M.~{McNeil Forbes}, E.~Gubankova, W.~V. Liu, F.~Wilczek, Stability criteria for
  breached-pair superfluidity, Phys. Rev. Lett. 94 (2005) 017001.

\bibitem{Takada-1969}
S.~Takada, T.~Izuyama, Superconductivity in a molecular field. {I}, Progr.
  Theor. Phys. 41 (1969) 635--663.

\bibitem{Eilenberger-1968}
G.~Eilenberger, {Transformation of Gorkov's equation for type II
  superconductors into transport-like equations}, Z. Phys. 214 (1968) 195--213.

\bibitem{Larkin-1968}
A.~I. Larkin, Y.~N. Ovchinnikov, Quasiclassical method in the theory of
  superconductivity, Sov. Phys. JETP 55 (1968) 1200--1205, [Zh. Eksp. Teor.
  Fiz. 55 (1968) 2262-2272].

\bibitem{Matsuo-1998}
S.~Matsuo, S.~Higashitani, Y.~Nagato, K.~Nagai, Phase diagram of the
  {Fulde-Ferrell-Larkin-Ovchinnikov} state in a three-dimensional
  superconductor, J. Phys. Soc. Jpn. 67 (1998) 280--289.

\bibitem{Mora-2005}
C.~Mora, R.~Combescot, {Transition to Fulde-Ferrell-Larkin-Ovchinnikov phases
  in three dimensions: A quasiclassical investigation at low temperature with
  Fourier expansion}, Phys. Rev. B 71 (2005) 214504.

\bibitem{Buzdin-1997}
A.~I. Buzdin, H.~Kachkachi, {Generalized Ginzburg-Landau theory for nonuniform
  FFLO superconductors}, Phys. Lett. A 225 (1997) 341--348.

\bibitem{Houzet-1999}
M.~Houzet, Y.~Meurdesoif, O.~Coste, A.~Buzdin, {Structure of the non-uniform
  Fulde-Ferrell-Larkin-Ovchinnikov state in 3D superconductors}, Physica C 316
  (1999) 89--96.

\bibitem{Combescot-2002}
R.~Combescot, C.~Mora, {Transition to Fulde-Ferrel-Larkin-Ovchinnikov phases
  near the tricritical point: an analytical study}, Eur. Phys. J. B 28 (2002)
  397--406.

\bibitem{Bowers-2002}
J.~A. Bowers, K.~Rajagopal, Crystallography of color superconductivity, Phys.
  Rev. D 66 (2002) 065002.

\bibitem{Machida-1984}
K.~Machida, H.~Nakanishi, Superconductivity under a ferromagnetic molecular
  field, Phys. Rev. B 30 (1984) 122--133.

\bibitem{Buzdin-1983}
A.~I. Buzdin, V.~V. Tugushev, Phase diagrams of electronic and superconducting
  transitions to soliton lattice states, Sov. Phys. JETP 58 (1983) 428--433,
  [Zh. Eksp. Teor. Fiz. 85 (1983) 735--745].

\bibitem{Buzdin-1987}
A.~I. Buzdin, S.~V. Polonskii, {Nonuniform state in quasi-1D superconductor},
  Sov. Phys. JETP 66 (1987) 422--429, [Zh. Eksp. Teor. Fiz. 93 (1987)
  747--761].

\bibitem{Bedaque-2003}
P.~F. Bedaque, H.~Caldas, G.~Rupak, Phase separation in asymmetrical fermion
  superfluids, Phys. Rev. Lett. 91 (2003) 247002.

\bibitem{Wu-2003}
S.-T. Wu, S.~Yip, Superfluidity in the interior-gap states, Phys. Rev. A 67
  (2003) 053603.

\bibitem{Son-2006}
D.~T. Son, M.~A. Stephanov, {Phase diagram of a cold polarized Fermi gas},
  Phys. Rev. A 74 (2006) 013614.

\bibitem{Iskin-2006}
M.~Iskin, C.~A.~R. S\'a~de Melo, Two-species fermion mixtures with population
  imbalance, Phys. Rev. Lett. 97 (2006) 100404.

\bibitem{Gubbels-2006}
K.~B. Gubbels, M.~W.~J. Romans, H.~T.~C. Stoof, Sarma phase in trapped
  unbalanced {Fermi} gases, Phys. Rev. Lett. 97 (2006) 210402.

\bibitem{Chen-2006}
Q.~Chen, Y.~He, C.-C. Chien, K.~Levin, Stability conditions and phase diagrams
  for two-component {Fermi} gases with population imbalance, Phys. Rev. A 74
  (2006) 063603.

\bibitem{Pao-2006}
C.-H. Pao, S.-T. Wu, S.-K. Yip, Superfluid stability in the {BEC-BCS}
  crossover, Phys. Rev. B 73 (2006) 132506, and Erratum {\em ibid.} 74 (2006)
  189901.

\bibitem{Pilati-2008}
S.~Pilati, S.~Giorgini, Phase separation in a polarized {Fermi} gas at zero
  temperature, Phys. Rev. Lett. 100 (2008) 030401.

\bibitem{Sheehy-2007}
D.~E. Sheehy, L.~Radzihovsky, {BEC-BCS} crossover, phase transitions and phase
  separation in polarized resonantly-paired superfluids, Ann. Phys. 322 (2007)
  1790--1924.

\bibitem{Pieri-2006}
P.~Pieri, G.~C. Strinati, Trapped fermions with density imbalance in the
  {Bose-Einstein} condensate limit, Phys. Rev. Lett. 96 (2006) 150404.

\bibitem{Yoshida-2007}
N.~Yoshida, S.-K. Yip, {Larkin-Ovchinnikov state in resonant Fermi gas}, Phys.
  Rev. A 75 (2007) 063601.

\bibitem{Pao-2009}
C.-H. Pao, S.-K. Yip, {Phase diagram of asymmetric Fermi gas across Feshbach
  resonance}, J. Phys. Conf. Ser. 150 (2009) 032078.

\bibitem{Shin-2008b}
Y.-I. {Shin}, C.~H. {Schunck}, A.~{Schirotzek}, W.~{Ketterle}, {Phase diagram
  of a two-component Fermi gas with resonant interactions}, Nature 451 (2008)
  689--693.

\bibitem{Kashimura-2012}
T.~{Kashimura}, R.~{Watanabe}, Y.~{Ohashi}, {Spin susceptibility and
  fluctuation corrections in the BCS-BEC crossover regime of an ultracold Fermi
  gas}, Phys. Rev. A 86 (2012) 043622.

\bibitem{Tartari-2011}
A.~Tartari, {Phase diagram of a two-component Fermi gas with density imbalance
  throughout the BCS-BEC crossover}, Ph.D. thesis, University of Camerino
  (2011).

\bibitem{Parish-2007}
M.~M. {Parish}, F.~M. {Marchetti}, A.~{Lamacraft}, B.~D. {Simons},
  {Finite-temperature phase diagram of a polarized Fermi condensate}, Nat.
  Phys. 3 (2007) 124--128.

\bibitem{Galitskii-1958}
V.~M. Galitskii, {The energy spectrum of a non-ideal Fermi gas}, Sov. Phys.
  JETP 7 (1958) 104--112, [Zh. Eksp. Teor. Fiz. 34 (1958) 151-162].

\bibitem{Vagov-2007}
A.~Vagov, H.~Schomerus, A.~Shanenko, {Generalized Galitskii approach for the
  vertex function of a Fermi gas with resonant interaction}, Phys. Rev. B 76
  (2007) 214513.

\bibitem{Enss-2012}
T.~Enss, {Quantum critical transport in the unitary Fermi gas}, Phys. Rev. A 86
  (2012) 013616.

\bibitem{Serene-1989}
J.~W. Serene, {Stability of two-dimensional Fermi liquids against pair
  fluctuations with large total momentum}, Phys. Rev. B 40 (1989) 10873--10877.

\bibitem{Mahan-2000}
G.~D. Mahan, {Many-Particle Physics}, 3rd Edition, Kluwer Academic/ Plenum
  Publishers, New York, 2000.

\bibitem{Urban-2014}
M.~Urban, P.~Schuck, {Occupation numbers in strongly polarized Fermi gases and
  the Luttinger theorem}, Phys. Rev. A 90 (2014) 023632.

\bibitem{Schuck-2003}
P.~Schuck, H.-J. Schulze, N.~Van~Giai, M.~Zverev, Two-dimensional electron gas
  in an improved random-phase approximation, Phys. Rev. B 67 (2003) 233404.

\bibitem{Haussmann-1994}
R.~Haussmann, {Properties of a Fermi liquid at the superfluid transition in the
  crossover region between BCS superconductivity and Bose-Einstein
  condensation}, Phys. Rev. B 49 (1994) 12975--12983.

\bibitem{Pieri-2000}
P.~Pieri, G.~C. Strinati, {Strong-coupling limit in the evolution from BCS
  superconductivity to Bose-Einstein condensation}, Phys. Rev. B 61 (2000)
  15370--15381.

\bibitem{Perali-2002}
A.~Perali, P.~Pieri, G.~C. Strinati, C.~Castellani, {Pseudogap and spectral
  function from superconducting fluctuations to the bosonic limit}, Phys. Rev.
  B 66 (2002) 024510.

\bibitem{Pieri-2004b}
P.~Pieri, L.~Pisani, G.~C. Strinati, {BCS-BEC crossover at finite temperature
  in the broken-symmetry phase}, Phys. Rev. B 70 (2004) 094508.

\bibitem{Pieri-2005b}
P.~Pieri, L.~Pisani, G.~C. Strinati, {Comparison between a diagrammatic theory
  for the BCS-BEC crossover and quantum Monte Carlo results}, Phys. Rev. B 72
  (2005) 012506.

\bibitem{Pieri-2004a}
P.~Pieri, L.~Pisani, G.~C. Strinati, {Pairing fluctuation effects on the
  single-particle spectra for the superconducting state}, Phys. Rev. Lett. 92
  (2004) 110401.

\bibitem{Andrenacci-2003}
N.~Andrenacci, P.~Pieri, G.~C. Strinati, {Evolution from BCS superconductivity
  to Bose-Einstein condensation: Current correlation function in the
  broken-symmetry phase}, Phys. Rev. B 68 (2003) 144507.

\bibitem{Prokofev-2008}
N.~Prokof'ev, B.~Svistunov, {Bold diagrammatic Monte Carlo: A generic
  sign-problem tolerant technique for polaron models and possibly interacting
  many-body problems}, Phys. Rev. B 77 (2008) 125101.

\bibitem{Janko-1997}
B.~Jank\'o, J.~Maly, K.~Levin, Pseudogap effects induced by resonant pair
  scattering, Phys. Rev. B 56 (1997) R11407--R11410.

\bibitem{Kosztin-1998}
I.~Kosztin, Q.~Chen, B.~Jank\'o, K.~Levin, {Relationship between the pseudo-
  and superconducting gaps: Effects of residual pairing correlations below
  ${T}_{c}$}, Phys. Rev. B 58 (1998) R5936--R5939.

\bibitem{Chen-1998}
Q.~Chen, I.~Kosztin, B.~Jank\'o, K.~Levin, Pairing fluctuation theory of
  superconducting properties in underdoped to overdoped cuprates, Phys. Rev.
  Lett. 81 (1998) 4708--4711.

\bibitem{Haussmann-2007}
R.~Haussmann, W.~Rantner, S.~Cerrito, W.~Zwerger, Thermodynamics of the
  {BCS-BEC} crossover, Phys. Rev. A 75 (2007) 023610.

\bibitem{Zimmermann-1985}
R.~{Zimmermann}, H.~{Stolz}, The mass action law in two-component {F}ermi
  systems revisited: Excitons and electron-hole pairs, Phys. Status Solidi B
  131 (1985) 151--164.

\bibitem{Hu-2008}
H.~Hu, X.-J. Liu, P.~D. Drummond, {Comparative study of strong-coupling
  theories of a trapped Fermi gas at unitarity}, Phys. Rev. A 77 (2008) 061605.

\bibitem{Chien-2010}
C.-C. Chien, H.~Guo, Y.~He, K.~Levin, {Comparative study of BCS-BEC crossover
  theories above ${T}_{c}$: The nature of the pseudogap in ultracold atomic
  Fermi gases}, Phys. Rev. A 81 (2010) 023622.

\bibitem{Combescot-2008}
R.~Combescot, S.~Giraud, Normal state of highly polarized fermi gases: Full
  many-body treatment, Phys. Rev. Lett. 101 (2008) 050404.

\bibitem{Engelbrecht-1997}
J.~R. Engelbrecht, M.~Randeria, C.~A.~R. S\'a~de Melo, {BCS to Bose crossover:
  Broken-symmetry state}, Phys. Rev. B 55 (1997) 15153--15156.

\bibitem{Mott-1969}
N.~F. Mott, Conduction in non-crystalline materials, Philos. Mag. 19 (1969)
  835--852.

\bibitem{Palestini-2012}
F.~Palestini, A.~Perali, P.~Pieri, G.~C. Strinati, {Dispersions, weights, and
  widths of the single-particle spectral function in the normal phase of a
  Fermi gas}, Phys. Rev. B 85 (2012) 024517.

\bibitem{Palestini-2010}
F.~Palestini, A.~Perali, P.~Pieri, G.~C. Strinati, Temperature and coupling
  dependence of the universal contact intensity for an ultracold {F}ermi gas,
  Phys. Rev. A 82 (2010) 021605(R).

\bibitem{Tsuchiya-2011}
S.~Tsuchiya, R.~Watanabe, Y.~Ohashi, {Pseudogap temperature and effects of a
  harmonic trap in the BCS-BEC crossover regime of an ultracold Fermi gas},
  Phys. Rev. A 84 (2011) 043647.

\bibitem{Kashimura-2014}
T.~Kashimura, R.~Watanabe, Y.~Ohashi, {Pseudogap phenomenon and effects of
  population imbalance in the normal state of a unitary Fermi gas}, Phys. Rev.
  A 89 (2014) 013618.

\bibitem{Marsiglio-2015}
F.~Marsiglio, P.~Pieri, A.~Perali, F.~Palestini, G.~C. Strinati, {Pairing
  effects in the normal phase of a two-dimensional Fermi gas}, Phys. Rev. B 91
  (2015) 054509.

\bibitem{Perali-2011}
A.~Perali, F.~Palestini, P.~Pieri, G.~C. Strinati, J.~T. Stewart, J.~P.
  Gaebler, T.~E. Drake, D.~S. Jin, Evolution of the normal state of a strongly
  interacting {Fermi} gas from a pseudogap phase to a molecular {Bose} gas,
  Phys. Rev. Lett. 106 (2011) 060402.

\bibitem{Gorkov-1961}
L.~P. Gor'kov, T.~K. Melik-Barkhudarov, Contribution to the theory of
  superfluidity in an imperfect {Fermi} gas, Sov. Phys. JETP 13 (1961)
  1018--1022, [Zh. Eksp. Teor. Fiz. 40 (1961) 1452-1458].

\bibitem{Thouless-1960}
D.~J. Thouless, Perturbation theory in statistical mechanics and the theory of
  superconductivity, Ann. Phys. 10 (1960) 553--588.

\bibitem{Hugenholtz-1959}
N.~M. Hugenholtz, D.~Pines, Ground-state energy and excitation spectrum of a
  system of interacting bosons, Phys. Rev. 116 (1959) 489--506.

\bibitem{Bulgac-2006}
A.~Bulgac, J.~E. Drut, P.~Magierski, Spin $1/2$ fermions in the unitary regime:
  A superfluid of a new type, Phys. Rev. Lett. 96 (2006) 090404.

\bibitem{Yu-2009a}
Z.-Q. Yu, K.~Huang, L.~Yin, {Induced interaction in a Fermi gas with a BEC-BCS
  crossover}, Phys. Rev. A 79 (2009) 053636.

\bibitem{Ruan-2013}
X.-X. Ruan, H.~Gong, L.~Du, W.-M. Sun, H.-S. Zong, {Effect of the induced
  interaction on the superfluid-transition temperature of ultracold Fermi gases
  within the $T$-matrix approximation}, Phys. Rev. A 87 (2013) 043608.

\bibitem{Floerchinger-2008}
S.~Floerchinger, M.~Scherer, S.~Diehl, C.~Wetterich, {Particle-hole
  fluctuations in BCS-BEC crossover}, Phys. Rev. B 78 (2008) 174528.

\bibitem{Floerchinger-2010}
S.~Floerchinger, M.~M. Scherer, C.~Wetterich, {Modified Fermi sphere, pairing
  gap, and critical temperature for the BCS-BEC crossover}, Phys. Rev. A 81
  (2010) 063619.

\bibitem{Tanizaki-2014}
Y.~Tanizaki, G.~Fej\H{o}s, T.~Hatsuda, Fermionic functional renormalization
  group approach to superfluid phase transition, Prog. Theor. Exp. Phys. 2014
  (2014) 043I01.

\bibitem{Pisani-2018}
L.~Pisani, A.~Perali, P.~Pieri, G.~C. Strinati, {Entanglement between pairing
  and screening in the Gorkov-Melik-Barkhudarov correction to the critical
  temperature throughout the BCS-BEC crossover}, Phys. Rev. B 97 (2018) 014528.

\bibitem{Nascimbene-2010}
S.~Nascimb{\`e}ne, N.~Navon, K.~J. Jiang, F.~Chevy, C.~Salomon, Exploring the
  thermodynamics of a universal {F}ermi gas, Nature 463 (2010) 1057--1060.

\bibitem{Ku-2012}
M.~J.~H. Ku, A.~T. Sommer, L.~W. Cheuk, M.~W. Zwierlein, Revealing the
  superfluid lambda transition in the universal thermodynamics of a unitary
  {F}ermi gas, Science 335 (2012) 563--567.

\bibitem{Burovski-2006a}
E.~Burovski, N.~Prokof'ev, B.~Svistunov, M.~Troyer, Critical temperature and
  thermodynamics of attractive fermions at unitarity, Phys. Rev. Lett. 96
  (2006) 160402.

\bibitem{Burovski-2008}
E.~Burovski, E.~Kozik, N.~Prokof'ev, B.~Svistunov, M.~Troyer, Critical
  temperature curve in {BEC-BCS} crossover, Phys. Rev. Lett. 101 (2008) 090402.

\bibitem{Goulko-2010}
O.~Goulko, M.~Wingate, {Thermodynamics of balanced and slightly spin-imbalanced
  Fermi gases at unitarity}, Phys. Rev. A 82 (2010) 053621.

\bibitem{Schulze-2001}
H.-J. Schulze, A.~Polls, A.~Ramos, Pairing with polarization effects in
  low-density neutron matter, Phys. Rev. C 63 (2001) 044310.

\bibitem{Heiselberg-2000}
H.~Heiselberg, C.~J. Pethick, H.~Smith, L.~Viverit, Influence of induced
  interactions on the superfluid transition in dilute {Fermi} gases, Phys. Rev.
  Lett. 85 (2000) 2418--2421.

\bibitem{Cao-2006}
L.~G. Cao, U.~Lombardo, P.~Schuck, Screening effects in superfluid nuclear and
  neutron matter within {Brueckner} theory, Phys. Rev. C 74 (2006) 064301.

\bibitem{Baldo-2000}
M.~Baldo, A.~Grasso, Dispersive effects in neutron matter superfluidity, Phys.
  Lett. B 485 (2000) 115--120.

\bibitem{Roepke-1982}
G.~R{\"o}pke, L.~M{\"u}nchow, H.~Schulz, Particle clustering and {Mott}
  transitions in nuclear matter at finite temperature, Nucl. Phys. A 379 (1982)
  536 -- 552.

\bibitem{Roepke-1983}
G.~{R{\"o}pke}, M.~{Schmidt}, L.~{M{\"u}nchow}, H.~{Schulz}, {Particle
  clustering and Mott transition in nuclear matter at finite temperature (II)
  Self-consistent ladder Hartree-Fock approximation and model calculations for
  cluster abundances and the phase diagram}, Nucl. Phys. A 399 (1983) 587--602.

\bibitem{Schmidt-1990}
M.~{Schmidt}, G.~{R{\"o}pke}, H.~{Schulz}, {Generalized Beth-Uhlenbeck approach
  for hot nuclear matter}, Ann. Phys. 202 (1990) 57--99.

\bibitem{Pantel-2014}
P.-A. Pantel, D.~Davesne, M.~Urban, {Polarized Fermi gases at finite
  temperature in the BCS-BEC crossover}, Phys. Rev. A 90 (2014) 053629, and
  Erratum {\em ibid.} 94 (2016) 019901.

\bibitem{Petrov-2004}
D.~S. Petrov, C.~Salomon, G.~V. Shlyapnikov, Weakly bound dimers of fermionic
  atoms, Phys. Rev. Lett. 93 (2004) 090404.

\bibitem{Petrov-2005}
D.~S. Petrov, C.~Salomon, G.~V. Shlyapnikov, Scattering properties of weakly
  bound dimers of fermionic atoms, Phys. Rev. A 71 (2005) 012708.

\bibitem{Brodsky-2005}
I.~V. Brodsky, A.~V. Klaptsov, M.~Y. Kagan, R.~Combescot, X.~Leyronas, Bound
  states of three and four resonantly interacting particles, JETP Lett. 82
  (2005) 273--278.

\bibitem{Brodsky-2006}
I.~V. Brodsky, M.~Y. Kagan, A.~V. Klaptsov, R.~Combescot, X.~Leyronas, Exact
  diagrammatic approach for dimer-dimer scattering and bound states of three
  and four resonantly interacting particles, Phys. Rev. A 73 (2006) 032724.

\bibitem{Elhatisari-2017}
S.~Elhatisari, K.~Katterjohn, D.~Lee, U.-G. Mei§ner, G.~Rupak, Universal
  dimer-dimer scattering in lattice effective field theory, Phys. Lett. B 768
  (2017) 337--344.

\bibitem{Deltuva-2017}
A.~Deltuva, Universality in fermionic dimer-dimer scattering, Phys. Rev. A 96
  (2017) 022701.

\bibitem{vonStecher-2008}
J.~von Stecher, C.~H. Greene, D.~Blume, {Energetics and structural properties
  of trapped two-component Fermi gases}, Phys. Rev. A 77 (2008) 043619.

\bibitem{D'Incao-2009}
J.~P. D'Incao, S.~T. Rittenhouse, N.~P. Mehta, C.~H. Greene, {Dimer-dimer
  collisions at finite energies in two-component Fermi gases}, Phys. Rev. A 79
  (2009) 030501.

\bibitem{Alzetto-2013}
F.~Alzetto, R.~Combescot, X.~Leyronas, Dimer-dimer scattering length for
  fermions with different masses: Analytical study for large mass ratio, Phys.
  Rev. A 87 (2013) 022704.

\bibitem{Wilczek-2012}
F.~Wilczek, {Origins of mass}, Open Physics 10 (2012) 1021--1037,
  arXiv:1206.7114.

\bibitem{Deltuva-2015}
A.~Deltuva, A.~C. Fonseca, Deuteron$-$deuteron scattering above four-nucleon
  breakup threshold, Phys. Lett. B 742 (2015) 285 -- 289.

\bibitem{Liu-2006}
X.-J. {Liu}, H.~{Hu}, {BCS-BEC crossover in an asymmetric two-component Fermi
  gas}, Europhys. Lett. 75 (2006) 364--370.

\bibitem{Kashimura-2013}
T.~Kashimura, R.~Watanabe, Y.~Ohashi, Magnetic properties and strong-coupling
  corrections in an ultracold {F}ermi gas with population imbalance, J. Low.
  Temp. Phys. 171 (2013) 355--361.

\bibitem{Maly-1996}
J.~Maly, K.~Levin, D.~Z. Liu, {Coulomb correlations and pseudogap effects in a
  preformed pair model for the cuprates}, Phys. Rev. B 54 (1996)
  R15657--R15660.

\bibitem{Ohashi-2002b}
Y.~Ohashi, On the {Fulde-Ferrell} state in spatially isotropic superconductors,
  J. Phys. Soc. Jap. 71 (2002) 2625--2628.

\bibitem{Shimahara-1998}
H.~Shimahara, {Phase fluctuations and Kosterlitz-Thouless transition in
  two-dimensional Fulde-Ferrell-Larkin-Ovchinnikov superconductors}, J. Phys.
  Soc. Jap. 67 (1998) 1872--1875.

\bibitem{Shimahara-1999}
H.~Shimahara, {Stability of Fulde-Ferrell-Larkin-Ovchinnikov state in type-II
  superconductors against the phase fluctuations}, Physica B 259-261 (1999)
  492--493.

\bibitem{Radzihovsky-2009}
L.~Radzihovsky, A.~Vishwanath, Quantum liquid crystals in an imbalanced {Fermi}
  gas: Fluctuations and fractional vortices in {Larkin-Ovchinnikov} states,
  Phys. Rev. Lett. 103 (2009) 010404.

\bibitem{Radzihovsky-2011}
L.~Radzihovsky, {Fluctuations and phase transitions in Larkin-Ovchinnikov
  liquid-crystal states of a population-imbalanced resonant Fermi gas}, Phys.
  Rev. A 84 (2011) 023611.

\bibitem{Pawel-2017}
P.~Jakubczyk, {Renormalization theory for the Fulde-Ferrell-Larkin-Ovchinnikov
  states at $T > 0$}, Phys. Rev. A 95 (2017) 063626.

\bibitem{Gubbels-2008}
K.~Gubbels, H.~Stoof, Renormalization group theory for the imbalanced {Fermi}
  gas, Phys. Rev. Lett. 100 (2008) 140407.

\bibitem{He-2007}
Y.~He, C.-C. Chien, Q.~Chen, K.~Levin, Thermodynamics and superfluid density in
  {BCS-BEC} crossover with and without population imbalance, Phys. Rev. B 76
  (2007) 224516.

\bibitem{Boettcher-2015}
I.~Boettcher, J.~Braun, T.~K. Herbst, J.~M. Pawlowski, D.~Roscher,
  C.~Wetterich, {Phase structure of spin-imbalanced unitary Fermi gases}, Phys.
  Rev. A 91 (2015) 013610.

\bibitem{Luttinger-1960}
J.~M. Luttinger, Fermi surface and some simple equilibrium properties of a
  system of interacting fermions, Phys. Rev. 119 (1960) 1153--1163.

\bibitem{Bloch-2008}
I.~Bloch, J.~Dalibard, W.~Zwerger, Many-body physics with ultracold gases, Rev.
  Mod. Phys. 80 (2008) 885--964.

\bibitem{Fano-1935}
U.~Fano, Sullo spettro di assorbimento dei gas nobili presso il limite dello
  spettro d'arco, Nuovo Cimento 12 (1935) 154--161.

\bibitem{Fano-1961}
U.~Fano, Effects of configuration interaction on intensities and phase shifts,
  Phys. Rev. 124 (1961) 1866--1878.

\bibitem{Feshbach-1962}
H.~Feshbach, A unified theory of nuclear reactions. {II}, Ann. Phys. 19 (1962)
  287--313.

\bibitem{Simonucci-2005}
S.~Simonucci, P.~Pieri, G.~C. Strinati, Broad vs. narrow {Fano-Feshbach}
  resonances in the {BCS-BEC} crossover with trapped {Fermi} atoms, Europhys.
  Lett. 69 (2005) 713--718.

\bibitem{Holland-2001}
M.~Holland, S.~J. J. M.~F. Kokkelmans, M.~L. Chiofalo, R.~Walser, Resonance
  superfluidity in a quantum degenerate {Fermi} gas, Phys. Rev. Lett. 87 (2001)
  120406.

\bibitem{Ohashi-2002a}
Y.~Ohashi, A.~Griffin, {BCS-BEC} crossover in a gas of {Fermi} atoms with a
  {Feshbach} resonance, Phys. Rev. Lett. 89 (2002) 130402.

\bibitem{Stoof-2004}
G.~M. Falco, H.~T.~C. Stoof, Crossover temperature of {Bose-Einstein}
  condensation in an atomic {Fermi} gas, Phys. Rev. Lett. 92 (2004) 130401.

\bibitem{Bruun-2004}
G.~M. Bruun, C.~J. Pethick, Effective theory of {Feshbach} resonances and
  many-body properties of {Fermi} gases, Phys. Rev. Lett. 92 (2004) 140404.

\bibitem{Stajic-2004}
J.~Stajic, J.~N. Milstein, Q.~Chen, M.~L. Chiofalo, M.~J. Holland, K.~Levin,
  {Nature of superfluidity in ultracold Fermi gases near Feshbach resonances},
  Phys. Rev. A 69 (2004) 063610.

\bibitem{Diehl-2006}
S.~Diehl, C.~Wetterich, Universality in phase transitions for ultracold
  fermionic atoms, Phys. Rev. A 73 (2006) 033615.

\bibitem{Pethick-2008}
C.~Pethick, H.~Smith, {Bose-Einstein Condensation in Dilute Gases}, Cambridge
  University Press, Cambridge, 2008.

\bibitem{Perali-2003}
A.~Perali, P.~Pieri, G.~C. Strinati, {Shrinking of a condensed fermionic cloud
  in a trap approaching the Bose-Einstein condensation limit}, Phys. Rev. A 68
  (2003) 031601.

\bibitem{Perali-2004b}
A.~Perali, P.~Pieri, L.~Pisani, G.~C. Strinati, {BCS-BEC} crossover at finite
  temperature for superfluid trapped {Fermi} atoms, Phys. Rev. Lett. 92 (2004)
  220404.

\bibitem{Haussmann-2008}
R.~Haussmann, W.~Zwerger, {Thermodynamics of a trapped unitary Fermi gas},
  Phys. Rev. A 78 (2008) 063602.

\bibitem{Shin-2007}
Y.~Shin, C.~H. Schunck, A.~Schirotzek, W.~Ketterle, Tomographic rf spectroscopy
  of a trapped {F}ermi gas at unitarity, Phys. Rev. Lett. 99 (2007) 090403.

\bibitem{Ho-2010}
T.-L. Ho, Q.~Zhou, Obtaining the phase diagram and thermodynamic quantities of
  bulk systems from the densities of trapped gases, Nat. Phys. 6 (2010)
  131--134.

\bibitem{Regal-2003b}
C.~A. Regal, C.~Ticknor, J.~L. Bohn, D.~S. Jin, Creation of ultracold molecules
  from a {Fermi} gas of atoms, Nature 424 (2003) 47--50.

\bibitem{Strecker-2003}
K.~E. Strecker, G.~B. Partridge, R.~G. Hulet, Conversion of an atomic {F}ermi
  gas to a long-lived molecular {B}ose gas, Phys. Rev. Lett. 91 (2003) 080406.

\bibitem{Cubizolles-2003}
J.~Cubizolles, T.~Bourdel, S.~J. J. M.~F. Kokkelmans, G.~V. Shlyapnikov,
  C.~Salomon, Production of long-lived ultracold {${\mathrm{L}\mathrm{i}}_{2}$}
  molecules from a {Fermi} gas, Phys. Rev. Lett. 91 (2003) 240401.

\bibitem{Jochim-2003b}
S.~Jochim, M.~Bartenstein, A.~Altmeyer, G.~Hendl, C.~Chin, J.~Hecker~Denschlag,
  R.~Grimm, Pure gas of optically trapped molecules created from fermionic
  atoms, Phys. Rev. Lett. 91 (2003) 240402.

\bibitem{Greiner-2003}
M.~Greiner, C.~A. Regal, D.~S. Jin, Emergence of a molecular {Bose}-{Einstein}
  condensate from a {Ferm}i gas, Nature 426 (2003) 537--540.

\bibitem{Jochim-2003a}
S.~Jochim, M.~Bartenstein, A.~Altmeyer, G.~Hendl, S.~Riedl, C.~Chin,
  J.~Hecker~Denschlag, R.~Grimm, Bose-{Einstein} condensation of molecules,
  Science 302 (2003) 2101--2103.

\bibitem{Zwierlein-2003}
M.~W. Zwierlein, C.~A. Stan, C.~H. Schunck, S.~M.~F. Raupach, S.~Gupta,
  Z.~Hadzibabic, W.~Ketterle, Observation of {Bose-Einstein} condensation of
  molecules, Phys. Rev. Lett. 91 (2003) 250401.

\bibitem{Regal-2004}
C.~A. Regal, M.~Greiner, D.~S. Jin, Observation of resonance condensation of
  fermionic atom pairs, Phys. Rev. Lett. 92 (2004) 040403.

\bibitem{Zwierlein-2004}
M.~W. Zwierlein, C.~A. Stan, C.~H. Schunck, S.~M.~F. Raupach, A.~J. Kerman,
  W.~Ketterle, Condensation of pairs of fermionic atoms near a {Feshbach}
  resonance, Phys. Rev. Lett. 92 (2004) 120403.

\bibitem{Perali-2005}
A.~Perali, P.~Pieri, G.~C. Strinati, Extracting the condensate density from
  projection experiments with {Fermi} gases, Phys. Rev. Lett. 95 (2005) 010407.

\bibitem{Zwierlein-2005a}
M.~W. Zwierlein, C.~H. Schunck, C.~A. Stan, S.~M.~F. Raupach, W.~Ketterle,
  Formation dynamics of a fermion pair condensate, Phys. Rev. Lett. 94 (2005)
  180401.

\bibitem{Chin-2004}
C.~Chin, M.~Bartenstein, A.~Altmeyer, S.~Riedl, S.~Jochim, J.~Hecker~Denschlag,
  R.~Grimm, Observation of the pairing gap in a strongly interacting {F}ermi
  gas, Science 305 (2004) 1128--1130.

\bibitem{Greiner-2005a}
M.~Greiner, C.~A. Regal, D.~S. Jin, Probing the excitation spectrum of a
  {F}ermi gas in the {BCS-BEC} crossover regime, Phys. Rev. Lett. 94 (2005)
  070403.

\bibitem{Schirotzek-2008}
A.~Schirotzek, Y.-i. Shin, C.~H. Schunck, W.~Ketterle, Determination of the
  superfluid gap in atomic {Fermi} gases by quasiparticle spectroscopy, Phys.
  Rev. Lett. 101 (2008) 140403.

\bibitem{Kinast-2004a}
J.~Kinast, S.~L. Hemmer, M.~E. Gehm, A.~Turlapov, J.~E. Thomas, Evidence for
  superfluidity in a resonantly interacting {F}ermi gas, Phys. Rev. Lett. 92
  (2004) 150402.

\bibitem{Bartenstein-2004b}
M.~Bartenstein, A.~Altmeyer, S.~Riedl, S.~Jochim, C.~Chin, J.~Hecker~Denschlag,
  R.~Grimm, Collective excitations of a degenerate gas at the {BEC-BCS}
  crossover, Phys. Rev. Lett. 92 (2004) 203201.

\bibitem{Kinast-2004b}
J.~Kinast, A.~Turlapov, J.~E. Thomas, {Breakdown of hydrodynamics in the radial
  breathing mode of a strongly interacting Fermi gas}, Phys. Rev. A 70 (2004)
  051401.

\bibitem{Altmeyer-2007a}
A.~Altmeyer, S.~Riedl, C.~Kohstall, M.~J. Wright, R.~Geursen, M.~Bartenstein,
  C.~Chin, J.~Hecker~Denschlag, R.~Grimm, Precision measurements of collective
  oscillations in the {BEC-BCS} crossover, Phys. Rev. Lett. 98 (2007) 040401.

\bibitem{Riedl-2008}
S.~Riedl, E.~{S\'anchez Guajardo}, C.~Kohstall, A.~Altmeyer, M.~Wright,
  J.~{Hecker Denschlag}, R.~Grimm, G.~Bruun, H.~Smith, {Collective oscillations
  of a Fermi gas in the unitarity limit: temperature effects and role of pair
  correlations}, Phys. Rev. A 78 (2008) 053609.

\bibitem{Altmeyer-2007b}
A.~Altmeyer, S.~Riedl, M.~J. Wright, C.~Kohstall, J.~Hecker~Denschlag,
  R.~Grimm, Dynamics of a strongly interacting {Fermi} gas: The radial
  quadrupole mode, Phys. Rev. A 76 (2007) 033610.

\bibitem{Wright-2007}
M.~J. Wright, S.~Riedl, A.~Altmeyer, C.~Kohstall, E.~R. S\'anchez~Guajardo,
  J.~Hecker~Denschlag, R.~Grimm, Finite-temperature collective dynamics of a
  {Fermi} gas in the {BEC-BCS} crossover, Phys. Rev. Lett. 99 (2007) 150403.

\bibitem{Tey-2013}
M.~K. Tey, L.~A. Sidorenkov, E.~R. S\'anchez~Guajardo, R.~Grimm, M.~J.~H. Ku,
  M.~W. Zwierlein, Y.-H. Hou, L.~Pitaevskii, S.~Stringari, Collective modes in
  a unitary {Fermi} gas across the superfluid phase transition, Phys. Rev.
  Lett. 110 (2013) 055303.

\bibitem{Kinast-2005}
J.~Kinast, A.~Turlapov, J.~E. Thomas, Q.~Chen, J.~Stajic, K.~Levin, Heat
  capacity of a strongly interacting {Fermi} gas, Science 307 (2005)
  1296--1299.

\bibitem{Luo-2007}
L.~Luo, B.~Clancy, J.~Joseph, J.~Kinast, J.~E. Thomas, Measurement of the
  entropy and critical temperature of a strongly interacting {F}ermi gas, Phys.
  Rev. Lett. 98 (2007) 080402.

\bibitem{Navon-2010}
N.~Navon, S.~Nascimb{\`e}ne, F.~Chevy, C.~Salomon, The equation of state of a
  low-temperature {Fermi} gas with tunable interactions, Science 328 (2010)
  729--732.

\bibitem{Horikoshi-2010}
M.~Horikoshi, S.~Nakajima, M.~Ueda, T.~Mukaiyama, Measurement of universal
  thermodynamic functions for a unitary {Fermi} gas, Science 327 (2010)
  442--445.

\bibitem{Partridge-2005}
G.~B. Partridge, K.~E. Strecker, R.~I. Kamar, M.~W. Jack, R.~G. Hulet,
  Molecular probe of pairing in the {BEC-BCS} crossover, Phys. Rev. Lett. 95
  (2005) 020404.

\bibitem{Greiner-2005b}
M.~Greiner, C.~A. Regal, J.~T. Stewart, D.~S. Jin, Probing pair-correlated
  fermionic atoms through correlations in atom shot noise, Phys. Rev. Lett. 94
  (2005) 110401.

\bibitem{Altman-2004}
E.~Altman, E.~Demler, M.~D. Lukin, Probing many-body states of ultracold atoms
  via noise correlations, Phys. Rev. A 70 (2004) 013603.

\bibitem{Zwierlein-2005b}
M.~W. Zwierlein, J.~R. Abo-Shaeer, A.~Schirotzek, C.~H. Schunck, W.~Ketterle,
  {Vortices and superfluidity in a strongly interacting Fermi gas}, Nature 435
  (2005) 1047--1051.

\bibitem{Shin-2006}
Y.~Shin, M.~W. Zwierlein, C.~H. Schunck, A.~Schirotzek, W.~Ketterle,
  Observation of phase separation in a strongly interacting imbalanced {Fermi}
  gas, Phys. Rev. Lett. 97 (2006) 030401.

\bibitem{Zwierlein-2006b}
M.~W. Zwierlein, C.~H. Schunck, A.~Schirotzek, W.~Ketterle, Direct observation
  of the superfluid phase transition in ultracold {Fermi} gases, Nature 442
  (2006) 54--58.

\bibitem{Shin-2008a}
Y.-i. Shin, A.~Schirotzek, C.~H. Schunck, W.~Ketterle, Realization of a
  strongly interacting {Bose-Fermi} mixture from a two-component {Fermi} gas,
  Phys. Rev. Lett. 101 (2008) 070404.

\bibitem{Olsen-2015}
B.~A. Olsen, M.~C. Revelle, J.~A. Fry, D.~E. Sheehy, R.~G. Hulet, {Phase
  diagram of a strongly interacting spin-imbalanced Fermi gas}, Phys. Rev. A 92
  (2015) 063616.

\bibitem{Joseph-2007}
J.~Joseph, B.~Clancy, L.~Luo, J.~Kinast, A.~Turlapov, J.~E. Thomas, Measurement
  of sound velocity in a {Fermi} gas near a {Feshbach} resonance, Phys. Rev.
  Lett. 98 (2007) 170401.

\bibitem{Sidorenkov-2013}
L.~A. Sidorenkov, M.~K. Tey, R.~Grimm, Y.-H. Hou, L.~Pitaevskii, S.~Stringari,
  Second sound and the superfluid fraction in a {Fermi} gas with resonant
  interactions, Nature 498 (2013) 78--81.

\bibitem{Miller-2007}
D.~Miller, J.~Chin, C.~Stan, Y.~Liu, W.~Setiawan, C.~Sanner, W.~Ketterle,
  Critical velocity for superfluid flow across the {BEC-BCS} crossover, Phys.
  Rev. Lett. 99 (2007) 070402.

\bibitem{Moritz-2015}
W.~Weimer, K.~Morgener, V.~P. Singh, J.~Siegl, K.~Hueck, N.~Luick, L.~Mathey,
  H.~Moritz, Critical velocity in the {BEC-BCS} crossover, Phys. Rev. Lett. 114
  (2015) 095301.

\bibitem{Schunck-2008}
C.~H. Schunck, Y.-i. Shin, A.~Schirotzek, W.~Ketterle, Determination of the
  fermion pair size in a resonantly interacting superfluid, Nature 454 (2008)
  739--743.

\bibitem{Stewart-2010}
J.~T. Stewart, J.~P. Gaebler, T.~E. Drake, D.~S. Jin, Verification of universal
  relations in a strongly interacting {F}ermi gas, Phys. Rev. Lett. 104 (2010)
  235301.

\bibitem{Kuhnle-2010}
E.~Kuhnle, H.~Hu, X.-J. Liu, P.~Dyke, M.~Mark, P.~Drummond, P.~Hannaford,
  C.~Vale, Universal behavior of pair correlations in a strongly interacting
  {Fermi} gas, Phys. Rev. Lett. 105 (2010) 070402.

\bibitem{Kuhnle-2011}
E.~Kuhnle, S.~Hoinka, P.~Dyke, H.~Hu, P.~Hannaford, C.~Vale, Temperature
  dependence of the universal contact parameter in a unitary {Fermi} gas, Phys.
  Rev. Lett. 106 (2011) 170402.

\bibitem{Sagi-2012}
Y.~Sagi, T.~E. Drake, R.~Paudel, D.~S. Jin, Measurement of the homogeneous
  contact of a unitary {F}ermi gas, Phys. Rev. Lett. 109 (2012) 220402.

\bibitem{Hoinka-2013}
S.~Hoinka, M.~Lingham, K.~Fenech, H.~Hu, C.~Vale, J.~Drut, S.~Gandolfi, Precise
  determination of the structure factor and contact in a unitary {Fermi} gas,
  Phys. Rev. Lett. 110 (2013) 055305.

\bibitem{Cao-2011}
C.~Cao, E.~Elliott, J.~Joseph, H.~Wu, J.~Petricka, T.~Sch{\"a}fer, J.~Thomas,
  Universal quantum viscosity in a unitary {Fermi} gas, Science 331 (2011)
  58--61.

\bibitem{Elliott-2014}
E.~Elliott, J.~Joseph, J.~Thomas, Anomalous minimum in the shear viscosity of a
  {Fermi} gas, Phys. Rev. Lett. 113 (2014) 020406.

\bibitem{Joseph-2015}
J.~A. Joseph, E.~Elliott, J.~E. Thomas, Shear viscosity of a unitary {Fermi}
  gas near the superfluid phase transition, Phys. Rev. Lett. 115 (2015) 020401.

\bibitem{Bluhm-2017}
M.~Bluhm, J.~Hou, T.~Sch\"afer, Determination of the density and temperature
  dependence of the shear viscosity of a unitary {Fermi} gas based on
  hydrodynamic flow, Phys. Rev. Lett. 119 (2017) 065302.

\bibitem{Kovtun-2005}
P.~K. Kovtun, D.~T. Son, A.~O. Starinets, Viscosity in strongly interacting
  quantum field theories from black hole physics, Phys. Rev. Lett. 94 (2005)
  111601.

\bibitem{Sanner-2011}
C.~Sanner, E.~Su, A.~Keshet, W.~Huang, J.~Gillen, R.~Gommers, W.~Ketterle,
  Speckle imaging of spin fluctuations in a strongly interacting {Fermi} gas,
  Phys. Rev. Lett. 106 (2011) 010402.

\bibitem{Hoinka-2012}
S.~Hoinka, M.~Lingham, M.~Delehaye, C.~J. Vale, Dynamic spin response of a
  strongly interacting {Fermi} gas, Phys. Rev. Lett. 109 (2012) 050403.

\bibitem{Lingham-2014}
M.~G. Lingham, K.~Fenech, S.~Hoinka, C.~J. Vale, Local observation of pair
  condensation in a {Fermi} gas at unitarity, Phys. Rev. Lett. 112 (2014)
  100404.

\bibitem{Riedl-2011}
S.~Riedl, E.~R. {S\'{a}nchez Guajardo}, C.~Kohstall, J.~{Hecker Denschlag},
  Superfluid quenching of the moment of inertia in a strongly interacting
  {F}ermi gas, New J. Phys. 13 (2011) 035003.

\bibitem{Husmann-2015}
D.~Husmann, S.~Uchino, S.~Krinner, M.~Lebrat, T.~Giamarchi, T.~Esslinger, J.-P.
  Brantut, Connecting strongly correlated superfluids by a quantum point
  contact, Science 350 (2015) 1498--1501.

\bibitem{Valtolina-2015}
G.~Valtolina, A.~Burchianti, A.~Amico, E.~Neri, K.~Xhani, J.~A. Seman,
  A.~Trombettoni, A.~Smerzi, M.~Zaccanti, M.~Inguscio, G.~Roati, {Josephson
  effect in fermionic superfluids across the BEC-BCS crossover}, Science 350
  (2015) 1505--1508.

\bibitem{Froehlich-2011}
B.~Fr\"ohlich, M.~Feld, E.~Vogt, M.~Koschorreck, W.~Zwerger, M.~K\"ohl,
  Radio-frequency spectroscopy of a strongly interacting two-dimensional
  {Fermi} gas, Phys. Rev. Lett. 106 (2011) 105301.

\bibitem{Sommer-2012}
A.~T. Sommer, L.~W. Cheuk, M.~J.~H. Ku, W.~S. Bakr, M.~W. Zwierlein, Evolution
  of fermion pairing from three to two dimensions, Phys. Rev. Lett. 108 (2012)
  045302.

\bibitem{Vogt-2012}
E.~Vogt, M.~Feld, B.~Fr\"ohlich, D.~Pertot, M.~Koschorreck, M.~K\"ohl, Scale
  invariance and viscosity of a two-dimensional {Fermi} gas, Phys. Rev. Lett.
  108 (2012) 070404.

\bibitem{Makhalov-2014}
V.~Makhalov, K.~Martiyanov, A.~Turlapov, Ground-state pressure of quasi-{2D
  Fermi and Bose} gases, Phys. Rev. Lett. 112 (2014) 045301.

\bibitem{Ong-2015}
W.~Ong, C.~Cheng, I.~Arakelyan, J.~E. Thomas, Spin-imbalanced
  quasi-two-dimensional {Fermi} gases, Phys. Rev. Lett. 114 (2015) 110403.

\bibitem{Ries-2015}
M.~G. Ries, A.~N. Wenz, G.~Z\"urn, L.~Bayha, I.~Boettcher, D.~Kedar, P.~A.
  Murthy, M.~Neidig, T.~Lompe, S.~Jochim, Observation of pair condensation in
  the quasi-{2D BEC-BCS} crossover, Phys. Rev. Lett. 114 (2015) 230401.

\bibitem{Murthy-2015}
P.~A. Murthy, I.~Boettcher, L.~Bayha, M.~Holzmann, D.~Kedar, M.~Neidig, M.~G.
  Ries, A.~N. Wenz, G.~Z\"urn, S.~Jochim, Observation of the
  {Berezinskii-Kosterlitz-Thouless} phase transition in an ultracold {Fermi}
  gas, Phys. Rev. Lett. 115 (2015) 010401.

\bibitem{Fenech-2016}
K.~Fenech, P.~Dyke, T.~Peppler, M.~G. Lingham, S.~Hoinka, H.~Hu, C.~J. Vale,
  Thermodynamics of an attractive {2D Fermi} gas, Phys. Rev. Lett. 116 (2016)
  045302.

\bibitem{Boettcher-2016}
I.~Boettcher, L.~Bayha, D.~Kedar, P.~A. Murthy, M.~Neidig, M.~G. Ries, A.~N.
  Wenz, G.~Z\"urn, S.~Jochim, T.~Enss, Equation of state of ultracold fermions
  in the {2D BEC-BCS} crossover region, Phys. Rev. Lett. 116 (2016) 045303.

\bibitem{Liao-2010}
Y.-a. Liao, A.~S.~C. Rittner, T.~Paprotta, W.~Li, G.~B. Partridge, R.~G. Hulet,
  S.~K. Baur, E.~J. Mueller, Spin-imbalance in a one-dimensional {Fermi} gas,
  Nature 467 (2010) 567--569.

\bibitem{Orso-2007}
G.~Orso, Attractive {Fermi} gases with unequal spin populations in highly
  elongated traps, Phys. Rev. Lett. 98 (2007) 070402.

\bibitem{Taylor-1972}
J.~R. Taylor, {Scattering theory: the quantum theory of nonrelativistic
  collisions}, Dover Publications, Mineola, N.Y., 1972,2006, Ch. 6-c.

\bibitem{Ho-2004}
T.-L. Ho, Universal thermodynamics of degenerate quantum gases in the unitarity
  limit, Phys. Rev. Lett. 92 (2004) 090402.

\bibitem{Perali-2004}
A.~Perali, P.~Pieri, G.~C. Strinati, Quantitative comparison between
  theoretical predictions and experimental results for the {BCS-BEC} crossover,
  Phys. Rev. Lett. 93 (2004) 100404.

\bibitem{Gehm-2003}
M.~E. Gehm, S.~L. Hemmer, S.~R. Granade, K.~M. O'Hara, J.~E. Thomas,
  {Mechanical stability of a strongly interacting Fermi gas of atoms}, Phys.
  Rev. A 68 (2003) 011401.

\bibitem{Bartenstein-2004a}
M.~Bartenstein, A.~Altmeyer, S.~Riedl, S.~Jochim, C.~Chin, J.~Hecker~Denschlag,
  R.~Grimm, Crossover from a molecular {Bose-Einstein} condensate to a
  degenerate {Fermi} gas, Phys. Rev. Lett. 92 (2004) 120401.

\bibitem{Nishida-2006}
Y.~Nishida, D.~T. Son, $\epsilon$ expansion for a {Fermi} gas at infinite
  scattering length, Phys. Rev. Lett. 97 (2006) 050403.

\bibitem{Nishida-2007a}
Y.~Nishida, D.~T. Son, {Fermi gas near unitarity around four and two spatial
  dimensions}, Phys. Rev. A 75 (2007) 063617.

\bibitem{Nishida-2009}
Y.~Nishida, {Ground-state energy of the unitary Fermi gas from the $\epsilon$
  expansion}, Phys. Rev. A 79 (2009) 013627.

\bibitem{Veillette-2007}
M.~Y. Veillette, D.~E. Sheehy, L.~Radzihovsky, {Large-$N$ expansion for unitary
  superfluid Fermi gases}, Phys. Rev. A 75 (2007) 043614.

\bibitem{Diehl-2007}
S.~Diehl, H.~Gies, J.~M. Pawlowski, C.~Wetterich, {Flow equations for the
  BCS-BEC crossover}, Phys. Rev. A 76 (2007) 021602.

\bibitem{Bartosch-2009}
L.~Bartosch, P.~Kopietz, A.~Ferraz, {Renormalization of the BCS-BEC crossover
  by order-parameter fluctuations}, Phys. Rev. B 80 (2009) 104514.

\bibitem{Nussinov-2004}
Z.~Nussinov, S.~Nussinov, {The BCS-BEC crossover in arbitrary dimensions}
  (2004).
\newblock \href {http://arxiv.org/abs/0410597} {\path{arXiv:0410597}}.

\bibitem{Nishida-2007b}
Y.~Nishida, {Unitary Fermi gas at finite temperature in the $\epsilon$
  expansion}, Phys. Rev. A 75 (2007) 063618.

\bibitem{Nikolic-2007}
P.~Nikoli\ifmmode~\acute{c}\else \'{c}\fi{}, S.~Sachdev, {Renormalization-group
  fixed points, universal phase diagram, and $1/N$ expansion for quantum
  liquids with interactions near the unitarity limit}, Phys. Rev. A 75 (2007)
  033608.

\bibitem{Metzner-2012}
W.~Metzner, M.~Salmhofer, C.~Honerkamp, V.~Meden, K.~Sch\"onhammer, Functional
  renormalization group approach to correlated fermion systems, Rev. Mod. Phys.
  84 (2012) 299--352.

\bibitem{Beth-1937}
E.~Beth, G.~E. Uhlenbeck, {The quantum theory of the non-ideal gas. II.
  Behaviour at low temperatures}, Physica 4 (1937) 915--924.

\bibitem{Liu-2009}
X.-J. Liu, H.~Hu, P.~D. Drummond, Virial expansion for a strongly correlated
  {Fermi} gas, Phys. Rev. Lett. 102 (2009) 160401.

\bibitem{Kaplan-2011}
D.~B. Kaplan, S.~Sun, New field-theoretic method for the virial expansion,
  Phys. Rev. Lett. 107 (2011) 030601.

\bibitem{Leyronas-2011}
X.~Leyronas, {Virial expansion with Feynman diagrams}, Phys. Rev. A 84 (2011)
  053633.

\bibitem{Rakshit-2012}
D.~Rakshit, K.~M. Daily, D.~Blume, {Natural and unnatural parity states of
  small trapped equal-mass two-component Fermi gases at unitarity and
  fourth-order virial coefficient}, Phys. Rev. A 85 (2012) 033634.

\bibitem{Ngampruetikorn-2015}
V.~Ngampruetikorn, M.~M. Parish, J.~Levinsen, {High-temperature limit of the
  resonant Fermi gas}, Phys. Rev. A 91 (2015) 013606.

\bibitem{Yan-2016}
Y.~Yan, D.~Blume, Path-integral {Monte Carlo} determination of the fourth-order
  virial coefficient for a unitary two-component {Fermi} gas with zero-range
  interactions, Phys. Rev. Lett. 116 (2016) 230401.

\bibitem{Carlson-2014}
J.~Carlson, S.~Gandolfi, {Predicting energies of small clusters from the
  inhomogeneous unitary Fermi gas}, Phys. Rev. A 90 (2014) 011601.

\bibitem{Blume-2007}
D.~Blume, J.~von Stecher, C.~H. Greene, Universal properties of a trapped
  two-component {Fermi} gas at unitarity, Phys. Rev. Lett. 99 (2007) 233201.

\bibitem{Yin-2015}
X.~Y. Yin, D.~Blume, {Trapped unitary two-component Fermi gases with up to ten
  particles}, Phys. Rev. A 92 (2015) 013608.

\bibitem{Chang-2007}
S.~Y. Chang, G.~F. Bertsch, {Unitary Fermi gas in a harmonic trap}, Phys. Rev.
  A 76 (2007) 021603.

\bibitem{Endres-2011}
M.~G. Endres, D.~B. Kaplan, J.-W. Lee, A.~N. Nicholson, {Lattice Monte Carlo
  calculations for unitary fermions in a harmonic trap}, Phys. Rev. A 84 (2011)
  043644.

\bibitem{Lee-2006}
D.~Lee, Ground-state energy of spin-$\frac{1}{2}$ fermions in the unitary
  limit, Phys. Rev. B 73 (2006) 115112.

\bibitem{Carlson-2003}
J.~Carlson, S.-Y. Chang, V.~R. Pandharipande, K.~E. Schmidt, Superfluid {Fermi}
  gases with large scattering length, Phys. Rev. Lett. 91 (2003) 050401.

\bibitem{Lee-2008}
D.~Lee, Ground state energy at unitarity, Phys. Rev. C 78 (2008) 024001.

\bibitem{Astrakharchik-2004}
G.~E. Astrakharchik, J.~Boronat, J.~Casulleras, S.~Giorgini, Equation of state
  of a {Fermi} gas in the {BEC-BCS} crossover: A quantum {Monte Carlo} study,
  Phys. Rev. Lett. 93 (2004) 200404.

\bibitem{Carlson-2005}
J.~Carlson, S.~Reddy, Asymmetric two-component fermion systems in strong
  coupling, Phys. Rev. Lett. 95 (2005) 060401.

\bibitem{Morris-2010}
A.~J. Morris, P.~L\'opez~R\'{\i}os, R.~J. Needs, {Ultracold atoms at unitarity
  within quantum Monte Carlo methods}, Phys. Rev. A 81 (2010) 033619.

\bibitem{Forbes-2011}
M.~McNeil~Forbes, S.~Gandolfi, A.~Gezerlis, Resonantly interacting fermions in
  a box, Phys. Rev. Lett. 106 (2011) 235303.

\bibitem{Li-2011}
X.~Li, J.~Koloren\ifmmode~\check{c}\else \v{c}\fi{}, L.~Mitas, {Atomic Fermi
  gas in the unitary limit by quantum Monte Carlo methods: Effects of the
  interaction range}, Phys. Rev. A 84 (2011) 023615.

\bibitem{Endres-2013}
M.~G. Endres, D.~B. Kaplan, J.-W. Lee, A.~N. Nicholson, {Lattice Monte Carlo
  calculations for unitary fermions in a finite box}, Phys. Rev. A 87 (2013)
  023615.

\bibitem{Carlson-2011}
J.~Carlson, S.~Gandolfi, K.~E. Schmidt, S.~Zhang, Auxiliary-field quantum
  {Monte Carlo} method for strongly paired fermions, Phys. Rev. A 84 (2011)
  061602.

\bibitem{OHara-2002}
K.~M. O{\textquoteright}Hara, S.~L. Hemmer, M.~E. Gehm, S.~R. Granade, J.~E.
  Thomas, Observation of a strongly interacting degenerate {Fermi} gas of
  atoms, Science 298 (2002) 2179--2182.

\bibitem{Bourdel-2004}
T.~Bourdel, L.~Khaykovich, J.~Cubizolles, J.~Zhang, F.~Chevy, M.~Teichmann,
  L.~Tarruell, S.~J. J. M.~F. Kokkelmans, C.~Salomon, Experimental study of the
  {BEC-BCS} crossover region in {Lithium} 6, Phys. Rev. Lett. 93 (2004) 050401.

\bibitem{Stewart-2006}
J.~T. Stewart, J.~P. Gaebler, C.~A. Regal, D.~S. Jin, Potential energy of a
  {$^{40}\mathrm{K}$ Fermi} gas in the {BCS-BEC} crossover, Phys. Rev. Lett. 97
  (2006) 220406.

\bibitem{Luo-2009}
L.~Luo, J.~E. Thomas, Thermodynamic measurements in a strongly interacting
  {Fermi} gas, J. Low Temp. Phys. 154 (2009) 1--29.

\bibitem{Zuern-2013}
G.~Z\"urn, T.~Lompe, A.~N. Wenz, S.~Jochim, P.~S. Julienne, J.~M. Hutson,
  Precise characterization of $^{6}\mathrm{Li}$ {Feshbach} resonances using
  trap-sideband-resolved {RF} spectroscopy of weakly bound molecules, Phys.
  Rev. Lett. 110 (2013) 135301.

\bibitem{Carlson-2008}
J.~Carlson, S.~Reddy, Superfluid pairing gap in strong coupling, Phys. Rev.
  Lett. 100 (2008) 150403.

\bibitem{Bulgac-2008a}
A.~Bulgac, J.~E. Drut, P.~Magierski, {Quantum Monte Carlo simulations of the
  BCS-BEC crossover at finite temperature}, Phys. Rev. A 78 (2008) 023625.

\bibitem{Drut-2012}
J.~E. Drut, T.~A. L\"ahde, G.~Wlaz\l{}owski, P.~Magierski, {Equation of state
  of the unitary Fermi gas: An update on lattice calculations}, Phys. Rev. A 85
  (2012) 051601.

\bibitem{Drut-2011}
J.~E. Drut, T.~A. L\"ahde, T.~Ten, Momentum distribution and contact of the
  unitary {Fermi} gas, Phys. Rev. Lett. 106 (2011) 205302.

\bibitem{Bulgac-2011}
P.~Magierski, G.~Wlaz\l{}owski, A.~Bulgac, Onset of a pseudogap regime in
  ultracold {Fermi} gases, Phys. Rev. Lett. 107 (2011) 145304.

\bibitem{Bulgac-2013a}
G.~Wlaz\l{}owski, P.~Magierski, J.~E. Drut, A.~Bulgac, K.~J. Roche, Cooper
  pairing above the critical temperature in a unitary {Fermi} gas, Phys. Rev.
  Lett. 110 (2013) 090401.

\bibitem{Wlazlowski-2012}
G.~Wlaz\l{}owski, P.~Magierski, J.~E. Drut, Shear viscosity of a unitary
  {Fermi} gas, Phys. Rev. Lett. 109 (2012) 020406.

\bibitem{Bulgac-2013b}
G.~Wlaz\l{}owski, P.~Magierski, A.~Bulgac, K.~J. Roche, {Temperature evolution
  of the shear viscosity in a unitary Fermi gas}, Phys. Rev. A 88 (2013)
  013639.

\bibitem{Bulgac-2015}
G.~Wlaz\l{}owski, W.~Quan, A.~Bulgac, {Perfect-fluid behavior of a dilute Fermi
  gas near unitary}, Phys. Rev. A 92 (2015) 063628.

\bibitem{Burovski-2006b}
E.~Burovski, N.~Prokof'ev, B.~Svistunov, M.~Troyer, {The Fermi-Hubbard model at
  unitarity}, New Jour. Phys. 8 (2006) 153.

\bibitem{Van-Houcke-2012}
K.~Van~Houcke, F.~Werner, E.~Kozik, N.~Prokof'ev, B.~Svistunov, M.~J.~H. Ku,
  A.~T. Sommer, L.~W. Cheuk, A.~Schirotzek, M.~W. Zwierlein, {Feynman diagrams
  versus Fermi-gas Feynman emulator}, Nat. Phys 8 (2012) 366--370.

\bibitem{Kozik-2015}
E.~Kozik, M.~Ferrero, A.~Georges, Nonexistence of the {Luttinger-Ward}
  functional and misleading convergence of skeleton diagrammatic series for
  {Hubbard}-like models, Phys. Rev. Lett. 114 (2015) 156402.

\bibitem{Rossi-2015}
R.~Rossi, F.~Werner, Skeleton series and multivaluedness of the self-energy
  functional in zero space-time dimensions, J. Phys. A 48 (2015) 485202.

\bibitem{Rossi-2016}
R.~Rossi, F.~Werner, N.~Prokof'ev, B.~Svistunov, Shifted-action expansion and
  applicability of dressed diagrammatic schemes, Phys. Rev. B 93 (2016) 161102.

\bibitem{Spuntarelli-2007}
A.~Spuntarelli, P.~Pieri, G.~C. Strinati, Josephson effect throughout the
  {BCS-BEC} crossover, Phys. Rev. Lett. 99 (2007) 040401.

\bibitem{Tan-2008a}
S.~Tan, Energetics of a strongly correlated {F}ermi gas, Ann. Phys. 323 (2008)
  2952--2970.

\bibitem{Tan-2008b}
S.~Tan, Large momentum part of a strongly correlated {F}ermi gas, Ann. Phys.
  323 (2008) 2971--2986.

\bibitem{Tan-2008c}
S.~Tan, Generalized virial theorem and pressure relation for a strongly
  correlated {F}ermi gas, Ann. Phys. 323 (2008) 2987--2990.

\bibitem{Braaten-2008a}
E.~Braaten, L.~Platter, Exact relations for a strongly interacting {F}ermi gas
  from the operator product expansion, Phys. Rev. Lett. 100 (2008) 205301.

\bibitem{Braaten-2008b}
E.~Braaten, D.~Kang, L.~Platter, {Universal relations for a strongly
  interacting Fermi gas near a Feshbach resonance}, Phys. Rev. A 78 (2008)
  053606.

\bibitem{Combescot-2009}
R.~Combescot, F.~Alzetto, X.~Leyronas, Particle distribution tail and related
  energy formula, Phys. Rev. A 79 (2009) 053640.

\bibitem{Braaten-2012}
E.~Braaten, {Universal relations for fermions with large scattering length},
  in: W.~Zwerger (Ed.), {The BCS-BEC Crossover and the Unitary Fermi Gas}, Vol.
  863 of Lecture Notes in Physics, Springer-Verlag, Berlin, 2012, pp. 193--231.

\bibitem{Pieri-2009}
P.~Pieri, A.~Perali, G.~C. Strinati, Enhanced paraconductivity-like
  fluctuations in the radiofrequency spectra of ultracold {F}ermi atoms, Nat.
  Phys. 5 (2009) 736--740.

\bibitem{Braaten-2010}
E.~Braaten, D.~Kang, L.~Platter, Short-time operator product expansion for rf
  spectroscopy of a strongly interacting {Fermi} gas, Phys. Rev. Lett. 104
  (2010) 223004.

\bibitem{Hu-2010}
H.~Hu, X.-J. Liu, P.~D. Drummond, Static structure factor of a strongly
  correlated {F}ermi gas at large momenta, EPL-Europhys. Lett. 91 (2010) 20005.

\bibitem{Werner-2008}
F.~Werner, L.~Tarruell, Y.~Castin, Number of closed-channel molecules in the
  {BEC-BCS} crossover, Eur. Phys. J. B 68 (2009) 401--415.

\bibitem{Zhang-2009}
S.~Zhang, A.~J. Leggett, Universal properties of the ultracold {Fermi} gas,
  Phys. Rev. A 79 (2009) 023601.

\bibitem{Van-Houcke-2013}
K.~Van~Houcke, F.~Werner, E.~Kozik, N.~Prokof'ev, B.~Svistunov, Contact and
  momentum distribution of the unitary {F}ermi gas by bold diagrammatic {Monte
  Carlo} (2013).
\newblock \href {http://arxiv.org/abs/1303.6245} {\path{arXiv:1303.6245}}.

\bibitem{Yu-2009b}
Z.~Yu, G.~M. Bruun, G.~Baym, Short-range correlations and entropy in
  ultracold-atom {F}ermi gases, Phys. Rev. A 80 (2009) 023615.

\bibitem{Sun-2015}
M.~Sun, X.~Leyronas, High-temperature expansion for interacting fermions, Phys.
  Rev. A 92 (2015) 053611.

\bibitem{Combescot-2006a}
R.~Combescot, X.~Leyronas, M.~Y. Kagan, {Self-consistent theory for molecular
  instabilities in a normal degenerate Fermi gas in the BEC-BCS crossover},
  Phys. Rev. A 73 (2006) 023618.

\bibitem{Pieri-2005a}
P.~Pieri, G.~C. Strinati, {Popov approximation for composite bosons in the
  BCS-BEC crossover}, Phys. Rev. B 71 (2005) 094520.

\bibitem{Enss-2011}
T.~Enss, R.~Haussmann, W.~Zwerger, {Viscosity and scale invariance in the
  unitary Fermi gas}, Ann. Phys. 326 (2011) 770 -- 796.

\bibitem{Hu-2011}
H.~Hu, X.-J. Liu, P.~D. Drummond, Universal contact of strongly interacting
  fermions at finite temperatures, New Jour. Phys. 13 (2011) 035007.

\bibitem{Boettcher-2013}
I.~Boettcher, S.~Diehl, J.~M. Pawlowski, C.~Wetterich, {Tan contact and
  universal high momentum behavior of the fermion propagator in the BCS-BEC
  crossover}, Phys. Rev. A 87 (2013) 023606.

\bibitem{Chen-2014}
Y.~Y. Chen, Y.~Z. Jiang, X.~W. Guan, Q.~Zhou, Critical behaviours of contact
  near phase transitions, Nat. Commun. 5 (2014) 5140.

\bibitem{Josephson-1962}
B.~Josephson, Possible new effects in superconductive tunnelling, Phys. Lett. 1
  (1962) 251-- 253.

\bibitem{Barone-1982}
A.~Barone, G.~Patern\`{o}, {Physics and Applications of the Josephson Effect},
  John Wiley $\&$ Sons, New York, 1982.

\bibitem{Abrikosov-1975}
A.~A. Abrikosov, L.~P. Gorkov, I.~E. Dzyaloshinski, {Methods of Quantum Field
  Theory in Statistical Physics}, Dover, New York, 1975.

\bibitem{Engels-2007}
P.~Engels, C.~Atherton, Stationary and nonstationary fluid flow of a
  {Bose-Einstein} condensate through a penetrable barrier, Phys. Rev. Lett. 99
  (2007) 160405.

\bibitem{Menotti-2002}
C.~Menotti, P.~Pedri, S.~Stringari, Expansion of an interacting {Fermi} gas,
  Phys. Rev. Lett. 89 (2002) 250402.

\bibitem{Stringari-1996}
S.~Stringari, Collective excitations of a trapped {Bose}-condensed gas, Phys.
  Rev. Lett. 77 (1996) 2360--2363.

\bibitem{Baranov-2000}
M.~A. Baranov, D.~S. Petrov, {Low-energy collective excitations in a superfluid
  trapped Fermi gas}, Phys. Rev. A 62 (2000) 041601.

\bibitem{Cozzini-2003}
M.~Cozzini, S.~Stringari, {Fermi} gases in slowly rotating traps: Superfluid
  versus collisional hydrodynamics, Phys. Rev. Lett. 91 (2003) 070401.

\bibitem{Dalfovo-1999}
F.~Dalfovo, S.~Giorgini, L.~P. Pitaevskii, S.~Stringari, {Theory of
  Bose-Einstein condensation in trapped gases}, Rev. Mod. Phys. 71 (1999)
  463--512.

\bibitem{Urban-2006}
M.~Urban, P.~Schuck, {Dynamics of a trapped Fermi gas in the BCS phase}, Phys.
  Rev. A 73 (2006) 013621.

\bibitem{LandauLifshitz10}
L.~P. Pitaevskii, E.~M. Lifshitz, Physical Kinetics, Vol.~10 of Landau-Lifshitz
  course of Theoretical Physics, Pergamon Press, 1981.

\bibitem{Bruun-2001}
G.~M. Bruun, B.~R. Mottelson, Low energy collective modes of a superfluid
  trapped atomic {Fermi} gas, Phys. Rev. Lett. 87 (2001) 270403.

\bibitem{Grasso-2005}
M.~Grasso, E.~Khan, M.~Urban, {Temperature dependence and finite-size effects
  in collective modes of superfluid-trapped Fermi gases}, Phys. Rev. A 72
  (2005) 043617.

\bibitem{Combescot-2004}
R.~Combescot, X.~Leyronas, Comment on ``{C}ollective excitations of a
  degenerate gas at the {BEC-BCS} crossover'', Phys. Rev. Lett. 93 (2004)
  138901.

\bibitem{Combescot-2006b}
R.~Combescot, M.~Y. Kagan, S.~Stringari, {Collective mode of homogeneous
  superfluid Fermi gases in the BEC-BCS crossover}, Phys. Rev. A 74 (2006)
  042717.

\bibitem{Bogoliubov-1959}
N.~Bogoliubov, V.~Tolmachev, D.~Shirkov, A New Method in the Theory of
  Superconductivity, Consultants Bureau, New York, 1959.

\bibitem{Anderson-1958}
P.~W. Anderson, Random-phase approximation in the theory of superconductivity,
  Phys. Rev. 112 (1958) 1900--1916.

\bibitem{Forbes-2013}
M.~McNeil~Forbes, R.~Sharma, {Validating simple dynamical simulations of the
  unitary Fermi gas}, Phys. Rev. A 90 (2014) 043638.

\bibitem{Martin-2014}
N.~Martin, M.~Urban, Collective modes in a superfluid neutron gas within the
  quasiparticle random-phase approximation, Phys. Rev. C 90 (2014) 065805.

\bibitem{Taylor-2005}
E.~Taylor, A.~Griffin, Two-fluid hydrodynamic modes in a trapped superfluid
  gas, Phys. Rev. A 72 (2005) 053630.

\bibitem{Betbeder-Matibet-1969}
O.~Betbeder-Matibet, P.~Nozi\`eres, Transport equations in clean
  superconductors, Ann. Phys. 51 (1969) 392--417.

\bibitem{Urban-2007}
M.~Urban, {Coupling of hydrodynamics and quasiparticle motion in collective
  modes of superfluid trapped Fermi gases}, Phys. Rev. A 75 (2007) 053607.

\bibitem{Urban-2008b}
M.~Urban, {Radial quadrupole and scissors modes in trapped Fermi gases across
  the BCS phase transition}, Phys. Rev. A 78 (2008) 053619.

\bibitem{Pantel-2015}
P.-A. Pantel, D.~Davesne, M.~Urban, {Numerical solution of the Boltzmann
  equation for trapped Fermi gases with in-medium effects}, Phys. Rev. A 91
  (2015) 013627.

\bibitem{Toschi-2003}
F.~{Toschi}, P.~{Vignolo}, S.~{Succi}, M.~P. {Tosi}, {Dynamics of trapped
  two-component Fermi gas: Temperature dependence of the transition from
  collisionless to collisional regime}, Phys. Rev. A 67 (2003) 041605.

\bibitem{Massignan-2005}
P.~{Massignan}, G.~M. {Bruun}, H.~{Smith}, {Viscous relaxation and collective
  oscillations in a trapped Fermi gas near the unitarity limit}, Phys. Rev. A
  71 (2005) 033607.

\bibitem{Chiacchiera-2009}
S.~Chiacchiera, T.~Lepers, D.~Davesne, M.~Urban, {Collective modes of trapped
  Fermi gases with in-medium interaction}, Phys. Rev. A 79 (2009) 033613.

\bibitem{Lepers-2010}
T.~Lepers, D.~Davesne, S.~Chiacchiera, M.~Urban, {Numerical solution of the
  Boltzmann equation for the collective modes of trapped Fermi gases}, Phys.
  Rev. A 82 (2010) 023609.

\bibitem{Chiacchiera-2011}
S.~Chiacchiera, T.~Lepers, D.~Davesne, M.~Urban, {Role of fourth-order
  phase-space moments in collective modes of trapped Fermi gases}, Phys. Rev. A
  84 (2011) 043634.

\bibitem{Bruun-2005}
G.~M. {Bruun}, H.~{Smith}, {Viscosity and thermal relaxation for a resonantly
  interacting Fermi gas}, Phys. Rev. A 72 (2005) 043605.

\bibitem{Bluhm-2014}
M.~{Bluhm}, T.~{Sch{\"a}fer}, {Medium effects and the shear viscosity of the
  dilute Fermi gas away from the conformal limit}, Phys. Rev. A 90 (2014)
  063615.

\bibitem{Farine-2000}
M.~{Farine}, P.~{Schuck}, X.~{Vi{\~n}as}, {Moment of inertia of a trapped
  superfluid gas of atomic fermions}, Phys. Rev. A 62 (2000) 013608.

\bibitem{Durand-1985}
M.~Durand, P.~Schuck, J.~Kunz, {Semiclassical description of currents in normal
  and superfluid rotating nuclei}, Nucl. Phys. A 439 (1985) 263.

\bibitem{Migdal-1959}
A.~B. Migdal, Superfluidity and the moments of inertia of nuclei, Nucl. Phys.
  13 (1959) 655--674.

\bibitem{Urban-2005}
M.~{Urban}, {Two-fluid model for a rotating trapped Fermi gas in the BCS
  phase}, Phys. Rev. A 71 (2005) 033611.

\bibitem{Urban-2003}
M.~{Urban}, P.~{Schuck}, {Slow rotation of a superfluid trapped Fermi gas},
  Phys. Rev. A 67 (2003) 033611.

\bibitem{Bausmerth-2008}
I.~{Bausmerth}, A.~{Recati}, S.~{Stringari}, Destroying superfluidity by
  rotating a {Fermi} gas at unitarity, Phys. Rev. Lett. 100 (2008) 070401.

\bibitem{Urban-2008a}
M.~Urban, P.~Schuck, {Pair breaking in rotating Fermi gases}, Phys. Rev. A 78
  (2008) 011601.

\bibitem{Simonucci-2015}
S.~Simonucci, P.~Pieri, G.~C. Strinati, {Vortex arrays in neutral trapped Fermi
  gases through the BCS-BEC crossover}, Nat. Phys. 11 (2015) 941--945.

\bibitem{Nozieres-1990}
P.~Nozi\`{e}res, D.~Pines, {The Theory of Quantum Liquids: Superfluid Bose
  Liquids}, Addison-Wesley, Reading, 1990.

\bibitem{Bertsch-1999}
G.~F. Bertsch, T.~Papenbrock, Yrast line for weakly interacting trapped bosons,
  Phys. Rev. Lett. 83 (1999) 5412--5414.

\bibitem{Bohr-1969}
A.~Bohr, B.~Mottelson, Nuclear structure. Vol. 1: Single-particle motion,
  Benjamin, New York, 1969.

\bibitem{Haidenbauer-1984}
J.~Haidenbauer, W.~Plessas, {Separable representation of the Paris
  nucleon-nucleon potential}, Phys. Rev. C 30 (1984) 1822--1839.

\bibitem{Lombardo-2001a}
U.~Lombardo, P.~Schuck, Size shrinking of deuterons in very dilute superfluid
  nuclear matter, Phys. Rev. C 63 (2001) 038201.

\bibitem{Andrenacci-2000}
N.~Andrenacci, P.~Pieri, G.~C. Strinati, {Size shrinking of composite bosons
  for increasing density in the BCS to Bose-Einstein crossover}, Eur. Phys. J.
  B 13 (2000) 637--642.

\bibitem{Itonaga-1970}
K.~Itonaga, H.~Band\=o, On the contraction of deuteron cluster in {Li}6, Prog.
  Theor. Phys. 44 (1970) 1232--1241.

\bibitem{Wiringa-1984}
R.~B. Wiringa, R.~A. Smith, T.~L. Ainsworth, Nucleon-nucleon potentials with
  and without {$\Delta$(1232)} degrees of freedom, Phys. Rev. C 29 (1984)
  1207--1221.

\bibitem{Stein-2014}
M.~Stein, A.~Sedrakian, X.-G. Huang, J.~W. Clark, {BCS-BEC crossovers and
  unconventional phases in dilute nuclear matter}, Phys. Rev. C 90 (2014)
  065804.

\bibitem{Yamaguchi-1954}
Y.~Yamaguchi, Two-nucleon problem when the potential is nonlocal but separable,
  Phys. Rev. 95 (1954) 1628--1634.

\bibitem{Taylor-1972b}
J.~R. Taylor, {Scattering theory: the quantum theory of nonrelativistic
  collisions}, Dover Publications, Mineola, N.Y., 1972,2006, Ch. 12-e.

\bibitem{Chomaz-2004}
P.~Chomaz, M.~Colonna, J.~Randrup, Nuclear spinodal fragmentation, Phys. Rep.
  389 (2004) 263--440.

\bibitem{Song-1990}
H.-Q. {Song}, G.-D. {Zheng}, R.-K. {Su}, {Critical phenomena in nuclear matter
  with Gogny interaction}, J. Phys. G: Nucl. Phys. 16 (1990) 1861--1871.

\bibitem{Ventura-1992}
J.~{Ventura}, A.~{Polls}, X.~{Vi{\~n}as}, S.~{Hernandez}, M.~{Pi},
  {Thermodynamic instabilities of nuclear matter at finite temperature with
  finite range effective interactions}, Nucl. Phys. A 545 (1992) 247--257.

\bibitem{Stein-1998}
H.~{Stein}, C.~{Porthun}, G.~{R{\"o}pke}, {Liquid-gas binodal anomaly for
  systems with pairing transition}, Eur. Phys. Jour. B 2 (1998) 393--398.

\bibitem{Su-1987}
R.~K. {Su}, S.~D. {Yang}, T.~T.~S. {Kuo}, {Liquid-gas and superconducting phase
  transitions of nuclear matter calculated with real time Green's function
  methods and Skyrme interactions}, Phys. Rev. C 35 (1987) 1539--1550.

\bibitem{Gardestig-2009}
A.~G{\aa}rdestig, {Extracting the neutron-neutron scattering length $-$ recent
  developments}, J. Phys. G 36 (2009) 053001.

\bibitem{Schwenk-2005}
A.~Schwenk, C.~J. Pethick, Resonant {Fermi} gases with a large effective range,
  Phys. Rev. Lett. 95 (2005) 160401.

\bibitem{Chamel-2008}
N.~Chamel, P.~Haensel, Physics of neutron star crusts, Living Rev. Relativity
  11 (2008) 10.
\newblock \href {http://dx.doi.org/10.1007/lrr-2008-10}
  {\path{doi:10.1007/lrr-2008-10}}.

\bibitem{Fortin-2010}
M.~Fortin, F.~Grill, J.~Margueron, D.~Page, N.~Sandulescu, Thermalization time
  and specific heat of the neutron stars crust, Phys. Rev. C 82 (2010) 065804.

\bibitem{Tamagaki-1968}
R.~Tamagaki, Potential models of nuclear forces at small distances, Prog.
  Theor. Phys. 39 (1968) 91--107.

\bibitem{Sun-2010}
B.~Y. Sun, H.~Toki, J.~Meng, {Relativistic description of BCS-BEC crossover in
  nuclear matter}, Phys. Lett. B 683 (2010) 134--139.

\bibitem{Maurizio-2014}
S.~Maurizio, J.~W. Holt, P.~Finelli, Nuclear pairing from microscopic forces:
  Singlet channels and higher-partial waves, Phys. Rev. C 90 (2014) 044003.

\bibitem{Drischler-2017}
C.~Drischler, T.~Kr\"uger, K.~Hebeler, A.~Schwenk, Pairing in neutron matter:
  New uncertainty estimates and three-body forces, Phys. Rev. C 95 (2017)
  024302.

\bibitem{Gandolfi-2008}
S.~Gandolfi, A.~Y. Illarionov, S.~Fantoni, F.~Pederiva, K.~E. Schmidt, Equation
  of state of superfluid neutron matter and the calculation of the
  {$^{1}S_{0}$} pairing gap, Phys. Rev. Lett. 101 (2008) 132501.

\bibitem{Schwenk-2004}
A.~Schwenk, B.~Friman, Polarization contributions to the spin dependence of the
  effective interaction in neutron matter, Phys. Rev. Lett. 92 (2004) 082501.

\bibitem{Schwenk-2003}
A.~{Schwenk}, B.~{Friman}, G.~E. {Brown}, {Renormalization group approach to
  neutron matter: quasiparticle interactions, superfluid gaps and the equation
  of state}, Nucl. Phys. A 713 (2003) 191--216.

\bibitem{Brueckner-1955}
K.~A. Brueckner, Two-body forces and nuclear saturation. {III}. {Details} of
  the structure of the nucleus, Phys. Rev. 97 (1955) 1353--1366.

\bibitem{Day-1967}
B.~D. Day, Elements of the {Brueckner-Goldstone} theory of nuclear matter, Rev.
  Mod. Phys. 39 (1967) 719--744.

\bibitem{Lombardo-2001c}
U.~Lombardo, H.-J. Schulze, Superfluidity in neutron star matter, in:
  D.~Blaschke, A.~Sedrakian, N.~K. Glendenning (Eds.), Physics of Neutron Star
  Interiors, Springer, Berlin, 2001, pp. 30--53.

\bibitem{Babu-1973}
S.~Babu, G.~E. Brown, Quasiparticle interaction in liquid $^3${H}e, Ann. Phys.
  78 (1973) 1--38.

\bibitem{Zhang-2016}
S.~S. Zhang, L.~G. Cao, U.~Lombardo, P.~Schuck, Medium polarization in
  asymmetric nuclear matter, Phys. Rev. C 93 (2016) 044329.

\bibitem{Bogner-2007}
S.~K. Bogner, R.~J. Furnstahl, S.~Ramanan, A.~Schwenk, Low-momentum
  interactions with smooth cutoffs, Nucl. Phys. A 784 (2007) 79--103.

\bibitem{Weinberg-1963}
S.~Weinberg, {Quasiparticles and the Born series}, Phys. Rev. 131 (1963)
  440--460.

\bibitem{Weiss-2015a}
R.~Weiss, B.~Bazak, N.~Barnea, Nuclear neutron-proton contact and the
  photoabsorption cross section, Phys. Rev. Lett. 114 (2015) 012501.

\bibitem{Levinger-1951}
J.~S. Levinger, The high energy nuclear photoeffect, Phys. Rev. 84 (1951)
  43--51.

\bibitem{Weiss-2016}
R.~Weiss, B.~Bazak, N.~Barnea, The generalized nuclear contact and its
  application to the photoabsorption cross-section, Eur. Phys. J. A 52 (2016)
  92.

\bibitem{Weiss-2015b}
R.~Weiss, B.~Bazak, N.~Barnea, Generalized nuclear contacts and momentum
  distributions, Phys. Rev. C 92 (2015) 054311.

\bibitem{Hen-2015}
O.~Hen, L.~B. Weinstein, E.~Piasetzky, G.~A. Miller, M.~M. Sargsian, Y.~Sagi,
  Correlated fermions in nuclei and ultracold atomic gases, Phys. Rev. C 92
  (2015) 045205.

\bibitem{Hen-2014}
O.~Hen, {\em et al.}, {Momentum sharing in imbalanced Fermi systems}, Science
  346 (2014) 614--617.

\bibitem{Nozieres-1982}
P.~Nozi{\`e}res, D.~Saint~James, {Particle vs. pair condensation in attractive
  Bose liquids}, J. Phys. (Paris) 43 (1982) 1133--1148.

\bibitem{Capponi-2008}
S.~Capponi, G.~Roux, P.~Lecheminant, P.~Azaria, E.~Boulat, S.~R. White,
  {Molecular superfluid phase in systems of one-dimensional multicomponent
  fermionic cold atoms}, Phys. Rev. A 77 (2008) 013624.

\bibitem{Sogo-2010a}
T.~Sogo, G.~R\"opke, P.~Schuck, Many-body approach for quartet condensation in
  strong coupling, Phys. Rev. C 81 (2010) 064310.

\bibitem{Schuck-2014}
P.~Schuck, Y.~Funaki, H.~Horiuchi, G.~R\"{o}pke, A.~Tohsaki, T.~Yamada, Theory
  for quartet condensation in {Fermi} systems with applications to nuclei and
  nuclear matter, J. Phys. Conf. Ser. 529 (2014) 012014.

\bibitem{Kamei-2005}
H.~Kamei, K.~Miyake, On quartet superfluidity of fermionic atomic gas, J. Phys.
  Soc. Jpn 74 (2005) 1911--1913.

\bibitem{Sogo-2009}
T.~Sogo, R.~Lazauskas, G.~R\"opke, P.~Schuck, Critical temperature for
  $\alpha$-particle condensation within a momentum-projected mean-field
  approach, Phys. Rev. C 79 (2009) 051301.

\bibitem{Malfliet-1969}
R.~A. Malfliet, J.~A. Tjon, Solution of the {Faddeev} equations for the triton
  problem using local two-particle interactions, Nucl. Phys. A 127 (1969)
  161--168.

\bibitem{Sogo-2010b}
T.~Sogo, G.~R\"opke, P.~Schuck, Critical temperature for $\alpha$-particle
  condensation in asymmetric nuclear matter, Phys. Rev. C 82 (2010) 034322.

\bibitem{Blin-1986}
A.~H. Blin, R.~W. Hasse, B.~Hiller, P.~Schuck, C.~Yannouleas, On the evaluation
  of semiclassical nuclear many-particle many-hole level densities, Nucl. Phys.
  A 456 (1986) 109--133.

\bibitem{Xu-2016}
C.~Xu, Z.~Ren, G.~R\"opke, P.~Schuck, Y.~Funaki, H.~Horiuchi, A.~Tohsaki,
  T.~Yamada, B.~Zhou, {$\ensuremath{\alpha}$-decay width of $^{212}\mathrm{Po}$
  from a quartetting wave function approach}, Phys. Rev. C 93 (2016) 011306.

\bibitem{Bulthuis-2016}
B.~Bulthuis, A.~Gezerlis, Probing mixed-spin pairing in heavy nuclei, Phys.
  Rev. C 93 (2016) 014312.

\bibitem{Gezerlis-2011}
A.~Gezerlis, G.~F. Bertsch, Y.~L. Luo, Mixed-spin pairing condensates in heavy
  nuclei, Phys. Rev. Lett. 106 (2011) 252502.

\bibitem{Pillet-2007}
N.~Pillet, N.~Sandulescu, P.~Schuck, {Generic strong coupling behavior of
  Cooper pairs on the surface of superfluid nuclei}, Phys. Rev. C 76 (2007)
  024310.

\bibitem{Fukushima-1978}
Y.~Fukushima, M.~Kamimura, Suppl. J. Phys. Soc. Jpn 44 (1978) 225.

\bibitem{Kamimura-1981}
M.~Kamimura, Transition densities between the $0_1^+$, $2_1^+$, $4_1^+$,
  $0_2^+$, $2_2^+$, $1_1^−$ and $3_1^−$ states in $^{12}\mathrm{C}$ derived
  from the three-alpha resonating-group wave functions, Nucl. Phys. A 351
  (1981) 456 -- 480.

\bibitem{Matsuo-2005}
M.~Matsuo, K.~Mizuyama, Y.~Serizawa, Di-neutron correlation and soft dipole
  excitation in medium mass neutron-rich nuclei near drip line, Phys. Rev. C 71
  (2005) 064326.

\bibitem{Pillet-2010}
N.~Pillet, N.~Sandulescu, P.~Schuck, J.-F. Berger, Two-particle spatial
  correlations in superfluid nuclei, Phys. Rev. C 81 (2010) 034307.

\bibitem{Hagino-2010}
K.~Hagino, H.~Sagawa, P.~Schuck, {Cooper pair sizes in {$^{11}$Li} and in
  superfluid nuclei: a puzzle?}, J. Phys. G 37 (2010) 064040.

\bibitem{Wiringa-2000}
R.~B. Wiringa, S.~C. Pieper, J.~Carlson, V.~R. Pandharipande, {Quantum Monte
  Carlo calculations of $A=8$ nuclei}, Phys. Rev. C 62 (2000) 014001.

\bibitem{Hoyle54}
F.~{Hoyle}, On nuclear reactions occuring in very hot stars. {I.} the synthesis
  of elements from {Carbon to Nickel}., Astrophys. J. Suppl. 1 (1954) 121--146.

\bibitem{Cook-1957}
C.~W. Cook, W.~A. Fowler, C.~C. Lauritsen, T.~Lauritsen, {B$^{12}$, C$^{12}$,
  and the red giants}, Phys. Rev. 107 (1957) 508--515.

\bibitem{Funaki-2009}
Y.~Funaki, H.~Horiuchi, W.~von Oertzen, G.~R\"opke, P.~Schuck, A.~Tohsaki,
  T.~Yamada, Concepts of nuclear $\ensuremath{\alpha}$-particle condensation,
  Phys. Rev. C 80 (2009) 064326.

\bibitem{Roepke-2014}
G.~R\"opke, P.~Schuck, Y.~Funaki, H.~Horiuchi, Z.~Ren, A.~Tohsaki, C.~Xu,
  T.~Yamada, B.~Zhou, {Nuclear clusters bound to doubly magic nuclei: The case
  of $^{212}$Po}, Phys. Rev. C 90 (2014) 034304.

\bibitem{Tohsaki-2001}
A.~Tohsaki, H.~Horiuchi, P.~Schuck, G.~R\"opke, Alpha cluster condensation in
  {$^{12}$C and $^{16}$O}, Phys. Rev. Lett. 87 (2001) 192501.

\bibitem{Tohsaki-1994}
A.~Tohsaki, New effective internucleon forces in microscopic
  \ensuremath{\alpha}-cluster model, Phys. Rev. C 49 (1994) 1814--1817.

\bibitem{Funaki-2006}
Y.~Funaki, A.~Tohsaki, H.~Horiuchi, P.~Schuck, G.~R\"opke, {Inelastic form
  factors to alpha-particle condensate states in $^{12}$C and $^{16}$O: What
  can we learn?}, Eur. Phys. J. A 28 (2006) 259--263.

\bibitem{Chernykh-2007}
M.~Chernykh, H.~Feldmeier, T.~Neff, P.~von Neumann-Cosel, A.~Richter, Structure
  of the {H}oyle state in $^{12}\mathrm{C}$, Phys. Rev. Lett. 98 (2007) 032501.

\bibitem{Schuck-2016}
P.~Schuck, Y.~Funaki, H.~Horiuchi, G.~R\"{o}pke, A.~Tohsaki, T.~Yamada, Alpha
  particle clusters and their condensation in nuclear systems, Phys. Scripta 91
  (2016) 123001.

\bibitem{Tohsaki-2017}
A.~Tohsaki, H.~Horiuchi, P.~Schuck, G.~R\"opke, {Status of $\alpha$-particle
  condensate structure of the Hoyle state}, Rev. Mod. Phys. 89 (2017) 011002.

\bibitem{Torma-2016}
P.~T\"orma, {Physics of ultracold Fermi gases revealed by spectroscopies},
  Phys. Scripta 91 (2016) 043006.

\bibitem{Salasnich-2016}
L.~Salasnich, F.~Toigo, Zero-point energy of ultracold atoms, Phys. Rep. 640
  (2016) 1--29.

\bibitem{Baym-1983}
G.~Baym, B.~L. Friman, J.-P. Blaizot, M.~Soyeur, W.~Czy\.{z}, Hydrodynamics of
  ultra-relativistic heavy ion collisions, Nucl. Phys. A 407 (1983) 541--570.

\bibitem{Yamamoto-2007}
N.~Yamamoto, M.~Tachibana, T.~Hatsuda, G.~Baym, {Phase structure, collective
  modes, and the axial anomaly in dense QCD}, Phys. Rev. D 76 (2007) 074001.

\bibitem{Nambu-1960}
Y.~Nambu, Quasi-particles and gauge invariance in the theory of
  superconductivity, Phys. Rev. 117 (1960) 648--663.

\bibitem{Yang-1962}
C.~N. Yang, Concept of off-diagonal long-range order and the quantum phases of
  liquid {He} and of superconductors, Rev. Mod. Phys. 34 (1962) 694--704.

\bibitem{Lubashevsky-2012}
Y.~Lubashevsky, E.~Lahoud, K.~Chashka, D.~Podolsky, A.~Kanigel, {Shallow
  pockets and very strong coupling superconductivity in FeSe$_x$Te$_{1-x}$},
  Nat. Phys. 8 (2012) 309--312.

\bibitem{Pieri-2007}
P.~Pieri, D.~Neilson, G.~C. Strinati, {Effects of density imbalance on the
  BCS-BEC crossover in semiconductor electron-hole bilayers}, Phys. Rev. B 75
  (2007) 113301.

\bibitem{Perali-2013}
A.~Perali, D.~Neilson, A.~R. Hamilton, High-temperature superfluidity in
  double-bilayer graphene, Phys. Rev. Lett. 110 (2013) 146803.

\bibitem{Li-2016}
J.~I.~A. Li, T.~Taniguchi, K.~Watanabe, J.~Hone, A.~Levchenko, C.~R. Dean,
  Negative {C}oulomb drag in double bilayer graphene, Phys. Rev. Lett. 117
  (2016) 046802.

\bibitem{Lee-2016}
K.~Lee, J.~Xue, D.~C. Dillen, K.~Watanabe, T.~Taniguchi, E.~Tutuc, Giant
  frictional drag in double bilayer graphene heterostructures, Phys. Rev. Lett.
  117 (2016) 046803.

\end{thebibliography}


\end{document}